\documentclass[twoside,12pt]{article}
\usepackage{epsfig,latexsym}

\def\Journal#1#2#3#4{{#1} {#2} (#4) #3 }

\def\NPA{{\em Nucl. Phys.} A}

\def\NPB{{\em Nucl. Phys.} B}

\def\PLB{{\em Phys. Lett.} B}

\def\PRL{\em Phys. Rev. Lett.}
\def\PREV{\em Phys. Rev.}
\def\PREP{\em Phys. Rep.}

\def\PRD{{\em Phys. Rev.} D}
\def\PRC{{\em Phys. Rev.} C}

\def\ZPC{{\em Z. Phys.} C}

\def\ANNP{\em Ann. Phys. (N.Y.)}
\def\RMP{{\em Rev. Mod. Phys.}}

\def\JETP{{\em J.E.T.P.}}
\def\JHEP{{\em JHEP}}

\newcommand{\be}{\begin{equation}}
\newcommand{\ee}{\end{equation}}
\newcommand{\bea}{\begin{eqnarray}}
\newcommand{\eea}{\end{eqnarray}}

\begin{document}

\title{ \vspace{1cm} The Quark-Gluon Plasma in Equilibrium}
\author{Dirk H.\ Rischke \\
\\
Institut f\"ur Theoretische Physik \\
Johann Wolfgang Goethe-Universit\"at Frankfurt am Main\\
Germany}
\maketitle
\begin{abstract} 
Our current knowledge of the quark-gluon plasma in thermodynamical equilibrium
is reviewed.
The phase diagram of strongly interacting matter is 
discussed, with emphasis on
the quark-hadron phase transition and the color-superconducting
phases of quark matter. Lattice QCD results on the order
of the phase transition, the thermodynamical functions,
the heavy quark free energy, mesonic spectral functions, and
recent results for nonzero quark chemical potential
are presented. Analytic attempts to compute the
thermodynamical properties of strongly interacting matter, such
as perturbation theory, quasiparticle models, ``hard-thermal-loop''
(HTL)-resummed perturbation theory, the Polyakov-loop model, as well as
linear sigma models are discussed.
Finally, color-superconducting quark matter is considered in
the limit of weak coupling. The gap equation and the excitation
spectrum are derived. The solution of the gap equation,
gap parameters in various color-superconducting phases, and
critical temperatures for the transition to
normal-conducting quark matter are presented.
A summary of gluon and photon properties in color superconductors is
given.
\end{abstract}

\section{Introduction and Summary}

Quantum chromodynamics (QCD) is the fundamental theory of the
strong interaction. QCD is an asymptotically free theory \cite{asympt},
i.e., interactions between quarks and gluons become weaker as the mutual
distance decreases or as the exchanged momentum increases. Consequently, at
very large temperatures and/or densities, the 
interactions which confine quarks and gluons inside hadrons
should become sufficiently weak to release them \cite{collins}.
The phase where quarks and gluons are deconfined is termed
the {\em quark-gluon plasma\/} (QGP). Lattice QCD calculations
have established the existence of such a phase of strongly interacting
matter at temperatures larger than $\sim 150$ MeV and zero net-baryon
density. Depending on the number of quark flavors and the
masses of the quarks, the transition between ordinary
hadronic matter and the QGP could be a thermodynamic phase transition
of first order, of second order, or simply a crossover transition.

The QGP was certainly present in the evolution of
the early universe. The early universe was very hot, but 
close to net-baryon free. In the opposite limit of
small temperature and large baryon density, the QGP may exist even nowadays
in the interior of compact stellar objects such as neutron stars.
The main effort in present-day nuclear physics
is to create the QGP under controlled conditions in the
laboratory via collisions of heavy nuclei at ultrarelativistic
energies \cite{QM}. The
temperatures and net-baryon densities reached
in nuclear collisions depend on the bombarding energy.
They interpolate between the extreme conditions
of the early universe on one side and compact stellar objects
on the other.

If at all, the QGP is only transiently created in a nuclear collision;
free quarks and gluons will not be released.
Therefore, deciding whether a QGP was formed or not is not easy.
Detectors in the laboratory can only measure hadrons, leptons, or photons.
The bulk of the particles emerging
from a nuclear collision are hadrons with transverse
momenta of order $\sim 1$ GeV. They carry information 
about the final stage of the collision after hadronization of
the QGP. The formation of the latter can only indirectly influence
this final stage, for instance by modifying
the collective dynamics of the system through a softening
of the equation of state in the hadronization transition \cite{soft}.
Very few hadrons are emitted with transverse momenta of the order 
of several GeV.
They arise from the fragmentation of jets and
may carry information also about the earlier stages of the collision. 
Of particular
interest is the situation where the jet has to traverse hot and dense matter
possibly formed in the collision and is quenched by multiple
rescattering in the medium \cite{quench}. 
From this ``jet-quenching'' process one may indirectly learn 
about the properties of the hot and dense environment.
Finally, leptons and photons only interact electromagnetically.
Once formed in the early stage of the collision, they leave the
system unimpededly and carry information about this 
stage to the detector \cite{photons}. The difficulty is to disentangle 
the thermal radiation from a hot, equilibrated QGP \cite{QGPphotons}
from the initial production of leptons and photons in the very first,
highly energetic partonic collisions and from the thermal
radiation of hot hadronic matter \cite{kapustaphotons}.

\begin{figure}[tb]
\begin{center}
\begin{minipage}[t]{14 cm}
\epsfig{file=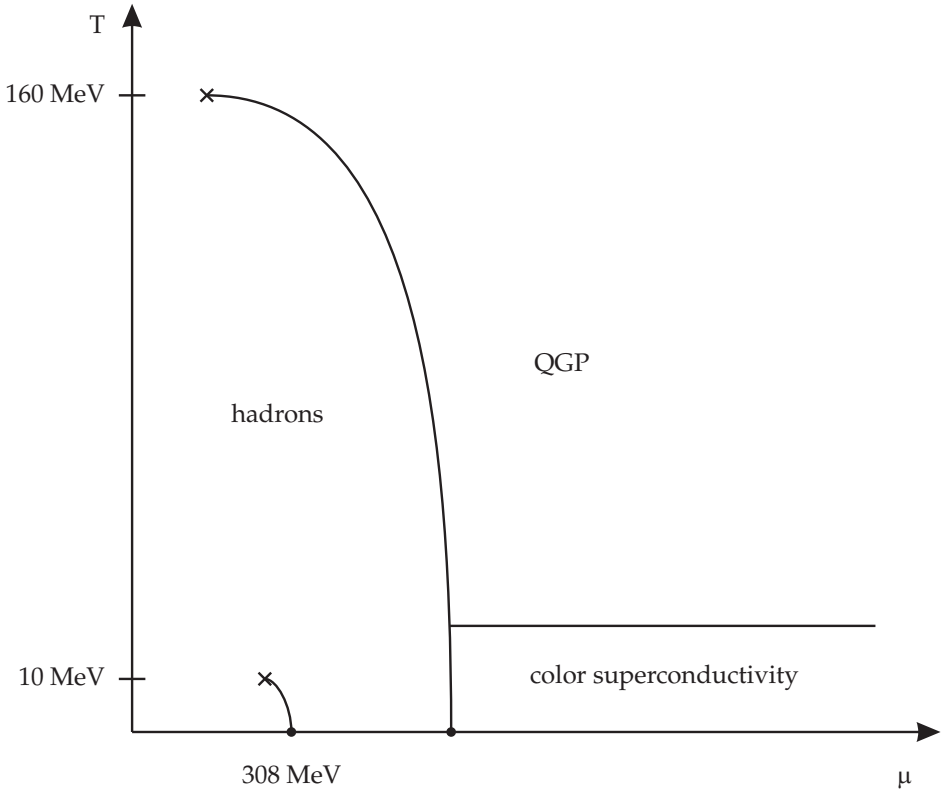,scale=0.8}
\end{minipage}
\begin{minipage}[t]{16.5 cm}
\caption{The phase diagram of strongly interacting matter (schematic).
\label{phasediagramNM}}
\end{minipage}
\end{center}
\end{figure}

In order to decide whether a QGP was formed, one has
to have detailed knowledge about its properties. Otherwise
it is impossible to find an unambiguous signature for QGP formation
in nuclear collisions.
In this review, I present an overview of 
the {\em thermodynamical properties\/} of the QGP. 
Section \ref{secphasediag} contains
a detailed discussion of the phase diagram of strongly interacting
matter. The present status of knowledge is shown schematically
in Fig.~\ref{phasediagramNM}. 
Depending on the temperature, $T$,
and the quark chemical potential, $\mu$, strongly interacting matter
may occur in three distinct phases: the hadronic phase, the QGP,
and color-superconducting quark matter. 
The ground state of (infinite) nuclear matter is at $(T, \mu)_0
= (0,308)$ MeV. There is a line of first-order phase transitions emerging
from this point and terminating in a critical endpoint
at a temperature of order $\sim 10$ MeV. At this point, the
transition is of second order.
This phase transition is the nuclear liquid-gas transition \cite{liquidgas}.
To the left of the line nuclear matter is in the gaseous phase,
and to the right in the liquid phase. Above the critical endpoint,
there is no distinction between these two phases.

For temperatures below $\sim 160$ MeV and quark chemical potentials
below $\sim 350$ MeV (corresponding to net-baryon densities
which are a few times the ground state density of nuclear matter), 
strongly interacting matter is in the hadronic phase. 
Quite similar to the liquid-gas
transition, there is a line of first-order phase transitions
which separates the hadronic phase from the QGP and terminates
in a critical endpoint where the transition is of second order.
This endpoint is approximately at $(T, \mu)\simeq (160, 240)$ MeV,
cf.~Sec.~\ref{nonzeromu}.
For smaller quark chemical potentials (smaller net-baryon densities), 
the transition becomes crossover,
and there is no real distinction between hadronic matter and the QGP.
As will be discussed in detail in Sec.~\ref{secphasediag},
the position of the critical endpoint depends on the value of
the quark masses. 
Finally, at large quark chemical potential (large baryon density)
and small temperature,
quark matter becomes a color superconductor. There can be multitude
of color-superconducting phases, depending on the symmetries of
the order parameter for condensation of quark Cooper pairs.
The discussion in Sec.~\ref{secphasediag} is qualitative and
is meant to give an overview of the phase structure of strongly
interacting matter at nonzero temperature and quark chemical
potential. The discussion in the following sections is
both more quantitative as well as technical and focusses on
the properties of the QGP and color-superconducting quark matter.

The early universe evolved close to the temperature axis in the phase
diagram of strongly interacting matter. Matter in the core of
compact stellar objects, like neutron stars, is close
to the quark chemical potential axis, 
at values of $\mu$ around 400 -- 500 MeV.
Nuclear collisions at
bombarding energies around $E_{\rm Lab} \sim 1$ AGeV
explore a region of temperatures and quark chemical potentials
around $(T, \mu) \sim (70, 250)$ MeV. Collisions at current RHIC
energies of $\sqrt{s} = 200$ AGeV are expected to excite 
matter in a region around and above $(T, \mu) \sim (170, 10)$ MeV.
Collision energies in between these two extremes cover the 
intermediate region and, in particular, may probe the
critical endpoint.

Section~\ref{LattQCD} presents a review of lattice QCD.
After a brief introduction to the basic
principles, results on
the order of the QCD phase transition, the equation of state
of strongly interacting matter, the heavy quark free energy, and
mesonic spectral functions are collected. For technical reasons, 
most lattice QCD calculations have been done at zero quark chemical
potential. An extension to nonzero values of $\mu$ is
difficult and has been started only fairly recently. First
results will also be discussed.

Lattice QCD is in principle an exact method to solve QCD.
If one had sufficiently large computer power, one could not only
decrease the lattice spacing and increase the size of the system
to come sufficiently close to the continuum and thermodynamic limit,
one could also sample over a sufficiently large number
of configurations to make the statistical errors arbitrarily small.
However, one still has to interpret the results in physical terms.
In this respect, analytic approaches to solve QCD 
have a certain advantage over lattice QCD. In an analytic approach,
one has complete control over the physical assumptions 
entering the calculation. Section~\ref{secIII} gives an overview
of what is known about the QGP from analytic calculations.

The most simple approach from a conceptual (albeit not
technical) point of view is to determine the 
thermodynamical properties of the QGP by a perturbative computation
of the QCD partition function in terms of a power series
in the strong coupling constant, $g$. This can be done
up to terms of order $O(g^6 \ln g)$. At order $O(g^6)$, the
perturbative series breaks down \cite{kapustaFTFT,linde},
and the remaining nonperturbative
contribution has to be determined, for instance, from a
lattice QCD calculation.
Those terms of the perturbative series, which are analytically
computable, are rapidly decreasing in magnitude at high temperatures
where the strong coupling constant is small, $g\ll 1$. 
This gives rise to the hope that only the first few terms
of the perturbative series are actually necessary to obtain
physically reasonable values for the QCD partition function. For
temperatures of order $\sim 150$ MeV, however, $g \sim 1$
and the perturbative series is obviously not convergent.
Therefore, one has tried
other ways to compute the partition function, either by expanding
around a nonperturbative ground state or by resumming certain classes
of diagrams to improve the convergence properties of the perturbative
series. In both approaches, quarks and gluons are considered
as quasiparticles with a dispersion relation which is modified
as compared to the one in the vacuum. 
Still another approach is to construct an effective theory
for QCD which can be solved exactly or at least within a many-body
approximation scheme. 
All these approaches will be reviewed in Sec.~\ref{secIII}.

Section~\ref{CSC} contains an introduction to color superconductivity
at large quark chemical potentials.
In this case, analytic calculations are well
under control, because corrections can be systematically 
computed in terms of powers of $g$. After a derivation of
the gap equation for the color-superconducting gap function,
the excitation spectrum in a color superconductor is
presented. The solution of the gap equation is discussed and
the magnitude of the gap parameter is determined. As
in ordinary superconductors, quark Cooper pairs break up, if the
thermal motion becomes too large. The critical temperature
for the transition between the normal- and the superconducting phase
is computed. Finally, the properties of gluons and photons
in color superconductors are discussed.
Section~\ref{concloutl} concludes this review with 
a brief summary of the material and an outlook towards future directions
of research in this area.

A lot of the material in this review can also be found in
other places. The standard review for properties of hot and dense,
strongly interacting matter is the seminal work of
Gross, Pisarski, and Yaffe \cite{grosspisarskiyaffe}.
The contents as well as more recent developments
have found their way into textbooks \cite{kapustaFTFT,lebellac}.
For early reviews focussing on the properties of the QGP, see 
Ref.~\cite{earlyQGP}. An introduction to lattice QCD
and recent results can be found in Ref.~\cite{schladming}.
The present status of lattice QCD is also reviewed in 
Ref.~\cite{laermannphilipsen}.
Resummation techniques which attempt to compute the
QCD partition function analytically are discussed in great detail in
Ref.~\cite{reviewBIR}. 

The present review tries to give a balanced overview of all
subfields concerned with the equilibrium properties of the QGP.
Therefore, the presentation is not as detailed
as in a more specialized review. On the other hand, 
I tried to explain the basics in somewhat greater detail than usually
found in the literature. My hope is that in this way, this review
will become useful for early-stage researchers working in
both theory as well as experiment, and
for all researchers who would like to get an overview of 
the theoretical activity related to equilibrium properties of the QGP.

The only somewhat more specialized and thus more technical part
is the section on color superconductivity.
This field has seen a lot of activity only fairly recently, but there
are already a couple of excellent reviews \cite{rajwil}. These,
however, focus mainly on the basics of the phenomenon of
color superconductivity and its phenomenological implications.
In contrast, Sec.~\ref{CSC} contains a very detailed discussion of
how to compute properties of the quasiparticle excitations in
a color superconductor in the weak-coupling limit.
By clarifying some of the technical details, I hope to remove the
obstacles that may prevent other researchers 
to enter this rapidly evolving and rather exciting new field of
strongly interacting matter physics.

Due to space restrictions this review is forced to omit many things
that could (and possibly, should)
also have been said about strongly interacting matter at
high temperature and/or density. 
Fortunately, most of these have already been covered in
review articles. These are, for instance, non-equilibrium
properties of the QGP \cite{litimmanuel} or the physics of instantons 
in non-Abelian gauge theories \cite{instantons}.
Another important topic which is not mentioned in this work, but
for which excellent reviews exist \cite{QGPexperiment},
are the experimental signatures for the QGP.
Recent developments in the field of color superconductivity
are mainly focussed on deriving effective theories for quarks 
around the Fermi surface. These greatly simplify calculations
and allow to systematically study effects of nonzero quark
masses, for details, see Ref.~\cite{CSCeffective}.
Finally, the list of references is, necessarily, far from
complete. I would like to apologize to all authors whose work
should have been (but was not) mentioned.

The units are $\hbar = c = k_B = 1$. I work in Euclidean
space-time at nonzero temperature $T$, i.e., space-time integrals are 
$\int_X \equiv \int_0^{1/T} {\rm d} \tau \int_V {\rm d}^3 {\bf x}$,
where $V$ is the 3-volume of the system. Energy-momentum integrals
are $\int_K \equiv T \sum_n \int {\rm d}^3 {\bf k}/(2 \pi)^3$;
$n$ labels the Matsubara frequencies $\omega_n^{\rm b} = 2 n \pi T$
for bosons and $\omega_n^{\rm f} = (2 n + 1) \pi T$ for fermions,
$n = 0, \pm 1, \pm 2 , \ldots$. 
I denote 4-vectors with capital letters, but unless mentioned
otherwise, retain a notation familiar from Minkowski space:
$X^\mu \equiv (t, {\bf x})$, where $t \equiv i \tau$, and
$K^\mu \equiv (k_0, {\bf k})$, where $k_0 = i \, \omega_n$,
with the metric tensor $g^{\mu \nu} = {\rm diag} (+,-,-,-)$. 
3-vectors ${\bf k}$ have
length $k \equiv |{\bf k}|$ and direction $\hat{\bf k} \equiv 
{\bf k}/k$.


\section{The QCD Phase Diagram} \label{secphasediag}

\subsection{\it Basics}

In order to understand the phase structure of strongly
interacting matter, one has to know its equation of state.
In the grand canonical ensemble, the equation of state
is determined by the grand partition function \cite{kapustaFTFT,lebellac}
\begin{equation} \label{Z}
{\cal Z}(T,V,\mu) = \int {\cal D}\bar{\psi} \,
{\cal D} \psi \, {\cal D}A^\mu_a \; \exp \left[ \int_X
\left( {\cal L} + \mu \, {\cal N} \right) \right]\,\, ,
\end{equation}
where $\mu$ is the quark chemical
potential associated with (net) quark number conservation.
The QCD Lagrangian is given by
\begin{equation} \label{LQCD}
{\cal L} = \bar{\psi}\, \left( i \gamma^\mu D_\mu - m\right)
\psi - \frac{1}{4} \, F^{\mu \nu}_a \, F^a_{\mu \nu} 
+ {\cal L}_{\rm gauge} \,\, .
\end{equation}
For $N_c$ colors and $N_f$ flavors, 
$\psi$ is the $4  N_c  N_f$-dimensional spinor of quark fields, 
$\bar{\psi} \equiv \psi^\dagger \gamma_0$ is the Dirac conjugate spinor, 
$\gamma^\mu$ are the Dirac matrices and $m$ is the quark mass matrix. The
covariant derivative is defined as 
$D_\mu = \partial_\mu - i g A_\mu^a \, T_a$, 
with the strong coupling constant $g$,
the gluon fields $A^\mu_a$, and the generators $T_a$ of the 
local $[SU(N_c)_c]$ symmetry. (Throughout this paper, 
I indicate local, i.e.~gauged, symmetries by square brackets.)
The latter are usually taken as $T_a \equiv \lambda_a/2$, where $\lambda_a$
are the Gell-Mann matrices. The gluonic field strength tensor is defined as
\begin{equation}
F^{\mu \nu}_a = \partial^\mu A^\nu_a - \partial^\nu A^\mu_a +
g f_{abc} \, A^\mu_b \, A^\nu_c \,\, ,
\end{equation}
where $f_{abc}$ are the structure constants of $[SU(N_c)_c]$.
The term ${\cal L}_{\rm gauge}$ in Eq.~(\ref{LQCD}) will not be
specified further. It comprises
gauge fixing terms and the contribution from Faddev-Popov ghosts.
The number density operator associated with the conserved (net) quark number 
is ${\cal N} \equiv \bar{\psi}\, \gamma_0 \, \psi$.

For any finite volume $V$ and nonzero temperature 
$T$, the partition function is defined for a compact Euclidean space-time
volume $V \times 1/T$. For the sake of simplicity (but without loss
of generality), assume that the spatial volume $V$ is
a box, $V = L^3$, with $L$ being the length of the box in one
spatial dimension. All fields are then usually taken
to be periodic in spatial directions, $\phi(\tau,0,y,z) = \phi(\tau,L,y,z)$,
where $\phi$ stands generically for $\psi$, $\bar{\psi}$, and $A^\mu_a$.
Bosonic fields, such as gluons, are periodic also in temporal direction,
$A^\mu_a(0,{\bf x}) = A^\mu_a(1/T,{\bf x})$, while fermionic fields, such as
quarks, are antiperiodic in temporal direction,
$\psi(0,{\bf x}) = - \psi(1/T,{\bf x})$.

From the grand partition function, one can derive other
thermodynamic quantities, for instance the pressure,
\begin{equation} \label{pressure}
p(T,\mu) = T \, \frac{\partial \ln {\cal Z}}{\partial V}\;\;
\rightarrow \;\; \frac{T}{V}\, \ln {\cal Z} \;\;(V \rightarrow \infty)\,\, .
\end{equation}
In the thermodynamic limit, $\ln {\cal Z}$ is an extensive quantity
($\sim V$) and the dependence of the pressure on $V$ drops out.

Phase transitions are determined by studying the derivatives of
the pressure with respect to $T$ and $\mu$
for a given point $(T,\mu)$ in the phase diagram of the independent
thermodynamic variables temperature and chemical potential.
For a phase transition of first order, the first derivatives
\begin{equation} \label{sn}
s = \left. \frac{\partial p}{\partial T} \right|_\mu
\;\;\;\; , \;\;\;\;\; 
n =\left. \frac{\partial p}{\partial \mu} \right|_T\,\, ,
\end{equation}
are discontinuous while the pressure $p$ is continuous at the
point $(T,\mu)$.
Here, $s$ is the entropy density and $n$ the (net) quark number density.
For a phase transition of second order, the second
derivatives are discontinuous, while $p$ and its first derivatives
are continuous. In this way, phase transitions of arbitrarily
high order can be defined. One speaks of a crossover transition, if
thermodynamic properties change rapidly within a narrow range
of values $T$ and $\mu$, but the pressure and all its derivatives 
remain continuous.
Usually, the points $(T,\mu)$ where a phase transition occurs are
continuously connected and thus determine a {\em line\/} of
phase transitions
in the phase diagram. These lines usually start on either the $T$ or
the $\mu$ axis. They may terminate for nonzero values of $T$ and
$\mu$. Two examples for this behavior, 
the liquid-gas transition in nuclear matter and the quark-hadron 
transition, have already been seen in the phase diagram of strongly
interacting matter, Fig.~\ref{phasediagramNM}, and will be 
discussed in more detail in the following.

\subsection{\it The Liquid-Gas Phase Transition} \label{LGTrans}

The liquid-gas transition in nuclear matter is a consequence of the fact
that nuclear matter assumes its ground state at a nonvanishing
baryon density $n_{B,0} \simeq 0.17\,{\rm fm}^{-3}$ at zero temperature
$T=0$. The underlying microscopic mechanism for this phenomenon
is a competition between attractive and repulsive forces 
among nucleons, with the
attraction winning at this particular value of the baryon density.
(This is good, because otherwise there would be no stable
atomic nuclei, precluding the existence of our universe 
as we know it.)
In infinite, isospin-symmetric nuclear matter, nucleons in the ground state
are bound by $-16$ MeV (if one neglects the Coulomb repulsion), 
i.e., the energy
per baryon is $(E/N_B)_0 \equiv (\epsilon/n_B)_0 = m_N - 16\, {\rm MeV}
\simeq 924$ MeV, where $\epsilon$ is the energy density,
$n_B$ the baryon density, and $m_N \simeq 939$ MeV is the rest mass of
the nucleon. Nuclear matter is mechanically stable in the ground state,
such that the pressure vanishes, $p = 0$.
From the fundamental relation of thermodynamics,
$\epsilon = T s + \mu n - p$, one then concludes that
the baryon chemical potential in the ground state is identical to
the energy per baryon, $\mu_{B,0} \equiv (\epsilon/n_B)_0
\simeq 924$ MeV. Since a baryon consists of three quarks,
$n_B = n/3$ and $\mu_B =3 \mu$. Hence, the ground state of nuclear
matter is the point $(T,\mu)_0 \simeq (0,308)$ MeV in the
nuclear matter phase diagram.

Obviously, it costs energy to compress nuclear matter to 
baryon densities $n_B > n_{B,0}$. Such an increase in energy
leads to an increase in pressure. 
At zero temperature, this can be immediately seen from the
identity $p = n_B^2\, {\rm d} (E/N_B)/{\rm d} n_B$.
Since the pressure is a monotonous function of the thermodynamic variables,
and since it vanishes in the ground state, there are only two
possibilities for the behavior of $p$ for densities $n_B < n_{B,0}$:
either the pressure remains zero as the density decreases, $p=0$, or
the pressure further decreases such that $p<0$. The latter possibility implies
that the system becomes mechanically unstable.
This can be prevented by fragmenting nuclear matter into
droplets. These droplets are mechanically stable, i.e.,
the density inside each droplet is equal to $n_{B,0}$ and
the pressure vanishes. The total baryon density
in the system can be further reduced by simply decreasing the droplet
density. The pressure in such a system remains zero down to arbitrarily small
densities, because compression just results in a decrease of the 
space between droplets. Thus, $p=0$ from $n_B = 0$ to $n_B = n_{B,0}$,
and then $p>0$ for $n_B> n_{B,0}$.

At small, but nonzero temperatures, this picture remains valid,
with the additional possibility to thermally evaporate single nucleons from
the surface of the droplets. At small temperatures and densities
below the ground state density, one thus has a mixed phase of nucleons
and droplets of nuclear matter. This is reminiscent of water which,
at room temperature and normal pressure,
consists of a mixed phase of water molecules
and water droplets. Changing the density
one can alter the relative fraction of molecules and droplets.
Beyond the density where droplets fill the entire volume
one enters the liquid phase, while below the density where
the last droplet fragments into molecules one enters the gas phase.
This behavior is typical for a first-order phase transition.
In this case, this is the so-called liquid-gas transition in water.

Nuclear matter shows a similar behavior, featuring a ``gaseous'' phase
of nucleons at small chemical potentials (densities) 
and a ``liquid'' phase of nuclear matter
at large chemical potentials (densities), cf.~Fig.~\ref{phasediagramNM}. 
At small temperatures
the transition between the two phases is of first order. Thus, in the
$(T,\mu)$ phase diagram there
is a line of first-order phase transitions extending from
the nuclear ground state $(0,308)$ MeV up towards larger values
of $T$ and smaller values of $\mu$. As for water, this line terminates
at a critical point where the transition becomes of second order.
The critical temperature is of the order of 10 MeV.
As for water, one cannot distinguish between the gaseous and the
liquid phase for temperatures above this critical temperature.
The existence of the liquid-gas phase transition has been confirmed
in heavy-ion collision experiments at BEVALAC and GSI energies
($E_{\rm Lab} \sim 1$ AGeV), although the precise value for
the critical temperature and the critical exponents remain a matter
of debate \cite{liquidgas}. 

The liquid-gas transition
is also a feature of phenomenological models for the nuclear
interaction, for instance, the Walecka model \cite{walecka}.
In the following section another phase transition in
strongly interacting matter is discussed,
which very much resembles the liquid-gas transition
in that it (most likely) is of first order for small temperatures
and terminates in a critical point where the transition becomes
of second order. This transition is the so-called quark-hadron
transition.

\subsection{\it The Quark-Hadron Phase Transition} \label{QHtransition}

\subsubsection{\it Qualitative arguments} \label{qualarg}

For a non-interacting, translationally invariant system a convenient 
basis of states are the single-particle momentum eigenstates.
Due to the Pauli principle, the density in a fermionic system
can only be increased by successively filling states with
higher momentum. The highest filled state 
defines the Fermi surface of the system, and
the corresponding momentum is the Fermi momentum, $k_F$.
For non-interacting systems at zero temperature, 
the single-particle density $n$
is given in terms of the Fermi momentum as
\begin{equation} \label{density}
n = \frac{d}{6 \pi^2}\, k_F^3\,\, ,
\end{equation}
where $d$ counts the internal degrees of freedom of the fermion
(like spin, color, flavor, etc.).
Thus, at large densities the Fermi momentum becomes large.

In a cold, dense fermionic system particles can only scatter
elastically if their momenta lie on the Fermi surface, as states below the
Fermi surface are not accessible due to the Pauli principle
(the so called ``Pauli-blocking'' phenomenon), and states above the Fermi
surface are not accessible due to energy conservation.
If the Fermi momentum exceeds the QCD scale parameter
$\Lambda_{\rm QCD} \sim 200 $ MeV, scattering events
between nucleons start to probe distances of the order
1 fm or less, i.e., the nucleonic substructure of quarks and gluons
becomes visible. 
The Fermi momentum in the ground state of nuclear matter
can be inferred from Eq.~(\ref{density}) to be
$k_{F,0} \simeq 250$ MeV. This is already of the order of $\Lambda_{\rm QCD}$.
Nevertheless, a description of nuclear matter in terms of nucleonic
degrees of freedom is certainly feasible around the
ground state. At which densities does
a description in terms of quark and gluon
degrees of freedom become more appropriate?
The ``volume'' occupied by a single nucleon can be estimated from
its charge radius to be $\sim 2\, {\rm fm}^3$. On the other hand,
the specific volume of the system in the ground state 
is $n_{B,0}^{-1} \sim 6\,{\rm fm}^3$. In this sense, nuclear matter
in the ground state is dilute. However, increasing the density
to about $3\, n_{B,0}$, the system becomes densely packed with nucleons.
At larger densities, they will even start to overlap.
Therefore, around densities of a few times nuclear matter ground state
density, one expects that a description of the system in terms
of quarks and gluons is more appropriate.

Similar arguments also apply to a system at nonzero temperature,
even when the net-baryon number density is small.
At nonzero temperature, nuclear matter consists not only of
nucleons but also of other, thermally excited hadrons.
For a non-interacting system
in thermodynamical equilibrium and neglecting 
quantum statistics, the hadron number densities are 
proportional to $n_i \sim m_i^2 \, T \, K_2(m_i/T) \, e^{\mu_i/T}$, 
where $i$ labels the hadron species, $m_i$ is their mass, 
$\mu_i$ is their chemical potential, and $K_2(x)$ is a
modified Bessel function of the second kind. 
For nonzero temperature and small net-baryon number density,
the lightest hadrons, the pions, are most abundant. 
At nonzero temperature and small baryon chemical potential,
the typical momentum scale for scattering events between hadrons
is set by the temperature $T$.
If the temperature is on the order of or larger than 
$\Lambda_{\rm QCD}$, scattering between
hadrons starts to probe their quark-gluon substructure.
Moreover, since the particle density increases with
the temperature, the hadronic
wave functions will start to overlap for large temperatures.
Consequently, above a certain temperature one expects a 
description of nuclear matter in terms
of quark and gluon degrees of freedom to be more
appropriate.

The picture which emerges from these considerations is the following:
for quark chemical
potentials $\mu$ which are on the order of 350 MeV or smaller, and
for temperatures $T < \Lambda_{\rm QCD} \sim 200$ MeV, nuclear matter is
a gas of hadrons. (At very small temperatures $T< 10$ MeV,
there is a gaseous and a liquid nucleonic phase, cf.\ Sec.~\ref{LGTrans}.)
On the other hand, for $T, \mu \gg \Lambda_{\rm QCD}$, 
nuclear matter consists of quarks and gluons.
The natural question which emerges is, whether the ``hadron phase''
and the ``quark-gluon phase'' (the QGP) are separated by a 
phase transition in the thermodynamic sense.
The rigorous way to decide this question is by identifying an
order parameter which is nonzero in one phase and zero in the other.
This will be discussed in more detail in the following.

\subsubsection{\it Pure gauge theory} \label{puregauge}

Let us first study the pure $[SU(N_c)]$ gauge theory, i.e., QCD
without dynamical quarks (sometimes also termed the $N_f=0$ case).
In this theory, there is
a phase transition between a low-temperature and a high-temperature
phase, cf.~Sec.~\ref{translatt}. 
The order parameter for this transition is the 
expectation value $\langle L({\bf x}) \rangle$
of the Polyakov loop (or Wilson line)
\begin{equation} \label{Polyakovloop}
L({\bf x}) = \frac{1}{N_c}\, {\rm Tr}\, \left\{ {\rm P}\, \exp \left[
i g \int_0^{1/T} {\rm d}\tau\, 
A_a^0(\tau, {\bf x}) \, T_a \right] \right\}\,\,,
\end{equation}
where ${\rm P}$ stands for path-ordering.
The expectation value of an operator ${\cal O}$ in the grand canonical
ensemble is defined as
\begin{equation}
\langle {\cal O} \rangle \equiv \frac{1}{{\cal Z}}
\int {\cal D}\bar{\psi} \,
{\cal D} \psi \, {\cal D}A^\mu_a \; {\cal O} \; \exp \left[ \int_X
\left( {\cal L} + \mu \, {\cal N} \right) \right]\,\, .
\end{equation}
The expectation value $\langle L({\bf x}) \rangle$ vanishes
in the low-temperature phase. If one interpretes this expectation
value as $\sim \exp(-F_Q/T)$, where $F_Q$ is
the free energy of an infinitely heavy quark 
\cite{mclerransvetitsky}, then $\langle L({\bf x}) \rangle = 0$ 
implies that the free energy is infinite, corresponding to confinement
of color charges.
In the high-temperature phase, $\langle L({\bf x}) \rangle \neq 0$,
which implies that the free energy of an infinitely heavy quark
is finite. This indicates
the liberation of colored degrees of freedom, i.e., deconfinement. 
The expectation value of the Polyakov loop is therefore the
order parameter for the deconfinement transition.

For an $[SU(N_c)]$ gauge theory the action has
a global $Z(N_c)$ symmetry: the action does not change when
multiplying all time-like links at a given spatial position ${\bf x}$
by an element $z = \exp(i 2\pi n/N_c)$ of the center $Z(N_c)$
of the gauge group $[SU(N_c)]$.
In the high-temperature phase, the
nonzero expectation value of the Polyakov loop 
breaks this symmetry spontaneously. In the low-temperature
phase, $\langle L({\bf x}) \rangle = 0$, and this symmetry is restored.
For two colors, $N_c=2$, the effective theory in the critical region
around the phase transition is given by a $Z(2)$ spin model,
i.e., it is in the same universality class as the Ising model
\cite{mclerransvetitsky}. 
This model has a second-order phase transition.
For $N_c=3$, the effective theory is that of a $Z(3)$ spin model
\cite{svetitskyyaffe},
i.e., it is in the universality class of the 3-state Potts model
which has a first-order phase transition
\cite{Potts}. The transition temperature was computed to be
$T_c \simeq 270$ MeV \cite{schladming,laermannphilipsen}, see also
Sec.~\ref{translatt}.

\subsubsection{\it Dynamical quarks} \label{dynquarks}

In the presence of dynamical quarks, $N_f >0$, the picture becomes
somewhat more complicated. The fermionic term in the
QCD Lagrangian~(\ref{LQCD}) breaks the $Z(N_c)$ symmetry explicitly, and thus
there is strictly speaking no order parameter for deconfinement.
Nevertheless, the QCD transition in the presence of massless dynamical quarks
has an order parameter, which is related to 
the chiral symmetry of QCD. While the QCD Lagrangian~(\ref{LQCD})
is chirally symmetric when $m=0$, the ground state of QCD is 
not, i.e., chiral symmetry is spontaneously broken. It is instructive to
review these arguments in more detail.

In the chiral limit, where the quark mass matrix is zero, $m=0$, 
the QCD Lagrangian~(\ref{LQCD})
is invariant under global chiral $U(N_f)_r \times U(N_f)_\ell$ rotations
of the quark fields. To see this, decompose the quark spinors into
right- and left-handed spinors,
\begin{equation}
\psi \equiv \psi_r + \psi_\ell
\;\; , \;\;\;\; 
\psi_{r,\ell} \equiv {\cal P}_{r,\ell} \, \psi
\;\; ,\;\;\;\;
{\cal P}_{r,\ell} \equiv \frac{1 \pm \gamma_5}{2} \,\, ,
\end{equation}
where ${\cal P}_{r,\ell}$ are chirality projectors. Then
perform a $U(N_f)_{r,\ell}$ 
transformation on the right/left-handed quark spinors,
\begin{equation} \label{Urot}
\psi_{r,\ell} \rightarrow U_{r,\ell}\, \psi_{r,\ell}
\;\; ,\;\;\;\; 
U_{r,\ell} \equiv \exp\left( i \sum_{a=0}^{N_f^2-1} \alpha^a_{r,\ell}\,
T_a \right) \in U(N_f)_{r,\ell} \,\, ,
\end{equation}
where $\alpha^a_{r,\ell}$ are
the parameters and $T_a$ the generators of $U(N_f)_{r,\ell}$.
The Lagrangian~(\ref{LQCD}) remains invariant under this transformation,
$ {\cal L}(\psi_r,\psi_\ell) \equiv {\cal L}(U_r \, \psi_r, 
U_\ell \, \psi_\ell)$.
The chiral group $U(N_f)_r \times U(N_f)_\ell$ is isomorphic to the
group $U(N_f)_V \times U(N_f)_A$ of unitary vector and axial
transformations, where $V \equiv r + \ell$, $A \equiv r - \ell$, i.e.,
$\alpha_V \equiv (\alpha_r + \alpha_\ell)/2$, $\alpha_A \equiv (\alpha_r -
\alpha_\ell)/2$. 
Any unitary group is the direct product
of a special unitary group and a complex phase,
$U(N_f) \cong SU(N_f) \times U(1)$. 
Thus, $U(N_f)_r \times U(N_f)_\ell \cong SU(N_f)_r \times SU(N_f)_\ell
\times U(1)_r \times U(1)_\ell \cong SU(N_f)_r \times SU(N_f)_\ell
\times U(1)_V \times U(1)_A$. The vector subgroup $U(1)_V$ of this
symmetry group corresponds to quark number conservation.
As physical states trivially conserve quark number,
this subgroup does not affect the chiral dynamics
and can be omitted from the further symmetry consideration.
This leaves an $SU(N_f)_r \times SU(N_f)_\ell \times U(1)_A$ symmetry.
The axial $U(1)_A$ symmetry is
broken explicitly by instantons (the so-called $U(1)_A$ anomaly of QCD)
\cite{tHooft}, leaving an $SU(N_f)_r \times SU(N_f)_\ell$ symmetry
which determines the chiral dynamics. 
Since instantons are screened in a hot and/or
dense medium \cite{grosspisarskiyaffe}, the $U(1)_A$ symmetry
may become effectively restored in matter. Then,
the chiral symmetry is again enlarged to
$SU(N_f)_r \times SU(N_f)_\ell \times U(1)_A$.

Nonzero quark masses break the chiral symmetry of the QCD Lagrangian
explicitly. The quark mass term in Eq.~(\ref{LQCD}) is
\begin{equation} \label{massterm}
\bar{\psi}^i \, m_{ij} \, \psi^j
\equiv \bar{\psi}^i_r \, m_{ij} \, \psi^j_\ell +
\bar{\psi}^i_{\ell} \, m_{ij} \, \psi^j_r \,\, ,
\end{equation}
where flavor indices $i,j=1, \ldots, N_f$ are explicitly written,
a sum over repeated indices is implied, and the properties
${\cal P}_{r,\ell} \gamma_0 = \gamma_0 {\cal P}_{\ell,r},\,
{\cal P}_r \, {\cal P}_\ell = {\cal P}_{\ell} \, {\cal P}_r = 0$ of
the chirality projectors were used. Now suppose all quark masses
were equal, $m_{ij} \equiv m \, \delta_{ij}$.
Performing chiral 
$SU(N_f)_r \times SU(N_f)_\ell \times U(1)_A$ rotations
of the quark fields, one observes that the mass term~(\ref{massterm}) 
preserves an $SU(N_f)_V$ symmetry.
All axial symmetries are explicitly broken.
If less than $N_f$ quark masses are equal, say $M < N_f$,
the preserved vector symmetry is $SU(M)_V$.
In nature, where $m_u \simeq m_d \ll m_s \ll m_c \ll m_b 
\ll m_t$, one only has the
well-known (approximate) $SU(2)_V$ isospin symmetry.
Consequently, exotic hadrons with strange, charm, bottom, or top
degrees of freedom are not degenerate in mass with
their non-strange counterparts.

The mass term $\bar{\psi}^i \, m_{ij} \, \psi^j$ in the QCD Lagrangian is
of the same form as the term ${\bf H} \cdot {\bf S}$ in spin models,
which couples the spin ${\bf S}$ to an external magnetic
field ${\bf H}$. Obviously, the operator 
$\bar{\psi}^i \, \psi^j$ corresponds to the spin ${\bf S}$, while 
the quark mass matrix $m_{ij}$ assumes the role of the external
magnetic field ${\bf H}$. 
Thus, the expectation value $\langle \bar{\psi}^i \, \psi^j \rangle$
is the analogue of the expectation value
of the spin, the magnetization ${\bf M} \equiv \langle {\bf S} \rangle$.
While the mass term explicitly breaks the chiral symmetry 
$SU(N_f)_r \times SU(N_f)_\ell \times U(1)_A$ of the 
QCD Lagrangian to an (approximate) $SU(2)_V$ symmetry, 
the external magnetic field in spin models explicitly 
breaks the rotational symmetry $O(3)$ to $O(2)$.

The analogy between QCD and spin models, however, extends further
than this. Even in the absence of external magnetic fields,
in spin models with ferromagnetic interactions 
rotational symmetry is spontaneously broken due to a nonvanishing
magnetization ${\bf M}\neq 0$ in the ferromagnetic phase.
Ana\-lo\-gously, in the QCD vacuum, chiral symmetry is spontaneously
broken by a nonvanishing expectation value 
$\langle \bar{\psi}^i \, \psi^j \rangle_{\rm vac.} \neq 0$.
Let us introduce the so-called chiral condensate $\Phi^{ij}$ 
and its complex conjugate, ${\Phi^{ij}}^\dagger$, via
\begin{equation} \label{chiralcond}
\Phi^{ij} \sim \langle \bar{\psi}^i_\ell \, \psi^j_r \rangle
\;\; , \;\;\;\;
{\Phi^{ij}}^\dagger \sim \langle \bar{\psi}^i_r \, \psi^j_\ell
\rangle \,\,.
\end{equation}
A nonvanishing expectation value 
$\langle \bar{\psi}^i \, \psi^j \rangle \neq 0$ is then
equivalent to $\Phi^{ij} + {\Phi^{ij}}^\dagger \neq 0$.
Just like the mass term in the QCD Lagrangian, a nonvanishing
chiral condensate breaks the chiral symmetry. In the chiral limit,
$m_{ij} \equiv 0$, nothing distinguishes one quark flavor
from another and, consequently, $\Phi^{ij}_{\rm vac.} = \phi_0\, \delta^{ij}$.
(In principle, there is another possibility how a chiral
condensate could break the chiral symmetry, for a more detailed
discussion see below and Ref.~\cite{pisarskistein}.)
This chiral condensate breaks the chiral
$U(N_f)_r \times U(N_f)_\ell$ symmetry spontaneously
to $U(N_f)_V$. To see this, note that
the chiral condensate is still
invariant under vector transformations,
$\Phi \rightarrow U_r\, \Phi \, U_\ell^\dagger
\equiv \Phi$, if $U_r = U_\ell \equiv U_V$, but not under axial
transformations, $\Phi \rightarrow U_r\, \Phi \, U_\ell^\dagger
\neq \Phi$, if $U_r =  U_\ell^\dagger \equiv U_A$.

According to Goldstone's theorem, the breaking
of the global chiral symmetry leads to the occurrence
of massless modes, the so-called Goldstone bosons. The
number of Goldstone bosons is identical to the number
of broken generators. In the QCD vacuum, where the $U(1)_A$ anomaly
is present, the breaking pattern is
$SU(N_f)_r \times SU(N_f)_\ell \rightarrow SU(N_f)_V$, i.e., 
in this case there are $N_f^2-1$ broken generators, corresponding to the
generators of the broken axial symmetry $SU(N_f)_A$.
For $N_f =1$, there is no global chiral symmetry that could
be broken, and thus no Goldstone boson.
For $N_f=2$, the Goldstone bosons are the three 
pions, the lightest hadronic
species. In nature, the pions are not completely massless,
because the chiral symmetry is explicitly broken by the
small, but nonzero quark mass term in the QCD Lagrangian.
This turns the Goldstone bosons into so-called pseudo-Goldstone bosons.
For $N_f=3$, the pseudo-Goldstone bosons correspond to
the pseudoscalar meson octet, comprising pions, kaons, and the eta
meson. Since chiral symmetry is more strongly broken by
the larger strange quark mass, the pseudo-Goldstone bosons
carrying strangeness are heavier than the pions. For
$N_f \geq 4$, the explicit symmetry breaking by the heavy exotic quark
flavors is so strong that the would-be Goldstone bosons are actually
heavier than the ordinary (i.e. non-Goldstone) non-strange bosons.

In spin models, rotational symmetry is restored above some
critical temperature and the magnetization vanishes. 
The magnetization is the order
parameter for this so-called ferromagnet-diamagnet phase transition.
By analogy, one expects a similar mechanism to occur in QCD, i.e., 
$\Phi^{ij}$ to vanish above some
critical temperature. The symmetry of the ground state
is then restored to the original
chiral symmetry, i.e., $SU(N_f)_r \times SU(N_f)_\ell$, if the
$U(1)_A$ anomaly is still present, or $SU(N_f)_r \times SU(N_f)_\ell
\times U(1)_A$, if instantons are sufficiently screened at the
transition in order to effectively restore the $U(1)_A$ symmetry.
Lattice QCD calculations show that this expectation is indeed fulfilled:
there is a phase transition between the low-temperature
phase where chiral symmetry is broken and the high-temperature
phase where it is restored, for details 
see Sec.~\ref{LattQCD}. Just like the magnetization in
spin models, the chiral condensate $\Phi^{ij}$ 
is the order parameter for this so-called
chiral phase transition.

In the case of massless quarks, $m_{ij} \equiv 0$, 
based on universality arguments one can analyse the
order of the chiral transition in the framework of a
linear sigma model for the order
parameter field $\Phi^{ij}$ \cite{pisarskiwilczek}. This linear
sigma model is an effective theory, i.e,  all terms 
allowed by the original chiral symmetry must in principle appear,
\begin{equation} \label{Leff}
{\cal L}_{\rm eff} = 
{\rm Tr} \left( \partial_0 \Phi^\dagger \partial^0 \Phi \right) 
- v^2 {\rm Tr} \left( \mbox{\boldmath$\nabla$} \Phi^\dagger \cdot 
\mbox{\boldmath$\nabla$} \Phi \right)
- V_{\rm eff}(\Phi) \,\, ,
\end{equation}
where the effective potential
\begin{equation} \label{Veff}
V_{\rm eff} (\Phi) =
 m^2\, {\rm Tr} \left(\Phi^\dagger \Phi \right)
 +\lambda_1 \, \left[ {\rm Tr} \left(\Phi^\dagger \Phi\right) \right]^2
 + \lambda_2 \, {\rm Tr} \left( \Phi^\dagger \Phi\right)^2 
 - c \, \left( {\rm det} \Phi + {\rm det} \Phi^\dagger \right)
+ \ldots \,\, 
\end{equation}
determines the ground state of the theory. In Eq.~(\ref{Leff})
it was assumed that the first term is canonically normalized.
However, since Lorentz symmetry is explicitly broken in a
medium at nonzero temperature, the coefficient $v^2$ in Eq.~(\ref{Leff})
may in general be different from one. In Eq.~(\ref{Veff}),
$\ldots$ denote higher-dimensional operators which are irrelevant
for the discussion of the order of the phase transition.
For $c \neq 0$, the chiral symmetry of
${\cal L}_{\rm eff}$ is $SU(N_f)_r \times SU(N_f)_\ell$,
while for $c=0$, it is $SU(N_f)_r \times SU(N_f)_\ell \times U(1)_A$.
Thus, the $U(1)_A$ anomaly is present for $c\neq 0$, and absent for
$c=0$. 
While these chiral symmetries are manifest in the Lagrangian~(\ref{Leff}), 
the ground state of the theory 
respects them only for $c = 0$ and $m^2 > 0$. For $c=0$ and $m^2 <0$ 
the chiral symmetry is spontaneously broken by a nonvanishing 
vacuum expectation value for the order parameter. 
Consequently, if the linear sigma model is to describe the chiral
transition in QCD, one has to ensure that $m^2 <0$ for $c=0$. 
There are still two possibilities how the order parameter can break the
symmetry. As shown in Ref.~\cite{pisarskistein}, if $\lambda_2 > 0$
the ground state is given by 
$\Phi^{ij}_{\rm vac.} = \phi_0 \, \delta^{ij}$, while for
$\lambda_2 < 0$ the ground state is given by
$\Phi^{ij}_{\rm vac.} = \phi_0 \, \delta^{i1}\, \delta^{j1}$.
(The choice of the 1-direction in right- and left-handed flavor space 
is arbitrary.)
Nature realizes the first possibility. 
For the case $c \neq 0$,
no general arguments can be made; whether the
ground state of the theory breaks chiral symmetry spontaneously
depends on the particular
values for the coupling constants $c, \lambda_1, \lambda_2$ and
the number of flavors $N_f$. 

For $N_f=1$, there is no difference between the two quartic invariants
in Eq.~(\ref{Veff}), and one may set $\lambda_1 +\lambda_2 \equiv \lambda$. 
In the presence of the $U(1)_A$ anomaly,
$c \neq 0$, there is also no chiral symmetry, and the transition is 
crossover, due to the linear term $\sim c$, which tilts
the effective potential such that the (thermal) ground state
$\langle \Phi \rangle_T \neq 0$.
If the $U(1)_A$ anomaly is absent, $c=0$,
the effective theory for the order parameter
falls in the same universality class as that of the $O(2)$ Heisenberg magnet,
and thus the transition is of second order. 

For $N_f=2$ and in the presence of the $U(1)_A$ anomaly, 
the chiral symmetry is $SU(2)_r \times SU(2)_\ell$, which is
isomorphic to $O(4)$. The effective theory for the order parameter
is in the universality class of the 
$O(4)$ Heisenberg magnet. Consequently, the transition is of
second order \cite{pisarskiwilczek}.  
If the $U(1)_A$ symmetry is effectively restored at the 
phase transition temperature, the symmetry group
is larger, $SU(2)_r \times SU(2)_\ell \times U(1)_A$, which is
isomorphic to $O(4) \times O(2)$, and the transition is of
first order. Lattice QCD calculations 
determine the transition temperature to be $T_c \simeq 172$ MeV
\cite{schladming,laermannphilipsen}.

For $N_f=3$, the chiral transition is of first order, both 
when the $U(1)_A$ symmetry is explicitly broken by instantons or
when it is effectively restored at the transition.
In the first case, the effective theory features a cubic invariant in the order
parameter field (the term $\sim {\rm det} \Phi + {\rm det} \Phi^\dagger$), 
which drives the chiral transition first order 
\cite{pisarskiwilczek}. In the second case,
the transition is fluctuation-induced of first order
\cite{pisarskiwilczek}. This also holds for
$N_f \geq 4$, irrespective of whether the $U(1)_A$ symmetry is 
explicitly broken or not. For $N_f = 3$, lattice QCD calculations
find the transition temperature to be $T_c \simeq 155$ MeV
\cite{laermannphilipsen}, cf.~Sec.~\ref{translatt}.
Note that nonvanishing quark masses 
can also be accounted for by adding a term
\begin{equation} \label{additional}
{\cal L}_H \equiv 
{\rm Tr} \left[ H \left( \Phi + \Phi^\dagger \right) \right]
\end{equation}
to the right-hand side of Eq.~(\ref{Leff}). As discussed above, this term
is the analogue of the term ${\bf H} \cdot {\bf S}$ in spin models.
Consequently, the external ``magnetic field'' $H_{ij}$ is proportional to the
quark mass matrix $m_{ij}$. 
\subsubsection{\it The quark-mass diagram} \label{quarkmassdiagram}

\begin{figure}[tb]
\begin{center}
\begin{minipage}[t]{9.5 cm}
\epsfig{file=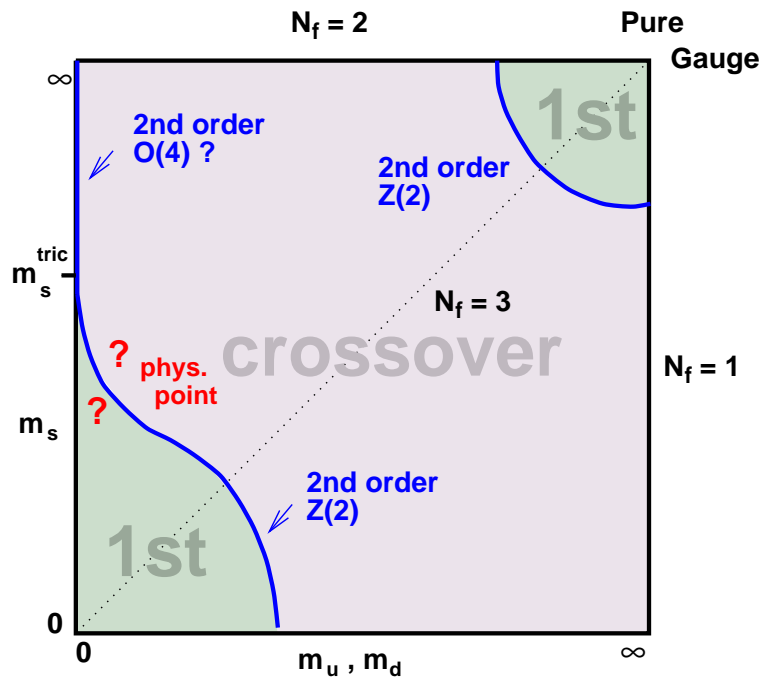,scale=0.6}
\end{minipage}
\begin{minipage}[t]{16.5 cm}
\caption{The quark-mass diagram (from Ref.~\cite{laermannphilipsen}).
\label{quark-mass}}
\end{minipage}
\end{center}
\end{figure}

Nonvanishing quark masses lead to the term~(\ref{additional}) 
in the effective theory for the order parameter field.
This term is linear in $\Phi$, such that the effective potential
is tilted. This may render a first or second order
phase transition a crossover transition (similar to the case
$N_f=1$ with $U(1)_A$ anomaly discussed in Sec.~\ref{dynquarks}, 
where a tilt in the potential is induced by the linear term $\sim c$).
For instance, the second-order transition for QCD with $N_f=2$ massless 
flavors is rendered crossover by a nonvanishing quark mass.
The first-order phase transition for QCD with $N_f=3$ massless flavors
can also become crossover, if the quark masses are sufficiently large.
In the real world, the up and down quark are approximately of
the same mass, while the strange quark is much heavier.
It is customary to put $m_q \equiv m_u \simeq m_d $ and
identify first-order regions, 
second-order lines, and crossover regions in an $(m_q, m_s)$ diagram,
see Fig.~\ref{quark-mass}. To simplify the following discussion, only 
the case where the $U(1)_A$ anomaly is present will be considered.

The origin in Fig.~\ref{quark-mass}
corresponds to the massless 3-flavor case, 
and the transition is of first order. 
The upper left corner corresponds to the massless 2-flavor case,
since the strange quark is infinitely heavy. Here, the transition
is of second order.
The lower right corner is the case of one
massless flavor. The transition is crossover.
The upper right corner, where all quark flavors are infinitely
heavy, corresponds to the pure gauge theory. At this point
the transition is of first order. 

The first-order regions around the
origin and the upper right corner extend somewhat into the $(m_q,m_s)$
plane and are bounded by critical lines where the transition
is of second order. Along these critical lines,
the second-order phase transitions 
are in the universality class of the Ising model, $Z(2)$.
In between the critical lines, the
transition is crossover. There is also a second-order phase transition
line (with a phase transition in the $O(4)$ universality class) 
extending downwards from the upper left corner along the
$m_s$ axis. There is a tricritical point where this line 
meets the second-order phase transition line bordering the first-order 
region around the origin \cite{pisarskigoksch}.
It is an interesting question whether the real world, where
$m_q  \sim 5\, {\rm MeV} \ll m_s \sim 100\, {\rm MeV}$, is
still inside the first-order phase transition region or already in the
crossover region. 
There are ongoing lattice QCD studies to decide this question, 
which at present favor the latter possibility \cite{laermannphilipsen},
see also Sec.~\ref{translatt} for more details.

\subsubsection{\it Nonzero quark chemical potential}

So far, the quark-hadron phase transition was studied at $\mu = 0$. 
Let us finally discuss the case of nonzero quark chemical
potential. 
For many years, lattice QCD studies at nonzero chemical potential 
were hampered by numerical problems related to the so-called sign
problem of the fermion determinant. Only recently there have
been attempts to compute the order of the phase transition, as
well as thermodynamical properties, at
nonzero quark chemical potential; for details,
see Sec.~\ref{nonzeromu}. So far, these calculations have been done
on fairly small lattices with rather heavy quarks.
Consequently, they show a crossover transition at $\mu=0$. This
crossover transition
extends to the point $(T,\mu)_{\rm cr} = (160 \pm 3.5,242 \pm 12)$ MeV, see 
also Fig.~\ref{endpoint}.

This point is a critical point where the transition is of second order. 
It is in the unversality class of the Ising model, i.e., $Z(2)$.
For smaller temperatures and larger chemical
potentials, the transition becomes of first order.
The critical point will move towards the
$T$ axis when the quark masses are decreased. From the discussion
in Sec.~\ref{quarkmassdiagram}
one cannot exclude the possibility that, for realistic quark masses,
the first-order phase transition line emerges
directly from the $T$ axis.

Finally, the question arises whether the line of first-order phase transitions
extends all the way down to $T=0$, and if so, at which point it
hits the $\mu$ axis. Renormalization group arguments \cite{hsuschwetz}
suggest that the behavior at zero temperature is very similar to
the one at nonzero temperature: the transition is of first order
for $N_f \geq 3$ as well as for 
$N_f =2$ in the absence of the $U(1)_A$ anomaly, 
while it could be of second order for $N_f=2$ in the
presence of the $U(1)_A$ anomaly.
In the latter case, however, it would remain of second order
along the whole phase transition line, and only the universality class
would change from $O(4)$ critical behavior 
at nonzero temperature to Gaussian critical behavior at
zero temperature. Since lattice QCD calculations \cite{fodorkatz} 
indicate that the transition becomes of first order for temperatures
below the critical point, and since it is hard to imagine that the transition
switches back to second order as the temperature decreases further,
this possibility can most probably be ruled out.
Note, however, that if quark-gluon matter is in a color-superconducting
phase to the right of the QCD phase transition line, other
possibilities emerge, for details, see Sec.~\ref{CSCphases}.

In any case, model calculations \cite{scavenius_NJL} within a 
Nambu--Jona-Lasinio (NJL) model \cite{NJL} support the picture
that the transition remains of first order below
the critical point all the way down to the $\mu$ axis. 
The value of $\mu$, where the first-order 
phase transition line meets the $\mu$ axis, depends sensitively
on the parameters used in these model calculations.
Its actual value should not be taken too seriously, because the
NJL model with quark degrees of freedom does not have
a phase where matter consists of nucleons. Instead, the
transition to quark matter happens at a quark chemical potential
which is of the order of the ground state quark chemical
potential, $\mu_0 \simeq 308$ MeV. Since we know that 
nucleonic matter exists, this behavior is clearly unphysical.

The critical point has recently received a fair amount of attention
\cite{shuryakrajagopalstephanov}.
For a second-order phase transition in the $Z(2)$ universality class, 
there must be one massless degree of freedom. The fact
that this degree of freedom is massless causes critical 
fluctuations.
These fluctuations were suggested to be an experimental signature
for the critical point in heavy-ion collisions.
Which physical particle does the massless degree of freedom correspond to?
For realistic quark masses, the pions are not massless in the vacuum, 
and it is unlikely that they become massless at the critical
point (the pions usually get more massive when the temperature is
increased). Moreover, since isospin is still a good symmetry at
the critical point, all pions would simultaneously become massless.
Then, one would have three massless modes instead of just
one. Consequently, the pions cannot assume the role of the massless mode. 
Since the critical point exists even when considering only $N_f=2$ quark
flavors and since, for $N_f=2$, there is only one other degree of freedom 
in the effective theory besides the pions, it must be this degree of
freedom that becomes massless: the scalar $\sigma$ meson 
\cite{scavenius_NJL,shuryakrajagopalstephanov}. In the vacuum, this
meson has a mass of about $400 - 1200$ MeV \cite{PDB}.
If it becomes massless, it can be copiously produced.
When these $\sigma$ mesons decouple
from the collision region, they assume their vacuum masses
and rapidly decay into pions. Besides critical fluctuations,
another signature for the
critical point would thus be the late emission of a large
amount of pions in a heavy-ion collision.

\subsection{\it The Color-Superconducting Phases of QCD} \label{CSCphases}

\subsubsection{\it Proof of existence of color superconductivity}

There are other phases in the phase diagram of 
nuclear matter, which have recently received much attention
in the literature, the so-called color-superconducting
phases in sufficiently cold and dense quark matter \cite{rajwil}. 
Color superconductivity occurs, because
there is an attractive interaction between two quarks at the Fermi
surface \cite{barrois,bailinlove}. 
Then, by Cooper's theorem, these quarks form 
Cooper pairs which condense in the new ground state of the system.

At least at asymptotically large quark chemical potentials,
the existence of an attractive interaction between quarks at the
Fermi surface, and thus the existence of color superconductivity, 
can be rigorously proven.
Due to asymptotic freedom
\cite{asympt}, when $\mu \gg \Lambda_{\rm QCD}$, the strong coupling constant
of QCD, evaluated at the scale $\mu$, becomes small, 
$g(\mu) \ll 1$, such that the
dominant interaction between quarks is given by single-gluon exchange.
The scattering amplitude for single-gluon exchange in
an $[SU(N_c)_c]$ gauge theory is proportional to
\begin{equation} \label{TaTa}
(T_a)_{ki} \, (T_a)_{lj} = - \frac{N_c+1}{4 N_c}\,
\left( \delta_{jk} \, \delta_{il} - \delta_{ik}\, \delta_{jl} \right) 
+ \frac{N_c-1}{4 N_c} \,
\left( \delta_{jk} \, \delta_{il} + \delta_{ik}\, \delta_{jl} \right) \,\, ,
\end{equation}
where $i,j$ are the fundamental colors of the two quarks in the
incoming channel, and $k,l$ their respective colors in the
outgoing channel.
Under the exchange of
the color indices of either the incoming or the outgoing quarks
the first term is antisymmetric,
while the second term is symmetric.
In group theoretical language, for $[SU(3)_c]$ Eq.~(\ref{TaTa}) represents
the coupling of two fundamental color triplets to an (antisymmetric)
color antitriplet and a (symmetric) color sextet,
\begin{equation}
[{\bf 3}]^c \times [ {\bf 3} ]^c = [ {\bf \bar{3} } ]_a^c 
+ [ {\bf 6} ]_s^c\,\, .
\end{equation}
The minus sign in front of the antitriplet contribution in
Eq.~(\ref{TaTa}) signifies the fact that this channel is
attractive, while the sextet channel is repulsive.
Therefore, one expects that quark Cooper pairs condense
in the color-antitriplet channel.

This argument holds rigorously at asymptotically large densities.
The highest densities of nuclear matter
that can be achieved in the laboratory through
heavy-ion collisions, or that occur in nature in the interior
of neutron stars, are of the order of ten times nuclear matter
ground state density. At these densities, the quark chemical
potential is still fairly small, $\mu \sim 0.5$ GeV.
For phenomenology it is therefore important 
to answer the question whether color superconductivity also
exists at the (comparatively moderate) densities occurring in nature.
There is no rigorous way to answer this question, as
an extrapolation of the above asymptotic argument becomes
unreliable when $g (\mu) \sim 1$.
Nevertheless, calculations in the framework of the
NJL model \cite{arw} 
show that color superconductivity does seem to occur also at
moderate densities. In this case, the attractive interaction
could be mediated by instanton (instead of single-gluon) exchange.

Color superconductivity is a much more complicated phenomenon
than ordinary superconductivity.
From a very qualitative point of view, in comparison to
electrons, quarks carry additional quantum numbers such as color and flavor.
The wave function of a Cooper pair has to be antisymmetric under the
exchange of the two fermions forming the pair. Consequently,
the possible color and
flavor representations of the two-fermion state have to be chosen 
in a way which respects this antisymmetry.
This requirement helps to classify all possible color-superconducting
condensates \cite{pr2,pisarskirischke}.
This classification will be presented in the following.

\subsubsection{\it Classification of color-superconducting phases}
\label{classcsc}

The baryon density in heavy-ion collisions or in neutron stars 
is at most of the order of ten times the nuclear ground state density.
Therefore, the quark chemical potential is unlikely to assume values 
beyond $\mu \sim 1$ GeV. At zero temperature, one can only have fermions
if the Fermi energy $\mu$ exceeds their mass. Thus,
only the three lightest quark flavors, up, down, and strange, play
any role; charm, bottom, and top quarks are too heavy.
For small temperatures, these heavy flavors can be thermally
excited, but their abundance is exponentially suppressed $\sim
\exp(-m_f/T)$, cf.~Sec.~\ref{qualarg}. Therefore, they will be excluded
from the following consideration.
For $\mu \sim 1$ GeV, up and down quarks 
can be treated as truly ultrarelativistic particles, as
$m_q/\mu \sim 10^{-3}$. To first approximation, also 
the strange quark will be considered to be massless. Corrections
due to the strange quark mass can be treated perturbatively,
the correction factor being of order $m_s/\mu \sim 10^{-1}$
\cite{CSCeffective}.

For ultrarelativistic particles, spin $S$ and angular momentum $L$
are not separately good quantum numbers, only the total
spin $J = L+S$ is. Therefore, possible Cooper pair
wave functions should be classified according to their total spin $J$.
Let us first focus on the spin-zero channel, $J=0$. The $J=0$ representation
of the $SU(2)_J$ spin group is totally antisymmetric. Therefore,
the remaining color and flavor part of the Cooper pair wave function
has to be symmetric under the simultaneous exchange of color
and flavor indices in order to fulfill the requirement of overall antisymmetry.
If one assumes that quarks pair in the antisymmetric color-antitriplet
channel, one has no choice but to also chose an antisymmetric flavor 
representation. For $N_f =1 $ flavor, this is impossible, as there
is no flavor symmetry group with an antisymmetric representation.

One therefore has to consider at least $N_f=2$ quark flavors
(for instance, up and down), where the most simple
representation is the antisymmetric flavor singlet $[{\bf 1}]_a^f$
representation of the $SU(2)_V$ flavor group. Therefore, 
the most simple $J=0$ quark Cooper pair condensate has the form
\begin{equation} \label{2SC}
\Phi^{fg}_{ij} = \epsilon_{ijk}\, \epsilon^{fg}\, \Phi_k \,\, .
\end{equation}
Here, $i,j=1,\ldots, N_c$ are the color indices of
the quarks forming the Cooper pair, while $f,g = 1, \ldots, N_f$
are the corresponding flavor indices. The two totally antisymmetric
tensors on the right-hand side ensure that the condensate belongs to the 
$[{\bf \bar{3}}]^c_a$ in color, as well as the $[{\bf 1}]^f_a$ representation
in flavor space. The
color-superconducting phase represented by the condensate~(\ref{2SC})
is commonly called the
``2SC'' phase (for ``${\bf 2}$-flavor color {\bf S}uper{\bf C}onductor'').

Condensation of quark Cooper pairs occurs if
the quantity $\Phi_k$ on the right-hand side of Eq.~(\ref{2SC})
is nonzero, $\Phi_k \neq 0$. Thus, the quark Cooper pair condensate
carries a fundamental color index $k$. This indicates that
the local $[SU(3)_c]$ color symmetry
is spontaneously broken by the quark Cooper pair condensate, similarly
to the spontaneous breaking of the global chiral symmetry 
$SU(N_f)_r \times SU(N_f)_\ell$ by the chiral condensate~(\ref{chiralcond}) in
the QCD vacuum discussed in Sec.~\ref{dynquarks}. In this sense,
$\Phi_k$ is the order parameter for color superconductivity.
It is zero in the phase of unbroken $[SU(3)_c]$ symmetry, and nonzero
in the broken phase where condensation of quark Cooper pairs occurs.

Of course, a local symmetry can never be truly broken 
spontaneously \cite{elitzur}.
However, after fixing the gauge, spontaneous breaking does occur, just like
in ordinary superconductivity or in the standard model of
electroweak interactions. In ordinary superconductivity, the
condensation of electron Cooper pairs breaks the $[U(1)_{\rm em}]$
gauge symmetry of electromagnetism, while in the standard model,
the Higgs field breaks $[SU(2)_\ell] \times [U(1)_Y]$ to $[U(1)_{\rm em}]$.
Analogously, the quark Cooper pair condensate~(\ref{2SC})
breaks the $[SU(3)_c]$ color gauge symmetry. 

Since quarks carry baryon and electric charge, the Cooper pair
condensate (\ref{2SC}) in principle also breaks the global $U(1)_V$ of baryon
number conservation and the local $[U(1)_{\rm em}]$ of electromagnetism.
In the discussion of chiral symmetry breaking, these symmetries
were never broken because the chiral condensate consists of a
quark and an antiquark which carry opposite baryon and electric charge.
The chiral condensate is thus a singlet under $U(1)_V$ and $[U(1)_{\rm em}]$
and consequently preserves these symmetries. This is different for
a color-superconducting condensate which consists of two quarks.
It turns out, however, that there exists a ``rotated'' baryon
number $\tilde{U}(1)$ symmetry and a ``rotated'' electromagnetic
$[\tilde{U}(1)]$ symmetry, which are formed from the original
baryon number and electromagnetic symmetries and the eighth generator
of $[SU(3)_c]$ \cite{schaferwilczek}. 
This is similar to electroweak symmetry breaking,
where the $[U(1)_{\rm em}]$ symmetry of electromagnetism is
a combination of the $[U(1)_Y]$ hypercharge symmetry and the
third generator of $[SU(2)_\ell]$. 
Thus, the $U(1)_V$ and $[U(1)_{\rm em}]$
symmetries are not really broken in the 2SC phase, but ``rotated''.
The rotation angle is the analogue of the Weinberg angle in the
standard model of electroweak interactions; for more details,
see Sec.~\ref{gluephoton}.

Spontaneous symmetry breaking in gauge theories does not lead to
Goldstone bosons. Rather, what would have been a Goldstone mode
will be ``eaten'' by a gauge boson which in turn becomes massive and 
thus acquires an additional longitudinal degree of freedom.
There are as many massive gauge bosons as there would have been
Goldstone modes due to spontaneous symmetry breaking.
In ordinary superconductors, the electromagnetic $[U(1)_{\rm em}]$ 
symmetry is broken, which has only one generator. 
Consequently, there is a single Goldstone mode which is 
``eaten'' by the single gauge boson present in this case, the photon. 
The photon acquires a so-called Meissner mass. What happens physically
is that magnetic fields are damped on length scales of the order
of the inverse Meissner mass, which in turn leads to the Meissner effect,
the expulsion of magnetic flux from the superconductor. In the
standard model of electroweak interactions, 
$[SU(2)_\ell] \times [U(1)_Y]$ is broken to $[U(1)_{\rm em}]$, i.e.,
there are three Goldstone modes which in turn lead to
the massive gauge bosons of the weak interaction, $W^\pm$ and $Z$.
The photon is massless. This is required, since 
it is the gauge boson of the
residual $[U(1)_{\rm em}]$ symmetry of electromagnetism.
Analogously, in a color superconductor
one expects some of the gluons to become massive. Exactly how many
gluons acquire a mass depends on the pattern of symmetry breaking.
For the condensate~(\ref{2SC}), one can clarify this via the
following argument.

By a global color rotation, one can always orient the 
order parameter $\Phi_k$ to point in the 3-direction in color space
(more precisely, the {\em anti\/}-3-direction, as condensation occurs
in the color {\em anti}-triplet channel),
\begin{equation} \label{2SCorient}
\Phi_k \equiv \delta_{k3} \, \Phi\,\, .
\end{equation}
Physically, this means that if we call color 1 red, color 2 green, and
color 3 blue,
red up (or down) quarks and green down (or up) quarks condense to 
form an anti-blue Cooper pair condensate.
Blue up and down quarks do not participate in condensation. 
The condensate~(\ref{2SCorient})
does not break the color $[SU(3)_c]$ gauge symmetry completely.
The residual symmetry is a local $[SU(2)_c]$ symmetry in the space
of the first two colors (in our conventions, red and green).
Including electromagnetism,
the symmetry breaking pattern for the condensate~(\ref{2SCorient})
is therefore 
$[SU(3)_c] \times [U(1)_{\rm em}] \rightarrow [SU(2)_c] \times [\tilde{U}(1)]$.
Consequently, there are $8-3 = 5$ broken generators, which lead
to five massive gluons. The remaining three gluons must remain
massless as they correspond to the gauge bosons of the residual
local $[SU(2)_c]$ symmetry. This is also borne out by an explicit
calculation of the gluon Meissner masses in the 2SC 
phase \cite{carter,rischke2SC}. The gauge boson of the local
$[\tilde{U}(1)]$ symmetry (the ``rotated'' photon) is also
massless. For more details, see Sec.~\ref{gluephoton}.

For $N_f = 3$ flavors, condensation of quark Cooper pairs becomes
considerably more interesting. First, to preserve the antisymmetry
of the Cooper pair wave function the two quarks have to be in
the $[{\bf \bar{3}}]_a^f$ representation of the global $SU(3)_V$
flavor symmetry. Consequently, the quark Cooper pair condensate has the form
\begin{equation}
\Phi^{fg}_{ij} = \epsilon_{ijk}\, \epsilon^{fgh} \, \Phi_k^h\,\, .
\end{equation}
The difference to Eq.~(\ref{2SC}) is that, to ensure antisymmetry
in flavor space, one is required to use the totally antisymmetric tensor
of rank 3, $\epsilon^{fgh}$, rather than its rank-2 counterpart.
Consequently, an additional flavor index $h$ appears in
the order parameter, $\Phi_k^h$. A nonvanishing order parameter
automatically implies that not only the local $[SU(3)_c]$ color, but also
the global $SU(3)_V$ flavor symmetry is broken.
The situation is not unlike the one encountered in superfluid helium-3
\cite{vollhardt}. Superfluid helium-3 forms Cooper pairs with
spin $S=1$ and angular momentum $L=1$. (Both spin and angular
momentum are good quantum numbers, as helium-3 is a non-relativistic
system.) Consequently, the order parameter breaks the global 
$SO(3)_S$ of spin as well as the global $SO(3)_L$ of angular momentum.
This breaking can occur in many possible ways, giving rise to a
plethora of phases in superfluid helium-3.

Similarly, one would expect many different phases to occur in a 3-flavor
color superconductor. However, in fact there are
only two possibilities, one of which is likely to be realized
in nature \cite{pisarskirischke}. To see this, note the formal
similarity between the order parameter $\Phi^h_k$ 
and the one encountered in chiral symmetry breaking, $\Phi^{ij}$, in
Sec.~\ref{dynquarks}. While $\Phi^{ij}$ transforms under
$SU(N_f)_r \times SU(N_f)_\ell$ (in the presence of the $U(1)_A$ anomaly),
$\Phi^h_k$ transforms under $[SU(3)_c] \times SU(3)_V$.
Consequently, the effective Lagrangian for $\Phi^h_k$ is of the
same form~(\ref{Leff}) as for $\Phi^{ij}$. The two possible
patterns of symmetry breaking occurring in such an effective
theory were already mentioned above. 

If the coupling constant $\lambda_2 > 0$, the order parameter
assumes the form
\begin{equation} \label{CFL}
\Phi^h_k =  \delta^h_k\, \Phi  \,\, .
\end{equation}
In contrast to the 2SC case, where blue quarks
remained unpaired, now all quark colors and flavors participate
in the pairing process. 
The order parameter~(\ref{CFL}) 
is similar to the one for chiral symmetry breaking, where
$\Phi^{ij} = \delta^{ij}\, \Phi$. (In the ground state, $\Phi_{\rm vac.}
\equiv \phi_0$.) Similar to the chiral symmetry breaking pattern 
$SU(N_f)_r \times SU(N_f)_\ell \rightarrow SU(N_f)_V$, $V = r + \ell$,
the condensate~(\ref{CFL}) breaks $[SU(3)_c] \times SU(3)_V$ to the vectorial
subgroup $SU(3)_{c+V}$. The condensate is still invariant
under vector transformations in color and flavor
space, or in other words, a transformation in color requires
a simultaneous transformation in flavor to preserve the
invariance of the condensate. Therefore, the 
discoverers \cite{arw2} of this 3-flavor color-superconducting
state termed it the ``{\bf C}olor-{\bf F}lavor-{\bf L}ocked'', or
short, ``CFL'' state. The residual $SU(3)_{c+V}$ symmetry is no
longer a local symmetry, so that there are no massless gluons.
The complete pattern of symmetry breaking, including the $U(1)_V$
symmetry of baryon number and the $[U(1)_{\rm em}]$ symmetry
of electromagnetism is 
$[SU(3)_c] \times SU(3)_V \times U(1)_V \times [U(1)_{\rm em}]
\rightarrow SU(3)_{c+V} \times [\tilde{U}(1)]$.
(This notation is slightly ambiguous: the $[U(1)_{\rm em}]$ symmetry
is generated by the quark charge operator $Q={\rm diag}(2/3,-1/3,-1/3)$
which is traceless. Thus, the {\em global\/} part of the $[U(1)_{\rm em}]$
symmetry is actually a subgroup of the {\em global\/} $SU(3)_V$ flavor
symmetry. However, the {\em local\/} part of $[U(1)_{\rm em}]$ is not.
Therefore, here and in the following I choose to explicitly denote the 
$[U(1)_{\rm em}]$ symmetry group.)
In the CFL case, unlike the 2SC case, baryon number is broken,
but a rotated electromagnetic $[\tilde{U}(1)]$ is again preserved.
The symmetry breaking pattern leads to nine Goldstone bosons, eight of
which are ``eaten'' by the gluons, i.e., all
gluons acquire a Meissner mass. There is one Goldstone boson
from the breaking of the $U(1)_V$ symmetry. 

In the chiral limit, the flavor symmetry
of QCD is actually not just $SU(3)_V$ but $SU(3)_r \times SU(3)_\ell$.
Assuming that also the $U(1)_A$ symmetry of QCD is effectively
restored at large quark densities,
quark Cooper pair condensation of the form~(\ref{CFL})
breaks $[SU(3)_c] \times SU(3)_r \times SU(3)_\ell \times U(1)_V \times
U(1)_A \times [U(1)_{\rm em}]$ to
$SU(3)_{c + V} \times [\tilde{U}(1)]$, 
i.e., not only the local color, but also the
global chiral symmetry is broken. In addition to the eight massive
gluons, there are also ten real Goldstone bosons, eight from the breaking 
of the $SU(3)_A$ chiral symmetry, and one each from
the breaking of $U(1)_V$ and $U(1)_A$.

Closer inspection \cite{continuity} shows that the excitation spectrum in the
CFL state bears a striking resemblance to the one in the hadronic phase.
Let us first focus on the fermionic sector.
In the CFL phase, there are nine gapped fermionic quasiparticles
(cf.~also Sec.~\ref{secexspec}), eight of which are degenerate in mass. These
correspond to the baryon octet in the QCD vacuum. (For this
argument we have to assume that there is no explicit
$SU(3)_V$ flavor-symmetry breaking in the QCD vacuum.)
The ninth quasiparticle is twice as heavy and does not have a counterpart in 
hadronic matter, but this is not a reason to worry, as such a particle
would have a large decay width into lighter particles.
In the bosonic sector, there are
nine Goldstone bosons from the breaking of the axial $SU(3)_A \times U(1)_A$
symmetry, which correspond to the pseudoscalar nonet in the hadronic phase.
Only the tenth Goldstone boson from the breaking of $U(1)_V$ does
not have a counterpart in the QCD vacuum. Such a boson exists, however, 
in dense nuclear matter, where a superfluid $\Lambda \, \Lambda$ condensate
may form, which also breaks the $U(1)_V$ baryon number symmetry.
The preceding arguments have led to the conjecture of ``continuity'' between 
hadron and quark matter \cite{continuity}. This conjecture
states that, since there is no difference in symmetry
between quark matter in the CFL state and ($SU(3)_V$ flavor-symmetric)
hadronic matter, 
there need not be any phase boundary between these two phases at all.
Of course, this requires that there is no other color-superconducting
phase, for instance the 2SC state, which separates 
CFL matter from hadronic matter. 

For $\lambda_2 < 0$, the order parameter is given by 
\begin{equation} \label{3SC}
\Phi^h_k =  \delta^{h3} \, \delta_{k3} \, \Phi\,\, ,
\end{equation}
where the 3-direction is arbitrary. This condensate breaks
$[SU(3)_c] \times SU(3)_V$ 
to $[SU(2)_c] \times SU(2)_V$. In this case,
$[SU(2)_c]$ is still a local symmetry.
Consequently, like in the 2SC case only five gluons become massive,
and blue up, down, and strange quarks do not participate in the
formation of Cooper pairs. Additionally, 
there are also five Goldstone bosons from the breaking of the 
flavor symmetry. 
There is a more technical and a more physical argument, why 
the CFL state is most likely realized in nature. From the 
more technical point of view, one can show \cite{pisarskirischke}
that, to one-loop order, in QCD $\lambda_1 = 0$ and $\lambda_2 > 0$.
From the more physical point of view, the CFL state is energetically
favored because {\em all\/} quark colors and flavors (instead of
just a few) acquire a gap at the Fermi surface. The gain in
condensation energy is thus expected to be
larger than for a state with a condensate of the form~(\ref{3SC}).

Although a single quark flavor cannot form Cooper pairs
with total spin $J=0$, it can pair in the $J=1$ channel.
(An exhaustive discussion of possible pairing channels
for a single quark flavor is given in Ref.~\cite{bowers}.)
This channel corresponds to the symmetric $[{\bf 3}]_s^J$
representation of the $SU(2)_J$ spin group. If one still
assumes pairing to occur in the color $[{\bf \bar{3}}]_a^c$
channel, the Cooper pair wave function is, as required, overall antisymmetric.
The condensate is a 3-vector in space which points in
the direction of the spin of the Cooper pair. It has 
the form \cite{bailinlove,pr2,schafer,schmitt}
\begin{equation}
\Phi_{ij}^a = \epsilon_{ijk} \, \Phi^a_k \,\, ,
\end{equation}
where $a = x,y,z$ denotes the spatial component of the spin vector.
Condensation breaks the local color $[SU(3)_c]$ symmetry and the
global $SO(3)_J$ spin symmetry. This is similar to superfluid helium-3
where condensation breaks $SO(3)_S \times SO(3)_L$ (see discussion
above). While many different phases arise, let us just mention two
which are quite similar to the ones discussed in the context
of three and two quark flavors, the so-called 
``{\bf C}olor-{\bf S}pin-{\bf L}ocked'' or CSL phase, where the
order parameter assumes the form
\begin{equation} \label{CSL}
\Phi^a_k =  \delta^a_k \, \Phi\,\, ,
\end{equation}
and the so-called polar phase, where
\begin{equation}
\Phi^a_k =  \delta_{k3}\, \delta^{az} \, \Phi \,\, .
\end{equation}
In the CSL phase, the order parameter~(\ref{CSL}) is
strikingly similar to the one in the CFL phase, cf.~Eq.~(\ref{CFL}).
All quark colors participate in the formation of Cooper pairs.
Also the symmetry breaking pattern is similar,
$[SU(3)_c] \times SO(3)_J \times U(1)_V \times
[U(1)_{\rm em}] \rightarrow SO(3)_{c+J}$.
The main difference is that now there is no rotated electromagnetism
$[\tilde{U}(1)]$, cf.~Sec.~\ref{gluephoton}.
Consequently, all eight gluons {\em and\/} the photon become massive 
\cite{schmitt2}.
In the polar phase, the order parameter resembles that of
Eq.~(\ref{3SC}). Neglecting electromagnetic interactions for the moment,
the symmetry breaking pattern is
$[SU(3)_c] \times SO(3)_J \times U(1)_V  
\rightarrow [SU(2)_c] \times SO(2)_J \times \tilde{U}(1)$.
Like in the 2SC phase, the residual $[SU(2)_c]$ color symmetry is
a local symmetry, and there are three massless and five massive gluons.
The breaking of the rotational $SO(3)_J$ symmetry to $SO(2)_J$ 
also leads to two real Goldstone bosons. Baryon number is not broken,
but merely rotated. Including the $[U(1)_{\rm em}]$ of electromagnetism
there is a small subtlety which is explained in more detail in 
Sec.~\ref{gluephoton}:
If there is only a single flavor present, or if all flavors 
carry the same electric charge, a rotated electromagnetic
$[\tilde{U}(1)]$ symmetry exists. If there are at least two flavors
which differ in charge, the $[U(1)_{\rm em}]$ symmetry is broken.
Table~\ref{table1} summarizes the results of this section for the
2SC, the CFL, the CSL, and the polar phase of color-superconducting quark
matter.
 
\begin{table}
\begin{center}
\begin{minipage}[t]{16.5 cm}
\caption{Color-superconducting phases in dense quark matter.
For the polar phase, there is an additional $[\tilde{U}(1)]$
symmetry, if all flavors in the system carry the same electric
charge.\vspace*{5mm}}
\label{table1}
\end{minipage}
\begin{tabular}{|l|c|c|c|}
\hline
      &                  &                    &                    \\[-2mm]
phase &  condensate      &    order parameter &  residual symmetry \\
      &                  &                    &                    \\[-2mm]
\hline
      &                  &                    &                    \\[-2mm]
2SC   &  $\Phi_{ij}^{fg} = \epsilon_{ijk}\, \epsilon^{fg} \,\Phi_k $ 
                         &  $\Phi_k = \delta_{k3}\, \Phi$ 
                                              & $[SU(2)_c] \times
           \tilde{U}(1) \times [\tilde{U}(1)] $                \\
      &                  &                    &                \\[-2mm]\hline
      &                  &                    &                    \\[-2mm]
CFL   &  $\Phi_{ij}^{fg} = \epsilon_{ijk}\, \epsilon^{fgh} \,\Phi_k^h $ 
                         &  $\Phi_k^h = \delta_{k}^h\, \Phi$ 
                                              & $SU(3)_{c+V} \times
                               [\tilde{U}(1)] $                \\
      &                  &                    &                \\[-2mm]\hline
      &                  &                    &                    \\[-2mm]
CSL   &  $\Phi_{ij}^{a} = \epsilon_{ijk} \,\Phi_k^a $ 
                         &  $\Phi_k^a = \delta_{k}^a\, \Phi$ 
                                              & $SO(3)_{c+J}$      \\
      &                  &                    &                \\[-2mm]\hline
      &                  &                    &                    \\[-2mm]
polar &  $\Phi_{ij}^{a} = \epsilon_{ijk}\,\Phi_k^a $ 
                         &  $\Phi_k^a = \delta_{k3}\, \delta^{az}\, \Phi$ 
                                              & $[SU(2)_c] \times
                               \tilde{U}(1)  $                \\
      &                  &                    &                \\[-2mm]\hline
\end{tabular}
\end{center}
\end{table}     

\subsubsection{\it Color-superconducting phases in the 
nuclear matter phase diagram} \label{cscphases}

How does color-superconductivity affect the phase diagram of
nuclear matter? Let us first assume that the temperature
is sufficiently small to favor a color-superconducting 
over the normal-conducting state.
As long as $\mu \gg m_s$, the CFL state is likely
to be the ground state of quark matter.
Since one has (approximately) equal numbers of up, down,
and strange quarks of colors red, green, and blue, the system is
(approximately) neutral with respect to color and electric charge.
However, when one extrapolates down to smaller quark chemical
potentials, say of the order of $\mu \sim 500$ MeV,
the strange quark mass is no longer negligibly small and causes,
for a given $\mu$, a mismatch in the Fermi surfaces 
between non-strange and strange quarks \cite{unlock}. 
In general, a nonzero strange quark mass reduces the number of
strange quarks as compared to the massless species. This, in turn, leads
to nonzero electric and color charge in the system.
Consequently, one is forced to introduce chemical potentials for electric and
color charge, which have to be tuned to again ensure overall electric and
color neutrality. The chemical potential for a quark species of
color $i$ and flavor $f$ thus reads
\begin{equation}
\mu^f_i = \mu - q^f \, \mu_e + t^3_i \, \mu_3 + t^8_i\, \mu_8
\,\, ,
\end{equation}
where $q^f$ is the electric charge of flavor $f$ ($q^u = 2/3$,
$q^{d,s} = - 1/3$), $\mu_e$ is the electron chemical potential,
$t^3_i$ and $t^8_i$ are the color charges associated with the third 
and eighth generator of $[SU(3)_c]$, respectively ($t^3_r = 1/2$,
$t^3_g = - 1/2$, $t^3_b = 0$, $t^8_{r,g} = 1/(2 \sqrt{3})$,
$t^8_b = - 1/\sqrt{3}$), and $\mu_3$, $\mu_8$ are the associated color
chemical potentials. (One could also introduce individual chemical potentials
for red, green, and blue quarks, but these can be written as
linear combinations of $\mu,\, \mu_3$, and $\mu_8$.)
The mismatch in the Fermi surfaces of different quark species
forming Cooper pairs is then
\begin{equation} \label{mismatch}
\delta {k_F}_{ij}^{fg} \equiv  {k_F}_i^f - {k_F}_j^g 
\simeq - (q^f - q^g) \, \mu_e + (t^3_i - t^3_j) \, \mu_3
+ (t^8_i - t^8_j) \, \mu_8 - \frac{m_f^2}{2 \mu_i^f} + \frac{m_g^2}{2 \mu_j^g}
\,\, ,
\end{equation}
where only the first correction in $m_f/\mu_i^f \ll 1$ was taken into account.
Equation~(\ref{mismatch}) shows that the mismatch in the Fermi surface
between different quark species is proportional to the electric
and color chemical potentials, as well as their mass difference.

The formation of Cooper pairs occurs at the Fermi surface. Typically,
a Cooper pair consists of
fermions with momenta which are in magnitude close to the Fermi momentum, 
but which have opposite directions, such that the total momentum
of the Cooper pair is zero. However,
when the mismatch $\delta {k_F}_{ij}^{fg}$ increases, it
becomes increasingly more difficult to form such pairs with zero
total momentum. For the
species with the smaller Fermi surface one may take a fermion
right at its Fermi surfaces, but in order to match the
momentum, one has to go deeper into the Fermi sea of the other
species. Pictorally speaking, 
forming a Cooper pair becomes energetically disfavored
once the cost of ``diving'' into the Fermi sea to find such a matching 
fermion is higher than the gain in condensation energy by forming
a Cooper pair. Whether this condition is fulfilled depends
on the magnitude of the color-superconducting gap at
the Fermi surface, $\phi_0$, compared to the
mismatch in the Fermi surfaces, $\delta {k_F}_{ij}^{fg}$. As
long as $\phi_0 \gg \delta {k_F}_{ij}^{fg}$, the Cooper-paired
state remains the ground state of the system. However,
when $\delta {k_F}_{ij}^{fg}$ becomes of the order
of $\phi_0$, or even considerably exceeds it,
the Cooper-paired state becomes energetically disfavored
as compared to normal-conducting state \cite{bedaque}.

It was recently realized, however, that instead of a transition to 
the normal-conducting state many other possibilities can be envisioned. 
For instance, imagine being in the CFL state and for the moment neglect
$\mu_e, \, \mu_3,$ and $\mu_8$ in Eq.~(\ref{mismatch}).
Then, the CFL state will become energetically disfavored when
$m_s^2/2 \mu$ exceeds $\phi_0$ \cite{unlock}. Nevertheless, quark
matter will not simply become normal-conducting, because there is
nothing to prevent the up and down quarks to form a 2SC state.
Of course, one cannot simply discard $\mu_e,\, \mu_3$, and $\mu_8$
from the consideration.
Taking these chemical potentials into account to ensure 
overall neutrality with respect
to color and electric charges, the 2SC state may become unstable
with respect to the formation of a gapless superconductor 
\cite{shovkovyhuang}, a crystalline color superconductor \cite{LOFF}, or
some other state with an even more exotic pairing scenario
\cite{muether,liuwilczek}. 

However, also a more conventional
pairing scenario is conceivable \cite{schmitt2}: the dominant terms in the
mismatch~(\ref{mismatch}) are the ones $\sim \mu_e$ and
the mass difference (the color chemical potentials $\mu_3$ and $\mu_8$
are parametrically of order $\phi_0^2/\mu \ll \phi_0$). Consequently, 
instead of realizing one of the more exotic pairing scenarios, it could be
energetically favorable to simply pair quarks with
the same charge and the same mass, i.e., 
of the same flavor. As discussed above, these Cooper pairs
must have spin one. Although spin-one gaps are
orders of magnitude smaller than spin-zero gaps 
\cite{schafer,schmitt,rockefeller}, the gain in condensation energy
$\Delta E_{\rm cond.}$
is parametrically larger than for some of the aforementioned exotic pairing
scenarios, for instance $\Delta E_{\rm cond.} \sim \mu^2 \, \phi_0^2$ 
for spin-one pairing vs.\ $\Delta E_{\rm cond.} \sim \mu^2 \phi_0^2
(\phi_{\rm LOFF}/\phi_0)^4$ for the crystalline color 
superconductor \cite{LOFF}.
(Here, $\phi_{\rm LOFF}$ is the value of the gap in the LOFF phase,
while $\phi_0$ is the gap in a superconductor with
equal Fermi surfaces for the particle species forming Cooper pairs.)
Whether, and if yes, which of 
these pairing scenarios are realized in nature, can only be decided
by a quantitative comparison of the pressure in the various cases. 
This has not been done so far and to draw definite conclusions
about the structure of the phase diagram of
nuclear matter at small temperatures and
chemical potentials of the order of $\mu \sim 500$ MeV
appears to be premature at this point.

On the other hand, it is far simpler to decide what happens
to a particular color-superconducting phase
when one increases the temperature at a given chemical potential.
Like in any other superconducting system,
thermal motion will break up quark Cooper pairs. In BCS theory,
the transition between superconducting and normal-conducting phases
is usually of second order and occurs at a temperature $T_c^{\rm BCS}$
proportional to the size of the superconducting gap
parameter $\phi_0$,
\begin{equation} \label{BCS}
T_c^{\rm BCS} = \frac{e^\gamma}{\pi}\, \phi_0 \simeq 0.567\, \phi_0\,\, ,
\end{equation}
where $\gamma \simeq 0.577$ is the Euler-Mascheroni constant.
At least in the mean-field approximation, in all color-superconducting
phases studied so far, $T_c$ either rigorously
obeys this relation or differs only by a factor of order one
from it, for details see Sec.~\ref{gapequation} and Ref.~\cite{schmitt}. 
The value of the color-superconducting
gap parameter $\phi_0$ is therefore of great importance in order to
locate the transition line between the normal and the color-superconducting 
quark matter phases in the nuclear matter phase
diagram. In Sec.~\ref{CSC} it will be discussed how to compute this gap 
parameter. Here it suffices to know that an extrapolation
of the result of solving a gap equation in weak coupling QCD
down to moderate densities suggests gap parameters of the order
of 10 MeV for pairing in the spin-zero channel.
NJL model calculations suggest somewhat larger
values around 100 MeV. With Eq.~(\ref{BCS}), this would lead
to transition temperatures of the order of 6 to 60 MeV.
In the spin-one channel, the gaps and critical temperatures
are typically smaller by two to three orders of magnitude 
\cite{schafer,schmitt,rockefeller}.


\section{Lattice QCD} \label{LattQCD}

\subsection{\it Basic Concepts} \label{basicslattice}

The most fundamental approach to compute thermodynamic properties of
strongly interacting matter and, in particular, its equation of state,
are lattice QCD calculations \cite{creutz}. 
In these calculations, one directly computes the grand partition
function~(\ref{Z}) on a discretized space-time lattice,
$V \times 1/T = (a_\sigma\, N_\sigma)^3 \, a_\tau \, N_ \tau$, where
$a_\sigma \equiv L/N_\sigma$ is the lattice spacing in spatial direction, 
$a_\tau \equiv 1/(N_\tau \, T)$ is
the lattice spacing in Euclidean time (i.e.~temperature) 
direction, and $N_\sigma$ and $N_\tau$ are the number of lattice points
in spatial and temporal direction, respectively.
Any space-time point in $V \times 1/T$ is then parametrized as
$x^\mu \equiv (\tau, {\bf x}) = (a_\tau \, l,\, a_\sigma \, i,\, 
a_\sigma \, j,\, a_\sigma \, k)$, with $0 \leq l \leq N_\tau$, 
and $0 \leq i,j,k \leq N_\sigma$. Commonly, one uses symmetric
lattices, where $a_\sigma = a_\tau \equiv a$. A space-time point
on the lattice, a lattice {\em site}, 
is then uniquely determined by the 4-vector $n^\mu = (l,{\bf n})$,
${\bf n} = (i,j,k)$.
Quantities with the dimension of energy are measured in 
units of the inverse lattice spacing $a^{-1}$,
and different lattices are simply characterized by their extension
$N_\sigma^3 \times N_\tau$.
The smallest length scale on a lattice is the lattice spacing $a$, 
corresponding to a maximum momentum scale $\Lambda_{\rm UV} \sim a^{-1}$.
This scale serves as ultraviolet cut-off which regulates the ultraviolet
divergences commonly appearing in quantum field theories.
The largest length scale on the lattice is the lattice extension
$a N_\sigma$. It determines the minimum momentum scale $\Lambda_{\rm IR}
\sim (a N_\sigma)^{-1}$. 

The next step is to define the QCD action $S \equiv \int_X {\cal L}$, 
with ${\cal L}$ given by 
Eq.~(\ref{LQCD}), on the discretized space-time lattice. As a first
guess, one could replace all derivatives with finite differences
between lattice points. For reasons explained below, 
this naive prescription is, however,
not particularly suitable, neither for the gauge field (gluon)
nor the matter (quark) part of the action. 
To find an alternative, note that the only condition a discretized
version of the QCD action has to fulfill is to
reproduce the continuum action in the limit $a \rightarrow 0$.
The choice of a discretized QCD action is therefore not unique.
This apparent shortcoming can, however, be turned into an advantage by 
choosing a form of the action which reduces or even completely eliminates
discretization errors (so-called {\em improved\/} or
{\em perfect\/} actions, respectively).

To find a suitable discretized version of the gauge field part of the action,
one first observes \cite{wilson} that, on a finite-size lattice, the 
gauge fixing term ${\cal L}_{\rm gauge}$ in Eq.~(\ref{LQCD}) is no 
longer necessary, as the integration over gauge fields becomes convergent.
Nevertheless, a naive discretization of the gauge field part of the
action is still not gauge-invariant, and will remain so even when
taking the continuum limit $a \rightarrow 0$.
It is therefore advantageous to formulate the gauge field part of the
action in a gauge-invariant form. A suitable choice was proposed
by Wilson \cite{wilson},
\begin{equation} \label{SA}
S_A =  \sum_n \sum_{0 \leq \mu < \nu \leq 3} 
\left[ 1 - \frac{1}{N_c}\, 
{\rm Re} {\rm Tr} \left( U_{n,\mu} \, U_{n+\hat{\mu},\nu} \,
U^\dagger_{n+\hat{\nu},\mu} \, U^\dagger_{n,\nu} \right) \right]\,\, .
\end{equation}
The sum over $n$ runs over all lattice sites $n^\mu$ and
the {\em link\/} variable $U_{n, \mu}$ is defined as
\begin{equation}
U_{n,\mu} = {\rm P} \, \exp \left[ ig \int_{x}^{x+ \hat{\mu}a}
{\rm d} y^\sigma \, A_\sigma^a(y) \, T^a \right]\,\, , \;\;\;\;
x^\mu \equiv a\, n^\mu \,\,.
\end{equation}
The link variable describes the parallel transport of the gauge field between 
two neighbouring lattice sites $n^\lambda$ and $n^\lambda
+ \hat{\mu}^\lambda $, where $\hat{\mu}^{\lambda} \equiv \delta^{\mu \lambda}$
is the 4-dimensional lattice unit vector pointing in $\mu$-direction.

Visualizing the product of the four link variables on the 
right-hand side of Eq.~(\ref{SA}), one realizes that this product
transports the gauge field around an elementary
lattice plaquette; it is therefore also called 
the {\em plaquette\/} operator. The trace of the plaquette operator,
and thus also the Wilson action~(\ref{SA}), is gauge-invariant.
The Polyakov loop~(\ref{Polyakovloop}) is related to the link variables via
\begin{equation} \label{Polyakovlatt}
L({\bf x}) \equiv \frac{1}{N_c}\, {\rm Tr} \prod_{l=1}^{N_\tau}
U_{n,0} \,\, , \;\;\;\; n^{\mu} = (l,{\bf n})\,\, ,
\;\;\;\; {\bf x} \equiv a\, {\bf n}\,\, .
\end{equation}
Expanding the Wilson action~(\ref{SA}) for small lattice spacing $a$, one
obtains the continuum limit
\begin{equation}
- \beta \, S_A \simeq   \int_X \left( -\frac{1}{4}\, 
F_{\mu \nu}^a F^{\mu \nu}_a \right) + O(a^2)\,\,,
\end{equation}
where $\beta \equiv 2N_c/g^2 \equiv 6/g^2$ 
is the so-called {\em bare coupling}.
The correction terms to the continuum result
are of order $O(a^2)$. The construction principle
behind an improved action~\cite{improvedaction} 
is to add further terms to $S_A$ in Eq.~(\ref{SA})
in order to eliminate corrections of order $O(a^2)$.
The improved action then reproduces the continuum limit
up to corrections of order $O(a^4)$. Repeating this procedure,
one can systematically eliminate discretization errors up to a
given power of $a^2$. Extending this procedure in order to
eliminate {\em all\/} corrections leads to so-called
{\em perfect\/} actions \cite{perfectaction}.

The naive discretization of the fermionic part of the QCD action
is not particularly suitable because of the so-called
{\em doubling problem\/} for massless fermions on the lattice \cite{creutz}.
Fermion doubler states originate from
the periodicity of the fermion dispersion relation within
the Brillouin zone. One obtains one extra
doubler state per space-time dimension, such that 
there are in total $2^4 = 16$ instead
of a single fermion species. One way out is to break chiral
symmetry explicitly by introducing a mass term. This leads
to the so-called {\em Wilson\/} fermion prescription \cite{wilson}. 
Wilson fermions eliminate the
doubler states completely, but they have the disadvantage that 
one can in principle no longer study the restoration
of chiral symmetry at the QCD transition.
Another possibility is to distribute components of the fermion Dirac
spinor over several lattice sites. These so-called {\em staggered\/}
or {\em Kogut-Susskind\/} fermions \cite{kogutsusskind} do not 
completely solve the fermion-doubling problem:
the number of doubler states is merely reduced to four. 
However, the solution to this problem is to
interpret the doubler fermions as different flavor states.
Hence, the standard staggered fermion action is interpreted as
describing QCD with $N_f=4$ flavors. 
The advantage of the staggered fermion prescription is that
it preserves a subgroup of the original chiral symmetry.
The chiral condensate is thus an order parameter for chiral
symmetry restoration at the QCD transition.
Other attempts have been made to 
solve the fermion-doubling problem, while at the same time
improving (or even preserving) the chiral symmetry of the lattice action.
To name a few, there are the so-called overlap \cite{overlap}, 
domain-wall \cite{domain}, fixed-point \cite{fixedpoint}
or chirally improved \cite{chirallyimproved} fermions.

Let us take a closer look into the staggered fermion
prescription, where the fermionic part of the QCD action reads
\begin{equation}
S_F^{KS} = \sum_{n,m} \bar{\psi}_n \, M_{n,m}^{KS} \, \psi_m \,\, ,
\end{equation}
with the inverse staggered fermion propagator
\begin{eqnarray} 
M_{n,m}^{KS} (\tilde{m},\tilde{\mu},U) & = & \frac{1}{2} \sum_{\mu = 1}^{3}
 (-1)^{n^0+\ldots+n^{\mu-1}} \left( \delta_{n+\hat{\mu},m}\, U_{n,\mu}
- \delta_{n,m+\hat{\mu}} \, U^\dagger_{m,\mu} \right) \nonumber \\
& +  &  \frac{1}{2} \left( \delta_{n+\hat{0},m}\, U_{n,0} \, e^{\tilde{\mu}}
- \delta_{n,m+\hat{0}}\, U_{m,0}^\dagger \, e^{-\tilde{\mu} } \right) 
+ \delta_{n,m} \, \tilde{m} \,\, . \label{KS}
\end{eqnarray}
Here, the fermion mass (in units of the inverse lattice
spacing) is denoted as $\tilde{m} \equiv a m$. 
This notation prevents confusion of the fermion mass
with the lattice site vector
$m \equiv m^\mu$. The chemical potential
(in units of the inverse lattice spacing) is $\tilde{\mu} \equiv a \mu$.
As shown in Ref.~\cite{karschhasenfratz}, the
correct prescription to introduce the chemical potential 
in the discretized fermion action is as indicated in Eq.~(\ref{KS}),
i.e., in exponential form on a temporal link.

The fermionic part of the QCD action is bilinear in the
Grassmann fields $\bar{\psi}$ and $\psi$. The fermion fields can thus be
integrated out exactly. The result for the QCD partition function
(\ref{Z}) on a discretized, 4-dimensional $N_\sigma^3 \times N_\tau$
lattice is
\begin{equation} \label{Zlatt}
{\cal Z}(N_\sigma, N_\tau,\beta,\tilde{m},\tilde{\mu})
= \int \prod_{n,\mu} {\rm d} U_{n,\mu} \, 
\left[ {\rm det} M^{KS} (\tilde{m},\tilde{\mu},U)
\right]^{N_f/4} \, e^{- \beta\, S_A}\,\, .
\end{equation}
The integration is over all link variables $U_{n,\mu}$.
The power $N_f/4$ of the fermion determinant takes into account
that, in the continuum limit, the standard
staggered fermion prescription leads to $N_f = 4$ fermion species. In 
order to obtain results with less than $N_f=4$ flavors, one has
to take the appropriate root in Eq.~(\ref{Zlatt}). In this way, one
can also obtain results for fermions with different masses.
For instance, in order to compute the partition function
for two light (say, up and down) and one heavy (say, strange) quark flavor
(also called the ``$2+1$'' flavor scenario) 
one replaces the fermion determinant
in Eq.~(\ref{Zlatt}) by the product
$[{\rm det} M^{KS}(\tilde{m}_q,\tilde{\mu}_q,U) ]^{1/2}
[{\rm det} M^{KS}(\tilde{m}_s,\tilde{\mu}_s,U) ]^{1/4}$.

For vanishing quark chemical potential, $\tilde{\mu}=0$, 
the fermion determinant in Eq.~(\ref{Zlatt}) is real and positive, 
and standard Monte Carlo methods \cite{creutz} can be applied
to evaluate the integral over the link variables $U_{n,\mu}$.
However, for nonzero quark chemical potential,
the fermion determinant becomes complex. It is clear that the
partition function itself cannot have an imaginary part, thus
the imaginary part of the fermion determinant has to cancel
when integrating over $U_{n,m}$. However, for a particular 
configuration of the gauge field, or equivalently, the
link variables $U_{n,\mu}$ on the space-time lattice,
the real part of the fermion determinant is no longer strictly
positive. This so-called {\em sign problem\/} prevents the
application of standard Monte Carlo techniques to evaluate the
partition function. For this reason, most lattice QCD calculations
have been performed at zero quark chemical potential, with data
reaching an impressive level of quality.
Results for the QCD phase transition and the equation of state, i.e.,
the pressure as a function of temperature,
are presented in Secs.~\ref{translatt} and \ref{EOSlatt}, respectively.
Only recently, attempts have been made to compute the partition
function also for nonzero values of the quark chemical potential.
This will be discussed in Sec.~\ref{nonzeromu}.

Finally, let us note that,
in order to extract continuum physics from lattice calculations,
one has to extrapolate the results to the case of vanishing
lattice spacing, $a \rightarrow 0$. 
In order to change the
value of $a$, one has to change the value of bare coupling
$\beta=6/g^2$. Since QCD is an asymptotic theory,
the strong coupling constant at the momentum scale 
$a^{-1}$ vanishes as $a$ goes to zero, $g(a) \rightarrow 0$ 
for $a\rightarrow 0$. This, in turn, implies that $\beta(a) \rightarrow \infty$
as $a \rightarrow 0$.
Asymptotically, the relation between $a$ and $\beta$ is given by
the leading-order renormalization group result
\begin{equation} \label{renorm}
a \, \Lambda_L \simeq \left( \frac{6 b_0}{\beta} \right)^{-b_1/(2b_0^2)}
\exp \left( -\frac{\beta}{12 b_0} \right)\,\, , \;\;\;\;
b_0 = \frac{11-2N_f/3}{16 \pi^2}\,\, , \;\;\;\;
b_1 = \frac{102 - 38N_f/3}{(16 \pi^2)^2} \,\, ,
\end{equation}
where $\Lambda_L$ is the lattice scale parameter that can be 
unambiguously related to the scale parameter in other regularization
schemes, for instance $\Lambda_{\overline{MS}}$ in
the $\overline{MS}$ scheme. 

In principle, this allows one to
convert the value of a physical quantity, say the pressure,
which on the lattice is computed in units of $a^{-4}$, i.e., as 
$\tilde{p} \equiv a^4 p $,
into physical units, i.e., MeV$^{4}$. However, as Eq.~(\ref{renorm})
is strictly valid only for asymptotically small values of $a$,
in practice one uses a different prescription. Consider
a physical quantity, for instance a hadronic 
mass $m_H$, which is well-known in the continuum. Compute
this quantity on the lattice, where its value is given in units of the 
inverse lattice spacing, $\tilde{m}_H = a m_H$. Then, any other 
quantity with the dimension of energy 
can be determined in units of $m_H$, say the temperature, which is 
$T/m_H = (aN_\tau)^{-1}/(\tilde{m}_H/a) \equiv 1/(\tilde{m}_H N_\tau)$.

When decreasing the lattice spacing $a$ (by increasing the value
of the bare coupling $\beta$), the temperature $T=1/(a N_\tau)$
increases, if one keeps the number of lattice points in temporal direction 
$N_\tau$ fixed.
(Simultaneously, for a fixed number of lattice points in spatial direction
$N_\sigma$, the volume $V=(a N_\sigma)^3$ decreases.)
Therefore, in order to determine the temperature dependence
of a quantity, one simply has to compute it on a lattice
with a fixed number of temporal points $N_\tau$, but for different
values of $a$, resp.\ $\beta$. Thus, one often finds
lattice data presented as a function of $\beta$ rather than as
a function of $T$.
Both presentations are equivalent, but note
that the temperature {\em increases\/} with the bare coupling $\beta$.
One should therefore never confuse the bare coupling $\beta$ with
the quantity $\beta \equiv (k_B T)^{-1}$ from
thermodynamics and statistical mechanics, which {\em decreases\/}
with temperature. (The way $\beta$ appears
in Eq.~(\ref{Zlatt}) certainly does not help to avoid this mistake.)

When extrapolating lattice results to the continuum limit $a \rightarrow 0$,
one does not simultaneously want to increase the temperature $T$ or 
decrease the volume $V$ of the system. Rather,
one has to ensure that these quantities are
kept fixed. In other words, the continuum limit $a \rightarrow 0$ 
is obtained by simultaneously
{\em increasing\/} the number of lattice points in space and time direction,
$N_\sigma,\, N_\tau \rightarrow \infty$, such that $a N_\sigma = V^{1/3}$
and $a N_\tau = 1/T$ are constant. This is obviously 
quite costly numerically. There is, however, also another 
problem of numerical nature with this limit. Consider, for
instance, a lattice computation of the pressure, which yields values 
for the quantity $\tilde{p} \equiv a^4 p \equiv (p / T^4) N_\tau^{-4}$. 
A given value for the physical temperature corresponds to
some value for the physical pressure, such that $p/T^4$ assumes
a certain value. Consequently, as $N_\tau \rightarrow
\infty$ the {\em numerical\/} value for $\tilde{p}$ on the lattice
rapidly decreases as $N_\tau^{-4}$ when $a \rightarrow 0$.
Since lattice QCD calculations are subject to statistical errors,
it therefore becomes increasingly more difficult to extract the 
physically relevant quantity from the statistical noise.
It is thus important to use improved actions (see discussion above), which
reduce the discretization errors and allow one to perform
calculations for moderate values of $N_\tau$ where $\tilde{p}$ is
still significantly larger than the statistical noise.

Finally, not only is one interested in the continuum limit for
a {\em finite\/} volume $V$, but one would like to extrapolate
to the thermodynamic limit $V = (a N_\sigma)^3 \rightarrow \infty$
as well. At a given nonzero temperature, however, $1/T = a N_\tau$
remains finite (in fact, it decreases as $T$ increases). Therefore,
simulations at nonzero temperature, which aim towards the thermodynamic
limit, require $N_\sigma \gg N_\tau$, which represents
another numerically expensive condition. Nowadays,
typical ``hot'' lattices have space-like extensions $N_\sigma \sim 16 - 32$
while the time-like extension is $N_\tau \sim 4 - 8$.
The only situation where one also has to have a large extension of the 
lattice in the time direction is the zero-temperature case, $T=0$.
``Cold'' lattices typically have $N_\sigma = N_\tau \sim 16 - 32$.

How close to the thermodynamic limit are present-day lattice QCD
calculations? Suppose one is doing a simulation
at the critical bare coupling, i.e., where the QCD transition
occurs (see Sec.~\ref{translatt}). Physical values for the
transition temperature are of the order of $T_c \sim 150$ MeV,
cf.~Sec.~\ref{dynquarks}. Consequently, the lattice spacing at
the critical bare coupling $\beta_c$ is $a_c = 1/(T_c N_\tau)$.
For temporal lattice extensions of the order of $N_\tau \sim 4- 8$,
this corresponds to values $a_c \sim 0.15 - 0.3$ fm. For
typical spatial lattice extensions $N_\sigma \sim 16 - 32$ 
on a ``hot'' lattice, the physical volume is then
$V \sim (2.5 - 10)^3\,\mbox{fm}^3 \sim (15 - 1000)\, \mbox{fm}^3$. 
Is such a system
sufficiently close to the thermodynamic limit? The answer is not
necessarily ``no'', as this depends on how large the system is in
comparison to the size of its constituents. 
The latter can be estimated via their Compton wavelength
$\lambda_C = m^{-1}$. For nucleons, the Compton wavelength
is $\lambda_C \sim 0.2$ fm, so that many nucleons would comfortably fit
into the system. (Of course, this is an optimistic estimate: taking
the nuclear charge radius $r \sim 0.8$ fm instead of the Compton
wavelength drastically worsens the situation.) 
For a pion, $\lambda_C \sim 1.4$ fm, such that
the lattice volume for these light particles appears to be on the verge
of being too small (unless the pion becomes much heavier at the 
phase transition, cf.~Sec.~\ref{linearsigmamodels}). In any case,
not more than a few pions would fit into the physical volume, which 
certainly casts doubts on whether one is able to reach the thermodynamic limit
with present-day lattice sizes.

In the following, results from lattice calculations
at zero and nonzero quark chemical potential will be reviewed. For $\mu = 0$,
a wealth of data is available; for the purpose of this
introductory review, I only focus
on the QCD phase transition, the equation of state,
the heavy quark free energy, and mesonic spectral functions.
The case $\mu \neq 0$ has only recently received a fair amount
of attention. The main activity is still to find solutions of (or
ways around) the sign problem of the fermion determinant. 
For more details, see Refs.~\cite{schladming,laermannphilipsen}.

\subsection{\it The QCD phase transition} \label{translatt}

\begin{figure}[tb]
\begin{center}
\begin{minipage}[t]{8 cm}
\epsfig{bbllx=105,bblly=220,bburx=460,bbury=595,clip=,
file=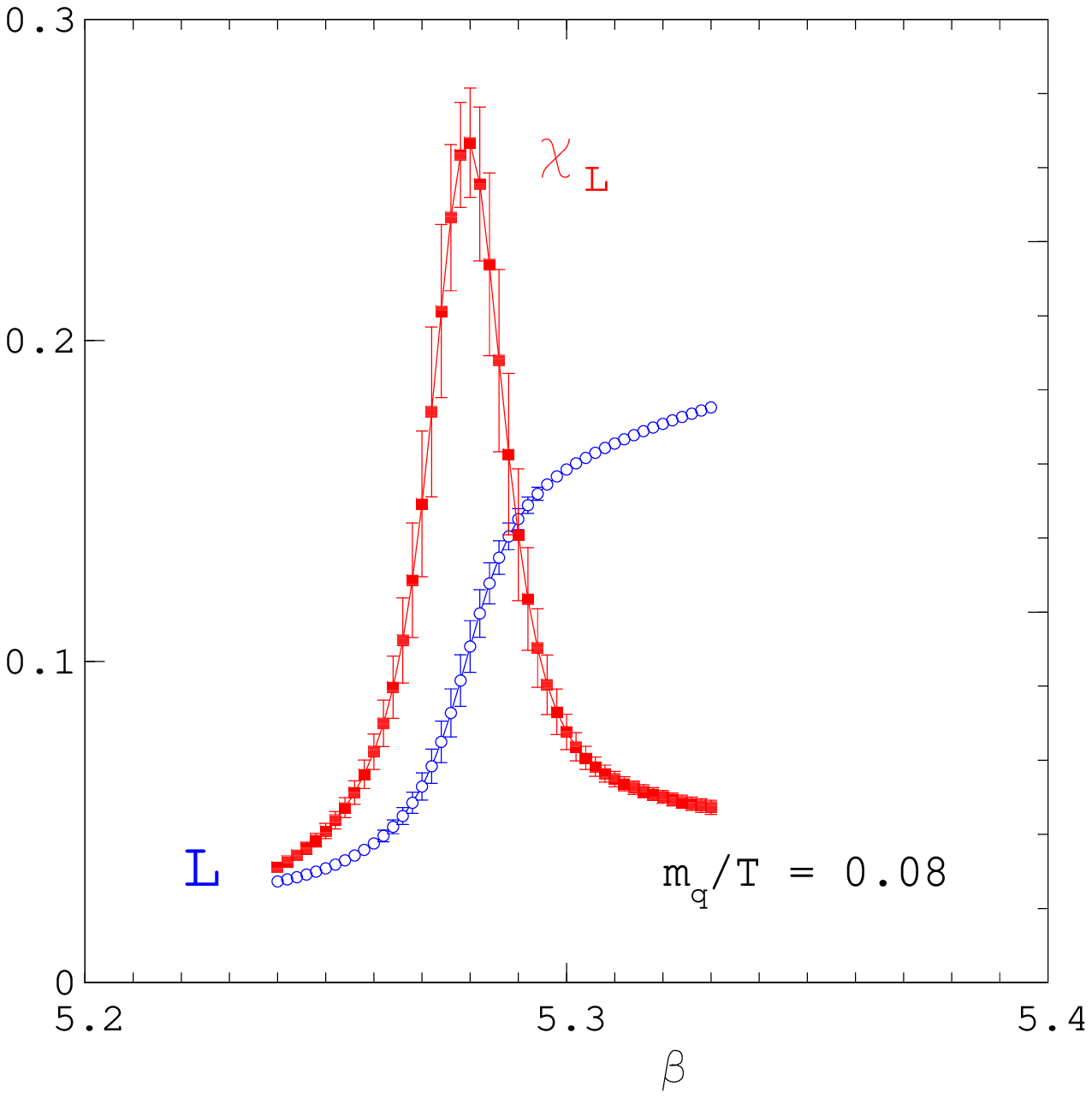,height=80mm}
\end{minipage}
\begin{minipage}[t]{8 cm}
\epsfig{bbllx=105,bblly=220,bburx=460,bbury=595,clip=,
file=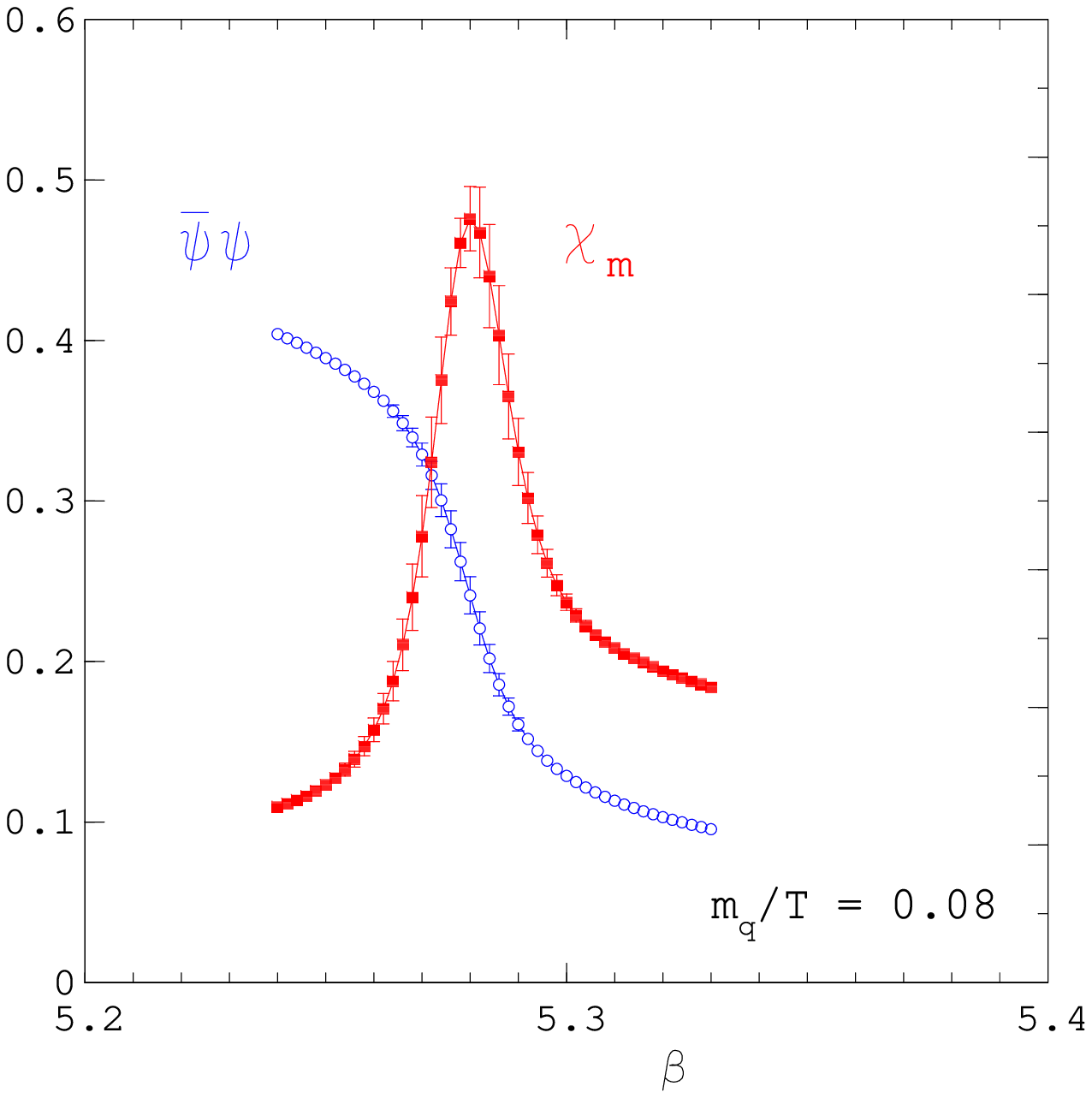,height=80mm}
\end{minipage}
\begin{minipage}[t]{16.5 cm}
\caption{Deconfinement and chiral symmetry restoration
in QCD with $N_f=2$ dynamical quark flavors.
(a) The expectation value of the Polyakov loop 
and the corresponding susceptibility as functions of the bare coupling.
(b) The chiral condensate and the corresponding susceptibility
as functions of the bare coupling. From 
Ref.~\cite{schladming}.\label{Ploop_and_Xcond}}
\end{minipage}
\end{center}
\end{figure}

As already discussed in Sec.~\ref{QHtransition}, lattice QCD calculations
have numerically established the existence of the quark-hadron transition.
Figure~\ref{Ploop_and_Xcond} (a) 
shows the expectation value of the Polyakov loop,
$\langle L({\bf x}) \rangle$, with $L({\bf x})$ as defined in
Eq.~(\ref{Polyakovlatt}), as a function of the bare coupling $\beta$
(i.e., as explained in Sec.~\ref{basicslattice}, as a function of 
temperature) for $N_f =2 $ quark flavors. 
For the pure gauge theory, i.e., for quark masses
$\tilde{m} \rightarrow \infty$, the Polyakov loop is
an order parameter for the deconfinement transition: it
changes its value from zero in the confined phase below $T_c$
to a nonzero value in the deconfined phase above
$T_c$, cf.~discussion in Sec.~\ref{puregauge}. 
However, the presence of dynamical quarks
in the calculation of Fig.~\ref{Ploop_and_Xcond} breaks the
$Z(3)$ symmetry of the pure gauge theory explicitly. Thus, 
the transition is no longer of first order,
but crossover. This is also observed in the data.

In Fig.~\ref{Ploop_and_Xcond} (b) the chiral condensate 
$\langle \bar{\psi}\psi \rangle$
is shown as a function of the bare coupling $\beta$. 
For vanishing quark masses, 
the chiral condensate serves as an order parameter for chiral symmetry
breaking: it is nonzero below and vanishes above $T_c$, cf.~discussion
in Sec.~\ref{dynquarks}.
Since the calculations of Fig.~\ref{Ploop_and_Xcond} have been done
for a nonzero quark mass, chiral symmetry is explicitly broken.
Consequently, the chiral transition is not of second order, as expected for
$N_f = 2$ flavors, but crossover, which is also seen in 
Fig.~\ref{Ploop_and_Xcond} (b).

Also shown in Figs.~\ref{Ploop_and_Xcond} (a) and (b) are the susceptibilities
corresponding to the Polyakov loop and the chiral condensate.
These are defined as
\begin{equation}
\chi_L \equiv N_\sigma^3 
\left( \langle L^2 \rangle - \langle L \rangle^2 \right) \,\, ,
\;\;\;\;
\chi_m \equiv \frac{ \partial}{\partial \tilde{m} } \,\langle \bar{\psi}\, 
\psi \rangle \,\, .
\end{equation}
These quantities have a maximum at the value of $\beta$ where the Polyakov 
loop and the chiral condensate change most rapidly. This value is 
the critical bare coupling $\beta_c$, which 
corresponds to the critical temperature $T_c$ for the QCD transition.
In this way, one can define a critical temperature, even if the
transition is not of first or second order, but only crossover.
The interesting observation one can make from 
Fig.~\ref{Ploop_and_Xcond} is that $\beta_c$
assumes the {\em same\/} value for 
the deconfinement transition as for the chiral symmetry restoration
transition. A possible explanation for this strong correlation
between deconfinement and chiral transition is provided by the
Polyakov-loop model of Ref.~\cite{pisarskidumitru}, see
also Sec.~\ref{Ploopmodel}.
Current results for the phase transition temperature in
the pure $[SU(3)]_c$ gauge theory, as well as in QCD with
different flavors, extrapolated to vanishing quark
mass (for details, see Ref.~\cite{peikert}) 
are summarized in Table~\ref{table2}.

\begin{table}
\begin{center}
\begin{minipage}[t]{16.5 cm}
\caption{The critical temperature $T_c$ for
QCD with different quark flavors (extrapolated to the chiral limit)
and the pure $[SU(3)_c]$
gauge theory ($N_f=0$). For the ``2+1'' case, $T_c$ is
close to the 2-flavor case. From Refs.~\cite{schladming,laermannphilipsen}.
\vspace*{5mm}}
\label{table2}
\end{minipage}
\begin{tabular}{|l|c|l|}
\hline
      &                  &                 \\[-2mm]
$N_f$ &  $T_c$           & Remarks          \\
      &                  &                 \\[-2mm]
\hline
      &                  &                 \\[-2mm]
2     &  $(171 \pm 4)$ MeV & Wilson fermions \\
      &                  &                 \\[-2mm]\hline
      &                  &                 \\[-2mm]
2     &  $(173 \pm 8)$ MeV & Kogut-Susskind fermions      \\
      &                  &                 \\[-2mm]\hline
      &                  &                 \\[-2mm]
3     &  $(154 \pm 8)$ MeV & Kogut-Susskind fermions    \\
      &                  &                 \\[-2mm]\hline
      &                  &                 \\[-2mm]
0     &  $(271 \pm 2)$ MeV &  pure gauge theory        \\
      &                  &                 \\[-2mm]\hline
\end{tabular}
\end{center}
\end{table}

\begin{figure}[tb]
\begin{center}
\begin{minipage}[t]{9 cm}
\epsfig{file=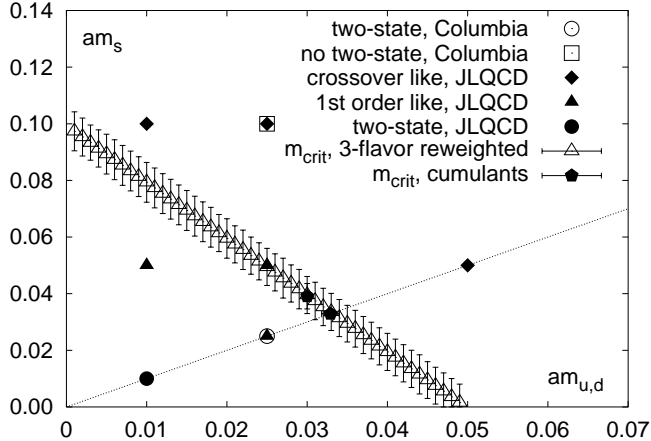,scale=0.7}
\end{minipage}
\begin{minipage}[t]{16.5 cm}
\caption{The quark-mass diagram as computed in lattice QCD.
Below the line of open triangles, the transition is of first order,
above it is crossover. The data points labelled
``Columbia'' are from Ref.~\cite{columbia}, the ones
labelled ``JLQCD'' are from Ref.~\cite{JLQCD}. The
other points are from Refs.~\cite{cschmidt,cschmidt2}.\label{quarkmasslatt}}
\end{minipage}
\end{center}
\end{figure}

Lattice QCD calculations have also begun to explore the quark-mass
diagram discussed in Sec.~\ref{quarkmassdiagram} in order to decide
the question about the order of the QCD transition. The present
knowledge is summarized in Fig.~\ref{quarkmasslatt}.
The open triangles are results from Ref.~\cite{cschmidt} and
correspond to the line of second-order transitions separating
the first-order from the crossover region in Fig.~\ref{quark-mass}.
The other data points 
confirm that the transition is of first order below the second-order
line and crossover above. It is somewhat difficult to locate
the physical point on this diagram.
Naively, one would think that it suffices to 
determine the lattice spacing $a$ in physical units,
after which one finds the physical point
in lattice units via multiplying the physical quark masses
by this value of $a$, 
$\tilde{m}_{q,s}^{\rm phys.} \equiv a\,m_{q,s}^{\rm phys.}$.
This deceptively simple method does not work in practice, because
the physical mass in lattice units $\tilde{m}_{q,s}^{\rm phys.}$ 
also receives contributions
from renormalization, which violate this simple relationship.
Present estimates seem to indicate, however,
that the physical point is deep in the crossover region \cite{privcomm}.

\subsection{\it Equation of state} \label{EOSlatt}

The equation of state is determined by the pressure $p(T,\mu)$ as a
function of temperature $T$ and chemical potential $\mu$.
Hence, according to Eq.~(\ref{pressure}) one has
to compute $(T/V) \, \ln {\cal Z}$. From the pressure, other thermodynamic
quantities can be derived via differentiation, cf.~Eq.~(\ref{sn}), 
and the fundamental relation of thermodynamics,
$\epsilon = Ts + \mu n - p$.
For any quantum field theory in the continuum as well as on the lattice, 
the calculation of the absolute value of $(T/V) \ln {\cal Z}$
is plagued by ultraviolet divergences~\cite{kapustaFTFT}.
These arise from vacuum fluctuations and have to be subtracted
in order to obtain a finite value for $p(T,\mu)$.
The simplest way to achieve this is to subtract the value
of $(T/V) \ln {\cal Z}$ in the vacuum, i.e., at $T= \mu =0$,
\begin{equation} \label{pnorm}
p(T,\mu) = \frac{T}{V}\,  \ln {\cal Z} - \left(\frac{T}{V} \, \ln{\cal Z}
\right)_{T=\mu=0}\,\, .
\end{equation}
In this way, the value of the pressure in the vacuum
is normalized to zero, $p(0,0) \equiv 0$.

A direct computation of the pressure using this formula
is still cumbersome, because it requires the calculation
of the absolute values of $ \ln {\cal Z}(T,V,\mu)$ and 
$ \ln {\cal Z}(0,V,0)$.
which then have to be subtracted from each other.
On the lattice, it is much simpler to compute average values of
quantities. Therefore, one uses the following method to determine
the pressure. First, note that 
\begin{equation} \label{pT4}
\frac{p}{T^4} \equiv  \frac{1}{T^4} \left[ \frac{T}{V} \, \ln {\cal Z}
- \left( \frac{T}{V} \, \ln {\cal Z} \right)_{T=\mu=0} \right]
\equiv N_\tau^4 \left( 
\frac{\ln{\cal Z}(N_\sigma,N_\tau,\beta,\tilde{m},\tilde{\mu})}{
N_\sigma^3 N_\tau}
-  \frac{\ln{\cal Z}(N_\sigma,N_\sigma,\beta,\tilde{m},\tilde{\mu})}{
N_\sigma^4} \right)\,\,.
\end{equation}
The assumption underlying this identity is that one
can approximate the vacuum subtraction 
by the value of $(T/V) \ln {\cal Z}$ computed
on a ``cold'' lattice with $N_\tau \equiv N_\sigma$, but at the same
value of the bare coupling $\beta$ (i.e., with the
same lattice spacing $a$) as for the ``hot'' lattice 
(where $N_\tau \ll N_\sigma$),
$(T/V) \ln {\cal Z} |_{T=\mu=0} \equiv (a N_\sigma)^{-4} 
\ln {\cal Z}(N_\sigma,N_\sigma,\beta,\tilde{m},\tilde{\mu})$.
Now introduce the expectation value 
of the (dimensionless) Wilson action {\em density\/} 
$\langle \tilde{s}_A \rangle \equiv a^4 \, \frac{T}{V} \langle S_A \rangle
\equiv (N_\sigma^3 N_\tau)^{-1} \langle S_A \rangle$,
\begin{equation} \label{SAexpect}
\langle \tilde{s}_A \rangle \equiv - 
\frac{1}{N_\sigma^3\, N_\tau}\, \frac{\partial}{\partial \beta}
\ln {\cal Z} (N_\sigma, N_\tau,\beta,\tilde{m},\tilde{\mu})\,\, ,
\end{equation}
and its zero-temperature value $\langle \tilde{s}_A \rangle_0$,
which is computed on a ``cold'' lattice, i.e., by
setting $N_\tau \equiv N_\sigma$
on the right-hand side of Eq.~(\ref{SAexpect}).
The quantity $p/T^4$ in Eq.~(\ref{pT4}) can now be obtained through
an integration of Eq.~(\ref{SAexpect}) with respect to the bare
coupling $\beta$,
\begin{equation}
\frac{p}{T^4}-\frac{p_1}{T_1^4} = -
N_\tau^4 \int_{\beta_1}^\beta {\rm d}\beta' \, 
\left( \langle \tilde{s}_A \rangle -
\langle \tilde{s}_A \rangle_0 \right) \,\, .
\end{equation}
In order to determine the second term on the left-hand side, $p_1/T_1^4$, 
one would like to choose a
rather low value for the temperature $T_1$. For temperatures $T_1 \ll m_H$,
where $m_H$ is the lightest hadronic particle,
the pressure is exponentially small, $p_1/T_1^4 \sim \exp(-m_H/T_1)$. 
This argument holds to very good approximation in the pure
gauge theory, since the lightest glueball state has a mass
of order 1 GeV. It does not hold in full QCD in the chiral limit, where there
are $N_f^2-1$ massless Goldstone particles, 
cf.~discussion in Sec.~\ref{dynquarks}.
For lattice QCD calculations, however, chiral symmetry is always broken
by a nonvanishing dynamical fermion mass,
thus $m_H$ is always positive. To very good approximation one may 
therefore set $p_1/T_1^4 \equiv 0$. 
Once the pressure is known, other thermodynamic quantities can be
determined from thermodynamic identities, for more details see 
Ref.~\cite{schladming}.

\begin{figure}[tb]
\begin{center}
\begin{minipage}[t]{8 cm}
\epsfig{file=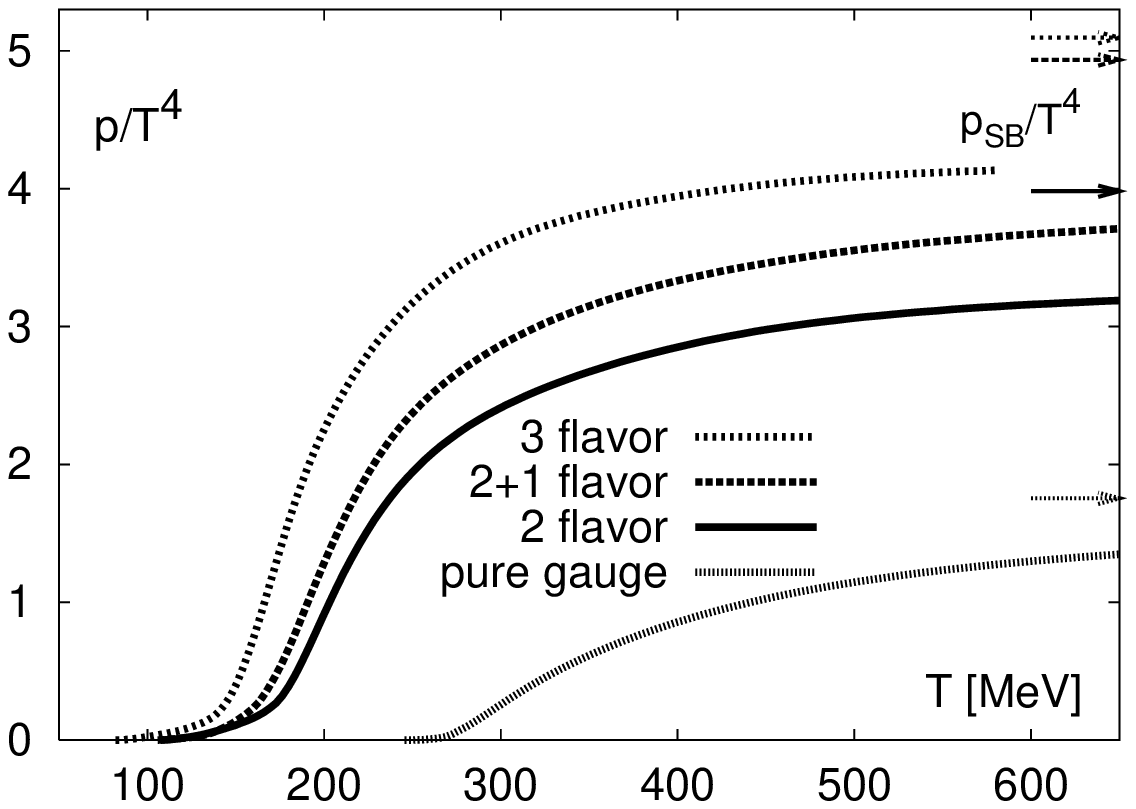,scale=0.6}
\end{minipage}
\begin{minipage}[t]{8 cm}
\epsfig{file=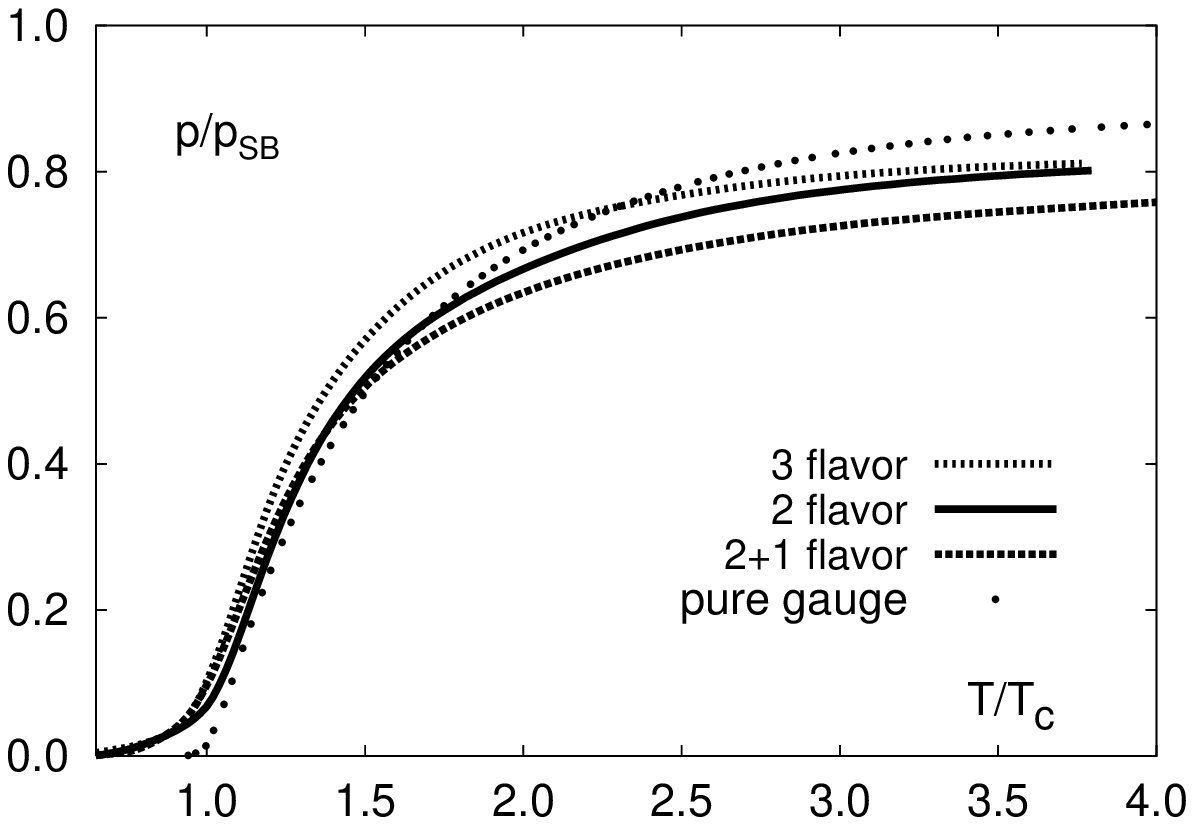,scale=0.6}
\end{minipage}
\begin{minipage}[t]{16.5 cm}
\caption{Left panel: Pressure (divided by $T^4$) as a function of temperature
for the pure gauge theory and for QCD with two and three light flavors, 
as well as with two light and a four times heavier quark flavor 
(curve labelled ``$2+1$''). 
Arrows denote the corresponding Stefan-Boltzmann values.
For the ``2+1'' case, the arrow is slightly below the three-flavor
case, due to the nonzero strange quark mass.
Right panel: Pressure normalized to its corresponding Stefan-Boltzmann
value as a function of temperature normalized to the corresponding
transition temperature $T_c$ for the four cases shown in the left panel.
From Ref.~\cite{schladming}.
\label{EOS}}
\end{minipage}
\end{center}
\end{figure}

As discussed in Sec.~\ref{basicslattice}, before one can
draw definite conclusions
about the thermodynamic properties of hot quark-gluon matter,
one has to extrapolate the lattice data to the
continuum limit (and hope that present-day lattices are sufficiently
large to be close to the thermodynamic limit). This has been done
in Fig.~\ref{EOS} which shows the pressure (normalized to $T^4$)
as a function of temperature (in physical units) for the case
of two light flavors, three light flavors, and the ``$2+1$'' case,
i.e., two light plus one heavy flavor~\cite{karsch}, in comparison
to the pressure for the pure $[SU(3)_c]$ gauge theory~\cite{boyd}.

One observes that the pressure is rather small at low temperatures.
This is to be expected, as the contributions from hadronic
resonances (or, in the pure gauge theory, from glueballs)
to the pressure are exponentially suppressed, 
$p_H/T^4 \sim \exp(-m_H/T)$ for a hadron (glueball) of mass $m_H$.
However, at the critical temperature $T_c$ for
the QCD transition (cf.~Table~\ref{table2}), the
pressure increases rapidly, approaching the so-called
Stefan-Boltzmann limit $p_{\rm SB}$ 
for a system of quarks and gluons as $T \rightarrow \infty$.
The Stefan-Boltzmann limit
is the pressure for an ideal (i.e., non-interacting) 
ultrarelativistic gas of particles.
For an ultrarelativistic gas at $\mu=0$, the temperature is the only
scale with the dimension of energy, consequently
$p_{\rm SB}/T^4 = const.$.
The value of this so-called Stefan-Boltzmann constant
only depends on the number of degrees of freedom
in the system. For an $[SU(N_c)]_c$ gauge theory with $N_f$ massless 
quark flavors one obtains
\begin{equation} \label{psb}
\frac{p_{\rm SB}}{T^4} 
= \left[ 2\, (N_c^2 -1)  + 2\, N_c\, N_f \, \frac{7}{4} \right] \,
\frac{\pi^2}{90}\,\,.
\end{equation}
Here, the first term in brackets is the contribution from the
gauge fields, while the second corresponds to that from the matter fields.
The factors of 2 in these terms arise from the spin degrees of
freedom of massless gauge fields and fermions. The factor $N_c^2-1$
counts the number of gauge fields which are in the adjoint representation
of the gauge group. The factor $N_c\,N_f$ counts the number of
colors and flavors of the fermions which are in the fundamental
representation of the gauge group. The factor 7/4 accounts
for the difference between Bose-Einstein and Fermi-Dirac statistics
and for the fact that at $\mu=0$ there are as many antifermions as
fermions. Finally, the factor $\pi^2/90$ is the value
of the (dimensionless) Bose-Einstein integral $(1/6 \pi^2)
\int_0^\infty {\rm d}x\, x^3 (e^x -1)^{-1}$ occurring in the
calculation of the pressure.

In all cases, the function $p/T^4$ approaches the
Stefan-Boltzmann value for a gas of quarks and gluons,
which indicates that there is indeed a transition from hadronic
degrees of freedom to quark and gluon degrees of freedom,
i.e., from hadronic matter to the QGP. From the behavior
of the Polyakov loop and the chiral condensate discussed in
Sec.~\ref{translatt}, in the QGP color charges become
deconfined and chiral symmetry is restored.
However, the approach of the pressure
to the corresponding Stefan-Boltzmann value is rather slow; even
at temperatures $\sim 3\, T_c$, deviations are typically of the order
of 20\%. This indicates that at such temperatures the QGP
cannot really be considered as a non-interacting gas of massless
quarks and gluons. 

In order to understand the deviations
from the Stefan-Boltzmann values, one has to resort to
analytic calculations of the pressure, taking into account
interactions between quarks and gluons. In an analytic approach,
deviations from the ideal-gas behavior are well under control and
can be physically interpreted. For instance, in a perturbative calculation
of the QCD pressure, deviations from the Stefan-Boltzmann limit
are due to corrections proportional to powers of the strong
coupling constant, see Sec.~\ref{pert}. Another
possible explanation for the deviation of the pressure from $p_{\rm SB}/T^4$ 
is that quarks and gluons are actually quasiparticles, i.e., 
they are not massless, but assume a thermal mass due
to interactions with the hot environment, 
see Secs.~\ref{quasiparticles} and \ref{HTLresum}.

An important step to understand the deviations from ideal-gas behavior
might be the observation that, when normalizing the
pressure to the corresponding Stefan-Boltzmann value and the 
temperature to the critical temperature, the curves
$p/p_{\rm SB}$ as a function of $T/T_c$ exhibit a universal behavior
for the pure gauge theory and for QCD with various dynamical
quark flavors, see right panel of Fig.~\ref{EOS}.
A possible explanation for this behavior is provided by the 
Polyakov-loop model of Ref.~\cite{pisarskidumitru}, see
Sec.~\ref{Ploopmodel}, where the dynamics of chiral symmetry restoration
is exclusively driven by the dynamics of the deconfinement transition.

\subsection{\it Heavy quark free energy} \label{heavyquarklatt}

\begin{figure}[tb]
\begin{center}
\begin{minipage}[t]{8 cm}
\epsfig{file=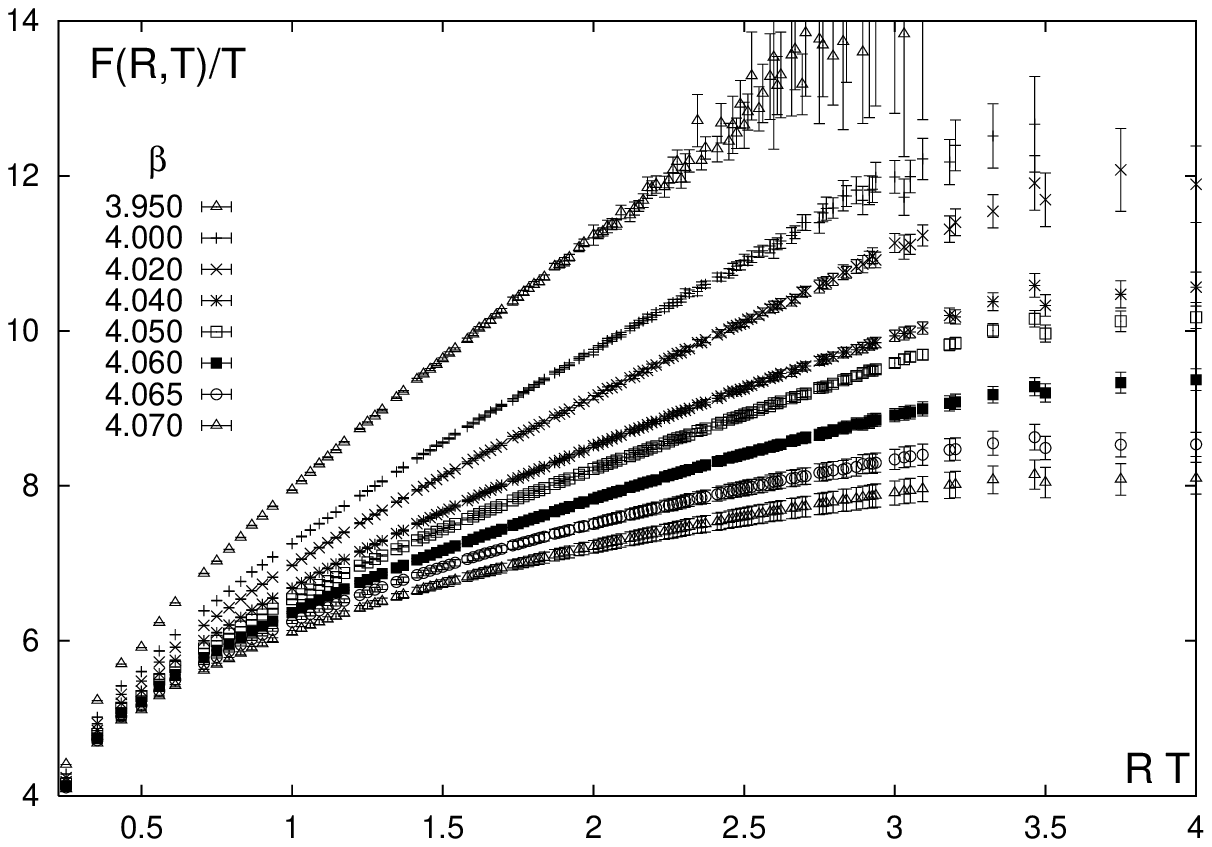,scale=0.65}
\end{minipage}
\begin{minipage}[t]{8 cm}\vspace*{-5.68cm}
\epsfig{file=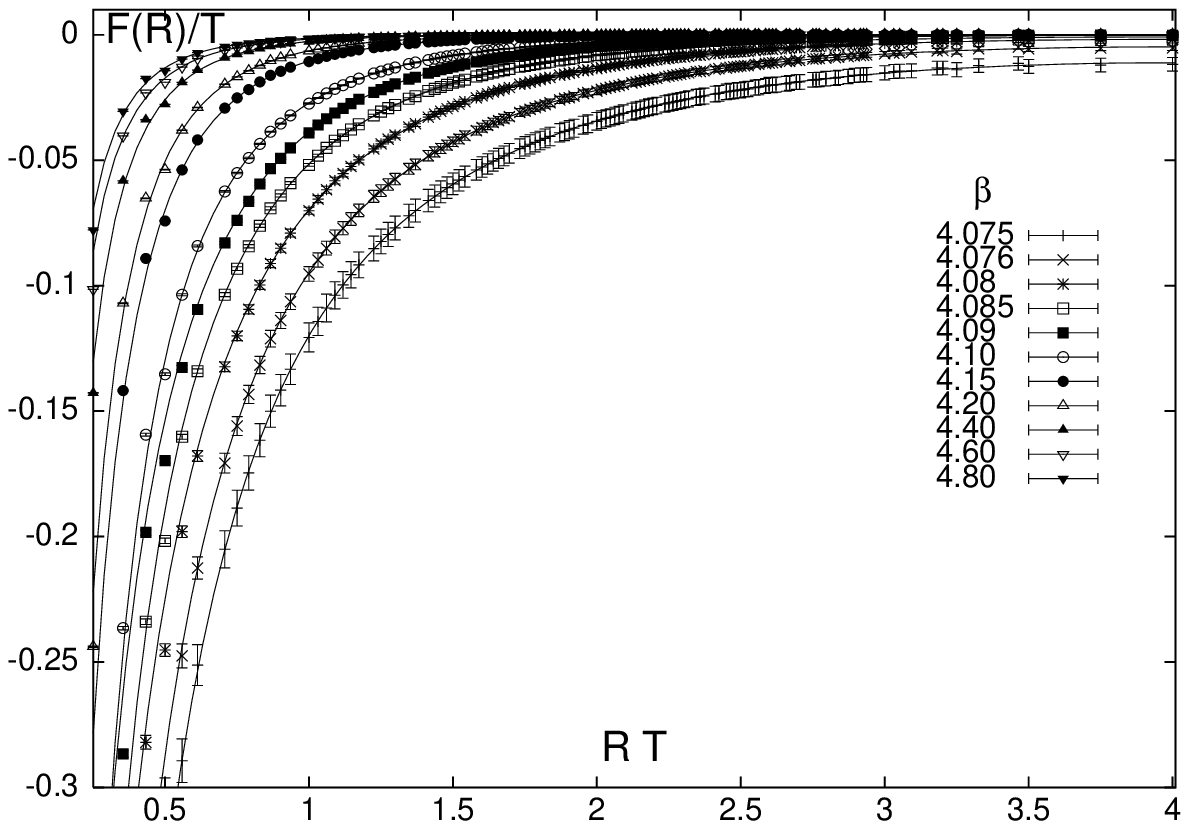,scale=0.65}
\end{minipage}
\begin{minipage}[t]{16.5 cm}
\caption{Left panel: The heavy quark free energy (in units of temperature)
as a function of the distance (in units of inverse temperature) for
various values of the bare coupling $\beta$ in the confined phase.
Right panel: The same for various values of $\beta$ 
in the deconfined phase. In this phase,
the Polyakov-loop correlation function of Eq.~(\ref{heavyquarkfreeenergy})
is normalized to $\langle |L(0)|^2 \rangle$.
Calculations are done for pure
$[SU(3)_c]$ gauge theory on a $32^3 \times 4$ lattice. The critical
bare coupling for this lattice is $\beta_c = 4.0729(3)$. From
Ref.~\cite{kaczmarek}.
\label{heavy_quark_pot}}
\end{minipage}
\end{center}
\end{figure}

The behavior of the heavy quark free energy as a function
of temperature is another indication for deconfinement in the
QGP. The heavy quark free energy $F_{\bar{Q}Q}(R,T)$ is the free energy of a
heavy quark and an antiquark, separated by a spatial
distance $R$, at a temperature $T$ \cite{mclerransvetitsky}. 
It is related to the Polyakov-loop correlation function via
\begin{equation} \label{heavyquarkfreeenergy}
\exp \left( - \frac{F_{\bar{Q}Q}(R,T)}{T} \right) 
= \langle L(0) \, L^\dagger({\bf x}) \rangle\,\, ,
\;\;\;\; R \equiv |{\bf x}|\,\, ,
\end{equation}
where $L({\bf x})$ is the Polyakov-loop operator defined in 
Eq.~(\ref{Polyakovlatt}). At $T=0$, the heavy quark free energy
is identical to the heavy quark potential, $V_{\bar{Q}Q}$, which
is expected to have a form motivated by the string model,
\begin{equation} \label{potTeq0}
V_{\bar{Q}Q}(R) =  -\frac{\alpha(0)}{R} +\sigma\, R + const.\,\, .
\end{equation}
The second term ensures confinement of color charge due
to the linear increase of $V_{\bar{Q}Q}$ with distance.
The constant $\sigma$ is the string tension.
The first term is an attractive Coulomb-like contribution
arising from fluctuations of the string.

Lattice QCD data \cite{kaczmarek} confirm these expectations,
cf.~Fig.~\ref{heavy_quark_pot}. Below $T_c$ (left panel of
Fig.~\ref{heavy_quark_pot}), at small distances
the heavy quark free energy is dominated by 
an attractive Coulomb-like part, while at large distances it linearly
rises with the distance, indicating confinement. The linear rise
becomes less pronounced with increasing bare coupling $\beta$, 
indicating that the string tension decreases with temperature.

\begin{figure}[tb]
\begin{center}
\begin{minipage}[t]{10 cm}
\epsfig{file=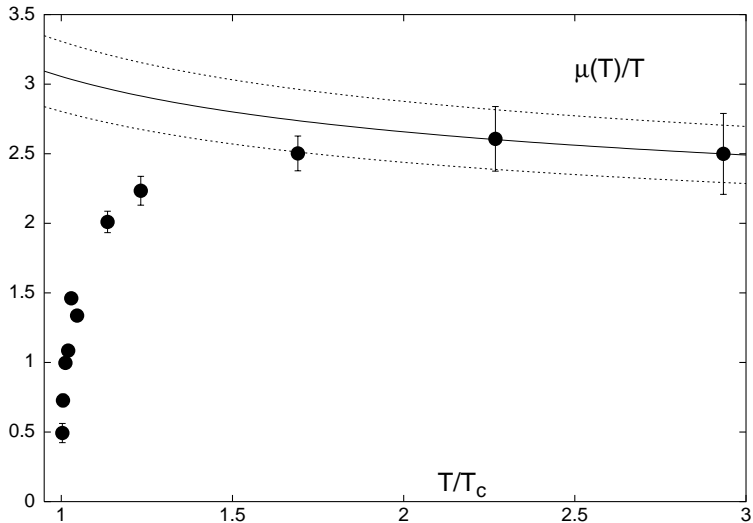,scale=0.8}
\end{minipage}
\begin{minipage}[t]{16.5 cm}
\caption{The temperature dependence of the
screening mass $\mu(T)$ (in units of temperature) 
as obtained from a fit of Eq.~(\ref{potfinT}) to the lattice 
data of Fig.~\ref{heavy_quark_pot}. The solid line is
a fit inspired by the perturbative expectation
$\mu_{\rm pert.}(T) = 2\, m_D(T)$. From Ref.~\cite{kaczmarek}.
\label{screeningmass}}
\end{minipage}
\end{center}
\end{figure}

At temperatures above $T_c$, color charges are deconfined, i.e.,
the linearly rising part of the potential in Eq.~(\ref{potTeq0})
has to vanish, leaving only a Coulomb-like part. The latter is,
however, screened due to the presence of a hot medium. 
This is confirmed by lattice QCD data above $T_c$ 
(right panel of Fig.~\ref{heavy_quark_pot}). It turns
out \cite{kaczmarek} that a fit to the numerically 
computed potential can be achieved by the formula
\begin{equation} \label{potfinT}
\frac{F_{\bar{Q}Q}(R,T)}{T} = - 
\frac{c(T)}{(RT)^{d(T)}} \, e^{-\mu(T)R} \,\, ,
\end{equation}
where $\mu(T)$ is the temperature-dependent screening mass 
(or inverse screening length).
This function is shown in Fig.~\ref{screeningmass}.

While the qualitative picture
of deconfinement and screening of color charges is certainly
applicable, the deconfined gluon-plasma phase cannot be described
perturbatively at temperatures in the range from $T_c$ to a few times $T_c$.
This is indicated by the fact that the fit function $d(T)$ in
Eq.~(\ref{potfinT}) is temperature-dependent \cite{kaczmarek} and
always smaller than $\simeq 1.5$ in the
range of temperatures considered here, while from perturbation theory one
expects $d_{\rm pert.} = 2$. Furthermore, the screening
mass $\mu(T)$ deviates from the perturbative value
$\mu_{\rm pert.}(T) = 2 \, m_D (T)$, where $m_D(T) = g T$ is 
the Debye mass in a hot gluonic medium. The solid line in
Fig.~\ref{screeningmass} represents a fit inspired by $\mu_{\rm pert.}(T)$
to the two data points corresponding to the highest temperatures,
for details see Ref.~\cite{kaczmarek}. One observes that while
the qualitative behavior of the data follows the perturbative
expectation at large temperature, near $T_c$ the data strongly
deviate from the perturbative result. They even suggest that
the screening mass goes to zero when $T \rightarrow T_c$.
This is an indication for critical behavior and is naturally
explained by the Polyakov-loop model of Ref.~\cite{pisarskidumitru},
see Sec.~\ref{Ploopmodel}.

More recent developments in the study of the heavy quark free energy
include a calculation in full QCD with dynamical quark flavors
\cite{peikert}. Below $T_c$, the string breaks when creation of dynamical
quark-antiquark pairs becomes energetically favorable. Consequently,
the heavy quark free energy saturates at larger distances
instead of increasing linearly. In another recent paper \cite{kaczmarek2}
the color-singlet and color-octet contributions to the heavy quark free energy 
were studied separately within the pure gauge theory, using
a novel prescription to renormalize the expectation value of
the Polyakov loop. It was found that
the singlet and octet contributions only deviate at smaller distances.
As expected, the color-octet channel is repulsive, while the
color-singlet channel is attractive.

\subsection{\it Mesonic spectral functions} \label{meson_spectral}

The correlation function of a mesonic state $G_H(\tau,{\bf x})$ is
defined as
\begin{equation} \label{corr}
G_H (\tau, {\bf x}) \equiv \langle \bar{\psi}(0) \, \Gamma_H \,
\psi(0) \, \bar{\psi}(\tau, {\bf x}) \, \Gamma_H \, \psi(\tau,{\bf x})
\rangle \,\, ,
\end{equation}
i.e., it is the overlap of a mesonic state with quantum numbers 
determined by the
$4 \times 4$ Dirac matrix $\Gamma_H$ at the origin with the same
mesonic state at $(\tau, {\bf x})$. 
Fourier-transforming Eq.~(\ref{corr}) with respect to 
the spatial variable, one obtains the mixed correlation function
$G_H(\tau,{\bf p})$, which has the spectral representation
\begin{equation} \label{spectralrep}
G_H(\tau,{\bf p}) = \int_0^\infty\frac{{\rm d}\omega}{2 \pi}\,
\sigma_H(\omega, {\bf p}) \, \frac{ \cosh \left[\omega 
\left(\tau - \frac{1}{2T} \right) \right] }{\sinh \left( \frac{\omega}{2T} 
\right)} \,\, .
\end{equation}
Here, $\sigma_H(\omega,{\bf p})$ is the spectral density in the
quantum number channel under consideration.
Suppose the spectral density is dominated by a single, stable, mesonic state 
with mass $m_H$. In this case, $\sigma_H (\omega,0) = \pi \lambda^2 
\delta(\omega - m_H)/\omega$, where $\lambda^2$ is
a constant with the dimension $[{\rm MeV}^4]$. Then, the susceptibility
\begin{equation}
\chi_H \equiv \frac{V}{T} \int_0^{1/T} {\rm d} \tau 
\int_V {\rm d}^3{\bf x}\, G_H(\tau, {\bf x})
\end{equation}
assumes the value $\chi_H = (V/T)\lambda^2 \, m_H^{-2}$, i.e.,
it is proportional to the inverse mass (squared) of the meson.
In a lattice QCD calculation, one can thus infer the mass of a mesonic
state in a given quantum number channel from the corresponding
susceptibility.

\begin{figure}[tb]
\begin{center}
\begin{minipage}[t]{13.5 cm}
\hspace*{2cm}
\epsfig{file=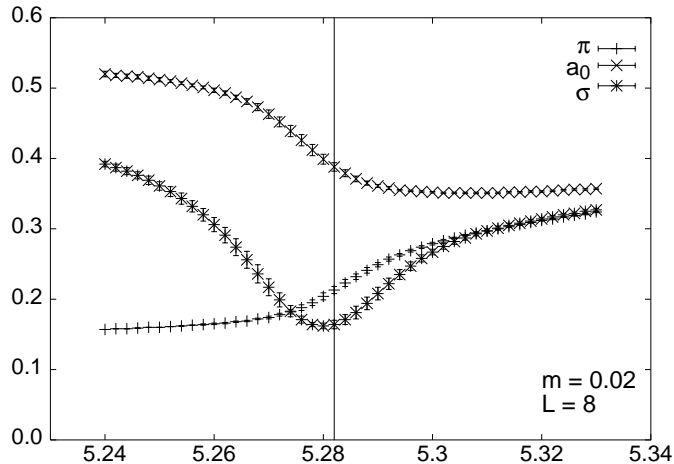,scale=0.7}
\end{minipage}
\begin{minipage}[t]{16.5 cm}
\caption{Mesonic masses as a function of the bare coupling,
computed on an $8^3 \times 4$ lattice
for QCD with $N_f=2$ dynamical flavors.
The vertical line indicates the critical bare coupling $\beta_c$
for the phase transition.
From Ref.~\cite{schladming}.\label{mesonmasslatt}}
\end{minipage}
\end{center}
\end{figure}

The masses for the pion, the $\sigma$ meson, and the
$a_0$ meson computed in this manner are shown in Fig.~\ref{mesonmasslatt}
as a function of the bare coupling (i.e., the temperature).
The results indicate restoration of chiral symmetry, i.e., the
mass of the pseudoscalar meson (pion) becomes degenerate
with those of the scalar mesons ($\sigma$ and $a_0$) 
at large temperature. The fact that the pion and the $\sigma$ meson become
degenerate in mass at smaller temperatures than 
the pion and the $a_0$ meson indicates that
the $SU(2)_r \times SU(2)_\ell$ symmetry is restored prior
to the $U(1)_A$ symmetry when increasing the temperature.

Instead of the susceptibility, one could also try 
to compute the complete spectral density
$\sigma_H(\omega,{\bf p})$ of a mesonic state from lattice QCD data. 
An important motivation for such a calculation is the fact that
the spectral density in the vector channel, $\sigma_V(\omega,{\bf p})$,
is directly proportional to the rate of dilepton emission
\cite{wetzorke}, which
is an experimentally observable quantity \cite{CERES},
\begin{equation} \label{rate}
\frac{{\rm d}N_{\ell^+ \ell^-}}{{\rm d}^4 X \, {\rm d}\omega \,
{\rm d^3}{\bf p} } = \frac{5 \alpha^2}{27 \pi^2} \,
\frac{1}{e^{\omega/T} - 1} \frac{\sigma_V(\omega,{\bf p})}{\omega^2} \,\, .
\end{equation}
To determine $\sigma_V(\omega,{\bf p})$, 
one would have to perform an inverse Laplace transformation
of Eq.~(\ref{spectralrep}).
This requires complete knowledge of the correlation function 
$G_V(\tau,{\bf p})$ on the left-hand side. 
On the lattice, however, this function
is only known at a few discrete points in $\tau$-direction. Moreover, its
value at these points is subject to statistical fluctuations.
Consequently, a computation of $\sigma_V(\omega,{\bf p})$
via inversion of Eq.~(\ref{spectralrep}) with lattice
data for $G_V(\tau,{\bf p})$ is impossible.

\begin{figure}[tb]
\begin{center}
\begin{minipage}[t]{9 cm}
\epsfig{file=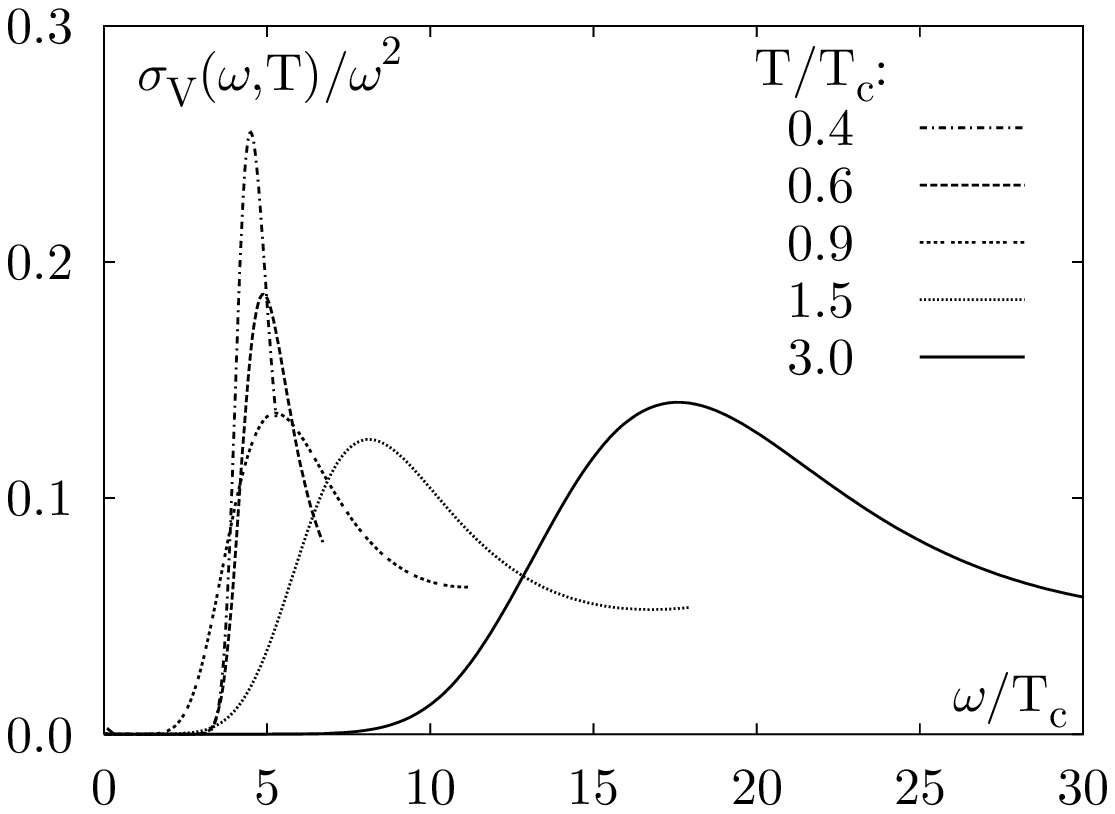,scale=0.7}
\end{minipage}
\begin{minipage}[t]{9 cm}
\vspace*{-6.1cm}
\hspace*{-6mm}
\epsfig{file=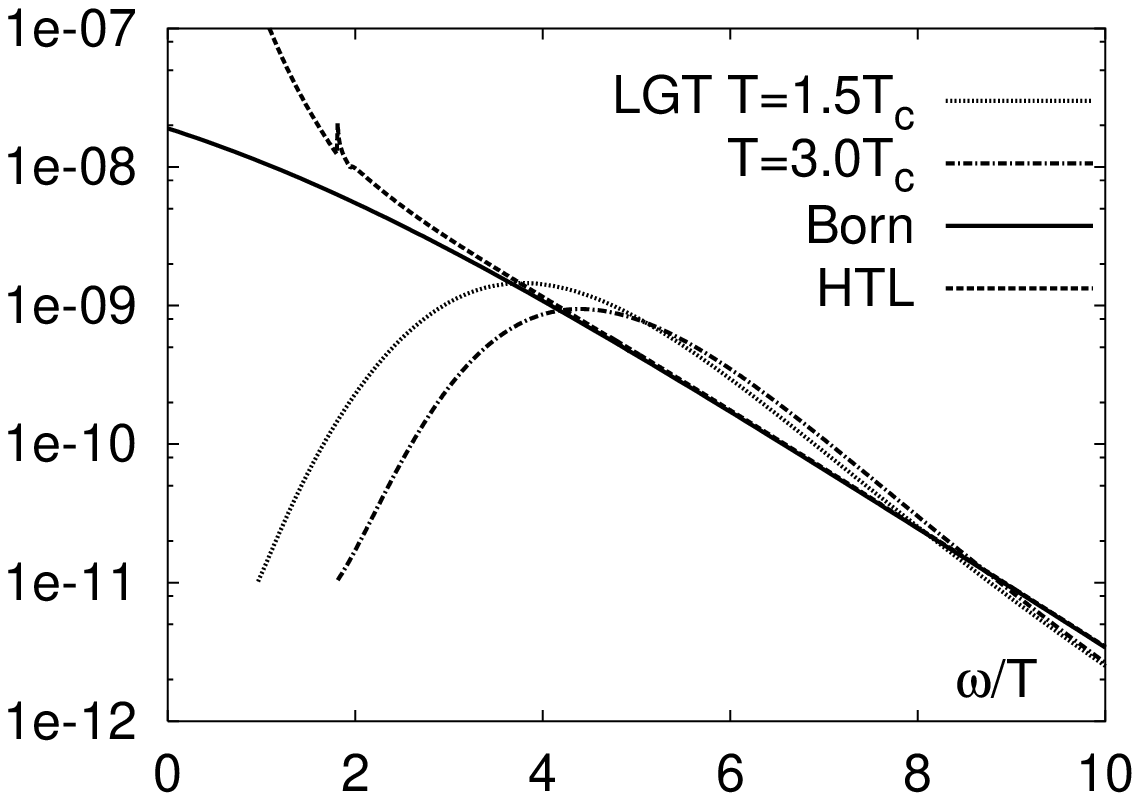,scale=0.7}
\end{minipage}
\begin{minipage}[t]{16.5 cm}
\caption{Left panel: The spectral density in the vector channel,
divided by $\omega^2$, for various temperatures.
Right panel: The dilepton rate computed in lattice QCD for
two different temperatures as a function
of energy (in units of temperature).
Also shown is the Born approximation and the result from
HTL-perturbation theory. From Ref.~\cite{laermannphilipsen}.
\label{spectraldens&dileptons}}
\end{minipage}
\end{center}
\end{figure}

Nevertheless, a solution of this problem is provided by 
the so-called ``Maximum Entropy Method'' (MEM). The basic idea
is to construct that particular spectral density $\sigma_H (\omega, {\bf p})$
under the integral in Eq.~(\ref{spectralrep}),
which is the {\em most probable\/} one to yield a given correlation function
$G_H(\tau, {\bf p})$ on the left-hand side of that equation;
for details see Ref.~\cite{asakawa}. 
Figure~\ref{spectraldens&dileptons} shows
the spectral function in the vector meson channel 
computed with this method \cite{wetzorke} (left panel) and the corresponding
dilepton rate~(\ref{rate}) (right panel). One observes that the peak in
the spectral density broadens and shifts towards larger energies as the
temperature increases. Consequently, the dilepton rate is
depleted for small dilepton energies. This behavior is in stark
contrast to the dilepton
emission rate computed in the Born approximation and in the so-called
``hard thermal loop'' (HTL-) resummation scheme, which are also shown
in Fig.~\ref{spectraldens&dileptons}. 

Finally, the low-energy behavior
of the spectral density determines the value of
transport coefficients in a hot medium \cite{aarts}. 
I do not elaborate further on this point, as it concerns the
non-equilibrium properties of the QGP, which are beyond the
scope of the present review.

\subsection{\it Nonzero Chemical Potential} \label{nonzeromu}

As discussed in Sec.~\ref{basicslattice}, 
for nonzero values of the quark chemical potential, $\mu \neq 0$,
a straightforward evaluation of the QCD partition function on
the lattice is not possible due to the sign problem of
the fermion determinant. However, for sufficiently small
$\mu$ progress has recently been made by applying methods which 
explicitly avoid the sign problem. Most notably among these
are multiparameter reweighting
\cite{fodorkatz}, Taylor expansion around $\mu=0$
\cite{allton}, and analytic continuation from imaginary values of $\mu$, 
where the fermion determinant is real-valued and positive, 
to real values of $\mu$ \cite{deForcrand}. For the sake of brevity,
here I only discuss the multiparameter-reweighting method; for a detailed
comparison of all approaches see the review~\cite{laermannphilipsen}.

The multiparameter-reweighting method is based on the so-called
Glasgow method \cite{glasgow}. The idea of the Glasgow method
is to treat the fermion determinant at nonzero $\mu$ in the partition function
(\ref{Zlatt}) as an {\em observable\/} rather than as a part of
the integration measure. The integration measure itself is computed with
a fermion determinant at $\mu = 0$, which is real-valued and positive
and thus causes no problems when applying standard Monte Carlo
methods to sample gauge field configurations,
\begin{equation} \label{Zlatt2}
{\cal Z}(N_\sigma, N_\tau,\beta,\tilde{m},\tilde{\mu})
= \int \prod_{n,\mu} {\rm d} U_{n,\mu} \, 
\frac{{\rm det} M(\tilde{m}, \tilde{\mu},U)}{
{\rm det} M (\tilde{m},0,U)} \,
{\rm det} M (\tilde{m},0,U) \, e^{- \beta\, S_A}
\equiv \left\langle \frac{{\rm det} M(\tilde{m}, \tilde{\mu},U)}{
{\rm det} M (\tilde{m},0,U)} \right\rangle_{\tilde{\mu} = 0}\,\, .
\end{equation}
Here, the expectation value $\langle {\cal O} \rangle_{\tilde{\mu} = 0}$ of
an operator ${\cal O}$ is defined with respect to an ensemble
of gauge fields and fermions at zero quark chemical potential.

This method is limited to small values of $\mu$. In order to understand this,
one has to remember the principle behind a Monte Carlo computation 
of the functional integral (\ref{Zlatt2}) \cite{creutz}.
A Monte Carlo computation assumes that in order to obtain
a reasonable approximate value of the
functional integral in Eq.~(\ref{Zlatt2}) it suffices to
sum only over (a few $10^2$ to $10^4$ of) the ``most probable''
gauge field configurations, rather than performing the integrals
over the link variables $U_{n,\mu}$ explicitly. The ``most probable''
gauge field configurations are obviously those which minimize
the action $S_A$. However, the most probable configurations at
$\mu \neq 0$ are not the same as the ones at $\mu = 0$.
Thus, approximating the partition function at $\mu \neq 0$ by
configurations obtained for $\mu = 0$ becomes increasingly
worse as $\mu$ increases. In other words, the ``overlap'' between
the ensemble of most probable configurations at $\mu = 0$ 
and the ensemble that consists of the configurations which
are actually most probable at $\mu \neq 0$ diminishes.

It has been recently realized \cite{fodorkatz} that a way to increase
this overlap is to also include the exponential of the action 
into the operator which is averaged over the ensemble,
\begin{equation} \label{Zlatt3}
{\cal Z}(N_\sigma, N_\tau,\beta,\tilde{m},\tilde{\mu})
\equiv \left\langle \frac{e^{-\beta S_A}\, {\rm det} 
M(\tilde{m}, \tilde{\mu},U)}{ e^{-\beta_0 S_A} \,
{\rm det} M (\tilde{m},0,U)} \right\rangle_{\tilde{\mu} = 0, \beta_0}\,\, ,
\end{equation}
i.e., the ensemble one averages over is generated at $\mu = 0$ and
a value $\beta_0$ for the bare coupling.
In this way, one not only reweights the ensemble
in the parameter $\tilde{\mu}$, as in the Glasgow approach 
(\ref{Zlatt2}), but also in the bare coupling $\beta$ 
(hence the name ``multiparameter reweighting'').

\begin{figure}[tb]
\begin{center}
\begin{minipage}[t]{13.5 cm}
\epsfig{file=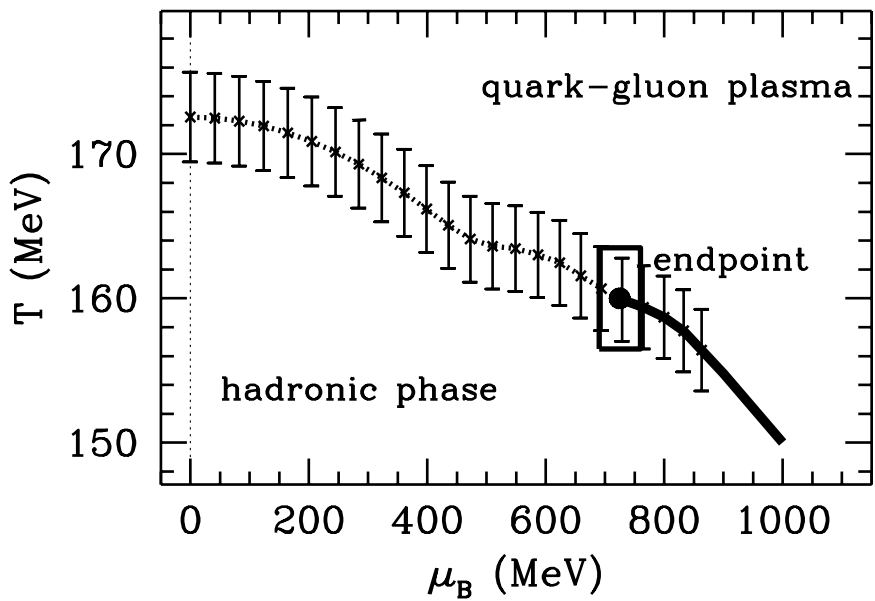,scale=0.6}
\end{minipage}
\begin{minipage}[t]{16.5 cm}
\caption{The QCD phase transition as computed in
lattice QCD \cite{fodorkatz}. To the left of the critical
point the transition is crossover, to the right it is of first
order. At the critical point, the transition is of second order and
in the universality class of the Ising model.
Note that $\mu_B = 3 \mu$.\label{endpoint}}
\end{minipage}
\end{center}
\end{figure}

How does one choose the second reweighting parameter $\beta$?
This depends on which physical question one asks. Suppose one wants
to compute the QCD phase transition line for nonzero values
of $\mu$. One first generates an ensemble
at $\tilde{\mu}=0$ and $\beta_0 \equiv \beta_c$. This
ensemble is ``maximally'' critical in the sense that it is
generated at the phase transition point $(\beta,\tilde{\mu}) = 
(\beta_c, 0)$ (which corresponds to the point $(T,\mu) = (T_c, 0)$
in the continuum).
For each nonzero value of $\tilde{\mu}$ one then determines $\beta$ such that
one remains on the phase transition line.

The criterion for ``remaining on the phase transition line''
is the position of the Lee-Yang zeroes 
$\beta^*_1, \beta^*_2, \ldots$ of the partition function 
${\cal Z}$ in the complex $\beta$-plane \cite{leeyang}. For a given 
set of parameters $N_\sigma,N_\tau,\tilde{m},\tilde{\mu}$, there 
are many Lee-Yang zeroes, i.e., roots of the equation
${\cal Z}(N_\sigma,N_\tau,\beta^*,\tilde{m},\tilde{\mu}) = 0$.
(In fact, the total number of Lee-Yang zeroes, $M$,
increases linearly with the volume of the system, $M \sim N_\sigma^3$.)
In the case of a first-order phase transition, one root, say
$\beta_1^*$, has a vanishing imaginary part, i.e., it lies on the
positive real $\beta$ axis. Then the value of $\beta$ 
which corresponds to the phase transition line in the $(\beta,\tilde{\mu})$
plane is $\beta \equiv {\rm Re} \beta_1^*$.

Note that in a finite system, such as the space-time lattice
considered in lattice QCD, all Lee-Yang zeroes have nonzero imaginary parts.
Then one has to study lattices of different sizes and
extrapolate to the infinite-volume limit. 
In Ref.~\cite{fodorkatz}, this is done via
linear extrapolation in the variable $1/V$, $\beta^*_1(V)
= \beta_1^*(\infty) + \alpha/V$.
In the case of a crossover transition, the imaginary parts of the
extrapolated Lee-Yang zeroes never vanish. In this case, the value of $\beta$
corresponding to the phase transition line is determined by
the real part of the Lee-Yang zero with the smallest imaginary part.

The phase transition line calculated in this way is shown in
Fig.~\ref{endpoint}. It agrees with the expectations discussed
in Secs.~\ref{quarkmassdiagram}, \ref{translatt}:
there is a line of first-order phase transitions, ending at the
point $(T, \mu)_{\rm cr} = (160 \pm 3.5,242 \pm 12)$ MeV, 
at which the transition is of second order. To the left of this point, the
transition is crossover. One should mention that the lattice QCD
calculation underlying Fig.~\ref{endpoint} was done on fairly small
lattice sizes, with probably unrealistically large quark masses.
As discussed in Sec.~\ref{quarkmassdiagram},
for smaller quark masses the endpoint should
move towards the temperature axis. For three massless flavors,
it should reach the temperature axis, since in this case the transition is
of first order. For realistic quark masses, however, 
as discussed in Sec.~\ref{translatt} the transition is crossover
at $\mu = 0$, and the line of
first-order transitions should always end at some nonzero value of $\mu$.

\begin{figure}[tb]
\begin{center}
\begin{minipage}[t]{11 cm}
\epsfig{file=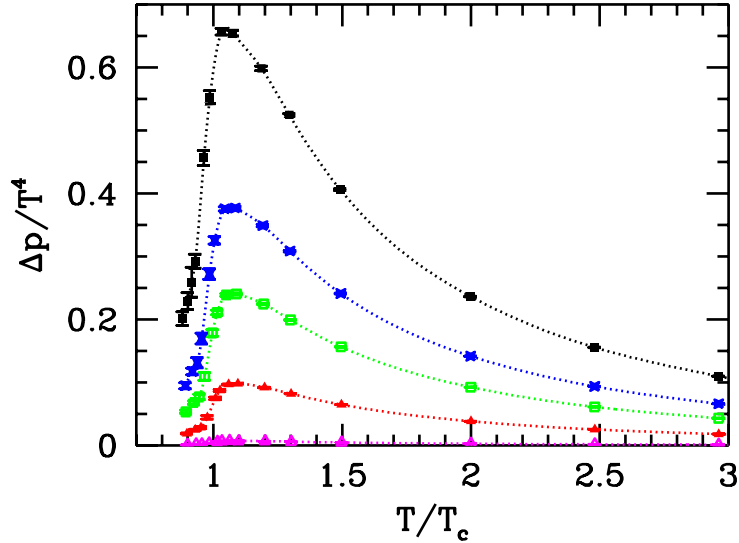,scale=0.5}
\end{minipage}
\begin{minipage}[t]{16.5 cm}
\vspace*{-2cm}
\caption{$\Delta p$ as a function of $T/T_c$ for various values
of $\mu_B$. From bottom to top, $\mu_B = 100, 210, 220, 310, 530$ MeV. 
From Ref.~\cite{fodorkatz2}.\label{finmuEOS}}
\end{minipage}
\end{center}
\end{figure}

The position of the phase transition line determined by multiparameter
reweighting is in good agreement with that computed by the 
other approaches mentioned previously, namely
the Taylor expansion method, and method of analytic continuation 
from imaginary values of $\mu$ \cite{laermannphilipsen}.
Recent developments \cite{fodorkatz2} are the application
of the multiparameter-reweighting method to
compute the equation of state at nonzero quark chemical potential.
Figure~\ref{finmuEOS} shows the results for the pressure
difference $\Delta p \equiv p(T,\mu) - p(T,0)$, normalized
to $T^4$, as a function of $T$ for various values of $\mu$.
There is a strong increase of $\Delta p$
around the phase transition temperature. This increase is larger
for larger values of $\mu$.

Multiparameter reweighting, as well as the other aforementioned methods,
is restricted to values of the quark chemical potential, which are not 
too large as compared to the temperature.
In order to compute the partition function of QCD for large
quark chemical potential at small or even zero temperature, and
possibly study the color-superconducting phases of quark-gluon matter,
one has to resort to other methods. A promising approach is the
so-called meron-cluster algorithm which has been shown to
solve the sign problem of the fermion determinant for the Hubbard
and the Potts model \cite{meron}. For QCD, as of yet no solution has
been found.

\begin{figure}[tb]
\begin{center}
\begin{minipage}[t]{11 cm}
\epsfig{file=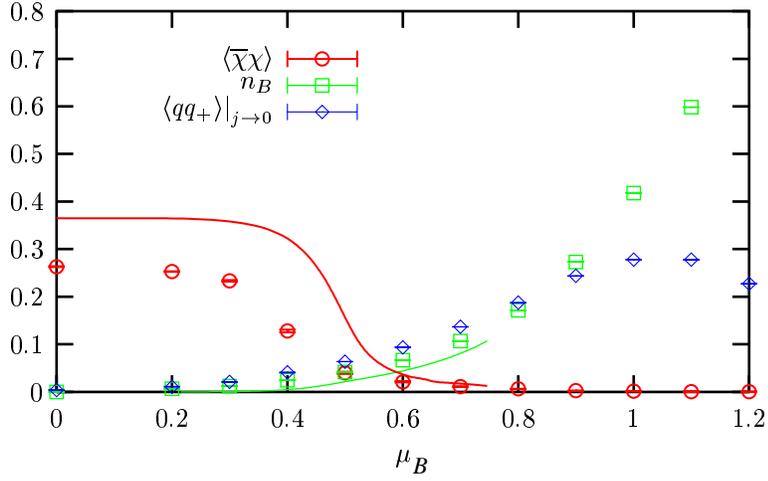,scale=1}
\end{minipage}
\begin{minipage}[t]{16.5 cm}
\caption{The chiral condensate, the baryon density, and the
diquark condensate as a function of baryon chemical potential
$\mu_B$. The solid line is the chiral condensate computed
analytically within the Hartree approximation. 
From Ref.~\cite{hands}.\label{NJLdiquark}}
\end{minipage}
\end{center}
\end{figure}

Another possibility is to study a model for QCD which does
not have the sign problem. Such a model is, for instance, 
the NJL model which has been investigated on the lattice at 
$T=0$ and $\mu \neq 0$ in Ref.~\cite{hands}. 
Although this model has no color gauge symmetry which could
be spontaneously broken, and thus strictly speaking cannot
exhibit color superconductivity, quarks can still form
Cooper pairs and the system may become superfluid.
Figure~\ref{NJLdiquark} shows the chiral condensate, the baryon density, and
the superfluid diquark condensate as a function of the
baryon chemical potential,
computed on the lattice and then extrapolated to the thermodynamic
limit. One observes that,
as the baryon density increases, the chiral condensate vanishes
and the diquark condensate increases, signalling the onset
of superfluid behavior. These results are in agreement with analytic
calculations \cite{arw} for the NJL model in the mean-field approximation,
which stimulated recent interest in color superconductivity.

\section{Analytic Approaches} \label{secIII}

\subsection{\it Perturbation Theory} \label{pert}

The QCD partition function (\ref{Z}) can be expanded in a power series
in the strong coupling constant $g$. In the following, I present
the general idea behind this approach, neglecting contributions from 
gauge fixing and from Fadeev-Popov ghosts. Of course,
these have to be properly accounted for in order to obtain the 
correct answer; for more details, see Ref.~\cite{kapustaFTFT}.
I also focus on the case $\mu = 0$.
The first step is to split the QCD Lagrangian (\ref{LQCD}) into two terms, 
the non-interacting part, ${\cal L}_0 \equiv {\cal L}_{g=0}$, and
the interaction part, ${\cal L}_I \equiv {\cal L} - {\cal L}_0$. 
Then, the QCD action 
$S \equiv \int_X {\cal L}$ can be written as 
\begin{eqnarray}
S & = & S_0 + S_I \equiv \int_X 
\left( {\cal L}_0 + {\cal L}_I \right)\,\, , \\ 
{\cal L}_0 & = & \bar{\psi}\, G_0^{-1} \, \psi +
\frac{1}{2} \, A_\mu^a \, {\Delta_0^{-1}}^{\mu \nu}_{ab} \, A_\nu^b
 \,\, , \\
{\cal L}_I & = &  g\, \bar{\psi}\, \gamma^\mu\, T^a \, \psi\, A_\mu^a
+ g \, f^{abc} \, \partial_\nu A_\mu^a \, A^\mu_b\, A^\nu_c
- \frac{g^2}{4} \, f^{abc} f^{ade} \, A_\mu^b\, A_\nu^c\, A^\mu_d \,
A^\nu_e \,\, .
\end{eqnarray}
Here $G_0^{-1} \equiv i \gamma^\mu \partial_\mu - m$ is
the free inverse quark propagator and
${\Delta_0^{-1}}^{\mu \nu}_{ab} \equiv \left(\Box g^{\mu \nu} - \partial^\mu
\partial^\nu\right) \delta_{ab}$ is the free inverse gluon propagator
(which will eventually receive another contribution from the
gauge fixing terms neglected here).
The next step is to introduce source terms for fermions and gauge fields
(which eventually have to be set to zero).
One can then replace all fields in $S_I$ in terms of functional
derivatives with respect to the sources, and thus
extract $e^{ S_I }$ from the functional integral. The functional
integration over the exponential of the
non-interacting part and the source terms is a Gaussian integral
and can be performed exactly. The result is
\begin{equation} \label{Z2}
{\cal Z} = {\cal Z}_0 \, \left.
\exp \left\{ S_I \left[ \frac{\delta}{\delta \bar{\eta}},
\frac{\delta}{\delta \eta}, \frac{\delta}{\delta J_\mu^a}\right]
\right\} \, \exp\left[  \int_X
\left( - \bar{\eta}\, G_0 \, \eta - \frac{1}{2} \,
J_\mu^a \, {\Delta_0}^{\mu \nu}_{ab}\, J_\nu^b 
\right) \right] 
\right|_{\bar{\eta}=\eta=J =0}\,\, ,
\end{equation}
where ${\cal Z}_0$ is the partition function for a system
of non-interacting quarks and gluons. Obviously, the pressure
$p_0 \equiv (T/V) \ln {\cal Z}_0$ is identical
to the Stefan-Boltzmann pressure defined through 
Eq.~(\ref{psb}), $p_0 \equiv p_{\rm SB}$.
The full pressure in QCD also receives contributions from
the remaining two terms in Eq.~(\ref{Z2}). After introducing Feynman
rules for propagators and vertices \cite{kapustaFTFT}, these
terms have a graphical representation as an infinite series of
diagrams with no external legs. The diagrams can be sorted according
to powers in the strong coupling constant $g$ associated with
the vertices. Thus, one obtains a perturbative series in powers
of $g$.

Inspecting the topology of these diagrams, one would 
naively conclude that this perturbative series is an expansion
in powers of $g^2$. In fact, it turns out that this is only true
at zero temperature \cite{freedman}.
At nonzero temperature, the expansion 
is in powers of $g$, rather than $g^2$, due to the different infrared
behavior of a field theory containing massless modes
(such as gauge fields) at nonzero temperature.
Roughly speaking, the difference arises from the infrared behavior of
single-particle phase space, which, at zero temperature, is
$\sim {\rm d}k\, k^3$, while at nonzero temperature it is
$\sim T\, {\rm d}k\, k^2$. The missing power of $k$ at
nonzero temperature leads to a completely different
infrared behavior as compared to the zero-temperature case.

At zero temperature the theory is well-behaved in the infrared
and the terms of the perturbative series are probably computable 
to all orders in $g^2$ \cite{pisarskirischke}. 
Freedman and McLerran computed the series up to terms of 
order $g^4$, for details see Ref.~\cite{freedman}. 
I do not present a more detailed discussion of
their results at this point, as quark matter
at zero temperature is a color superconductor, cf.~Secs.~\ref{cscphases} 
and \ref{CSC}. Color superconductivity is a nonperturbative phenomenon,
which cannot be described in a purely perturbative
calculation of the pressure. 

At nonzero temperature, the infrared behavior
of the theory leads to terms proportional to odd powers of $g$
in the perturbative expansion (\ref{Z2}) of the partition function.
Technically, they arise from a resummation of an infinite subset of
diagrams describing the screening of long-range electric fields.
Moreover, there are infinitely many diagrams 
at order $O(g^6)$, and the perturbative expansion breaks down
\cite{kapustaFTFT}. This is sometimes called 
the {\em Linde problem\/} of QCD, after its discoverer \cite{linde}.
Nevertheless, what is perturbatively computable has been evaluated. 
These are all terms up to $O(g^5)$,
and the terms of order $O(g^6 \ln g)$. How to obtain the latter is
discussed in greater detail in Sec.~\ref{dimred}. 
While the terms which are
genuinely of order $O(g^6)$ cannot be computed perturbatively, 
they can in principle be evaluated via a lattice calculation.

At zero chemical potential, the pressure assumes the form
\begin{equation} \label{ppert}
p = T^4 \, \left[ c_0 + c_2 \, g^2 + c_3\, g^3 + 
\left(c_4'\, \ln g + c_4 \right)\, g^4 +  c_5 \, g^5
+ c_6\, g^6 \right] \,\, .
\end{equation}
The coefficient $c_0$ is equal to the Stefan-Boltzmann
constant (\ref{psb}).
The coefficient $c_2$ arises from the lowest-order perturbative
correction to the pressure of an ideal gas. It consists
of two-loop diagrams, and was first computed by Shuryak \cite{shuryak}
\begin{equation}
c_2 = - \frac{N_c^2-1}{144} \,
\left( N_c + \frac{5}{4}\, N_f \right)
\,\, .
\end{equation}
The computation of the coefficient $c_3$ 
requires a nonperturbative resummation of plasmon ring diagrams
in the infrared limit. This was first done correctly by
Kapusta \cite{kapusta}, with the result
\begin{equation}
c_3 =  \frac{N_c^2-1}{36 \,\sqrt{3}\, \pi} \,
\left( N_c + \frac{1}{2}\,N_f \right)^{3/2}
\,\, .
\end{equation}
The coefficient $c_4'$ has been computed by Toimela \cite{toimela},
\begin{equation}
c_4' =  \frac{N_c^2-1}{48 \pi^2}\, N_c \,
\left( N_c + \frac{1}{2}\, N_f \right)
\,\, .
\end{equation}
The coefficient $c_4$ is due to three-loop diagrams
and has been computed by Arnold and Zhai \cite{arnoldzhai},
\begin{eqnarray}
c_4 & = & - \frac{N_c^2-1}{(48 \pi)^2}\,  \left\{ 
- 24\, N_c \, \left(N_c + \frac{1}{2}\,N_f\right) \,
\ln \left( \frac{N_c + N_f/2}{12  \pi^2} \right) \right . \nonumber \\
&    & \hspace*{2cm} + 
N_c^2 \, \left[ \frac{22}{3}\, \ln \frac{\bar{\mu}}{4\pi T}
+ \frac{38}{3}\, \frac{\zeta'(-3)}{\zeta(-3)}
- \frac{148}{3}\, \frac{\zeta'(-1)}{\zeta(-1)}
- 4 \gamma_E + \frac{64}{5} \right] \nonumber \\
&    &  \hspace*{2cm} + 
N_c N_f\, \left[ \frac{47}{6} \, \ln \frac{\bar{\mu}}{4\pi T}
+ \frac{1}{6}\, \frac{ \zeta'(-3)}{\zeta(-3)}
- \frac{37}{3}\, \frac{\zeta'(-1)}{\zeta(-1)}
- 4 \gamma_E + \frac{1759}{120} 
+ \frac{37}{10}\, \ln 2 \right] \nonumber \\
&    &  \hspace*{2cm} + 
N_f^2\, \left[- \frac{5}{3} \, \ln \frac{\bar{\mu}}{4\pi T}
+ \frac{2}{3}\, \frac{ \zeta'(-3)}{\zeta(-3)}
- \frac{4}{3}\, \frac{\zeta'(-1)}{\zeta(-1)}
- \gamma_E - \frac{1}{12} 
+ \frac{22}{5}\, \ln 2 \right] \nonumber \\
&    &  \hspace*{2cm} + \left.
\frac{N_c^2-1}{N_c}\, N_f \left[- \frac{105}{16} 
+ 6\, \ln 2 \right] \right\}
\,\, . \label{c4}
\end{eqnarray}
Here, $\zeta(x)$ is Riemann's zeta function, 
$\gamma_E$ is the Euler-Mascheroni constant, and
$\bar{\mu}$ is the renormalization scale in the $\overline{MS}$ scheme, 
for details see Ref.~\cite{arnoldzhai}.
The coefficient $c_5$ arises from corrections
to the three-loop diagrams due to Debye screening of electric gluons. 
It has been computed by Zhai and Kastening \cite{zhaikastening} 
and by Braaten and Nieto \cite{braatennieto},
\begin{eqnarray}
c_5 & = & \frac{N_c^2-1}{9216\, \sqrt{3}\, \pi^3}\, 
\left( N_c + \frac{1}{2}\, N_f \right)^{1/2}\,
 \left[ N_c^2 \, \left( 176\, \ln \frac{\bar{\mu}}{4\pi T}
+ 176 \gamma_E - 24 \pi^2 - 494 + 264\, \ln 2 \right) \right. \nonumber \\
&    &  \hspace*{5.3cm} + 
N_c N_f\, \left( 56 \, \ln \frac{\bar{\mu}}{4\pi T}
+ 56 \gamma_E + 36 - 64\, \ln 2 \right) \nonumber \\
&    &  \hspace*{5.3cm} + 
N_f^2\, \left(- 16 \, \ln \frac{\bar{\mu}}{4\pi T}
- 16 \gamma_E + 8 - 32\, \ln 2 \right) \nonumber \\
&    &  \hspace*{5.3cm} - \left. 36\,
\frac{N_c^2-1}{N_c}\, N_f \right]
\,\, . \label{c5}
\end{eqnarray}
The coefficient $c_6$ contains terms $\sim \ln g$
and constant terms. The former contribute to order
$O(g^6 \ln g)$ and can be evaluated perturbatively,
while the latter are genuinely of order $O(g^6)$ and can
only be computed e.g.\ via a lattice calculation, for
more details, see Sec.~\ref{dimred}.
At finite temperature and nonzero chemical potential $\mu$,
the contributions of order $O(1)$, $O(g^2)$, and $O(g^4 \ln g)$
to the pressure have been computed by Toimela \cite{toimela2}.

\begin{figure}[tb]
\begin{center}
\begin{minipage}[t]{8 cm}
\epsfig{file=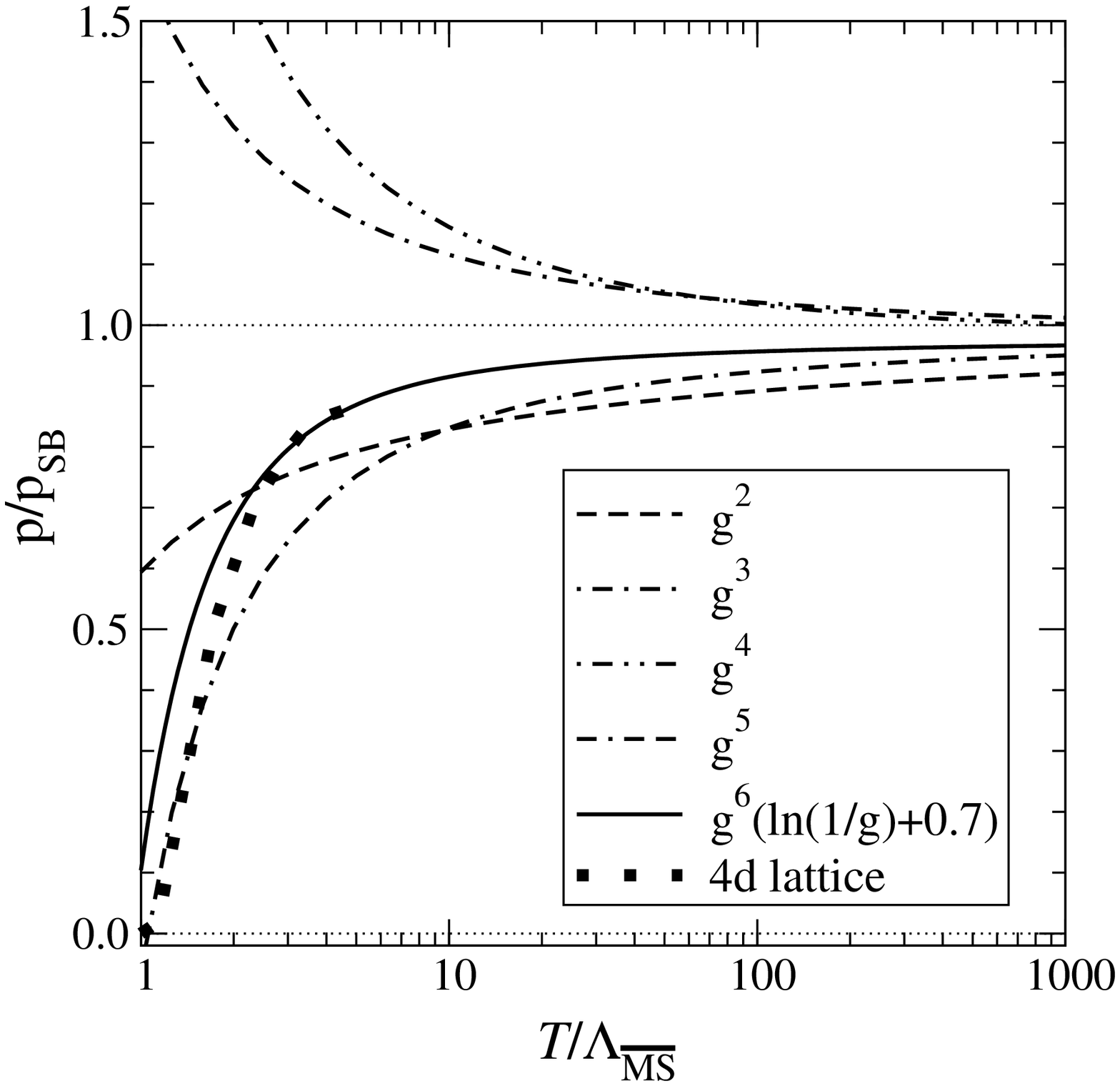,scale=0.45}
\end{minipage}
\begin{minipage}[t]{8 cm}
\epsfig{file=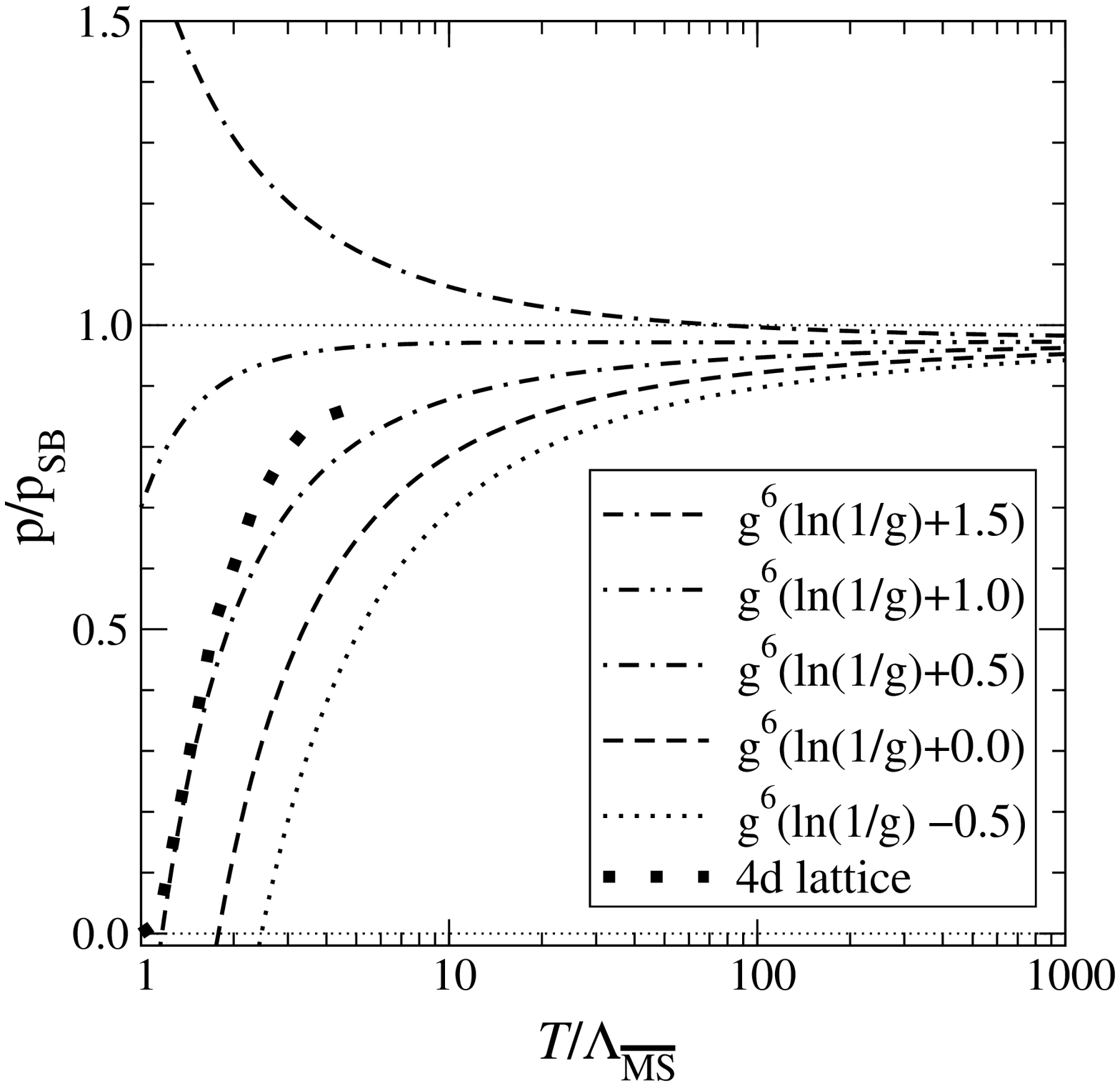,scale=0.45}
\end{minipage}
\begin{minipage}[t]{16.5 cm}
\caption{Left panel: Pressure (divided by $T^4$) as a function of temperature
for the pure $[SU(3)_c]$ gauge theory. Several perturbative contributions 
(up to a given power in $g$) are shown, as well as lattice QCD
data from Ref.~\cite{boyd}.
Right panel: Sensitivity of the pressure 
to the value of the constant $\delta$ in the 
term $\sim g^6$. From Ref.~\cite{schroder}.
\label{EOSpert}}
\end{minipage}
\end{center}
\end{figure}

To be more explicit, consider pure $[SU(3)_c]$ gauge theory, 
i.e., $N_c=3$, $N_f = 0$.
The pressure up to terms of a given order in $g$ is shown
in the left panel of Fig.~\ref{EOSpert}. The strong coupling constant $g$
is taken to be running and evaluated at the scale $\bar{\mu}$.
After applying the principle of fastest apparent convergence
to minimize the two-loop corrections to the running of $g$,
this scale is chosen as $\bar{\mu} \simeq 6.742\, T$;
for more details see \cite{schroder}.
The scale $\bar{\mu}$ also enters 
under the logarithms in Eqs.~(\ref{c4}), (\ref{c5}). 
In principle, the complete result for the pressure, being a physically
observable quantity, must be independent of the renormalization scale 
$\bar{\mu}$. The way this works out is that $\bar{\mu}$ under some
logarithm, such as occurring in Eqs.~(\ref{c4}), (\ref{c5}),
is cancelled by a similar logarithm from the running of the coupling constant
in a lower-order contribution. Nevertheless, while terms 
$\sim \ln \bar{\mu}$ must cancel, there still exist physical terms
$\sim \ln g$, and here $g$ has to be evaluated at the scale $\bar{\mu}$.
The cancellation of the $\bar{\mu}$-dependence holds
for the complete result for $p$, but this does not happen if one
terminates the perturbative expansion at some given order.
This is the reason why, for instance,
the $O(g^2)$ contribution to the pressure in Fig.~\ref{EOSpert} is
not flat. Here the curvature arises from the logarithmic running
of the strong coupling constant with the scale $\bar{\mu} \simeq
6.742\, T$.

The perturbative series (\ref{ppert}) converges badly.
The second-order term $\sim c_2 g^2$ gives a negative
contribution to the Stefan-Boltzmann
pressure, which is less than 10\% of $p_{\rm SB}$ at 
$T \sim 10^3 \, \Lambda_{\overline{MS}}$
and at most 40\% of $p_{\rm SB}$ 
at $T \sim \Lambda_{\overline{MS}}$ ($\sim T_c$).
However, the next contribution $\sim c_3 g^3$ is positive and so large
that the pressure overshoots $p_{\rm SB}$ up to the largest values of $T$ shown
in Fig.~\ref{EOSpert}. The terms of order $g^4$ are again 
small, but also positive, such that, to order $O(g^4)$, 
the pressure is larger than $p_{\rm SB}$.
The terms of order $O(g^5)$ are negative and so large in magnitude, 
that the pressure even vanishes at $T \simeq \Lambda_{\overline{MS}}$.
Thus, naive perturbation theory is clearly not applicable for
temperatures of order $T_c$. 

In the sections following Sec.~\ref{dimred},
several ways to improve the situation will be explained. All of them
are based on the observation that the {\em odd\/} powers of $g$ in the 
perturbative expansion (\ref{ppert})
are responsible for the bad convergence properties, i.e., the latter are
caused by the {\em infrared\/} properties of QCD.
Note that there have also been attempts to improve the convergence properties
of perturbation theory by using mathematical devices such
as Pad\'{e} approximates \cite{pade} and
Borel resummation \cite{borel}.
Here, I do not discuss these methods in more detail, because
the physical problem of improving the description of the infrared 
sector of QCD cannot be solved in this way.
For the sake of completeness, one should also mention Ref.~\cite{letessier},
where a phenomenological solution to the problem of convergence
of the perturbative series was presented.

\subsection{\it Dimensional Reduction} \label{dimred}

Consider a quantum field theory at nonzero temperature
in the limit $T \rightarrow \infty$.
In this limit, the Euclidean time interval in the partition function
(\ref{Z}) shrinks to zero, $1/T \rightarrow 0$. Consequently,
the original 3+1-dimensional theory 
reduces to a theory in three spatial dimensions.
This is called {\em dimensional reduction\/} \cite{appelquistpisarski}.
What are the degrees of freedom in the dimensionally reduced
theory? Recall that the compactification of the Euclidean time
interval $[0,1/T]$ at nonzero temperature leads to discrete
energies for the field modes, the so-called Matsubara frequencies
\cite{kapustaFTFT}. Bosonic degrees of freedom have periodic
boundary conditions in Euclidean time, and thus their Matsubara
frequencies are even multiples of $\pi T $,
$\omega_n^{\rm b} = 2 n \pi T$, $n = 0, \pm 1, \pm 2, \ldots$. 
On the other hand, fermionic degrees of freedom are
antiperiodic in Euclidean time, and consequently their
Matsubara frequencies are odd multiples of $\pi T$,
$\omega_n^{\rm f} = (2 n+1) \pi T$, $n = 0, \pm 1, \pm 2, \ldots$.
As $T \rightarrow \infty$, all modes with nonzero Matsubara
frequency become infinitely heavy, and are thus removed from
the spectrum of physical excitations. These are {\em all\/} fermionic
modes, and all {\em non-static\/} bosonic modes. Consequently,
dimensional reduction leads to a theory of static bosonic fields
in three spatial dimensions.

The dimensionally reduced theory can be viewed as an {\em effective\/} theory
at energy scales much less than the temperature. 
Consider, for example, QCD in weak coupling, $g \ll 1$, where there
is a distinct separation of energy scales, $g^2 T \ll g T \ll T$.
The dimensionally reduced theory is then the effective theory
for modes at energies of order $O(gT)$ which one obtains from
the underlying theory, i.e., QCD, by integrating out modes at
energy scales of order $O(T)$. One can then take this idea one step
further and integrate out modes at energies of order $O(gT)$ and
obtain an effective theory at an energy scale of order $O(g^2T)$.
In Ref.~\cite{braatennieto2} it was suggested to apply this
principle of constructing a series of effective theories
to compute the pressure in QCD. This task was recently carried out
to order $O(g^6)$ in a succession of papers \cite{schroder}. 
In the following, I outline the idea
and discuss the results, which are also shown in Fig.~\ref{EOSpert}.
Note that the idea of constructing an effective theory valid on
a certain energy scale has also been applied to non-Abelian
transport theories \cite{transport}. As transport theory
concerns non-equilibrium situations, a discussion
of these aspects are beyond the scope of the present review.

After the first step of integrating out modes at energy scales
of order $O(T)$, the pressure in QCD takes the form (at $\mu = 0$)
\begin{equation} \label{presseff1}
p(T) = p_T(T) + \frac{T}{V}\, \ln \left[ 
\int {\cal D} A_i^a \, {\cal D} A_0^a \, \exp \left( - S_E \right) 
\right]\,\, ,
\end{equation}
where $p_T(T)$ is the pressure of the modes at energy scales
of order $O(T)$ and the remaining term is the contribution
from modes at energy scales of order $O(gT)$. 
The argument of the logarithm is the partition function of
the effective theory for these modes. Since the energy scale
$gT$ is that of the Debye mass, $m_D = gT$,
which determines the screening length of static color-electric fields, 
quantities appearing in this partition function will be labelled
with a subscript ``{\it E\/}''.
The action $S_E$ of the effective theory is that of a
three-dimensional non-Abelian gauge theory (i.e., consisting of
the color-magnetic fields of the original theory) coupled to a Higgs 
field in the adjoint representation of the gauge group
(corresponding to the static color-electric fields of the original theory)
\cite{schroder},
\begin{eqnarray}
S_E & = & \int_V {\rm d}^3 {\bf x} \, {\cal L}_E \,\, , \\
{\cal L}_E & = & \frac{1}{2} \, {\rm Tr}\, {\cal F}_{ij}^2
 + {\rm Tr}\, [D_i, {\cal A}_0]^2 + m_E^2 \, {\rm Tr}\, {\cal A}_0^2
 + \lambda_E^{(1)} \, \left( {\rm Tr} \,{\cal A}_0^2 \right)^2
 + \lambda_E^{(2)} \, {\rm Tr} \,{\cal A}_0^4 + \ldots\,\, . 
\label{L_E}
\end{eqnarray}
Here, ${\cal F}_{ij} \equiv (i/g_E) [D_i,D_j] = F_{ij}^a T^a$, 
$D_i = \partial_i - i g_E {\cal A}_i$, and ${\cal A}_\mu \equiv A_\mu^a T^a$. 
There are five unknown quantities on the right-hand side of 
Eq.~(\ref{presseff1}): the pressure $p_T$, the mass
$m_E$ of the adjoint Higgs field ${\cal A}_0$ and the
coupling constants $g_E, \lambda_E^{(1)},\lambda_E^{(2)}$.
Their values have to be determined by ``matching'' the
effective theory to the original theory at some matching energy scale.
At this point, however, one can already determine their scaling behavior
from counting dimensions. The pressure $p_T$ is that of a 3+1-dimensional
theory of modes with momenta of order $O(T)$. 
Thus, at nonzero temperature (see above)
$p_T \sim T \int {\rm d} k\, k^2 \sim T^4$.
Since the action $S_E$ has to be dimensionless, one
can deduce the dimensionality of the fields ${\cal A}_i$ from
the kinetic term $\sim {\cal F}_{ij}^2$. As lengths have dimension
$T^{-1}$, the fields have to scale as ${\cal A}_i \sim T^{1/2}$.
The adjoint Higgs field must have the same dimension, 
${\cal A}_0 \sim T^{1/2}$. From
this one deduces that the mass term scales as $m_E \sim m_D = g T$, 
the coupling constant $g_E$ scales $\sim g T^{1/2}$, and
the four-point couplings behave as $\lambda_E^{(i)} \sim g^4 T$.
The dots in Eq.~(\ref{L_E}) denote higher-dimensional operators. One can
show by power counting that they are not relevant if one is
interested in a calculation of the pressure to order $O(g^6)$ \cite{schroder}.

The next step consists of integrating out modes at energy scales
of order $O(gT)$. Since physics at this scale is determined by static
color-electric fields, or in other words, by the adjoint Higgs field
${\cal A}_0$, one has to integrate out this field,
\begin{equation} \label{presseff2}
\frac{T}{V}\, \ln \left[ 
\int {\cal D} A_i^a \, {\cal D} A_0^a \, \exp \left( - S_E \right) 
\right] = p_E(T) + \frac{T}{V}\, \ln \left[ 
\int {\cal D} A_i^a \, \exp \left( - S_M \right) \right]\,\, .
\end{equation}
The term $p_E$ is the pressure of modes with energy of order
$O(gT)$. The argument of the logarithm on the right-hand side defines 
the partition function of an effective theory at energy scales
of order $O(g^2T)$. Since the energy scale $g^2T$ is associated with
the scale of the magnetic screening mass $m_M \sim g^2 T$ in 
non-Abelian gauge theories at nonzero temperature, 
quantities appearing in this partition function are labelled with 
a subscript ``{\it M\/}''.
The action $S_M$ entering the partition function of the effective
theory at an energy scale $O(g^2T)$ is simply that for
a three-dimensional non-Abelian field theory for color magnetic fields,
\begin{eqnarray}
S_M = \int {\rm d}^3 {\bf x}\, {\cal L}_M\,\, , \\
{\cal L}_M = \frac{1}{2}\, {\rm Tr}\, {\cal F}_{ij}^2 + \ldots\,\, ,
\label{L_M}
\end{eqnarray}
where ${\cal F}_{ij} \equiv (i/g_M) [D_i, D_j]$,
$D_i = \partial_i -i g_M {\cal A}_i$, and 
${\cal A}_i \equiv A_i^a T^a$. The two constants $p_E$, $g_M$ on the
right-hand side of Eq.~(\ref{presseff2}) have to be determined
by matching the effective theory at the energy scale $g^2T$ to that
at the energy scale $gT$. However, their scaling behavior can
already be determined by power counting. The pressure $p_E$ is 
again $\sim T \int {\rm d} k \, k^2$, but now the integral runs only
over modes with momenta of order $O(gT)$, thus $p_E \sim (g T)^3 T \sim
m_E^3 T$. The dimensionality of the fields ${\cal A}_i$ is the
same as in the previous effective theory, thus
$g_M \sim g T^{1/2} \sim g_E$. 
The dots in Eq.~(\ref{L_M}) denote higher-dimensional operators
which are again
irrelevant if one is interested in a computation of the pressure to
order $O(g^6)$.

The final step is to compute the pressure of modes with energies
of order $O(g^2T)$,
\begin{equation}
p_M(T) \equiv \frac{T}{V}\, \ln \left[ 
\int {\cal D} A_i^a \, \exp \left( - S_M \right) \right]\,\,.
\end{equation}
From power counting one deduces that
$p_M \sim T \int {\rm d}k\, k^2 \sim (g^2 T)^3 T \sim g^6 T^4 \sim g_M^6 T$, 
since the integral runs over modes with momenta of order $O(g^2T)$.
Due to the Linde problem, this contribution cannot be 
obtained perturbatively. What one can evaluate \cite{schroder} is the
contribution of order $O(g^6 \ln g)$ to $p_M$, since this arises
from ultraviolet divergences $\sim \ln(\bar{\mu}/{m_M})$ and not
from the nonperturbative infrared sector which yields a genuine
$O(g^6)$ contribution to the pressure.
The latter has to be evaluated e.g.\ via a lattice calculation.

The final answer for the pressure in QCD is then 
$p(T) = p_T(T) + p_E(T) + p_M(T)$. If one wants to
determine the pressure to order $O(g^6 \ln g)$, one 
has to compute all terms appearing to this order in $p_T$,
$p_E$, and $p_M$. This can be done perturbatively. The contributions
to $p_T$ constitute a power series in $g^2$, and not in $g$. 
They are needed explicitly only to order $O(g^4)$, since the 
full $O(g^6)$ contribution
to the pressure is nonperturbative in any case.
One then evaluates all four-loop diagrams in the
effective theory at scales $gT$, in order to determine
$p_E$ up to order $O(g^6 \ln g)$ \cite{schroder}.
As expected from power counting (see above), the lowest-order terms
in $p_E$ are $\sim m_E^3 T \sim g^3 T^4$.
Finally one adds everything to the $O(g^6 \ln g)$ term from
$p_M$. One obtains a well-defined expression
for the pressure up to order $O(g^6 \ln g)$.
The term which is genuinely of order $O(g^6)$ remains unknown.
The result for the pressure in QCD 
is then given by Eq.~(\ref{ppert}) with the $O(g^6)$ contribution
($N_c = 3, \, N_f = 0$)
\begin{equation}
c_6 = N_c^3\, \frac{ N_c^2 -1}{(4 \pi^4)} 
\left[ \left( \frac{215}{12} - \frac{805}{768}\,\pi^2 \right)
\,\ln \frac{1}{g} + 8 \, \delta \right]\,\, ,
\end{equation}
where $\delta$ is an unknown constant. 

In the right panel of Fig.~\ref{EOSpert} the result for the
pressure as a function of temperature is shown for various values of 
$\delta$. Comparing to lattice QCD data for the pure $[SU(3)_c]$ gauge
theory, the optimum value appears to be $\simeq 0.7$, since then
the perturbative calculation nicely matches onto the results from
the lattice computation, see left panel of Fig.~\ref{EOSpert}.
One also observes that for the optimum value, the pressure
up to $O(g^6)$ is rather close to the result to order $O(g^2)$, unless
the temperature is very close to $T_c$. This provides a certain amount
of confidence that this perturbative evaluation of the pressure
is reasonable. 

Finally, note that the above framework of constructing a sequence
of effective theories via dimensional reduction
was recently extended to include quark degrees of freedom at 
nonzero chemical potential \cite{vuorinen}.
At nonzero temperature, the quark Matsubara frequencies 
$\omega_n^{\rm f} = (2n+1) \pi T$ are always of order $O(T)$, consequently
quark degrees of freedom have to integrated out in the first step
(\ref{presseff1}) in the
construction of the sequence of effective theories.

\subsection{\it Quasiparticle Models} \label{quasiparticles}

In Sec.~\ref{dimred}, the pressure of QCD was computed by evaluating
the partition functions of various effective theories.
This considerably improved the somewhat unsatisfactory situation 
of a purely perturbative evaluation of the pressure up to terms
of order $O(g^5)$ as discussed in Sec.~\ref{pert}.
Another way to improve the situation is based on the following
observation.
The results of Sec.~\ref{heavyquarklatt} suggest that
nonperturbative effects still influence the physics
at temperatures in the range from $T_c$ to a few times $T_c$.
Consequently, a perturbative expansion of the pressure 
around the {\it perturbative\/} vacuum in terms of {\it massless\/} 
quarks and gluons seems inappropriate.
What is obviously missing in a perturbative description
of the QCD partition function are nonperturbative effects
which, when decreasing the temperature from $T > T_c$, are
responsible for the phase transition at $T_c$.
This was realized a while ago and attempts were made to
incorporate them into the properties of the
physical degrees of freedom.
In the following, I discuss two such attempts, the
so-called ``cut-off model'' and a model which treats 
quarks and gluons as massive quasiparticles.

\subsubsection{\it The cut-off model}

Let us again restrict the discussion to the pure $[SU(3)_c]$
gauge theory. The cut-off model is motivated by the fact that
QCD is an asymptotically free theory \cite{asympt}, i.e.,
only gluons with large momenta can be considered to be perturbative, 
while those with small momenta are subject to confinement.
Quite similar to the effective-theory approach discussed
in Sec.~\ref{dimred}, one then introduces a cut-off momentum
$\Lambda$ to separate these two regions \cite{cutoff1,cutoff2}.
Gluons with momenta larger than $\Lambda$ are treated perturbatively,
while gluons with momenta smaller than $\Lambda$ are assumed to 
remain bound inside colorless objects
(glueballs in the case of pure $[SU(3)_c]$ gauge theory).
The dispersion relation for perturbative gluons
then changes from the one for free massless particles, 
$\omega({\bf k}) = |{\bf k}| = k$ to
\begin{equation} \label{cut}
\omega({\bf k}) = \Theta(k-\Lambda)\, k\,\,.
\end{equation}
Gluons with momenta $k<\Lambda$ are bound inside glueballs. The
glueball mass scale $M$ is of the order of 1 GeV. The contribution
of glueballs to the thermodynamic functions is then exponentially suppressed
$\sim \exp(-M/T)$, and can thus be neglected for the temperature
range of interest.

The leading-order contribution to the pressure arises from
non-interacting gluons with momenta $k>\Lambda$,
\begin{equation}
p_0^{\rm cut}(T) = - 2(N_c^2-1)\, T \int \frac{{\rm d}^3{\bf k}}{(2 \pi)^3}\,
\Theta(k-\Lambda)\, \ln \left[1-\exp\left(-\frac{k}{T}\right) \right] \,\, .
\end{equation} 
One can also compute perturbative corrections to this leading-order
result. To this end, one has to evaluate the standard diagrams 
of the perturbative expansion of the pressure as discussed in 
subsection~\ref{pert}, but with additional theta-functions 
like in Eq.~(\ref{cut}) to restrict the phase space of the 
internal gluon lines. In Ref.~\cite{cutoff2} this has been done
up to order $O(g^2)$. Due to the restricted phase space in the loop
integrals, the perturbative corrections 
become relatively small compared to the zeroth-order contribution.
In this sense, the perturbative series for
particles with the dispersion relation~(\ref{cut})
is better behaved than the original perturbative series when 
the cut-off $\Lambda=0$.

Besides the cut-off $\Lambda$, the cut-off model has another
parameter, the MIT bag constant $B$ \cite{MIT},
which describes the energy difference
between the perturbative and the nonperturbative vacuum.
Fitting the parameters of the cut-off model to lattice QCD data for the pure
$[SU(3)_c]$ gauge theory, quite reasonable agreement could be obtained. 
I do not explicitly show results from
Ref.~\cite{cutoff2}, because lattice data at that time were not
yet extrapolated to the continuum limit. Consequently, the values for
$\Lambda$ and $B$ obtained previously will quantitatively
change once continuum-extrapolated data are used for the fit.

Nevertheless, on a qualitative level, the values for $\Lambda$ necessary to
fit the data were on the order a typical glueball mass $\Lambda \sim
M \sim 1$ GeV. This value for $\Lambda$ nicely confirms the consistency of
the assumption underlying the cut-off model, namely that
gluons with momenta $k < \Lambda \sim M$ are bound into glueballs
of mass $M$. Moreover, if one interprets the cut-off model as
an effective theory in the sense of Sec.~\ref{dimred},
the physics cannot depend on the precise value of $\Lambda$.
In other words, if one properly matches the effective theory 
for gluons with momenta $k > \Lambda$ to the effective theory for 
glueballs (i.e., gluons with momenta $k < \Lambda$), the cut-off 
would drop out.
This matching procedure has not been done for the cut-off model.
However, one may expect that a proper matching calculation
would just confirm the result $\Lambda \sim M$ obtained from
the fit to lattice QCD data.
One may thus simply replace the unphysical parameter 
$\Lambda$ by the physical value of the glueball mass $M$
to obtain a model which is independent of the arbitrary (and thus
unphysical) cut-off scale $\Lambda$.
The cut-off model has not been applied to lattice data for
full QCD, because then the assumption that all colorless
objects are heavy and are negligible when computing thermodynamic
functions breaks down (the pion mass is of the order of $T_c$).

The gluon dispersion relation (\ref{cut}) can be interpreted
in the way that gluons with momenta below the cut-off momentum $\Lambda$
have infinite mass, while those with momenta above $\Lambda$ have
zero mass. It is hard to believe that 
the true dispersion relation of gluons as computed in Yang-Mills theory 
would sustain the oversimplified and rather
radical assumptions of the cut-off model. 
A more conservative model to improve
our understanding of the thermodynamic properties of the QGP
is explained in the following section.

\subsubsection{\it Models with massive quasiparticles} \label{massiveglue}

In a hot and dense medium, particles attain a self-energy
$\Pi(\omega,{\bf k})$, which (due to the breaking of
Lorentz invariance) depends separately on energy $\omega$ and
3--momentum ${\bf k}$, as well as on the properties of the medium (i.e.,
its temperature $T$ and chemical potential $\mu$).
If the imaginary part of the self-energy on the dispersion
branch of the physical excitations is not too large
compared to the real part, these excitations are
called quasiparticles \cite{fetwal}. The simplest situation
is when the self-energy is independent of energy and momentum, i.e.,
constant and real, corresponding to a mass term, 
which depends on $T$ and $\mu$, but not on the energy and the momentum
of the particle. In this case, the dispersion relation for 
quasiparticles of mass $m$ reads
\begin{equation} \label{massdisp}
\omega({\bf k}) = \sqrt{{\bf k}^2 + m^2}\,\,.
\end{equation}

The self-energies of quarks and gluons in QCD are certainly
not constant (see also Sec.~\ref{HTLresum}). Nevertheless, 
one can still simply {\it assume\/} that they are constant
and explore the consequences.
These so-called ``massive quasiparticle models''
have been investigated in great detail in the literature
as a means to describe and interpret lattice QCD data on 
thermodynamic functions of QCD \cite{satz,rischkePR,peshier1,levai,schneider}.
The advantage of these models is that it is straightforward
to extend them to nonzero quark chemical potential \cite{peshier2}.

Here, as in the previous section
let us only focus on a quasiparticle model for pure $[SU(3)_c]$ gauge
theory. For the generalization to QCD with dynamical quarks,
see Refs.~\cite{levai,schneider,peshier2}. 
For the pure gauge theory, there are only massive gluon degrees of freedom
and the pressure reads
\begin{equation} \label{pmqp}
p^{\rm mqp}(T) = - D\, T \int \frac{{\rm d}^3{\bf k}}{(2 \pi)^3}\,
\ln \left[1-\exp\left(-\frac{\omega({\bf k})}{T}\right) \right] 
- B\,\, ,
\end{equation}
with $\omega({\bf k})$ given by Eq.~(\ref{massdisp}). As in the
cut-off model, an MIT bag constant is introduced to account for
the difference between the perturbative and the nonperturbative
vacuum. 

Two comments regarding the pressure (\ref{pmqp}) are in order.
First, while massless gluons have two transverse polarizations,
massive gluons have an additional longitudinal polarization degree of 
freedom. Therefore, one would naively argue that
the constant $D$ in Eq.~(\ref{pmqp})
should assume the value $D= 3 (N_c^2 -1)$, instead of
$D= 2 (N_c^2 -1)$, as for massless gluons.
This is, however, not quite correct.
First, the pressure (\ref{pmqp}) has to approach
the correct Stefan-Boltzmann limit when $T \rightarrow \infty$.
For three polarization degrees of freedom, the Stefan-Boltzmann
pressure is a factor $3/2$ larger than the correct value.
Another reason why this is incorrect is the following. In
Sec.~\ref{HTLresum} we shall see that gluons indeed 
acquire a longitudinal degree of freedom in a hot or dense medium, but
that the respective longitudinal spectral density 
vanishes rapidly for energies and momenta larger than $gT$. 
Thus, in a calculation of the pressure, which is
dominated by modes with momenta of order $T$, one should 
in principle not count the longitudinal degrees of freedom.
We shall therefore set $D= 2 (N_c^2 -1)$ in the following.
(Note that Ref.~\cite{levai} also investigated a scenario where
$D$ is a function of temperature.)

\begin{figure}[tb]
\begin{center}
\begin{minipage}[t]{8 cm}
\epsfig{bbllx=105,bblly=200,bburx=600,bbury=700,
file=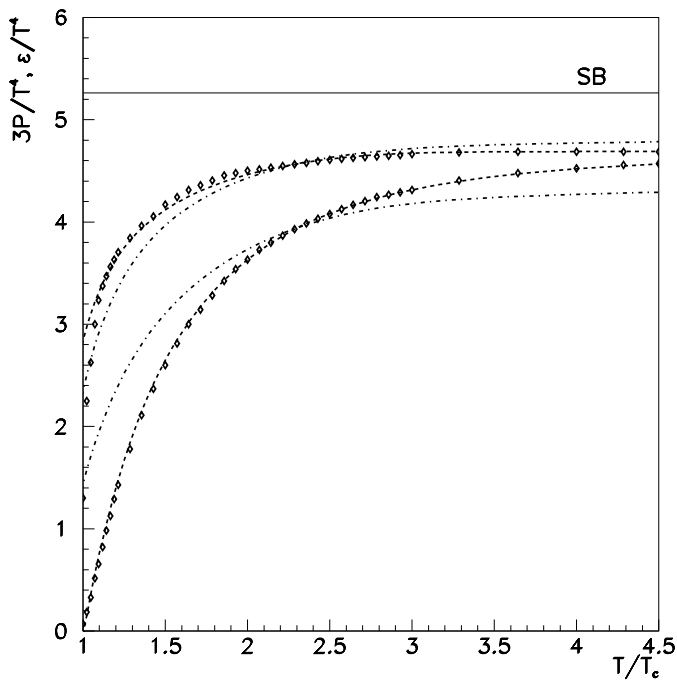,height=90mm}
\end{minipage}
\begin{minipage}[t]{16.5 cm}
\caption{The temperature dependence of
$3\,p/T^4$ and $\epsilon/T^4$ in pure $[SU(3)_c]$ gauge theory. 
Symbols are lattice QCD data from \cite{boyd}. The dashed lines
are a fit within the massive gluon model. The dash-dotted lines
represent the contribution of the kinetic term in Eq.~(\ref{pmqp}).
The horizontal line is the Stefan-Boltzmann limit. (Since
for an ultrarelativistic ideal gas
$\epsilon \equiv 3\, p$, this limit is
the same for the functions $3p/T^4$ and $\epsilon/T^4$.)
From Ref.~\cite{levai}.
\label{levaifit}}
\end{minipage}
\end{center}
\end{figure}

The second comment concerns the possibility
to fit lattice QCD data with Eq.~(\ref{pmqp}). It turns out that
for constant mass and bag parameters, the quality of the fit is
not satisfactory. Consequently, one needs to generalize
the model (\ref{pmqp}) to allow for a temperature-dependent
gluon mass, $m \rightarrow m(T)$. 
A convenient parametrization is motivated by the dispersion
relation for transverse gluons at large momenta, which takes the form
$\omega_t(k) = \sqrt{k^2 + m_{t\,\infty}^2}$, see Eq.~(\ref{transglueasym}).
Consequently,
\begin{equation}
m(T) = \frac{g(T)\, T}{\sqrt{2}}\,\, , \;\;\;\;
g^2(T) = \frac{8 \pi^2}{ 11\, \ln 
\left[ F(T/T_c,T_c/\Lambda_{\overline{MS}}) \right] }\,\, ,
\;\;\;\;
F(x,y) = K(x)\, x \, y\,\,.
\end{equation}
For a fixed value of $T_c/\Lambda_{\overline{MS}} \simeq 1.03 \pm 0.19$
for pure $[SU(3)_c]$ gauge theory \cite{fingberg},
one only needs to know the function $K(T/T_c)$ in order to
determine the mass, $m(T)$, and thus the kinetic term in
the pressure (\ref{pmqp}). In Ref.~\cite{levai} the function $K(T/T_c)$
is simply fit to reproduce lattice QCD data. A surprisingly good
fit is obtained with the functional form 
$K(x) = 18/[18.4\, \exp(-0.5\,x^2) +1]$, see Fig.~\ref{levaifit}.
Once the gluon mass is a function of temperature, 
thermodynamical consistency requires
the bag parameter to depend on the temperature as well, $B \rightarrow B(T)$.
The functional form of $B(T)$ can be uniquely determined from
$m(T)$; for details see Ref.~\cite{levai}.

\subsection{\it HTL/HDL-Resummed Perturbation Theory} \label{HTLresum}

Apparently, the idea that quarks and gluons are quasiparticles 
works rather well to describe the thermodynamic functions of the QGP.
Therefore, it seems appropriate to put this concept onto a
more formal basis. In fact, the quasiparticle excitations in the QGP are well 
known in the weak-coupling limit, $g \ll 1$,
and for temperatures and/or chemical potentials much larger than the
quasiparticles' energies and momenta.
At nonzero temperature, these quasiparticles form the basis of the
so-called ``hard thermal loop'' (HTL-) resummation scheme 
\cite{lebellac,braatenpisarski}. 
At zero temperature, but large quark
chemical potential, there is an equivalent approach,
the so-called ``hard dense loop'' (HDL-) resummation scheme
\cite{lebellac,HDL}. 
From the quasiparticle excitation spectrum, one can also
construct the equation of state. All this will be discussed in detail
in the following.

\subsubsection{\it The excitation spectrum in a hot and dense medium}
\label{HTLexcite}

How does one determine the spectrum of physical excitations
in a hot and/or dense medium? I shall illustrate this explicitly
for the case of gluons. The case of quarks can be considered analogously,
I only briefly report the results at the end of this section.
The outline of the procedure is the following. First, one
computes the 
gluon self-energy, $\Pi^{\mu \nu}_{ab} (\omega, {\bf p})$.
From the self-energy, one then determines the full gluon propagator,
$\Delta^{\mu \nu}_{ab} (\omega, {\bf p})$. 
From the gluon propagator, one then deduces the spectral density,
which provides all information about the excitation spectrum
in a hot and/or dense medium.
Quite surprisingly, it turns out that one can follow 
this procedure in complete generality, without
actually specifying $\Pi^{\mu \nu}_{ab}$ until the very end.
Although the derivation is somewhat formal, 
it is nevertheless a rather instructive exercise and will
therefore be discussed in more detail \cite{rischkeshovkovy}.
Note that, in a medium, Lorentz symmetry is explicitly broken, and
all quantities depend separately on energy $\omega$ and momentum
${\bf p}$. Nevertheless, to abbreviate the notation, I shall
frequently use the 4-vector $P^{\mu} \equiv (\omega, {\bf p})$ to characterize
this dependence. Note also that, at nonzero temperature, one usually
computes in Euclidean space time, i.e., all energies are 
discrete Matsubara frequencies on the {\em imaginary\/}
energy axis. However, in order to determine the
physical excitation spectrum, one has to analytically continue to 
{\em real\/} energies, $i \omega_n \rightarrow \omega + i \eta$. 
This $i \eta$ prescription produces the {\em retarded\/} Greens functions
\cite{fetwal}. When writing real energies $\omega$ in the following,
the $i \eta$ prescription will be suppressed.

The self-energy $\Pi^{\mu \nu}_{ab}$
can be decomposed according to its tensor structure.
First of all, if the color $[SU(3)_c]$ gauge symmetry is not
broken, one may assume that the gluon self-energy is diagonal in adjoint color,
$\Pi^{\mu \nu}_{ab} \equiv \delta_{ab}\, \Pi^{\mu \nu}$.
(In a color superconductor, this is in general no longer the case,
for examples see Sec.~\ref{gluephoton}.) It thus suffices to consider
$\Pi^{\mu \nu}$.
The next step is to decompose $\Pi^{\mu \nu}$ in terms
of tensors, multiplied by scalar functions of $\omega$ and
${\bf p}$ \cite{lebellac,heinz}. Let us define
\begin{equation}
{\rm E}^{\mu \nu} \equiv \frac{P^\mu \, P^\nu}{P^2}\
\end{equation}
as the projector onto the subspace parallel to $P^{\mu}$.
Then one chooses a vector orthogonal to $P^\mu$, for instance
\begin{equation}
N^\mu \equiv \left( \frac{\omega\, p^2}{P^2}, \frac{\omega^2\, {\bf p}}{P^2}
\right) \equiv \left(g^{\mu \nu} - {\rm E}^{\mu \nu}\right)\, f_\nu\,\, ,
\end{equation}
with $f^\mu = (0,{\bf p})$. Now one defines the tensors
\begin{equation} \label{BCA}
{\rm B}^{\mu \nu} = \frac{N^\mu\, N^\nu}{N^2}\,\,\,\, , \,\,\,\,\,
{\rm C}^{\mu \nu} = N^\mu \, P^\nu + P^\mu\, N^\nu \,\,\,\, , \,\,\,\,\,
{\rm A}^{\mu \nu} = g^{\mu \nu} - {\rm B}^{\mu \nu} - {\rm E}^{\mu \nu} \,\, .
\end{equation}
With the help of these tensors one can write the gluon self-energy as
\begin{equation} \label{decomp}
\Pi^{\mu \nu} = \Pi^{\rm a} \, {\rm A}^{\mu \nu}
+ \Pi^{\rm b} \, {\rm B}^{\mu \nu} + \Pi^{\rm c}\, 
{\rm C}^{\mu \nu} + \Pi^{\rm e}\, {\rm E}^{\mu \nu}\,\, .
\end{equation}
The scalar functions $\Pi^{\rm a,b,c,e}$ can be obtained by
suitable projections of $\Pi^{\mu \nu}$ onto the respective
tensor structures. 

Using the explicit form of $N^\mu$, one convinces oneself
that the tensor ${\rm A}^{\mu \nu}$ projects onto the spatially transverse
subspace orthogonal to $P^\mu$, 
\begin{equation}
{\rm A}^{00} = {\rm A}^{0i}=0\,\,\,\, , \,\,\,\,\,
{\rm A}^{ij} = - \left(\delta^{ij} - \hat{p}^i \, \hat{p}^j \right)\,\, .
\end{equation}
This means that the self-energy function $\Pi^{\rm a}$ determines
the excitation spectrum of the spatially transverse gluon fields
\begin{equation}
{A_\perp}^a_\mu(P) \equiv {\rm A}_{\mu}^{\nu}\, A_\nu^a(P)\,\, .
\end{equation}
As ${\rm A}^{\mu \nu}$
projects onto a two-dimensional subspace, there are two degrees of freedom
associated with ${A_\perp}^a_\mu$. In the vacuum, these are the only
physical degrees of freedom, since gluons are massless.

The tensor ${\rm B}^{\mu \nu}$ projects onto the spatially
longitudinal subspace orthogonal to $P^\mu$,
\begin{equation}
{\rm B}^{00} = - \frac{p^2}{P^2} \,\,\,\, , \,\,\,\,\,
{\rm B}^{0i} = - \frac{\omega\, p^i}{P^2}\,\,\,\, ,\,\,\,\,
{\rm B}^{ij} = - \frac{\omega^2}{P^2}\,\hat{p}^i\,\hat{p}^j\,\, .
\end{equation}
Consequently, the polarization function $\Pi^{\rm b}$ determines
the excitation spectrum of the longitudinal gluon degree of freedom, 
\begin{equation}
A_N^a(P) \equiv \frac{N^\mu\, A_\mu^a (P)}{ N^2}\,\,,
\end{equation} 
which becomes a physical degree of freedom in a medium.

The spatially transverse and spatially longitudinal gluon fields
${A_\perp}^a_\mu$ and $A_N^a$ 
are the only physical degrees of freedom.
There is an unphysical degree of freedom associated with
the projection onto $P^\mu$, 
\begin{equation}
A_\parallel^a(P) \equiv 
\frac{P^\mu \, A^a_\mu (P)}{P^2}\,\,.
\end{equation}
The scalar function
$\Pi^{\rm e}$ is the self-energy of this unphysical degree of 
freedom. It will be seen that gauge fixing ultimately removes this
degree of freedom from the theory.
A nonvanishing $\Pi^{\rm c}$ indicates that the 
spatially longitudinal, physical gluon degree of freedom 
$A^a_N$ mixes with the unphysical degree of freedom $A^a_\parallel$. 
Before extracting the physical excitation spectrum, one has to 
remove this mixing term, as will be discussed below.

Now use the tensor decomposition (\ref{decomp}) to determine
the full gluon propagator $\Delta^{\mu \nu}_{ab}$. Since 
the free inverse gluon propagator
\begin{equation}
{\Delta_0^{-1}}^{\mu \nu}_{ab}  \equiv
\delta_{ab} (P^2 g^{\mu \nu} - P^{\mu} P^{\nu})
\equiv \delta _{ab} \, P^2 ({\rm A}^{\mu \nu} + {\rm B}^{\mu \nu})
\end{equation}
is diagonal in adjoint colors, so is the full inverse gluon
propagator, ${\Delta^{-1}}^{\mu \nu}_{ab} \equiv \delta_{ab}\,
{\Delta^{-1}}^{\mu \nu}$. One obtains
\begin{equation}
{\Delta^{-1}}^{\mu \nu} 
\equiv {\Delta_0^{-1}}^{\mu \nu} + \Pi^{\mu \nu}
\equiv \left( P^2 + \Pi^{\rm a} \right)\, {\rm A}^{\mu \nu}
+ \left( P^2 + \Pi^{\rm b} \right) \, {\rm B}^{\mu \nu}
+ \Pi^{\rm c}\,{\rm C}^{\mu \nu} + \Pi^{\rm e}\, {\rm E}^{\mu \nu}\,\, .
\end{equation}
In an effective action, this inverse propagator
is the coefficient of the term quadratic in the gauge fields.
In momentum space (and choosing a normalization such that
$A_\mu^a(P)$ retains dimensions of energy),
\begin{eqnarray}
S_2 & = &  -\frac{1}{2}\, \frac{V}{T} \sum_P \sum_{a=1}^{N_c^2-1}
\left\{ \frac{}{} A_{\perp\, \mu}^a(-P)\, \left[ P^2 + \Pi^{\rm a}(P)
\right]\, {\rm A}^{\mu \nu}\, A_{\perp\,\nu}^a(P)  
- A_N^a(-P) \, \left[ P^2 + \Pi^{\rm b}(P) \right] \, N^2 \, A_N^a(P) 
\right.  \nonumber \\
&   & \hspace*{2.6cm} - \, A_\parallel^a(-P)\,
\Pi^{\rm c}(P) \,N^2 P^2 \, A_N^a(P) 
- A_N^a(-P) \, \Pi^{\rm c}(P) \,N^2 P^2 \, A_\parallel^a(P) 
\nonumber \\
&   & \hspace*{2.6cm}
 - \left. A_\parallel^a(-P)\, \Pi^{\rm e}(P) \, P^2 \, 
A_\parallel^a(P) \frac{}{} \right\}\,\,, \label{S2}
\end{eqnarray}
where $\sum_P \equiv (V/T) \int_P$.
The physical excitation spectrum can be most easily extracted
from a diagonalized inverse gluon propagator, i.e., 
the term which mixes the physical field component
$A_N^a$ with the unphysical component $A_\parallel^a$ has to be
eliminated. This can be done as follows. Remember that
in the partition function one functionally integrates the exponential
of the action (a part of which is $S_2$ in Eq.~(\ref{S2}))
over all gauge field components, including $A_\parallel^a$. 
For the purpose of diagonalizing the inverse gluon propagator
one may simply redefine the integration variable
\begin{equation}
A_\parallel^a(P) \rightarrow \hat{A}^a_\parallel (P)
\equiv A_\parallel^a(P) + 
\frac{\Pi^{\rm c}(P)\,  N^2}{\Pi^{\rm e}(P)} \, A_N^a(P)\,\, .
\end{equation}
This redefinition does not change the physics (as it only
involves the unphysical component of the gauge field) and
diagonalizes the action $S_2$ in the components of the gauge field,
\begin{eqnarray}
S_2 & = &  -\frac{1}{2}\, \frac{V}{T} \sum_P \sum_{a=1}^{N_c^2-1}
\left\{ \frac{}{} A_{\perp\, \mu}^a(-P)\, \left[ P^2 + \Pi^{\rm a}(P)
\right]\, {\rm A}^{\mu \nu}\, A_{\perp\,\nu}^a(P)  
- A_N^a(-P) \, \left[ P^2 + \hat{\Pi}^{\rm b}(P) \right] \, N^2 \, A_N^a(P) 
\right.  \nonumber \\
&   & \hspace*{2.6cm}
 - \left. \hat{A}_\parallel^a(-P)\, \Pi^{\rm e}(P) \, P^2 \, 
\hat{A}_\parallel^a(P) \frac{}{} \right\}\,\,, \label{S2diag}
\end{eqnarray}
where
\begin{equation}
\hat{\Pi}^{\rm b}(P) \equiv \Pi^{\rm b}(P) - \frac{\left[ \Pi^{\rm c}(P)
\right]^2 \, N^2 \, P^2 }{\Pi^{\rm e}(P)}\,\,.
\end{equation}
From Eq.~(\ref{S2diag}) one can read off the inverse gluon propagator
in diagonal form. However, in order to be able to invert it, it is
necessary to fix the gauge. (To see this, consider the case where
$\Pi^{\rm e}=0$. Then, the inverse propagator has a zero eigenvalue 
and is not invertible). 
For the sake of simplicity, let us choose
covariant gauge, where the gauge fixing term in the action 
only involves the unphysical components of the gauge field,
\begin{equation}
S_{\rm gf} = \frac{1}{2 \, \lambda} \, \frac{V}{T} \sum_P \sum_{a=1}^{N_c^2-1}
\hat{A}_\parallel^a(-P) \, P^2 \, P^2 \, \hat{A}_\parallel^a(P) \,\,.
\end{equation}
Adding $S_{\rm gf}$ to $S_2$, one reads off the (gauge-fixed)
inverse gluon propagator 
\begin{equation}
{\Delta^{-1}}^{\mu \nu}(P) = 
\left[ P^2 + \Pi^{\rm a}(P) \right]\, {\rm A}^{\mu \nu}
+ \left[ P^2 + \hat{\Pi}^{\rm b}(P) \right] \, {\rm B}^{\mu \nu}
+ \left[ \frac{1}{\lambda} \, P^2 + \Pi^{\rm e}(P) \right]\,
{\rm E}^{\mu \nu}\,\, ,
\end{equation}
which can be straightforwardly inverted, since
${\rm A}^{\mu \nu}$, ${\rm B}^{\mu \nu}$, and ${\rm E}^{\mu \nu}$ are
projectors,
\begin{equation} 
\label{glueprop}
\Delta^{\mu \nu} (P) = \Delta_t(P)\,
{\rm A}^{\mu \nu} - \Delta_\ell(P) \frac{p^2}{P^2}
\, {\rm B}^{\mu \nu} + \frac{\lambda}{P^2 + \lambda\,
\Pi^{\rm e}(P)} \, {\rm E}^{\mu \nu}\,\,,
\end{equation}
where the transverse and longitudinal propagators are defined as
\begin{equation} \label{translongprop}
\Delta_t(P) \equiv \frac{1}{P^2 + \Pi^{\rm a}(P)}
\equiv \frac{1}{P^2 - \Pi_t(P)} \,\, , \;\;\;\;
\Delta_\ell(P) \equiv - \frac{P^2}{p^2} \, \frac{1}{
P^2 + \hat{\Pi}^{\rm b}(P)} = - \frac{1}{p^2 - \Pi_{\ell}(P)} \,\, ,
\end{equation}
with the transverse and
longitudinal polarization functions
$\Pi_t \equiv - \Pi^{\rm a}$ and 
$\Pi_{\ell} \equiv - (p^2/P^2) \hat{\Pi}^{\rm b}$.
The last term in Eq.~(\ref{glueprop}) can be removed from the spectrum
of physical excitations by the gauge choice $\lambda = 0$.
The physical excitations are described by the transverse and
longitudinal propagators defined in Eq.~(\ref{translongprop}).
From these one can derive the corresponding spectral densities in the usual way
\cite{lebellac}
\begin{equation} \label{specdens}
\rho_t (\omega,{\bf p}) \equiv \frac{1}{\pi}\, {\rm Im} 
\, \Delta_t(\omega + i \eta, {\bf p}) \,\, , \;\;\;\;
\rho_\ell (\omega,{\bf p}) \equiv \frac{1}{\pi}\, {\rm Im} 
\, \Delta_\ell(\omega + i \eta, {\bf p})
\,\, .
\end{equation}

\begin{figure}[tb]
\begin{center}
\begin{minipage}[t]{15 cm}
\epsfig{file=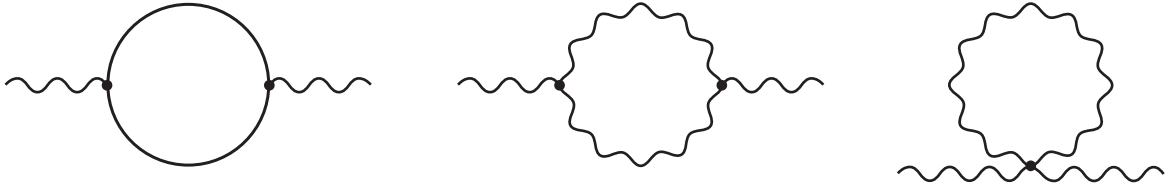,scale=1}
\end{minipage}
\begin{minipage}[t]{16.5 cm}
\caption{The one-loop diagrams contributing to the
gluon self-energy (possible ghost contributions are not shown).
\label{gluonselfenergy}}
\end{minipage}
\end{center}
\end{figure}

All that is left is to actually specify the form of the
polarization functions $\Pi_t$ and $\Pi_\ell$.
In the weak-coupling limit, $g \ll 1$, this can be done via
a one-loop calculation of these functions.
To this end, one has to compute the diagrams shown in
Fig.~\ref{gluonselfenergy} using standard methods \cite{kapustaFTFT,lebellac}.
Note, however, that the result depends in general
on the choice of gauge and thus cannot determine the physical
excitation spectrum, which is by definition
independent of the choice of gauge.
However, it was noticed many years ago \cite{weldon}, that the
high-temperature limit $T \gg \omega, p$ 
of the polarization functions is actually independent
of the choice of gauge,
\begin{eqnarray}
{\rm Re}\, \Pi_t(\omega,{\bf p}) & = & \frac{3}{2}\, m_g^2\, \left[
\frac{\omega^2}{p^2} + \left( 1- \frac{\omega^2}{p^2} \right) \,
\frac{\omega}{2\, p} \, \ln \left| \frac{\omega + p}{\omega - p} \right| \,
\right] \,\, , \label{RePit} \\
{\rm Im}\, \Pi_t(\omega,{\bf p}) & = & - \pi\, \frac{3}{4}\, m_g^2 \, 
\frac{\omega}{p}\, \left(1- \frac{\omega^2}{p^2} \right) \, 
\Theta(p-\omega)\,\, ,
\\
{\rm Re}\, \Pi_\ell(\omega,{\bf p}) & = & - 3\, m_g^2\, \left( 1-
\frac{\omega}{2\, p} \, \ln \left| \frac{\omega + p}{\omega - p} \right| \,
\right) \,\, , \\
{\rm Im}\, \Pi_\ell(\omega,{\bf p}) & = & - \pi\, \frac{3}{2}\, m_g^2 \, 
\frac{\omega}{p} \, \Theta(p-\omega)\,\, , \label{ImPil}
\end{eqnarray}
where the gluon mass parameter at a given $T$ and $\mu$ is
\begin{equation} \label{m_g}
m_g^2 = g^2 \, \left[ \frac{2\,N_c + N_f}{18}\, T^2
+  \frac{N_f}{6\,\pi^2}\, \mu^2 \right]\,\, .
\end{equation}
This result is not restricted to the high-temperature
limit, in fact it holds as long as either $T$ or $\mu$ is
much larger than $\omega$ and $p$. Thus, it
also describes gluonic quasiparticle excitations
at $T=0$ and high density. It is only this gauge-invariant
high-$T$ (or high-$\mu$) limit of the one-loop polarization functions,
which is relevant for the HTL- (or HDL-) resummation 
scheme discussed in Sec.~\ref{HTLHDLresum}.

What is the physical meaning of the result (\ref{RePit}) -- (\ref{ImPil})?
The condition $T,\, \mu \gg \omega,\, p$ implies that there
is a separation of scales, just like in the construction of the
effective theories in Sec.~\ref{dimred}. The temperature
(or the chemical potential) sets a ``hard'' energy scale, while
the external energy and momentum are ``soft''. As will be seen
in Sec.~\ref{HTLHDLresum}, this separation
of scales forms the basis of the HTL- (or HDL-) resummation scheme.
In this scheme, energies of order $gT$ (or $g \mu$,
in the HDL-resummation scheme) are ``soft'', while energies of
order $T$ (or $\mu$, respectively) are ``hard''.
In the loops in Fig.~\ref{gluonselfenergy}, one integrates over the
internal loop momentum, ${\bf k}$,
but the result must be finite, as there are no other 
ultraviolet divergences at nonzero $T$ and/or $\mu$ than those 
already known from the vacuum. 
Therefore, the ultraviolet regularization must be
provided by the distribution functions of quarks and gluons, which
decrease exponentially with temperature. (At $T=0$ and $\mu \neq 0$,
the gluon distribution function vanishes, while the quark
distribution function is a step function $\sim \Theta(\mu - k)$, which
cuts off momenta $k > \mu$). On the other hand, the
phase space in the loop integral grows $\sim \int {\rm d}k k^2$.
One thus expects that
the {\em dominant\/} contribution to the loop integral comes from
``hard'' momenta of order $T$ (or, at $T=0$ and $\mu \neq 0$, from
momenta close to the Fermi surface, $k \sim \mu$).
For dimensional reasons, and including factors of $g$ from
the vertices, $\Pi \sim g^2 T^2$ (or $ \sim g^2 \mu^2$,
at $T=0$ and $\mu \neq 0$). This gives rise to the prefactor
$\sim m_g^2$ in Eqs.~(\ref{RePit}) -- (\ref{ImPil}).
The ``soft'' external energy and momentum cannot
significantly alter the kinematics in the loop, where the
dominant contribution comes from ``hard''  momenta. 
In fact, it suffices to expand the integrands of the loop integrals 
to leading order in these external 
quantities. This gives rise to the particular dependence on $\omega$ and $p$
of the result (\ref{RePit}) -- (\ref{ImPil}).
The essential approximation which leads to this result is
the assumption that internal momenta are exclusively ``hard''. Therefore,
the loops computed under this assumption are called
``{\em hard thermal loops\/}'' (or ``{\em hard dense loops\/}'', 
at $T=0$ and $\mu \neq 0$).

\begin{figure}[tb]
\begin{center}
\begin{minipage}[t]{13.5 cm}
\epsfig{file=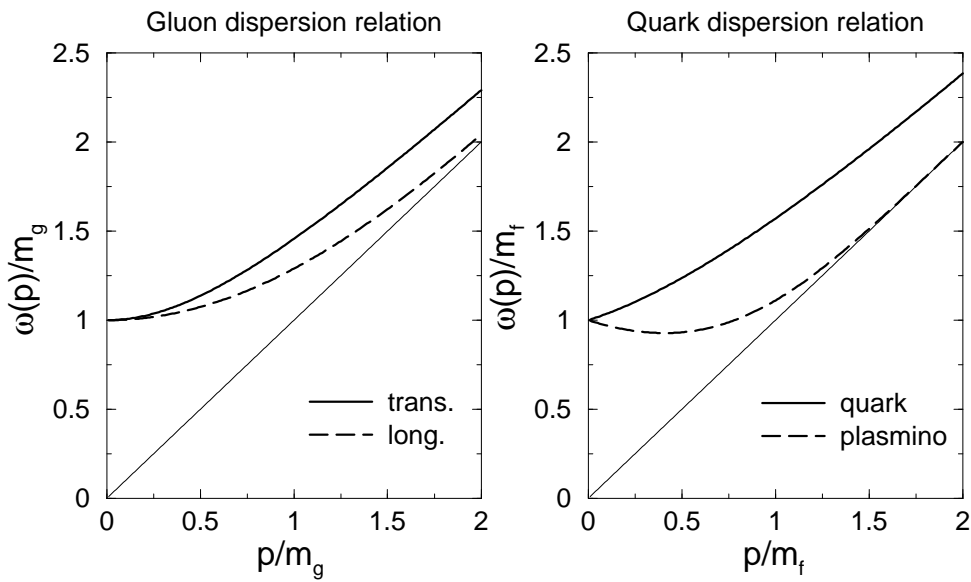,scale=0.7}
\end{minipage}
\begin{minipage}[t]{16.5 cm}
\caption{The excitation branches for gluons (left panel) 
and quarks (right panel) in a hot and/or dense medium.
For gluons, energy and momenta are shown in units of the
gluon mass parameter, $m_g$. For quarks, energy and momenta
are shown in units of the fermion mass parameter, $m_f$.
\label{excite}}
\end{minipage}
\end{center}
\end{figure}

From the self-energies one can construct the propagators 
(\ref{translongprop}) and the spectral densities (\ref{specdens}), 
respectively. Following Ref.~\cite{pisarskiphysica} it turns out
that the propagators have
poles above the light-cone, $\omega > p$, and a cut below, $\omega < p$.
In the spectral densities, the poles become $\delta$-functions.
These determine the excitation branches $\omega(p)$. Since
a $\delta$-function has no width, the quasiparticles corresponding
to these excitation branches are {\em stable}, i.e., they
have an infinite lifetime. (This changes if one computes
beyond one-loop order.) The excitation branches are above
the light-cone, i.e., they correspond to time-like, propagating gluons. 
In the left panel of Fig.~\ref{excite} they are shown for 
transverse and longitudinal gluon modes.
For large momenta, the longitudinal mode has an exponentially
vanishing residue \cite{pisarskiphysica}. In contrast, the
transverse mode has a finite residue and 
a dispersion relation which approaches the form
\begin{equation} 
\label{transglueasym}
\omega_t(p) \rightarrow \sqrt{p^2 + m_{t\, \infty}^2}\,\, , \;\;\;\;
m_{t \, \infty}^2 \equiv \frac{3}{2}\, m_g^2\,\, .
\end{equation}
The cuts in the propagator 
become continuous distributions in the spectral density.
They provide Landau-damping for space-like gluons.

Performing a similar exercise for quarks \cite{weldon2},
one obtains the spectrum of fermionic quasiparticle excitations.
It turns out that there are twice as many excitation branches as
expected. There are in fact {\em two\/} solutions for positive energies,
$\omega_\pm (p)$, determined by the equation
\begin{equation}
\omega_\pm (p) = \pm p \pm \frac{m_f^2}{p}
\, \left[ 1 - \frac{\omega_\pm(p) \mp p}{2\, p}\, 
\ln \left( \frac{\omega_\pm(p) + p}{\omega_\pm(p) - p} \right) \right]
\,\, ,
\end{equation} 
where the fermionic mass parameter (squared) is
\begin{equation}
m_f^2 = g^2 \ \frac{N_c^2-1}{16 \, N_c} \, \left( T^2 + \frac{\mu^2}{\pi^2}
\right)\,\, .
\end{equation}
These two solutions are shown in the right panel of
Fig.~\ref{excite}. (The two solutions for negative energies
mirror the above solutions below the $p$ axis.)
The solution $\omega_-(p)$ corresponds to a quasiparticle with the opposite
chirality than the one associated with the solution $\omega_+(p)$.
This peculiar quasiparticle is commonly called
the ``plasmino''. Note that, while the ordinary quasiparticle
dispersion branch $\omega_+(p)$ has
a positive group velocity ${\rm d} \omega_+(p)/ {\rm d} p >0$,
the plasmino branch has negative group velocity at small momenta. 
For large momenta, the residue of the plasmino branch becomes
exponentially small, while the one for the ordinary quasiparticle
remains finite and its dispersion relation approaches the form
\begin{equation}
\omega_+ (p) \rightarrow \sqrt{p^2 + m_{f \, \infty}^2}\,\, , \;\;\;\;
m_{f\, \infty}^2 = 2 \, m_f^2\,\, .
\end{equation}
The quark propagator also features a cut below the light-cone,
which gives rise to Landau damping.
Finally, note that the fermionic quasiparticle spectrum 
in Fig.~\ref{excite} is shown for energies and momenta much smaller
than either temperature and/or chemical potential. 
However, at $T=0$ and large $\mu$, this kinematic region is irrelevant,
as it reflects the situation at the bottom of the Fermi sea.
At $T=0$ and large $\mu$, the {\em relevant\/} 
fermionic excitations are those around the Fermi surface, where $p \simeq \mu$.
The excitation spectrum for this case 
will be discussed in more detail in Sec.~\ref{secexspec}.

\subsubsection{\it HTL/HDL-resummation scheme} \label{HTLHDLresum}

The HTL-resummation scheme and its counterpart at $T=0$ and
nonzero $\mu$, the HDL-resummation scheme, are explained in
great detail in textbooks on field theory at nonzero
temperature and density (see, for instance, Ref.~\cite{lebellac}).
Therefore, for more details, I refer the interested reader to the literature 
and restrict myself here to a short discussion of
the general idea behind these methods.

As already discussed in Sec.~\ref{pert}, 
due to the infrared behavior of gauge theories at
nonzero temperature, naive perturbation theory breaks down.
We have seen in Sec.~\ref{HTLexcite} that loop calculations
involve propagators of the form
\begin{equation} \label{scalarprop}
\Delta(\omega,{\bf p}) \sim  \frac{1}{\omega^2 - p^2 + 
\Pi(\omega,{\bf p})}\,\, .
\end{equation}
The leading-order terms 
in the one-loop self-energy arise from
``hard'' particles with momenta $k \sim T$ inside the loop.
Together with factors of the coupling constant arising from
the vertices, the self-energy is $\Pi \sim g^2 T^2$.
Therefore, as long as either $\omega$ {\em or\/} 
$p$ are ``hard'', i.e., of order $T$, the self-energy $\Pi$
in the propagator (\ref{scalarprop}) can be {\em neglected\/}.
However, for ``soft'' $\omega$ {\em and\/} $p$ of order $gT$, the
self-energy is of the same order of magnitude as the first two
terms in the denominator and {\em cannot\/} be neglected.

This observation forms the basis of the so-called HTL-resummation scheme
\cite{lebellac,braatenpisarski} in field theories at nonzero temperature.
In simple words it states that whenever the energy and the momentum
of a propagator in a given diagram is ``soft'', one has to use the 
``dressed'' propagator 
(\ref{scalarprop}) including the self-energy $\Pi$, and if either
energy or momentum is ``hard'', one may use the ``bare'' propagator
$\Delta_0(\omega,{\bf p}) = 1/(\omega^2 - p^2)$ without the self-energy $\Pi$.
How does the name ``HTL-resummation scheme'' arise?
The dressed propagator is the solution of the Dyson-Schwinger equation 
\begin{equation}
\Delta = \Delta_0 - \Delta_0 \, \Pi \, \Delta\,\, .
\end{equation}
Iterating this equation, one realizes that
it stands for an infinite series of diagrams; consequently
the solution (\ref{scalarprop}) is a {\em resummed\/} propagator.
As explained in Sec.~\ref{HTLexcite},
the quantity which is resummed is the self-energy $\Pi$ computed
in the HTL approximation, i.e., one resums HTL's. 

It is now also easy to see why naive perturbation theory breaks down
at nonzero temperature.
Imagine a diagram with $n$ vertices, such that naive perturbation
theory would tell us that this diagram is of order $O(g^n)$.
Now imagine that there is a loop in this diagram 
with propagators of the type (\ref{scalarprop}) and
that the dominant contribution to this loop
arises not from the ``hard'' region of phase space, 
i.e., from momenta of order $T$, as in HTL's, but from
the ``soft'' region, i.e., from momenta $k \sim g T$.
The contribution of the propagator (\ref{scalarprop}) to the
diagram is then $\sim 1/(g^2 T^2)$, instead of $\sim 1/T^2$.
This cancels two powers of the coupling
constant in the naive perturbative counting scheme. The diagram
is thus actually of order $O(g^{n-2})$ 
(or even of lower order, if other propagators
contribute additional powers of $g^{-2}$). The occurrence of
the additional energy scale $T$ (or $\mu$, at $T=0$)
compared to the vacuum invalidates the naive perturbative
counting scheme.

For gauge theories, the Ward identities require to
extend the HTL-resummation scheme from propagators to vertices as well.
For physical quantities which are determined by computing diagrams
with at least one loop, depending on whether the dominant contribution 
in the loop arises from the hard or the soft region of phase space,
one may be required to use only bare propagators, 
resummed propagators, or both resummed
propagators and resummed vertices. The major success of the
HTL-resummation scheme was the proof that the leading-order
result for the gluon-damping rate is independent of the gauge
and positive \cite{gluondamping}.

\subsubsection{\it The equation of state for quasiparticles
in HTL/HDL approximation} \label{HTLEOS}

The objective of this section is to apply our knowledge 
from Sec.~\ref{HTLexcite} about
the quasiparticle excitation spectrum in QCD to determine
the equation of state at high temperature and/or density.
What one obviously needs is a thermodynamically consistent way
to construct the pressure including information about the
spectral density of the quasiparticles. Obviously, classical statistical
mechanics, such as applied in Sec.~\ref{massiveglue}, is of no use;
we need a field-theoretical approach. The method of choice
is the so-called ``Cornwall-Jackiw-Tomboulis'' (CJT) formalism \cite{CJT}.
The CJT formalism determines the effective action
of a theory as a functional of the one- and
and two-point functions. The stationary values of the effective action yield
the expectation value of the field and the full propagator.
The stationarity conditions are
Dyson-Schwinger equations for these quantities.
The CJT formalism is particularly useful for theories with spontaneously
broken symmetries, see Sec.~\ref{linearsigmamodels}.
For unbroken symmetries, the CJT formalism is equivalent
to the so-called $\Phi$-functional approach \cite{baym}.

Let us elaborate on this in somewhat greater detail.
In the CJT formalism the effective action of a theory with
bosonic fields $\phi$ and corresponding propagators $\Delta$,
as well as fermionic fields $\bar{\psi}, \psi$ with propagators $G$ reads
\begin{eqnarray}
\Gamma[\phi,\bar{\psi}, \psi, \Delta,G] & = & 
I[\phi,\bar{\psi},\psi] - \frac{1}{2}\, {\rm Tr} \, \ln \Delta^{-1}
- \frac{1}{2}\, {\rm Tr} \,\left( D^{-1}\,\Delta - 1 \right) \nonumber \\
& + & {\rm Tr}\, \ln G^{-1} + {\rm Tr}\, \left( S^{-1} \, G - 1 \right)
+ \Gamma_2[\phi,\bar{\psi},\psi,\Delta,G] \,\, . \label{effact}
\end{eqnarray}
Here, $I[\phi,\bar{\psi},\psi]$ 
is the classical action and all traces are taken in the
functional sense. The quantities $D^{-1}$ and $S^{-1}$ are
the inverse tree-level propagators for bosons and fermions, respectively,
\begin{equation}
D^{-1}(X,Y) \equiv - \frac{\delta I[\phi,\bar{\psi},\psi]}{\delta 
\phi(X) \, \delta \phi(Y)} \,\, , \;\;\;\;
S^{-1}(X,Y) \equiv - \frac{\delta I[\phi,\bar{\psi},\psi]}{\delta 
\bar{\psi}(X) \, \delta \psi(Y)} \,\, .
\end{equation}
The functional $\Gamma_2$ is the sum of all two-particle irreducible
(2PI) diagrams without external legs and with internal lines given by
the propagators $\Delta$ and $G$. The stationarity conditions which
determine the expectation value of the bosonic field, $\varphi$, as
well as the full propagators for bosons, ${\cal D}$, and for
fermions, ${\cal G}$, read
\begin{equation} 
\frac{\delta \Gamma[\phi,\bar{\psi},\psi,\Delta,G]}{\delta \phi(X)} 
= \frac{\delta \Gamma[\phi,\bar{\psi},\psi,\Delta,G]}{\delta \Delta(X,Y)} 
= \frac{\delta \Gamma[\phi,\bar{\psi},\psi,\Delta,G]}{\delta G(X,Y)} 
= 0
\label{statio}
\end{equation}
In principle, there are also stationarity conditions for the 
fermionic fields $\bar{\psi}$ and $\psi$, but their solution is
always trivial, as fermionic fields are Grassmann-valued and thus cannot
assume a nonzero expectation value.

Inserting the explicit form of $\Gamma$ from Eq.~(\ref{effact}) 
into the last two equations of (\ref{statio}), 
one obtains the Dyson-Schwinger equations for the full propagators
for bosons, ${\cal D}$, and fermions, ${\cal G}$,
\begin{equation} \label{DS}
{\cal D}^{-1} = D^{-1} + \Pi\,\, , \;\;\;\;
{\cal G}^{-1} = S^{-1} + \Sigma \,\, ,
\end{equation}
where the bosonic and fermionic self-energies are
\begin{equation} \label{selfenergy}
\Pi \equiv - 2 \, \frac{\delta 
\Gamma_2[\phi,\bar{\psi},\psi,\Delta,G]}{\delta \Delta} 
\,\, , \;\;\;\;
\Sigma \equiv  \frac{\delta 
\Gamma_2[\phi,\bar{\psi},\psi,\Delta,G]}{\delta G} 
\,\, .
\end{equation}
The right-hand sides of these equations
have to be taken at the stationary point
$\phi = \varphi,\, \bar{\psi}=\psi=0, \, \Delta = {\cal D}, \, G = {\cal G}$.
According to their definition (\ref{selfenergy}), the self-energies are 
obtained from the set of 2PI-diagrams $\Gamma_2$ by opening one
internal line. 

For translationally invariant systems, $\phi(X) \equiv \phi = const.$,
$\Delta(X,Y) \equiv \Delta(X-Y)$, $ G(X,Y) \equiv  G(X-Y)$, and
it is advantageous to work in energy-momentum space instead of
in space-time.
The effective action is, up to a sign and 
a factor of the four-dimensional volume of the system, $V/T$,
equal to the effective potential,
\begin{equation}
V [\phi,\bar{\psi},\psi,\Delta,G] \equiv - \frac{T}{V}\, 
\Gamma[\phi,\bar{\psi},\psi,\Delta,G] \,\, .
\end{equation}
At the stationary point, the effective potential is, again up to a sign,
equal to the thermodynamic pressure. Utilizing the stationarity
conditions (\ref{DS}) and the definitions (\ref{selfenergy}),
\begin{equation} \label{pCJT}
p \equiv - V[\varphi,{\cal D},{\cal G}]
= - U(\varphi)  - \frac{1}{2}\, {\rm Tr} \, \ln {\cal D}^{-1}
+ \frac{1}{2}\, {\rm Tr} \, \Pi\,{\cal D} 
 +  {\rm Tr}\, \ln {\cal G}^{-1} - {\rm Tr}\, \Sigma\, {\cal G}
- V_2[\varphi,{\cal D},{\cal G}] \,\, .
\end{equation}
Here, $U(\varphi)$ is the tree-level potential, $V_2 \equiv - (T/V) \Gamma_2$,
and ${\rm Tr} \, A \equiv \int_K \,
{\rm tr} \, A(\omega_n,{\bf k})$, where ${\rm tr}$ runs over possible
internal indices (Lorentz, Dirac, color, flavor, etc.) of $A$.

Equation~(\ref{pCJT}) is the desired result. All one has to do is to
apply it to QCD. In QCD, there is no spontaneously broken symmetry
(unless one considers color superconductivity, see Sec.~\ref{CSC}),
and thus $U(\varphi) \equiv 0$. The bosons are the gluons
and the fermions are the quarks.
Of course, one also has to account for ghost degrees 
of freedom. These look like another fermion contribution in Eq.~(\ref{pCJT}),
but the corresponding Matsubara sum in ${\rm Tr}$ has to run over {\em even\/}
multiples of $\pi T$ \cite{kapustaFTFT}.

\begin{figure}[tb]
\begin{center}
\begin{minipage}[t]{14 cm}
\epsfig{file=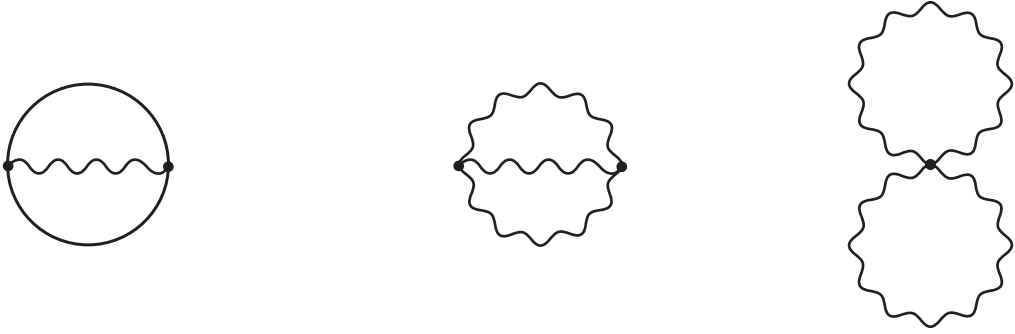,scale=1}
\end{minipage}
\begin{minipage}[t]{16.5 cm}
\caption{The two-loop approximation to
$\Gamma_2$ (possible ghost contributions are not shown).
\label{gamma2fig}}
\end{minipage}
\end{center}
\end{figure}

This sounds much simpler than it is to realize in practice.
The difficulty obviously lies in solving the Dyson-Schwinger
equations for the gluon and quark propagators.
It is clear that,
as $V_2$ consists of an infinite set of diagrams, one can never 
aspire to solve the problem exactly.
However, a great advantage of the CJT formalism is that 
the solutions of the Dyson-Schwinger equations (\ref{DS}) are
self-consistent and conserving, even if one truncates this infinite set.
Any truncation defines a meaningful many-body approximation scheme
(for a well-known example, the Hartree approximation,
see Sec.~\ref{linearsigmamodels}).
Let us therefore imagine we only take a finite subset of all diagrams
in $V_2$, for instance the one consisting of the two-loop
diagrams shown in Fig.~\ref{gamma2fig}.
This subset is particularly interesting, because 
when computing the self-energies according to Eq.~(\ref{selfenergy}) one 
obtains these self-energies to one-loop order. 
As discussed in Sec.~\ref{HTLexcite},
the high-temperature (high-density) limit of the one-loop self-energies
defines the HTL (HDL) approximation, which is already known to
provide a physically interesting, meaningful, and gauge-invariant 
quasiparticle excitation spectrum.
Another advantage of restricting $V_2$ to a set of two-loop diagrams
is that the entropy density $s \equiv \partial p/\partial T$ and the
quark number density $n \equiv \partial p / \partial \mu$
assume the particularly simple form \cite{vanderheyden}
\begin{eqnarray} \label{sCJT}
\!\!\!\! s &\!\! = & \!\! - {\rm tr} \int \frac{{\rm d}^4 K}{(2 \pi)^4} \,
\frac{\partial n(\omega)}{\partial T} \left[ {\rm Im} \ln {\cal D}^{-1}
 - {\rm Im} \Pi\, {\rm Re} {\cal D} \right]
 -2\, {\rm tr} \int \frac{{\rm d}^4 K}{(2 \pi)^4} \,
\frac{\partial f(\omega)}{\partial T} \left[ {\rm Im} \ln {\cal G}^{-1}
 - {\rm Im} \Sigma\, {\rm Re} {\cal G} \right], \\
\!\!\!\! n &\!\! = & \!\! -2\, {\rm tr} \int \frac{{\rm d}^4 K}{(2 \pi)^4} \,
\frac{\partial f(\omega)}{\partial \mu} \left[ {\rm Im} \ln {\cal G}^{-1}
 - {\rm Im} \Sigma\, {\rm Re} {\cal G} \right] , \label{nCJT}
\end{eqnarray}
where $K^\mu \equiv (\omega, {\bf k})$ and $n(\omega) \equiv 
(e^{\omega/T}-1)^{-1}$ is the Bose-Einstein distribution function,
while $f(\omega) \equiv [e^{(\omega-\mu)/T} + 1]^{-1}$ is the
Fermi-Dirac distribution function. While one has
to compute two-loop diagrams to obtain the pressure (\ref{pCJT}),
the entropy and quark number densities (\ref{sCJT},\ref{nCJT})
are essentially one-loop quantities and thus much simpler to calculate.

However, even when restricting $V_2$ to the simple subset of
Fig.~\ref{gamma2fig}, the solution of the Dyson-Schwinger
equations is highly nontrivial (see also Sec.~\ref{linearsigmamodels}).
Let us suppose, however, that it can be achieved.
Then there is still the issue whether
the propagators thus obtained obey the Ward identities.
In general, this is not the case \cite{smit}, and a cure 
of this problem would most likely
require a self-consistent calculation of
the three- and four-point functions on top of the two-point functions.
An extension of the CJT formalism
in this direction was proposed in Ref.~\cite{nortoncornwall}, but
since that work considers only scalar field theories, 
it is not clear whether this approach also applies
to gauge theories in a way which preserves the Ward identities.
Another question is the gauge invariance of the solution. 
Fortunately, as shown
in Ref.~\cite{smit}, the gauge dependence always enters at
an order which is higher than the truncation order.
Finally, there is the issue of renormalizability. 
For scalar theories, renormalizability was demonstrated in
Ref.~\cite{vanhees}. For gauge theories, it still remains an open 
question.

In order to make progress, one has to simplify the solution of
the problem. Instead of a completely self-consistent solution of
the Dyson-Schwinger equations (\ref{DS}), as a first
educated guess one might simply use the self-energies in the
HTL (or HDL) approximation to compute the propagators
entering the pressure (\ref{pCJT}). Since this approximation
is independent of the choice of gauge, gauge invariance is not
an issue anymore. There is also another advantage:
since the HTL- (HDL-) approximated
self-energies are just the high-$T$ (high-$\mu$) limit
of the {\em one-loop\/} self-energies, Eqs.~(\ref{sCJT},\ref{nCJT}) for the
entropy and quark number densities apply.
These expressions are completely ultraviolet safe
and thus do not require normalization. They are one-loop
quantities and thus simpler to compute than the pressure
which contains two-loop diagrams. From the entropy 
and quark number densities one
can always deduce the pressure via an integration with respect
to $T$ and $\mu$. 

\begin{figure}[tb]
\begin{center}
\begin{minipage}[t]{14cm}
\vspace*{-4cm}
\epsfig{bbllx=105,bblly=220,bburx=500,bbury=650,
file=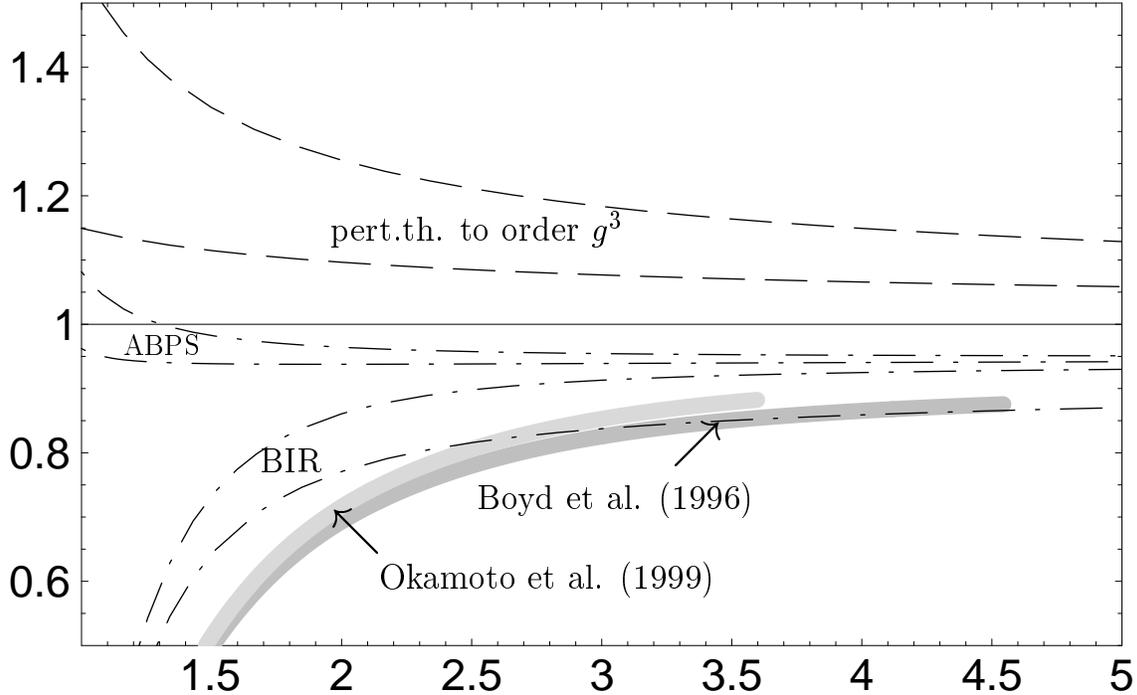,height=140mm}
\end{minipage}
\begin{minipage}[t]{16.5 cm}
\caption{The pressure, normalized to the Stefan-Boltzmann limit,
as a function of temperature, normalized to the critical temperature.
Different resummation methods are compared to lattice QCD results
for pure $[SU(3)_c]$ gauge theory. The band labelled 
``pert.th.\ to order $g^3$'' is the order $O(g^3)$ perturbative 
result computed in Sec.~\ref{pert} and shown in Fig.~\ref{EOSpert}. 
The band labelled ``ABPS'' is the
two-loop HTL-resummed pressure from Ref.~\cite{andersen}.
The band labelled ``BIR'' is the result from Ref.~\cite{BIR}.
Bands arise from varying the renormalization
scale $\bar{\mu}$ within certain limits.
Lattice data from Refs.~\cite{boyd,okamoto}
are shown as grey bands.
\label{HTLcomp}}
\end{minipage}
\end{center}
\end{figure}

This approach has been followed in Ref.~\cite{BIR},
where a detailed discussion of the results and their interpretation
can be found. In particular, since the self-consistent approach
is based on a partial resummation of a subset of perturbative diagrams,
i.e., a reorganization of the perturbative series,
it is possible to compare with standard perturbation theory 
(Sec.~\ref{pert}) by re-expanding the results 
in powers of $g$; for more details, see Ref.~\cite{reviewBIR,BIR}.
Here, I restrict myself to a discussion of one of the main results,
namely the pressure for pure $[SU(3)]_c$ gauge theory as computed
in this approach in comparison to lattice QCD data, cf.~Fig.~\ref{HTLcomp}.
One observes that the agreement at large temperatures is quite good,
and that deviations occur only below $T \simeq 2.5\, T_c$.
The quality of reproducing lattice QCD data is
comparable to that in the perturbative approach of Sec.~\ref{dimred}. However, 
a distinct advantage of the self-consistent approximation scheme is
that the physical interpretation of the results
in terms of a gas of quasiparticles is much more appealing.

Note that a very similar approach has been pursued in Ref.~\cite{peshierHTL},
utilizing directly the expression (\ref{pCJT}) for the pressure,
with propagators in the HTL approximation and a suitably modified
$V_2$. Another related approach is so-called
``HTL-resummed perturbation theory'' \cite{andersen}.
Here, the pressure of the pure gauge theory is computed to two-loop order
with HTL-resummed propagators. A variational procedure like that of
Ref.~\cite{OPT} is applied to determine the gluon mass parameter.
The pressure computed from this approach is closer to
the Stefan-Boltzmann limit and fails to agree with
lattice QCD data in the region of temperatures from $T_c$ to
several times $T_c$, cf.~Fig.~\ref{HTLcomp}. The reason for the differences
between this approach and the one of Ref.~\cite{BIR}
lies in the way the perturbative series is reorganized when
resumming certain subsets of diagrams. The
differences can be explained by comparing to
a dimensionally reduced version of HTL-resummed perturbation theory
\cite{BIR2}. Finally, note that the CJT formalism \cite{BIR} as well as
HTL-resummed perturbation theory \cite{baierredlich} were also applied 
to compute thermodynamic properties of the QGP at $T=0$ and nonzero $\mu$.

\subsection{\it Polyakov-Loop Model} \label{Ploopmodel}

The pressure, normalized to the appropriate Stefan-Boltzmann limit,
as a function of temperature, normalized to the appropriate
critical temperature, shows a universal behavior, see right panel
of Fig.~\ref{EOS}. Since the normalized pressure for full QCD with dynamical
fermions looks the same as for the pure gauge theory (i.e., without dynamical
fermions), a natural conclusion would be to assume that the 
dynamics of the gluons drives the QCD phase transition, not that of 
the fermions.
Consequently, the order parameter for the transition in full QCD should
be the same as in the pure gauge theory, i.e., the Polyakov loop, see
Sec.~\ref{puregauge}. For an $[SU(N_c)_c]$ gauge symmetry,
the Polyakov loop (\ref{Polyakovloop})
is invariant under $[SU(N_c)_c]$ gauge transformations, up to an element
of the center of the gauge group, $Z(N_c)$,
$L({\bf x}) \rightarrow \exp(2\pi i n/N_c) \, L({\bf x})$.
The effective theory for the Polyakov loop 
consists of all possible terms invariant under 
$Z(N_c)$ transformations \cite{pisarskidumitru},
\begin{equation}
{\cal L}_{\rm eff} = c_0 \, |\mbox{\boldmath$\nabla$} L|^2
+ c_2\, |L|^2 + c_3\, \left[L^3+ (L^*)^3 \right] 
- c_4 \, \left(|L|^2\right)^2  + \ldots\,\, ,
\label{LeffPLM}
\end{equation}
where the dots denote higher-order terms (which will be neglected in
the following).
For $N_c=2$, one would have $c_3=0$, because $L^3$ is not $Z(2)$-symmetric.
For $N_c =3$, it is precisely this cubic term which
drives the transition first order \cite{svetitskyyaffe}, 
see also Sec.~\ref{puregauge}. 

The ground state of the theory, $L_0 \equiv \langle L({\bf x}) \rangle$
(which is assumed to be real), is determined by the global minimum
of the effective potential 
\begin{equation} \label{VeffPLM}
V_{\rm eff}(L) = - c_2\, |L|^2  - c_3\, \left[L^3
+ (L^*)^3 \right] + c_4 \, \left(|L|^2\right)^2 \,\,.
\end{equation}
The Polyakov loop $L$ is dimensionless, therefore
all coupling constants in Eq.~(\ref{VeffPLM})
carry dimension $[\mbox{energy}]^4$. Since the only dimensionful scale
is set by the temperature, one may pull out a factor $T^4$ from
the right-hand side of Eq.~(\ref{VeffPLM}) and after appropriately
renaming the constants write
\begin{equation}  \label{VeffPLM2}
V_{\rm eff}(L) = b_4\, T^4 \left\{ - b_2\, |L|^2  - b_3\, \left[L^3
+ (L^*)^3 \right] +  \left(|L|^2\right)^2 \right\}\,\, ,
\end{equation}
where the new coupling constants are dimensionless.
In order to have a potential that is bounded from below for large
values of $L$, one has to assume $b_4 > 0$. For the moment
take $b_3 = 0$ (for $N_c=2$, $b_3 \equiv c_3/(b_4 T^4)$ must always vanish). 
Then the sign of $b_2$ drives the transition. 
Below the transition, $T < T_c$, there is confinement, 
which can only be achieved if the curvature of $V_{\rm eff}$ at the
origin is positive, $b_2  < 0$, such that the global minimum is
at $L_0 = 0$. Above the transition, $T > T_c$, there is deconfinement, which is
achieved by a negative curvature of $V_{\rm eff}$ at the origin, 
$b_2 > 0$, giving rise to a nonvanishing $L_0 = \pm \sqrt{b_2/2}$. 
The constant $b_2$ is therefore a function of temperature,
$b_2 = b_2(T)$, while
in the simplest version of the Polyakov-loop model, $b_3$ and
$b_4$ are assumed to be constant. Through the temperature
dependence of $b_2$, $L_0$ also becomes a function of temperature,
$L_0 \equiv L_0(T)$.
At $T=T_c$, one must have $b_2(T_c) = 0$,
i.e., $L_0$ vanishes continuously at $T_c$, and one has a
second-order phase transition. A nonvanishing $b_3$ (which is
allowed for $N_c=3$) turns
this second-order transition into a first-order transition.
Lattice QCD data indicate that the transition for pure $[SU(3)_c]$ gauge
theory has a comparatively small latent heat, i.e., the transition
is only {\em weakly\/} of first order. This suggests that the
constant $b_3$ is small.

\begin{figure}[tb]
\begin{center}
\begin{minipage}[t]{11cm}
\epsfig{file=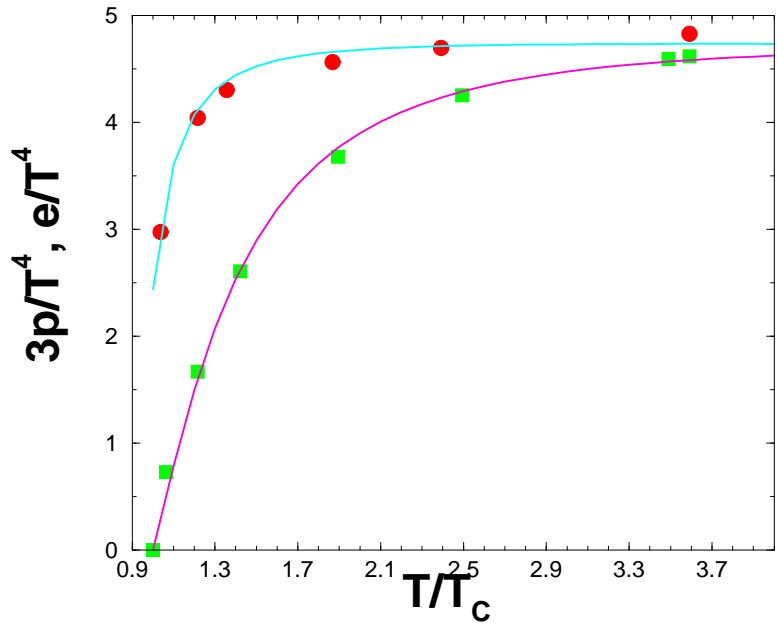,scale=0.6}
\end{minipage}
\begin{minipage}[t]{16.5 cm}
\caption{The functions $3p/T^4$ (lower curve) and $\epsilon/T^4$
(upper curve) as a function of $T/T_c$. Fit of the thermodynamic
functions of the Polyakov-loop model
to lattice data (shown as circles and boxes)
for the pure $[SU(3)_c]$ gauge theory.
\label{PLMpress}}
\end{minipage}
\end{center}
\end{figure}

Now remember that the pressure is, up to sign, equal to the
value of the effective potential at its minimum
\begin{equation} \label{pPLM}
\frac{p}{T^4} \equiv - \frac{V_{\rm eff}(L_0)}{T^4} 
=  \frac{b_4}{2}\, L_0^2 \left( b_2 + b_3\, L_0 \right) \,\, .
\end{equation}
For $N_c=2$, where $b_3= 0$, one has $p/T^4 = b_4 b_2(T)^2 /4$, where
the explicit value of $L_0$ was used.
The physical interpretation of this result is quite astonishing: 
the pressure in QCD, as calculated on the lattice, cf.~Fig.~\ref{EOS},
is {\em not\/} determined by the {\em kinetic\/} energy
of a gas of weakly interacting quasiparticles,
as advocated in Secs.~\ref{quasiparticles}, \ref{HTLresum},
but is simply given by the {\em potential\/} for the Polyakov loop
(at the respective minimum). All that one has to know is
$L_0(T)$ (or, equivalently, $b_2(T)$, under the assumption that
this is the only coupling constant which depends on temperature), 
and the pressure in QCD can be immediately computed
from Eq.~(\ref{pPLM}). Comparing Figs.~\ref{Ploop_and_Xcond} and
\ref{EOS}, this conjecture is at least qualitatively correct.
Quantitatively, it is not as simple: not only is the absolute
value of $L_0$ as calculated on the lattice subject to renormalization,
also the constants $b_2, b_3,$ and $b_4$ are not known.
In the meantime, one can at least convince oneself that
there are no principal obstacles for this interpretation:
one may insert the value of $L_0$ (which is only
a function of $b_2$ and $b_3$) into Eq.~(\ref{pPLM}) and simply fit
$b_3$, $b_4$, as well as $b_2(T)$ to lattice QCD data. 
The result of this fit is $b_3 = 4/3$, $b_4 \simeq 0.1515$,
$b_2(x) = 2 (1-1.11/x)(1+0.265/x)^2(1+0.300/x)^3 - 0.974$,
where $x\equiv T/T_c$. The pressure and the energy density obtained
from this fit are shown in Fig.~\ref{PLMpress}.

The Polyakov-loop model also allows to predict other
physical observables, for instance, the ratio of screening masses
related to the correlation function of the real and the
imaginary part of the Polyakov loop \cite{pisarskidumitru}.
Decompose the Polyakov loop into its real and imaginary part,
$L \equiv R + i I$. In analogy to Eq.~(\ref{heavyquarkfreeenergy}),
the correlation functions for the real and imaginary part are then 
defined as $\langle R(0) \, R({\bf x}) \rangle$ and
$\langle I(0) \, I({\bf x}) \rangle$, respectively.
In the deconfined phase, one expects them to decrease exponentially
with the distance,
\begin{equation}
\langle R(0) \, R({\bf x}) \rangle \sim \exp \left( - m_R |{\bf x}| \right)
\,\, , \;\;\;\;
\langle I(0) \, I({\bf x}) \rangle \sim \exp \left( - m_I |{\bf x}| \right)
\,\, ,
\end{equation}
where $m_R$, $m_I$ are the corresponing screening masses.
In weak coupling, one can make a definite prediction for these
masses. To this end, expand $L$ in powers of $g$ to 
leading nontrivial order. One obtains
$R \sim 1 - g^2 /(T^2 N_c) \, {\rm Tr} {\cal A}_0^2$.
Thus, the correlation function for the real part involves
the exchange of two static ${\cal A}_0$ fields, which are Debye screened.
One therefore expects
$\langle R(0) \, R({\bf x}) \rangle \sim \exp( - 2m_D |{\bf x}|)$.
Analogously, $\langle I(0) \, I({\bf x}) \rangle \sim \exp( - 3m_D |{\bf x}|)$.
Thus, the perturbative prediction for the screening masses is
$m_R = 2 m_D$, $m_I = 3 m_D$, and $m_I/m_R = 3/2$.

In general, because of the renormalization of the Polyakov loop
it is not as simple to obtain an answer for the absolute values
of $m_R$ and $m_I$. However, in the ratio $m_I/m_R$ one expects
unknown factors to cancel out \cite{pisarskidumitru}.
The Polyakov-loop model makes a definite prediction for this ratio.
First, write the potential (\ref{VeffPLM}) in
terms of $R$ and $I$. As usual, the curvature of the effective potential in
the ground state provides the masses. Consequently, computing the
curvature in $R$- and in $I$-direction,
one obtains
\begin{equation} \label{massesPLM}
m_R^2 \sim - b_2 - 6 \, b_3 \, L_0 + 6 \, L_0^2\,\, , \;\;\;\;
m_I^2 \sim - b_2 + 6 \, b_3 \, L_0 + 2 \, L_0^2 \,\, .
\end{equation}
I have refrained from writing the constant of proportionality,
which also provides the correct dimension for $m_{R,I}$,
because this constant will drop out anyway in the ratio $m_I/m_R$.
Let us compute the ratio close to $T_c$. In the case of a first-order
transition, at $T= T_c$ the effective
potential $V_{\rm eff}$ has two minima, one at the origin and one
at $L_0 \neq 0$, which are degenerate and
separated by an energy barrier. The condition $V_{\rm eff}(0) = 
V_{\rm eff} (L_0)$ gives $L_0 = - b_2 / b_3$, which together with
the condition that $L_0$ is the nontrivial minimum of $V_{\rm eff}$,
$L_0^2 = (3 b_3 L_0 + b_2)/ 2$, allows to express all
quantities in Eq.~(\ref{massesPLM}) in terms of, say $b_3 L_0$.
This yields $m_I / m_R = 3$.
This value differs from the perturbative expectation by a factor of two,
and is a definite prediction of the Polyakov-loop model 
which can be tested, for instance, on the lattice.

Another particularly appealing aspect of the Polyakov-loop model
is that it correctly describes the behavior of the screening masses
when approaching $T_c$ from above, cf.~Fig.~\ref{screeningmass}.
Near $T_c$, the temperature dependence of $b_2$ and $L_0$ cause
the screening masses $m_R, m_I$ to decrease, in agreement with
lattice QCD data.
The fact that the perturbative prediction 
completely fails to reproduce this behavior was already mentioned in 
Sec.~\ref{heavyquarklatt}.

Finally, let us remark that the Polyakov-loop model may also 
naturally explain
why the chiral symmetry restoration transition occurs
at the same temperature as the deconfinement transition.
There is nothing to prevent the effective theory (\ref{LeffPLM}) from
having a term $\sim c |L|^2 {\rm Tr} \, \Phi^\dagger \Phi$, 
where $\Phi$ is the chiral condensate and $c>0$ a constant. 
Thus, when $|L|$ condenses (in the deconfined phase),
this term suppresses condensation of $\Phi$ on account
of the positive coupling constant $c$, effectively restoring
the chiral symmetry. Thus, chiral symmetry restoration and
deconfinement occur at the same temperature.

\subsection{\it Linear Sigma Models with Hadronic Degrees of Freedom}
\label{linearsigmamodels}

In Sec.~\ref{translatt} we have seen that the QCD transition is
most likely crossover for small values of the chemical potential
$\mu$. Therefore, there is no real distinction between
hadronic degrees of freedom on the one side and quark and gluon
degrees of freedom on the other side of the transition.
Consequently, there is no compelling reason 
why the thermodynamic properties of QCD around the phase transition 
should be described in terms of quarks and gluons; a hadronic description
should be equally adequate. (For a more complete discussion of this
argument, see Ref.~\cite{mullerQM91}.)

Let us assume that, somewhat contrary
to the idea behind the Polyakov-loop model discussed in 
Sec.~\ref{Ploopmodel}, the QCD transition is driven by
chiral symmetry restoration instead of deconfinement.
In this case, the order parameter for the transition is the
chiral condensate, see Sec.~\ref{dynquarks}, and the effective
theory is given by the linear sigma model (\ref{Leff}). 
The degrees of freedom in this effective theory are
the fluctuations of the order parameter field
$\Phi$ around the ground state, $\langle \Phi \rangle$. Physically,
these fluctuations correspond to the scalar and pseudoscalar mesons.
Following the above line of arguments, this effective theory could 
therefore not only serve to understand
the dynamics of chiral symmetry restoration,
it could equally well be used to describe
the thermodynamic properties of QCD around the phase transition.

Of particular interest is the question how chiral symmetry restoration
is exhibited in the meson mass spectrum. In the framework of lattice
QCD, this was discussed in Sec.~\ref{meson_spectral}.
Here, I explain how to answer this question using
the effective theory (\ref{Leff}) at nonzero temperature.
The standard definition of a particle mass is via the pole of the propagator
at zero momentum. Consequently, the goal is to determine the 
mesonic propagators. The method of choice is obviously the
CJT formalism discussed in Sec.~\ref{HTLEOS}, because this
approach allows to compute the full propagator from the stationary points of
the effective action $\Gamma$, cf.~Eq.~(\ref{statio}), or in other
words, from the Dyson-Schwinger equations (\ref{DS}). 
Of course, the solution of these equations
requires $\Gamma_2$, i.e., the complete set of 2PI diagrams.
For practical purposes, a solution of the Dyson-Schwinger equations
is therefore impossible.
However, as already discussed in Sec.~\ref{HTLEOS}, one may truncate
$\Gamma_2$ at some given order. Such a truncation
defines a many-body approximation, within which
a solution of the Dyson-Schwinger equations becomes feasible.

\begin{figure}[tb]
\begin{center}
\begin{minipage}[t]{5cm}
\epsfig{file=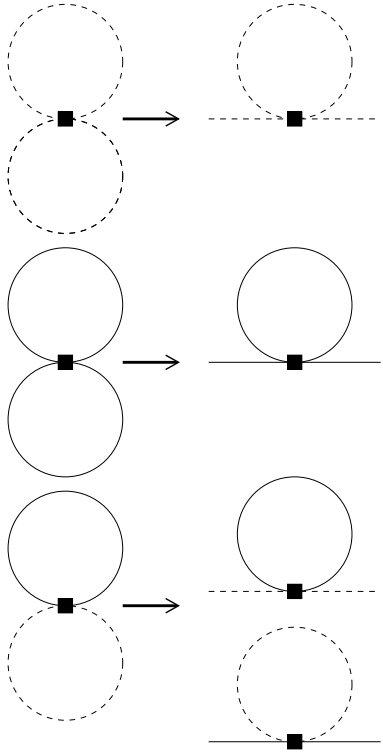,scale=0.6}
\end{minipage}
\begin{minipage}[t]{16.5 cm}
\caption{Double-bubble diagrams (left-hand side),
where full lines denote scalar
mesons and dashed lines denote pseudoscalar mesons. Cutting the
bubbles produces the tadpole-diagram contributions to the self-energies 
shown on the right-hand side. The tadpole diagrams constitute
the so-called Hartree approximation.
\label{doublebubble}}
\end{minipage}
\end{center}
\end{figure}

The most simple, nontrivial truncation of $\Gamma_2$ is to
include only the set of double-bubble
diagrams shown on the left in Fig.~\ref{doublebubble}.
The self-energies (\ref{selfenergy}) computed from these
diagrams consist only of the tadpole diagrams shown on the right-hand
side of Fig.~\ref{doublebubble}. 
This is the well-known Hartree approximation \cite{fetwal}.
In general, all particle species in a particular effective theory
contribute via tadpole diagrams to the self-energy of a given
particle species.
Since the tadpole diagrams do not have any dependence
on the external momentum, the self-energies are (temperature-dependent)
constants in the Hartree approximation. They simply shift the
meson masses as compared to their vacuum values. In principle, 
the Dyson-Schwinger equations (\ref{DS}) for the propagators
are coupled integral equations, but in the Hartree approximation,
they reduce to a system of coupled fix-point equations for the
meson masses. While numerically much simpler than solving
integral equations, the solution can still be a formidable task, if
the underlying chiral symmetry group is large.
For an $O(4)$ chiral symmetry, this problem was
solved in Refs.~\cite{petro,jtldhr}.
The $U(3)_r \times U(3)_\ell$ case was discussed in Ref.~\cite{jtljsbdhr}.
The cases $U(2)_r \times U(2)_\ell$ as well as $U(4)_r \times U(4)_\ell$
were investigated in Ref.~\cite{drjrdhr}. (The $U(1)_V$ of baryon
number is always trivially
respected in these models, cf.~the discussion in Sec.~\ref{dynquarks}.
Nevertheless, in order to simplify the notation I include it in 
characterizing the symmetry of a particular chiral effective theory.)

\begin{figure}[tb]
\begin{center}
\begin{minipage}[t]{10cm}
\vspace*{-1cm}
\epsfig{file=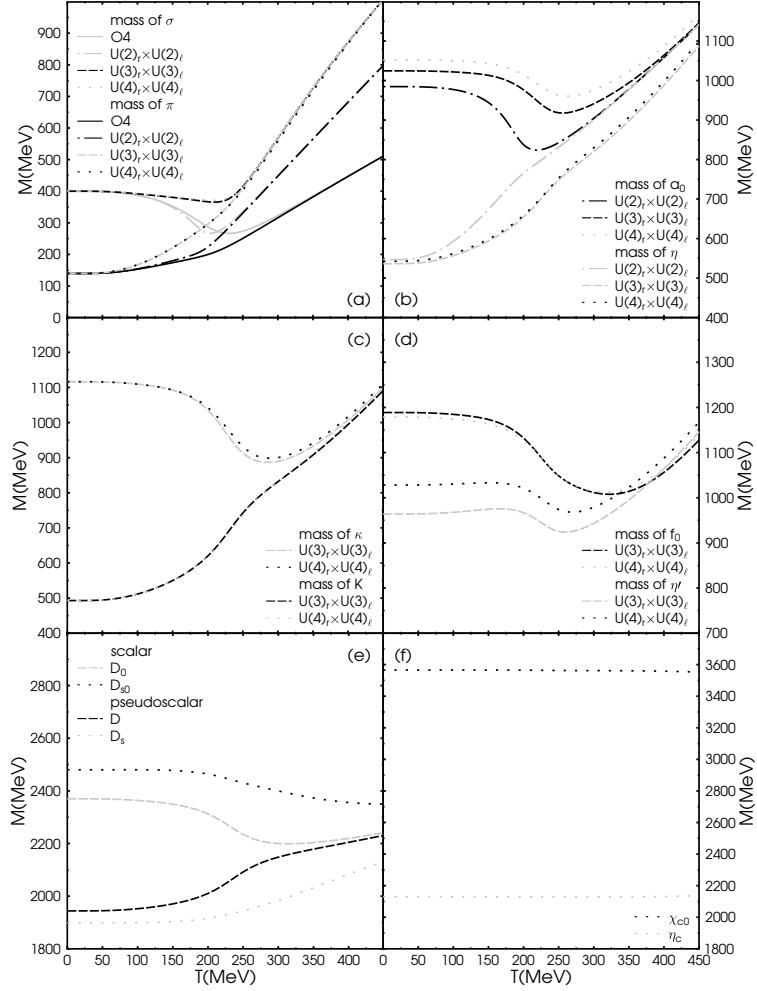,scale=0.75}
\end{minipage}
\begin{minipage}[t]{16.5 cm}
\vspace*{-0.5cm}
\caption{Scalar and pseudoscalar meson masses as a function of temperature
in the Hartree approximation,
for effective theories models with different chiral symmetry:
$O(4)$, $U(2)_r \times U(2)_\ell$, $U(3)_r \times U(3)_\ell$,
and $U(4)_r \times U(4)_\ell$. Chiral partners are shown in
the same panel to demonstrate that their masses become degenerate
at high temperature. From Ref.~\cite{drjrdhr}.
\label{mesonmasses}}
\end{minipage}
\end{center}
\end{figure}

In Fig.~\ref{mesonmasses} the masses for the
scalar and pseudoscalar mesons are shown as a function of temperature,
calculated within the framework of linear sigma models with
$O(4)$, $U(2)_r \times U(2)_\ell$, $U(3)_r \times U(3)_\ell$, and
$U(4)_r \times U(4)_\ell$ chiral symmetry. 
The different subpanels show the masses of the respective chiral
partners. In the non-strange sector, these are
(a) the $\sigma$ meson and the pion, and (b) the $a_0$ and the
$\eta$ meson. In the strange sector, these are (c) 
the $\kappa$ meson and the kaon, and (d) the $f_0$ 
and the $\eta'$ meson. Finally, in the charmed meson sector
one has (e) the scalar $D_0$ and $D_{s0}$ meson and the 
pseudoscalar $D$ and $D_s$ meson, and (f) the $\eta_c$ and
$\chi_{c0}$ meson. One observes that in all cases the
chiral transition occurs at temperatures of the
order of $200 - 300$ MeV. For the results shown in Fig.~\ref{mesonmasses}
realistic, nonzero values for the quark masses were assumed, 
and consequently the transition is crossover, cf.~the discussion in
Sec.~\ref{quarkmassdiagram}. Figure~\ref{mesonmasses} also allows to
compare the results for models with different chiral symmetry.
For instance, Fig.~\ref{mesonmasses} (a) nicely demonstrates the effect
of enlarging the symmetry group. In general, the higher the 
symmetry, the more particle species contribute via tadpole diagrams 
to the mass of a particular species, and consequently the larger 
is its mass at a given temperature. Furthermore, one can
learn from all subpanels that the difference between the $U(3)_r \times
U(3)_\ell$ and the $U(4)_r \times U(4)_\ell$ model is at best marginal.
The reason is that the additional charmed meson degrees of freedom in
the latter are so heavy that the contribution of the respective
tadpole diagrams is exponentially suppressed $\sim \exp(- m/T)$.
Consequently, they only minimally influence the behavior of
the non-charmed mesons in the temperature range considered here.
In turn, in this temperature range 
the charmed meson masses change little from their vacuum values,
cf.~Fig.~\ref{mesonmasses} (e), (f).

\begin{table}
\begin{center}
\begin{minipage}[t]{16.5 cm}
\caption{The critical temperature $T_c$ for the chiral
transition, computed for various chiral effective theories 
in the chiral limit with and without $U(1)_A$ anomaly \cite{drjrdhr}, 
in comparison to results from lattice QCD, extrapolated to the chiral limit.
For $N_f=2$ flavors, the QCD results 
for Wilson and Kogut-Susskind fermions from Table~\ref{table2}
have been averaged, assuming uncorrelated statistical errors. 
\vspace*{5mm}}
\label{tableTc}
\end{minipage}
\begin{tabular}{|c|c|c|c|}
\hline
       &                          &                            &\\[-2mm]
Chiral model  &  $T_c$ with $U(1)_A$ anomaly   &  
                            $T_c$ without $U(1)_A$ anomaly  &  LQCD    \\
       &                          &                            &\\[-2mm]
\hline
       &                          &                            &\\[-2mm]
$O(4)$ &  $(159.5 \pm 0.2)$ MeV   &  N/A                       &
                                                    $(172 \pm 9)$ MeV \\
       &                          &                            &\\[-2mm]\hline
       &                          &                            &\\[-2mm]
$U(2)_r \times U(2)_\ell$  &  $(154.5 \pm 0.2)$ MeV & 
           $(149.2 \pm 0.2)$ MeV    &             $ (172 \pm 9)$ MeV  \\
       &                          &                            &\\[-2mm]\hline
       &                          &                            &\\[-2mm]
$U(3)_r \times U(3)_\ell$  &  $(165.5 \pm 0.2)$ MeV & 
           $(147.5 \pm 0.2)$ MeV     &   $(154 \pm 8)$ MeV  \\
       &                          &                            &\\[-2mm]\hline
\end{tabular}
\end{center}
\end{table}

Lattice QCD calculation cannot be done in the chiral limit,
because as the quark mass decreases, 
the computation of the fermion determinant becomes more and more 
difficult. The only way to make predictions about
the chiral limit is to extrapolate data obtained for
nonzero quark masses. This is computationally expensive, as it
requires calculations at several different values of
the quark mass.
In contrast, in the framework of linear sigma models, taking the chiral limit 
actually simplifies the calculation.
A comparison between quantities computed in lattice QCD and extrapolated
to the chiral limit with the corresponding values obtained in linear sigma
models is therefore straightforward.
Let us dicuss two examples. The first is the critical temperature
for the chiral transition. The way to identify the critical temperature
is the following.
For a first-order phase transition, look for the temperature
where two minima of the effective potential
become degenerate. For a second-order
transition, the criterion is a vanishing second derivative of the 
potential in one direction in order parameter space 
(indicating a massless degree of freedom and critical behavior).
It should be mentioned that the Hartree approximation does not always
reproduce the correct order of the transition as predicted from
universality arguments. (Aside from that, 
being a mean-field type approximation, it
always fails to predict the correct critical exponents.)
For instance, for the $O(4)$ model and the
$U(2)_r \times U(2)_\ell$ model with $U(1)_A$ anomaly, the transition
should be of second order, but the Hartree approximation predicts
a first-order transition. For the $U(2)_r \times U(2)_\ell$ model
without $U(1)_A$ anomaly, as well as for the $U(N_f)_r \times U(N_f)_\ell$
models with $N_f \geq 3$, the Hartree
approximation predicts a first-order transition in agreement with
universality arguments. 
In chiral effective theories, one can simply ``switch off'' the
$U(1)_A$ anomaly by setting the constant $c=0$ in Eq.~(\ref{Veff}).
This is not possible for lattice QCD: the $U(1)_A$ anomaly 
may or may not be present, depending on how strongly instantons
are screened at $T_c$. This question has been studied on the lattice
\cite{alles}, with the result that the $U(1)_A$ anomaly is not
completely absent at the critical temperature, but at least rapidly
decreasing. 

The critical temperatures obtained
in various chiral models, with and without the $U(1)_A$
anomaly, are shown in Table~\ref{tableTc} in comparison
to the results from lattice QCD, cf.~Table~\ref{table2}.
Two things are noteworthy. First, the values of the critical temperature
obtained in chiral models are rather close to the ones
computed in lattice QCD. This is surprising given the fact
that chiral effective theories only contain a certain subset
of the degrees of freedom of full QCD, do not describe confinement, and
moreover are treated here in the framework of a very
simple many-body approximation. Second,
the ordering of the critical temperatures 
as a function of quark flavors 
is the same in lattice QCD and chiral models {\em without\/}
$U(1)_A$ anomaly, but the opposite in chiral models {\em with\/}
$U(1)_A$ anomaly. This also lends support to the above mentioned
results \cite{alles} regarding the rapid decrease of the $U(1)_A$ 
anomaly near $T_c$.

\begin{figure}[tb]
\begin{center}
\begin{minipage}[t]{12cm}
\epsfig{file=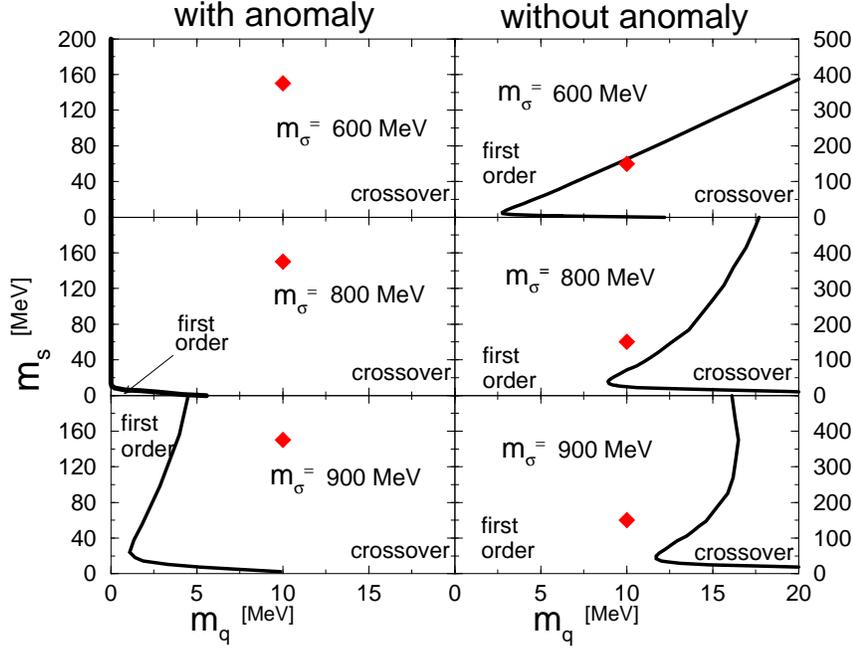,scale=0.6}
\end{minipage}
\begin{minipage}[t]{16.5 cm}
\vspace*{-0.5cm}
\caption{The quark-mass diagram computed in a
$U(3)_r \times U(3)_\ell$ chiral model \cite{jtl},
with or without $U(1)_A$ anomaly for different values of the 
$\sigma$ meson mass in vacuum. The physical point $(m_q, m_s) = (10,150)$ MeV
is shown as a diamond.
\label{quarkmasschiral}}
\end{minipage}
\end{center}
\end{figure}

The second example is the investigation of the quark-mass diagram. 
Figure~\ref{quarkmasschiral}
shows this diagram as determined within a $U(3)_r \times U(3)_\ell$
linear sigma model \cite{jtl}.
Of course, the linear sigma model does not have quark degrees of freedom.
Nonzero quark masses correspond to a term of the form (\ref{additional}) in
the Lagrangian, where the matrix $H$ is proportional to the quark mass matrix.
Since the vacuum expectation value of $\Phi$ is a diagonal
matrix, the matrix $H$ must also be diagonal. 
If one assumes $SU(2)_V$ isospin symmetry, $H = h_0 T_0 + h_8 T_8$.
Consequently, $m_q = a(h_0 + h_8/\sqrt{2})$, $m_s = b(h_0 - \sqrt{2} h_8)$.
The fields $h_0$, $h_8$ can be determined from the vacuum values
for the pion and kaon masses, as well as the pion and kaon decay
constants. Then, setting $m_q = 10 $ MeV, $m_s = 150$ MeV
fixes the values for the constants of proportionality $a$ and $b$; for
details see Ref.~\cite{jtl}.

One observes in Fig.~\ref{quarkmasschiral} that the position of
the line of second-order
phase transitions between the first-order region around the origin
and the crossover region depends sensitively on the value of
the $\sigma$ meson mass in vacuum. However, the physical point is always in
the crossover region, if the $U(1)_A$ anomaly is present.
Without the $U(1)_A$ anomaly, the transition could be of first
order, if the $\sigma$ meson is sufficiently heavy.

Recently, there have been attempts to go beyond the Hartree
approximation by including diagrams with more
complicated topologies in $\Gamma_2$ \cite{vanhees,baacke}. 
In this case, a self-consistent 
calculation becomes technically rather difficult, because these topologies
lead to momentum-dependent self-energies, so that the
Dyson-Schwinger equations turn from simple fix-point equations for
the masses into integral equations for the full propagators.
Moreover, momentum-dependent self-energies 
have in general nonzero imaginary parts, such that quasiparticles, which
are stable in the Hartree approximation, develop a finite decay width.
The spectral density carries the complete information of
the spectral properties of the quasiparticles in the system. It is
therefore natural to solve the Dyson-Schwinger equations 
for the full propagators as self-consistency
equations for the respective spectral densities. 
Work in this direction is in progress \cite{jrdhr}.


\section{Color Superconductivity} \label{CSC}

\subsection{\it Derivation of the Gap Equation}

As discussed in Sec.~\ref{cscphases}, 
quark matter at high density and sufficiently 
low temperature is most probably a color superconductor \cite{bailinlove,arw}. 
While the color quantum numbers of a Cooper pair
are determined by the attractive gluon interaction in the color-antitriplet 
channel, there are still many different ways to combine 
flavor and spin quantum numbers, giving rise to a plethora of possible
color-superconducting phases. In general, the energetically most favorable
phase will prevail at a given $T$ and $\mu$. In order to decide which
is the most favorable phase one has to compute the gain in condensation
energy when forming Cooper pairs. To this end, one has to calculate
the color-superconducting gap parameter. In general, the larger the
gap, and the more degrees of freedom participate in
forming Cooper pairs, the larger the gain condensation energy. 
The gap is computed from a so-called {\em gap equation}. 
In this section, I shall outline the derivation of this equation.

Consider the QCD action
\begin{equation}
S \equiv S_A + S_F + g \int_X \bar{\psi}(X) \, \gamma^\mu \,
T_a\, \psi(X) \, A_\mu^a (X)\,\, ,
\end{equation}
where $S_A$ is the gauge field action (including ghost and gauge
fixing contributions) and
\begin{equation}
S_F = \int_X
\bar{\psi}(X) \, (i \gamma \cdot \partial_X + \mu \gamma_0 -m ) 
   \, \psi(X)
\end{equation}
is the action for non-interacting fermion fields in the presence
of a chemical potential. For reasons which will become clear
in the following, $S_F$ and the quark-gluon coupling are rewritten
in terms of ordinary quark fields, $\bar{\psi}$,
$\psi$, {\em and\/} their {\em charge-conjugate\/} counterparts,
$\bar{\psi}_C \equiv \psi^T C$, $\psi_C \equiv C \bar{\psi}^T$,
where $C \equiv i \gamma^2 \gamma_0$ is the charge conjugation matrix.
To this end, introduce the so-called Nambu-Gor'kov basis, 
with the $2  \cdot 4 \, N_c N_f$-dimensional quark spinors
\begin{equation}
\Psi \equiv \left( \begin{array}{c} \psi \\ \psi_C 
                   \end{array} \right) \,\, , \;\;\;\;
\bar{\Psi} \equiv (\bar{\psi}, \bar{\psi}_C)\,\, .
\end{equation}
For the fermion action one then obtains
\begin{equation}
S_F = \frac{1}{2} \int_{X,Y} \bar{\Psi}(X) \,
{\cal S}_0^{-1}(X,Y)\, \Psi(Y) \,\, , 
\end{equation}
where 
\begin{equation}
{\cal S}_0^{-1}  \equiv \left( \begin{array}{cc}
                             [G_0^+]^{-1} &  0  \\
                              0 & [G_0^-]^{-1} \end{array} \right)\,\, ,
\;\;\;\; 
[G_0^\pm]^{-1} (X,Y) \equiv -i (i \gamma \cdot \partial_X \pm \mu \gamma_0
-m )\, \delta^{(4)}(X-Y)\,\, ,
\end{equation}
is the free inverse fermion propagator in the Nambu-Gor'kov basis.
The quark-gluon coupling becomes
\begin{equation}
\bar{\psi}(X) \, \gamma^\mu\, T_a \,\psi(X)\, A_\mu^a(X) =
\frac{1}{2} \, \bar{\Psi}(X) \, \Gamma^\mu_a \,\Psi(X)\, A_\mu^a(X)\,\, ,
\end{equation}
where
\begin{equation}
\Gamma^\mu_a \equiv \left( \begin{array}{cc}
                           \gamma^\mu T_a & 0 \\
                            0 & - \gamma^\mu T_a^T \end{array} \right)
\end{equation}
is the Nambu-Gor'kov matrix which couples the corresponding
quark spinors to gluon fields.

Now add a bilocal source term to the QCD action,
\begin{equation}
S[{\cal K}] \equiv S + 
\int_{X,Y} \bar{\Psi}(X)\, {\cal K}(X,Y)\,
\Psi(Y) \,\, ,
\end{equation}
where, in the Nambu-Gor'kov basis,
\begin{equation} \label{K}
{\cal K} \equiv \left( \begin{array}{cc}
                             \sigma^+ &  \varphi^-  \\
                             \varphi^+ & \sigma^- \end{array} \right)\,\, .
\end{equation}
Here, $\sigma^+$ and $\sigma^-$ are sources
which couple to adjoint quark spinors and quark spinors,
while $\varphi^+$ couples to two quark spinors 
(remember that $\bar{\psi}_C \sim \psi^T$)
and $\varphi^-$ couples to two adjoint quark spinors, respectively
($\psi_C \sim \bar{\psi}^T$). Charge-conjugation invariance of the action
relates the sources $\sigma^+$ and $\sigma^-$:
$\sigma^- \equiv C \, [ \sigma^+ ]^T\, C^{-1}$.
The action must also be real-valued, which leads to the
condition
$\varphi^- \equiv \gamma_0 [ \varphi^+ ]^\dagger \gamma_0$.

The next step is to derive the effective action from
the partition function of QCD in the presence of the external
source ${\cal K}$,
\begin{equation}
{\cal Z}[{\cal K}] = 
\int {\cal D} A_\mu^a \, {\cal D} \bar{\Psi} \, {\cal D} \Psi\,
\exp S[{\cal K}] \,\, .
\end{equation}
The details of this derivation are beyond the scope of this 
review, but the interested reader can readily convince himself
that, in the presence of bilocal sources, this problem 
is solved precisely by the CJT formalism \cite{CJT} discussed in
Sec.~\ref{HTLEOS}.
Consequently, if one takes into account that 
there is no nonvanishing expectation value for a 
fermionic one-point function, and
if one adapts Eq.~(\ref{effact}) to the notation of this section,
the effective action reads
\begin{equation}
\Gamma[A,\Delta,{\cal S}] = I[A] - \frac{1}{2}\, {\rm Tr} \, \ln \Delta^{-1}
- \frac{1}{2}\, {\rm Tr} \,\left( \Delta_0^{-1}\,\Delta - 1 \right) 
 + \frac{1}{2}\, {\rm Tr}\, \ln {\cal S}^{-1} + \frac{1}{2}\,{\rm Tr}\, 
       \left( {\cal S}_0^{-1} \, {\cal S} - 1 \right) 
 + \Gamma_2[\Delta,{\cal S}] \,\, ,
\end{equation}
where $\Delta_0$ is the free and $\Delta$ the full gluon propagator,
and ${\cal S}$ is the full fermion propagator in the Nambu-Gor'kov
basis. The factors of $1/2$ in front of the fermionic
terms account for the double-counting of
fermionic degrees of freedom in the Nambu-Gor'kov formalism. 
The full propagators for the physical situation (where
${\cal K} =0$) are obtained from the stationarity
conditions (\ref{statio}). These conditions are
Dyson-Schwinger equations of the type (\ref{DS}),
\begin{equation} \label{DSNG}
\Delta^{-1} = \Delta^{-1}_0 + \Pi\,\, ,\;\;\;\;
{\cal S}^{-1}  =  {\cal S}_0^{-1} + \Sigma \,\, .
\end{equation}
(In a slight abuse of notation, we
also denote the full gluon propagator
at the stationary point of $\Gamma[A,\Delta,{\cal S}]$ by
$\Delta$ and, similarly, the full quark propagator 
by ${\cal S}$.)

\begin{figure}[tb]
\begin{center}
\begin{minipage}[t]{12cm}
\epsfig{file=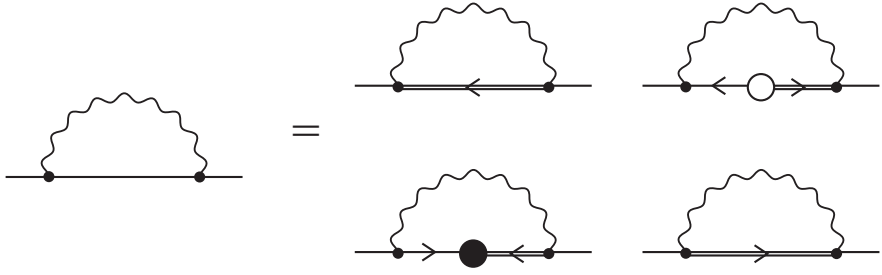,scale=0.7}
\end{minipage}
\begin{minipage}[t]{16.5 cm}
\caption{The quark self-energy. 
Writing out the individual components in the Nambu-Gor'kov
basis one obtains the diagrams on the right-hand side.
\label{quarkselfenergy}}
\end{minipage}
\end{center}
\end{figure}
 
In order to proceed one has to make an approximation for
$\Gamma_2$. As in Sec.~\ref{HTLEOS}, 
the discussion will be restricted to the set of
diagrams shown in Fig.~\ref{gamma2fig}, however, with
the quark propagators in the quark loop now given
by the Nambu-Gor'kov propagator ${\cal S}$.
The gluon self-energy is computed as 
$\Pi= - 2 \,\delta \Gamma_2 / \delta \Delta$, i.e., by cutting
a gluon line in the diagrams of Fig.~\ref{gamma2fig}. Thus,
$\Pi$ is given by the diagrams shown in Fig.~\ref{gluonselfenergy}
(with Nambu-Gor'kov propagators ${\cal S}$ in the quark loop).
The quark self-energy in the Nambu-Gor'kov basis is
\begin{equation} \label{NGselfenergy}
\Sigma \equiv \left( \begin{array}{cc} 
              \Sigma^+ & \Phi^- \\
              \Phi^+ & \Sigma^- \end{array} \right) 
  \equiv 2\, \frac{\delta \Gamma_2 [\Delta,{\cal S}]}{\delta {\cal S}}\,\, .
\end{equation}
It is given by the diagram shown on the left-hand side
in Fig.~\ref{quarkselfenergy}. 
The diagonal components $\Sigma^\pm$ correspond to the
ordinary self-energies for particles and charge-conjugate particles.
In space-time, $\Sigma^+(X,Y)$ ($\Sigma^-(X,Y)$) 
has a particle (charge-conjugate particle) entering at $X$ and
another particle (charge-conjugate particle) emerging at $Y$.
On the other hand, the off-diagonal components $\Phi^\pm$
have to interpreted as follows: a particle (charge-conjugate particle)
enters $\Phi^+(X,Y)$ ($\Phi^-(X,Y)$) at $X$ and a {\em
charge-conjugate particle\/} (an {\em ordinary particle\/}) emerges at
$Y$. This is typical for systems with a fermion-fermion condensate
in the ground state \cite{fetwal}. The self-energies $\Phi^\pm$
symbolize this condensate. Note that the two off-diagonal
components are related in the same way as the bilocal sources
$\varphi^\pm$ in Eq.~(\ref{K}), i.e.,
$\Phi^- \equiv \gamma_0 [\Phi^+]^\dagger \gamma_0$.
In the following, also the term {\em gap matrix\/} will be used
for $\Phi^+$.

In momentum space, the self-energy is given by
\begin{equation}
\Sigma(K) = - g^2 \, \int_Q
\Gamma^\mu_a \, {\cal S}(Q) \, \Gamma^\nu_b \, \Delta_{\mu \nu}^{ab}(K-Q)\,\, .
\end{equation}
This equation can be decomposed in terms of its
Nambu-Gor'kov components. To this end, let us first determine
the full Nambu-Gor'kov quark propagator by formally inverting the
Dyson-Schwinger equation (\ref{DSNG}) for the quark propagator
\cite{manuellifetime},
\begin{equation} \label{SNG}
{\cal S} =  \left( \begin{array}{cc}
                  G^+ & \Xi^- \\
                  \Xi^+  & G^- \end{array} \right) \,\, ,
\end{equation}
where
\begin{eqnarray} \label{Gplusminus}
G^\pm & \equiv & \left\{ [ G_0^\pm ]^{-1} + \Sigma^\pm
 - \Phi^\mp \, \left( [ G_0^\mp ]^{-1} + \Sigma^\mp \right)^{-1}\,
\Phi^\pm \right\}^{-1} \,\, , \\
\Xi^\pm & \equiv & - \left( [G_0^\mp ]^{-1} + 
\Sigma^\mp \right)^{-1}\, \Phi^\pm \, G^\pm \,\, . \label{Xiplusminus}
\end{eqnarray}
Here $G^+$ ($G^-$) is the propagator for quasiparticles (charge-conjugate
quasiparticles).
Besides these quantities describing the ordinary propagation
of quasiparticles, there are also off-diagonal, or
``anomalous'' propagators $\Xi^\pm$ in Eq.~(\ref{SNG}).
These anomalous propagators are typical for superconducting
systems \cite{fetwal} and account for the possibility that in the
presence of a Cooper-pair condensate, symbolized by
$\Phi^\pm$, a fermion can always be absorbed
in the condensate, while its charge-conjugate counterpart is emitted
from the condensate and continues to propagate.

In terms of its Nambu-Gor'kov components, the quark self-energy is
\begin{eqnarray}
\Sigma^+(K) & = & - g^2 \, \int_Q
\gamma^\mu T_a \, G^+(Q) \, \gamma^\nu T_b \, \Delta_{\mu \nu}^{ab}(K-Q)
\,\, , \\
\Sigma^-(K) & = & - g^2 \, \int_Q
\gamma^\mu T_a^T \, G^-(Q) \, \gamma^\nu T_b^T \, \Delta_{\mu \nu}^{ab}(K-Q)
\,\, , \\
\Phi^+(K) & = &  g^2 \, \int_Q
\gamma^\mu T_a^T \, \Xi^+(Q) \, \gamma^\nu T_b \, \Delta_{\mu \nu}^{ab}(K-Q)
\,\, , \label{Phi+}\\
\Phi^-(K) & = &  g^2 \, \int_Q
\gamma^\mu T_a \, \Xi^-(Q) \, \gamma^\nu T_b^T \, \Delta_{\mu \nu}^{ab}(K-Q)
\,\, . \label{Phi-}
\end{eqnarray}
The integrals in these equations are
shown diagrammatically on the right-hand side of Fig.~\ref{quarkselfenergy}.
In these diagrams, a normal full propagator $G^+$ ($G^-$) is denoted as
a double line with an arrow pointing to the left (right).
According to Eq.~(\ref{Xiplusminus}), the anomalous 
propagators $\Xi^+$ and $\Xi^-$ consist of a combination of
propagators $\left([G_0^\pm]^{-1} + \Sigma^\pm\right)^{-1}$, 
gap matrices $\Phi^\pm$, and full propagators $G^\pm$. This
combination is explicitly drawn on the right-hand side
of Fig.~\ref{quarkselfenergy}, with propagators
$\left([G_0^\pm]^{-1} + \Sigma^\pm\right)^{-1}$ 
as single lines with arrows pointing left/right, and gap matrices
$\Phi^\pm$ as full/empty circles.
Inserting these self-energies (and the corresponding one for
the gluons) into the Dyson-Schwinger equations (\ref{DSNG})
one obtains a coupled set of integral equations which has
to be solved self-consistently.
In particular, the Dyson-Schwinger equations for the off-diagonal
components $\Phi^\pm$ of the inverse propagator ${\cal S}^{-1}$,
i.e., Eqs.~(\ref{Phi+}) and (\ref{Phi-}), are
the {\em gap equations\/} for the color-superconducting
condensate.

A completely self-consistent solution of the Dyson-Schwinger equations
is technically too difficult to be feasible, 
and one has to make certain approximations. It turns
out that in weak coupling a well-controlled, approximate solution
is possible. This will be demonstrated in Sec.~\ref{gapequation}.
Prior to that, I shall discuss the excitation spectrum
in a superconductor, which
follows from the poles of the quark propagator (\ref{SNG}).

\subsection{\it Excitation spectrum} \label{secexspec}

In order to find the quasiparticle excitations, one has
to determine the poles of $G^\pm$ and $\Xi^\pm$. For an arbitrary
quark mass $m$, this is a formidable task, see Ref.~\cite{fugleberg}.
For our purpose it is sufficient to consider the
ultrarelativistic limit, $m=0$, where the situation simplifies
considerably \cite{prscalar}. Let us also focus exclusively on $G^+$;
the poles of $G^-$ and $\Xi^\pm$ can be determined accordingly
(in fact, they have precisely the same poles). In order to proceed, we
need to specify the color, flavor, and Dirac structure
of the inverse free propagators $[G_0^\pm]^{-1}$, and of the
self-energies $\Sigma^\pm$ and $\Phi^\pm$. 
The inverse propagators $[G_0^\pm]^{-1}$ are diagonal in color and
flavor space. To determine the Dirac structure, it is convenient
to Fourier transform them into energy-momentum space and then
to expand them in terms of projectors onto states of positive
or negative energies,
$\Lambda^e_{\bf k} \equiv (1 + e\, \gamma_0 \, \mbox{\boldmath$\gamma$}
\cdot {\bf k})/2$, $e = \pm$,
\begin{equation}
[G_0^\pm]^{-1} (K) = \gamma \cdot K \pm \mu\, \gamma_0
= \gamma_0 \sum_{e=\pm} \left[ k_0 - (ek \mp \mu) \right] \, 
\Lambda^e_{\bf k} \,\,.
\end{equation}
For the weak-coupling solution discussed in Sec.~\ref{gapequation}
it turns out that it suffices to compute the self-energies $\Sigma^\pm$ 
neglecting effects from
the breaking of the $[SU(3)_c]$ color symmetry due to condensation
of Cooper pairs. Then, the self-energies are diagonal 
in color and flavor space, and one only needs to know their Dirac structure.
One may expand them in terms of Dirac matrices,
\begin{equation}
\Sigma^\pm = s_0^\pm \, \gamma^0 + {\bf s}^\pm \cdot \mbox{\boldmath$\gamma$}
+ \ldots\,\, .
\end{equation}
Other Dirac structures are simply abbreviated by dots. It turns out
that, in weak coupling, the dominant contribution to the quark self-energy
arises from the exchange of almost static (i.e., Landau-damped)
magnetic gluons and, in momentum space, is \cite{manuelwavefct}
\begin{equation}
\Sigma^+(K) = \Sigma^-(K) \simeq \gamma_0 \, \bar{g}^2 \,
\left( k_0 \, \ln \frac{M^2}{k_0^2} + i \pi |k_0| \right)\,\, ,
\end{equation}
where $\bar{g} \equiv g / (3 \sqrt{2} \pi)$ and $M^2 \equiv (3 \pi/4) m_g^2$.
The imaginary part gives rise to a finite lifetime of quasiparticle
excitations off the Fermi surface \cite{manuellifetime}, but in
weak coupling this contributes to the color-superconducting gap
only at sub-subleading order. (How to count orders in weak coupling will be
discussed in detail in Sec.~\ref{gapequation}). 
There are other contributions at
sub-subleading order which have not been reliably computed so far. 
Only the subleading order terms are under complete analytic control.
They determine the prefactor of the color-superconducting gap to order
$O(1)$. The sub-subleading contributions contribute to
the prefactor at order $O(g)$. Therefore, in the following
the imaginary part will be neglected and only the real part
will be kept,
which, for $\ln (M/k_0) \sim 1/g$ contributes at subleading order
to the color-superconducting gap \cite{qwangdhr}.
Defining the wave function renormalization factor
\begin{equation} \label{wavefctrenfac}
Z(k_0) \equiv \left( 1 + \bar{g}^2 \, \ln\frac{M^2}{k_0^2} \right)^{-1}\,\, ,
\end{equation}
the effect of the quark self-energy is to shift the poles
in the propagator, $k_0 \rightarrow k_0/Z(k_0)$. Note that the logarithm
renders a normal-conducting system a non-Fermi liquid \cite{boyanovsky}.

The color, flavor, and Dirac structure of the off-diagonal
self-energies $\Phi^\pm$ is less trivial. As they symbolize
the Cooper-pair condensate, they must have the quantum
numbers of the particular channel where condensation occurs. 
For the purpose of illustration, I specify this structure for the 
four color-superconducting phases already discussed in Sec.~\ref{cscphases}.
Let us furthermore only consider parity-even spin-zero and
spin-one condensates. In this case, one may expand the
gap matrix $\Phi^+$ (in momentum space) as follows \cite{schmitt},
\begin{equation} \label{Phi+decomp}
\Phi^+ (K) = \sum_{e = \pm} \phi^e(K)\, {\cal M}_{\bf k} \, \Lambda^e_{\bf k}
\,\,,
\end{equation}
where $\phi^e$, the so-called {\em gap function},
is a scalar function of 4-momentum and
${\cal M}_{\bf k}$ is a matrix in color, flavor, and Dirac space, which
is determined by the symmetries of the color-superconducting order parameter 
(cf.~Sec.~\ref{cscphases}). An important property is
that it commutes with the energy projectors, 
$[{\cal M}_{\bf k}, \Lambda^e_{\bf k}] = 0$. 
 
\begin{table}
\begin{center}
\begin{minipage}[t]{16.5 cm}
\caption{The structure of the matrices ${\cal M}_{\bf k}$
and $L_{\bf k}$ in various color-superconducting phases,
$(\tau_2)^{fg} \equiv i \epsilon^{fg}$, 
$(J_k)_{ij} \equiv - i \epsilon_{ijk}$,
$(I^h)^{fg} \equiv - i \epsilon^{fgh}$,
and $\mbox{\boldmath$\gamma$}_\perp({\bf k}) \equiv 
\mbox{\boldmath$\gamma$} - \mbox{\boldmath$\gamma$} \cdot {\bf k} \, {\bf k}$.
For the matrix $L_{\bf k}$ in the CSL phase, the second
term is a matrix in color space, formed by the dyadic product
of the two vectors $\hat{\bf k} + \mbox{\boldmath$\gamma$}_\perp({\bf k})$
and $\hat{\bf k} - \mbox{\boldmath$\gamma$}_\perp({\bf k})$,
and a matrix in Dirac space, formed by the product of the
two $\mbox{\boldmath$\gamma$}_\perp({\bf k})$ matrices.
The last two columns show the two eigenvalues $\lambda_r$ of $L_{\bf k}$ and
their degeneracy $d_r$ (counting color-flavor degrees of freedom
in the 2SC and CFL phases, and color-Dirac degrees of freedom in
the two spin-one phases).
\vspace*{5mm}}
\label{tableM}
\end{minipage}
\begin{tabular}{|l|c|c|c|c|}
\hline
      &                     &                 &    &        \\[-2mm]
phase &  ${\cal M}_{\bf k}$ &   $L_{\bf k}$   & $\lambda_1\; (d_1)$
                                        & $\lambda_2\; (d_2)$ \\
      &                     &                 &    &        \\[-2mm]
\hline
      &                     &                 &    &        \\[-2mm]
2SC   &  $\gamma_5 \, \tau_2 \, J_3$ 
                     &  $(J_3)^2 \,(\tau_2)^2$  & 1 (4) & 0 (2) \\
      &                     &                 &    &        \\[-2mm]\hline
      &                     &                 &    &        \\[-2mm]
CFL   &  $\gamma_5 \, {\bf I} \cdot {\bf J} $  
                     & $({\bf I} \cdot {\bf J})^2$  &  4 (8) & 1 (1)   \\
      &                     &                 &    &        \\[-2mm]\hline
      &                     &                 &    &        \\[-2mm]
CSL   &  ${\bf J} \cdot \left[ \hat{\bf k} + \mbox{\boldmath
            $\gamma$}_\perp({\bf k}) \right]$      
& ${\bf 1} + [\hat{\bf k} + \mbox{\boldmath$\gamma$}_\perp({\bf k})]\,
             [\hat{\bf k} - \mbox{\boldmath$\gamma$}_\perp({\bf k})] $
                                                    & 4 (4) &  1 (8) \\
      &                     &                &     &       \\[-2mm]\hline
      &                     &                &     &       \\[-2mm]
polar &  $J^3 \left[ \hat{k}^z + 
               \gamma^z_\perp({\bf k})\right]$ & $(J_3)^2$ 
                                                    & 1 (8) & 0 (4) \\
      &                     &                &     &         \\[-2mm]\hline
\end{tabular}
\end{center}
\end{table}

In Table~\ref{tableM} the explicit expressions for
${\cal M}_{\bf k}$ are listed for the four phases considered here.
For these expressions, I have used the fact
that a color antitriplet is totally antisymmetric and thus
has a representation 
in terms of the antisymmetric Gell-Mann matrices $\lambda_2$,
$\lambda_5$, and $\lambda_7$. These Gell-Mann matrices 
form an $SO(3)$ subgroup of $SU(3)$, and are thus identical to the generators
of $SO(3)$, $(J_1,J_2,J_3) \equiv {\bf J}$. These matrices were finally used 
in Table~\ref{tableM} to parametrize that the
gap matrix $\Phi^+$ is a color-antitriplet.
Similarly, $\tau_2$ (the only Pauli matrix which is antisymmetric) 
symbolizes that the gap in the 2SC phase is a flavor singlet.
The Dirac matrix $\gamma_5$ in the 2SC and CFL cases 
is necessary to obtain even parity. 
For the CFL case, the flavor-antitriplet
nature is represented by another set of generators $(I^1, I^2, I^3) \equiv
{\bf I}$ of $SO(3)$. Color-flavor locking is obtained by
a scalar product of ${\bf I}$ with ${\bf J}$, cf.~also Table~\ref{table1}.
For the two spin-one phases, the CSL phase and the polar phase,
the order parameter is a 3-vector,
see Sec.~\ref{cscphases}, but the gap matrix is a scalar.
This requires it to be proportional to 
a scalar product of the order parameter with another
3-vector. There are only two other 3-vectors (in momentum space),
the direction of momentum of the quark in the Cooper pair,
$\hat{\bf k}$, and the vector $\mbox{\boldmath$\gamma$}$.
Consequently, $\Phi^+$ has to be proportional to a linear
combination of these two 3-vectors. It is convenient \cite{pr2} to use
the projection of $\mbox{\boldmath$\gamma$}$ onto the subspace
orthogonal to $\hat{\bf k}$, $\mbox{\boldmath$\gamma$}_\perp({\bf k})$,
because then ${\cal M}_{\bf k}$ commutes with the energy projectors.
Finally, color-spin locking
requires a scalar product of ${\bf J}$ with
this linear combination of 3-vectors. In the polar phase,
one may independently choose a direction for the gap
in color space and in space-time. Conveniently, one chooses
the 3- (anti-blue) direction in color space and the $z$-direction
in space-time.

In order to proceed, however, one does not require the explicit form
of ${\cal M}_{\bf k}$ in the various phases. The existence
of the decomposition (\ref{Phi+decomp}) and the commutation property of
${\cal M}_{\bf k}$ with the energy projectors is sufficient to
derive the quasiparticle spectrum as a function of the absolute
magnitude of the gap function, $|\phi^e(K)|$. To see this, compute 
$\Phi^- \equiv \gamma_0 [\Phi^+]^\dagger  \gamma_0$ and, together
with $\Phi^+$ and $[G_0^-]^{-1}+ \Sigma^-$, the quantity
\begin{equation}
\Phi^- \, \left( [G_0^-]^{-1} + \Sigma^- \right)^{-1} \, \Phi^+\,
\left([G_0^-]^{-1} + \Sigma^- \right)
= \sum_{e= \pm} |\phi^e(K) |^2 \, L_{\bf k}\, \Lambda_{\bf k}^{-e}\,\, ,
\end{equation}
where 
\begin{equation}
L_{\bf k} \equiv \gamma_0\, {\cal M}_{\bf k}^\dagger\,
{\cal M}_{\bf k} \, \gamma_0 
\end{equation}
is another central quantity for the quasiparticle excitation spectrum.
Expressions for the matrix $L_{\bf k}$
in the various phases are also listed in Table~\ref{tableM}.
Note that also $L_{\bf k}$ commutes with the energy projectors,
$[L_{\bf k}, \Lambda_{\bf k}^e] = 0$. Since $L_{\bf k}$ is hermitian,
it has real eigenvalues, $\lambda_r$, and can be expanded in terms
of a complete set of orthogonal projectors, ${\cal P}_{\bf k}^r$,
\begin{equation}
L_{\bf k} = \sum_r \lambda_r\, {\cal P}_{\bf k}^r\,\, .
\end{equation}
In the four phases considered here, there are only two distinct
eigenvalues and therefore two distinct projectors. The eigenvalues
are also listed in Table~\ref{tableM}, and the projectors can be
expressed in terms of $L_{\bf k}$ via
\begin{equation}
{\cal P}_{\bf k}^{1,2} = \frac{L_{\bf k} - \lambda_{2,1}}{\lambda_{1,2}
-\lambda_{2,1}}\,\, .
\end{equation}
Obviously, also these projectors commute with the energy projectors,
$[{\cal P}_{\bf k}^{1,2}, \Lambda_{\bf k}^e] = 0$.
It is now straightforward to compute the full propagator
by inverting the term in curly brackets in Eq.~(\ref{Gplusminus}),
since the projectors ${\cal P}^r \, \Lambda_{\bf k}^e$ form a 
complete, orthogonal set in color, flavor, and Dirac space,
\begin{equation} \label{G+}
G^+(K) = \left[ [G_0^-]^{-1}(K) + \Sigma^-(K) \right]
\sum _{e,r} {\cal P}_{\bf k}^r\, \Lambda_{\bf k}^{-e}\, 
\frac{1}{[k_0/Z(k_0)]^2 - \left[ \epsilon^e_{{\bf k},r}(\phi^e) \right]^2}
\,\, ,
\end{equation}
where
\begin{equation}
\epsilon^e_{{\bf k},r}(\phi^e) = \left[ (ek - \mu)^2 + \lambda_r\,
|\phi^e |^2 \right]^{1/2}\,\, .
\end{equation}
Obviously, the poles of the full propagator are located
at $k_0 \equiv \pm Z(k_0) \, \epsilon_{{\bf k},r}^e (\phi^e)$. 
Because of the $k_0$ dependence of the gap
function $\phi^e(K)$, this is a condition which has
to be solved self-consistently.

\begin{figure}[tb]
\begin{center}
\begin{minipage}[t]{12cm}
\vspace*{1cm}
\epsfig{file=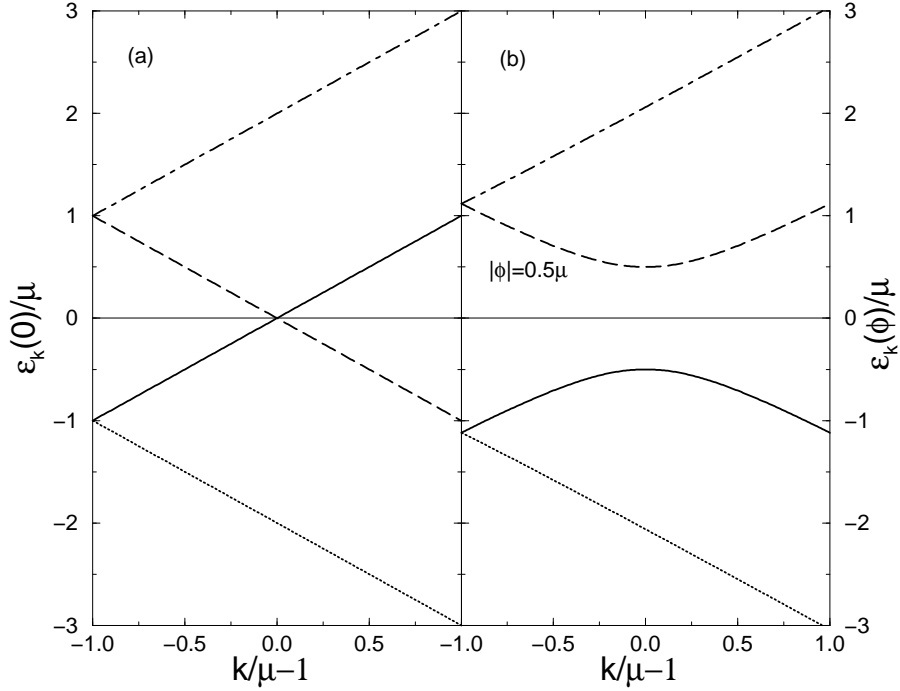,height=120mm,angle=-90}
\end{minipage}
\begin{minipage}[t]{16.5 cm}
\caption{The excitation spectrum for (a) non-interacting massless
particles and (b) quasiparticles in a superconductor.
The value of the gap function $\phi^e(K)$ is assumed to be constant, 
$\phi = 0.5 \, \mu$. The excitation energies for non-interacting
particles are $\epsilon_{\bf k}(0) = k-\mu$ for particles (solid),
$\epsilon_{\bf k}(0) = \mu -k$ for holes (dashed),
$\epsilon_{\bf k}(0) = - (k + \mu)$ for antiparticles (dotted),
and $\epsilon_{\bf k}(0) = k + \mu$ for antiparticle-holes (dash-dotted).
The excitation energies for quasiparticles
are $\epsilon_{\bf k}(\phi) = -\sqrt{(k-\mu)^2 + \phi^2}$
(solid), $\epsilon_{\bf k}(\phi) = \sqrt{(k-\mu)^2 + \phi^2}$
for quasiparticle-holes (dashed),
$\epsilon_{\bf k}(\phi) = -\sqrt{(k+\mu)^2 + \phi^2}$ for 
quasi-antiparticles (dotted),
and $\epsilon_{\bf k}(\phi) = \sqrt{(k+\mu)^2 + \phi^2}$ for 
quasi-antiparticle-holes (dash-dotted)
\cite{prscalar}.
\label{excitationspectrum}}
\end{minipage}
\end{center}
\end{figure}

In order to get an impression what the excitation spectrum of
the quasiparticles in a superconductor looks like, let us
for the moment approximate $\phi^e(K) \equiv \phi = const.$, and
also set $Z(k_0) \equiv 1$ (corrections are of order $O(\bar{g})$).
Let us also neglect the fact that there are two different sets of excitation
branches depending on the value of $\lambda_r$.
In Fig.~\ref{excitationspectrum}
the excitation spectrum is shown for non-interacting massless fermions
as well as for quasiparticles in a superconductor. 
The dispersion branches for
the quasiparticle excitations corresponding to negative energies,
$e = -$, i.e., the quasi-antiparticles and quasi-antiparticle-holes, 
hardly differs from the non-interacting antiparticle or antiparticle-hole
branches. As we shall see in Sec.~\ref{gapequation}, in weak coupling, 
$\phi \ll \mu$, such that to very good approximation
$\epsilon^-_{{\bf k}, r} \simeq k + \mu$.
On the other hand, the dispersion branches for the quasiparticle
excitations corresponding to positive energies, $e = +$, differ
considerably from the non-interacting particle or hole branches. The most
notable feature is an energy gap at the Fermi surface, $k = \mu$, 
between the quasiparticle
and quasiparticle-hole branches. This indicates that, in a superconductor,
it costs an energy $2\, \phi$ to excite quasiparticle--quasiparticle-hole
pairs at the Fermi surface. In contrast, in a non-interacting system
it costs no energy to excite particle-hole pairs at the Fermi surface.
The superconducting state is thus energetically favored compared
to the normal-conducting (non-interacting) state. 
As a rule
of thumb, the more fermionic excitations branches develop a gap 
(the more fermions form Cooper pairs),
and the larger the associated gap, the lower is the ground state energy,
and the more energetically favored is the particular superconducting
state.

In the CFL and CSL case, there are two different excitation 
branches with two different gaps, $\epsilon_{\mu,1}^+(\phi) = 
\sqrt{\lambda_1} \, \phi \equiv 2 \, \phi$ and $\epsilon_{\mu,2}^+(\phi) = 
\sqrt{\lambda_2} \, \phi \equiv \phi$.
Consequently, it costs twice the amount of energy to
excite quasiparticle excitations from the first branch than from
the second. 
In the 2SC and polar phases, there also two different excitation
branches, but the one corresponding to unpaired fermionic
excitations is gapless, $\epsilon_{\mu,2}^+ = 0$.
In the next section, the gap equation (\ref{Phi+})
for the gap function $\phi^e(K)$ will be solved, which allows us to
determine the magnitude of the gap at the Fermi surface.

\subsection{\it Solution of the Gap Equation} \label{gapequation}

In order to solve the gap equation (\ref{Phi+}), insert
Eqs.~(\ref{Phi+decomp}) and (\ref{G+}) into Eq.~(\ref{Xiplusminus}) for
$\Xi^+$,
\begin{equation} 
\label{Xi+}
\Xi^+(K) = - \sum_{e,r} \gamma_0 \, {\cal M}_{\bf k}\,\gamma_0\,
{\cal P}_{\bf k}^r\, \Lambda_{\bf k}^{-e}\, \frac{\phi^e(K)}{
[k_0/Z(k_0)]^2 -  \left[ \epsilon^e_{{\bf k},r}(\phi^e) \right]^2} \,\, .
\end{equation}
(By the way, this result demonstrates the claim made earlier that $G^+$ and
$\Xi^+$ have the same poles.) Now insert this equation into the
gap equation (\ref{Phi+}), multiply from the right with 
${\cal M}_{\bf k}^\dagger \, \Lambda_{\bf k}^e$, and trace over color,
flavor, and Dirac space. The result is an equation for the scalar
gap function $\phi^e(K)$,
\begin{equation}  \label{GEphie}
\phi^e(K) = g^2\, T\sum_n \int \frac{{\rm d}^3 {\bf q}}{(2 \pi)^3}
\sum_{e',s} \frac{\phi^{e'}(Q)}{\left[q_0/Z(q_0)\right]^2-
\left[\epsilon_{{\bf q},s}^{e'}(\phi^{e'})\right]^2} \, 
\Delta_{\mu\nu}^{ab}(K-Q) \,
{\cal T}^{\mu\nu, ee'}_{ab,s}({\bf k},{\bf q}) \,\, ,
\end{equation}
where 
\begin{equation} \label{T2SC}
{\cal T}^{\mu\nu, ee'}_{ab,s}({\bf k},{\bf q})=-\frac{ {\rm Tr}
\left[\gamma^\mu \, T_a^T \, \gamma_0 \,{\cal M}_{\bf q}\, \gamma_0
\,  {\cal P}_{\bf q}^s \, 
\Lambda_{\bf q}^{-e'}\, \gamma^\nu \, T_b\, {\cal M}^\dagger_{\bf k} \, 
\Lambda_{\bf k}^e\right]}{{\rm Tr}
\left[{\cal M}_{\bf k}\,  {\cal M}^\dagger_{\bf k} \, 
\Lambda_{\bf k}^e\right]} \,\, .
\end{equation}
The form of this gap equation is the same 
for all color-superconducting phases considered here.
The difference lies in the excitation spectrum and
the structure of the term ${\cal T}^{\mu\nu, ee'}_{ab,s}({\bf k},{\bf q})$.

At this point it is convenient to explain the power counting scheme
in weak coupling, $g \ll 1$. The right-hand side of
Eq.~(\ref{GEphie}) has a prefactor $g^2$. Consequently, 
in order to satisfy the equality, after performing the integral 
there have to be terms $\sim \phi/g^2$, which together with
the prefactor produce a term $\sim O(\phi)$, i.e., which is of the same order
as the left-hand side. These are the so-called terms of
{\em leading order\/} in the gap equation.
Then there are terms of so-called {\em subleading order}.
These enter the right-hand side of the gap equation
at order $O( g \phi)$. The terms of so-called {\em sub-subleading order\/}
are $\sim O(g^2 \phi)$. It turns out that only the terms of leading and 
subleading order can be reliably calculated in weak coupling.

In order to proceed, one has to make further approximations.
As shown in Ref.~\cite{dhr2SCselfenergy},
to leading and subleading order
one does not need the fully self-consistent gluon propagator; it suffices
to employ the gluon propagator in HDL approximation,
cf.~Sec.~\ref{HTLexcite}. 
The HDL gluon propagator is diagonal in adjoint colors,
$\Delta_{\mu \nu}^{ab} \equiv \delta^{ab} \Delta_{\mu \nu}$.
For the sake of definiteness,
I choose pure Coulomb gauge, where
\begin{equation}
\Delta_{00}(P) = \Delta_\ell(P)\,\, , \;\;\;\;
\Delta_{0i}(P) = 0 \,\, , \;\;\;\;
\Delta_{ij} (P) = \Delta_t(P) \, \left( \delta_{ij} - \hat{p}_i \, \hat{p}_j
\right) \,\, ,
\end{equation}
with $P \equiv K-Q$ and
the longitudinal and transverse propagators $\Delta_{\ell, t}$
introduced in Eq.~(\ref{translongprop}). 
In fact, it is not even necessary to use the full form of the
HDL propagator. In weak coupling, power counting along the lines of
argument given above reveals \cite{pr2,rockefeller,sonetal} that
the dominant, leading-order contribution to the gap equation
comes from almost static, Landau-damped magnetic gluons. 
Their propagator may be approximated by 
\begin{equation}
\Delta_t^{\rm LDM}(P) \simeq \frac{p^4}{p^6 + M^4 \, \omega^2}\, 
\Theta(M-p)\,\, ,
\end{equation}
where $M^2 = (3 \pi/4) m_g^2$.
To subleading order, there are two contributions, from non-static
magnetic gluons,
\begin{equation}
\Delta_t^{\rm NSM}(p) \simeq \frac{1}{p^2}\, \Theta(p-M)\,\, ,
\end{equation}
and from static electric gluons,
\begin{equation}
\Delta_{00}^{\rm SE}(p) \simeq - \frac{1}{p^2 + 3 m_g^2}\,\, .
\end{equation}
All other terms in the HDL gluon propagator contribute to sub-subleading
order.

In pure Coulomb gauge, one only
needs the $00$-component, ${\cal T}^{00, ee'}_{ab,s}({\bf k},{\bf q})$,
and the transverse projection of the $ij$-components of the
tensor ${\cal T}^{\mu\nu, ee'}_{ab,s}({\bf k},{\bf q})$,
\begin{equation}
{\cal T}^{t,ee'}_{ab,s}({\bf k},{\bf q})\equiv
- \left( \delta_{ij} - \hat{p}_i \, \hat{p}_j
\right) \, {\cal T}^{ij, ee'}_{ab,s}({\bf k},{\bf q})\,\, ,
\end{equation}
where the extra minus sign is convention. In Ref.~\cite{schmitt} it
was shown that one can write these components in terms of a power
series in $p^2/(kq)$, where $p \equiv |{\bf k} - {\bf q}|$, with
coefficients $\eta_{2m}^{\ell,t}$ which depend on $k$, $q$ and 
the product $e e'$, and an overall normalization factor $a_s$ which
is the same for the $00$- and the transverse component. 
The power series over $m$ start at $m = -1$ and, 
for spin-zero and spin-one gaps,
terminate at $m=2$. The normalization factors satisfy the
constraint $\sum_s a_s=1$. 

Performing the Matsubara sum in Eq.~(\ref{GEphie})
one then obtains
\begin{eqnarray}
\phi^e(\epsilon_{{\bf k},r}^e,k)& = &\frac{g^2}{16\pi ^2 k} 
\int_{\mu-\delta}^{\mu+\delta}
{\rm d} q \, q \sum_{e',s} a_s\,  Z(\epsilon_{{\bf q},s}^{e'})\,
\frac{\phi^{e'}(\epsilon_{{\bf q},s}^{e'},q)}{\epsilon_{{\bf q},s}^{e'}}\,
\tanh\left(\frac{\epsilon_{{\bf q},s}^{e'}}{2T}\right)
\sum_m  \int_{|k-q|}^{k+q} {\rm d}p\,p
 \left(\frac{p^2}{kq}\right)^m \nonumber \\
 && \hspace*{-2cm}  \times  \left\{- 2\, \Delta_{00}^{\rm SE}(p)\, 
\eta_{2m}^\ell 
+  \left[\;2\, \Delta_t^{\rm NSM}(p)
+ \Delta_t^{\rm LDM}\left(\epsilon_{{\bf q},s}^{e'}+\epsilon_{{\bf k},r}^e,p
\right)
+ \Delta_t^{\rm LDM}\left(\epsilon_{{\bf q},s}^{e'}-\epsilon_{{\bf k},r}^e,p
\right)
\right] \eta_{2m}^t \right\} .
\label{a1} 
\end{eqnarray}
Several approximations have been made to obtain this result.
First, the integration over $q$ has been restricted to a narrow
interval of length $2\, \delta$ around the Fermi surface, $\delta \sim M$.
It turns out that this approximation is good to subleading order;
due to the momentum dependence of the gap function, the value
of the cut-off affects the color-superconducting
gap parameter only at sub-subleading order \cite{pr2}.
Second, when evaluating the Matsubara sum via contour integration, 
to subleading order the value of the poles
$q_0 = \pm Z(q_0) \, \epsilon_{{\bf q},s}^{e'}$ may be approximated
by $q_0 \simeq \pm \epsilon_{{\bf q},s}^{e'}$ everywhere except
in the residue of the contour integral. 
The single factor of $Z(\epsilon_{{\bf q},s}^{e'})$ 
under the integral arises from the residue. 
In the argument of this factor, one has also made 
the approximation $q_0 \simeq \pm \epsilon_{{\bf q},s}^{e'}$, 
since the logarithm in $q_0$ in Eq.~(\ref{wavefctrenfac})
gives at most a subleading contribution to the integral
\cite{schmitt,qwangdhr}.
Third, the gap function was assumed to be an even function of
energy, $\phi^e(k_0) \equiv \phi^e(-k_0)$ \cite{pr2}.
Fourth, the gap function was assumed to be  isotropic in momentum
space, $\phi^e({\bf k}) \equiv \phi^e(k)$ \cite{schmitt}.
Finally, the imaginary part of $\phi^e$ was neglected \cite{pr2}.

How do the leading, $\sim O(\phi)$, and subleading, $\sim O(g \phi)$, 
terms arise? To this end, one has to power-count the different
contributions to the integral in Eq.~(\ref{a1}). 
One uses the fact (which will be confirmed below) that
in weak coupling, $\phi \sim \mu \exp(-1/g)$.
Neglecting all subtleties regarding different excitation branches,
and setting $\epsilon_{{\bf q},s}^{e'} \equiv \epsilon_{\bf q}^{e'}
\equiv \sqrt{(e'q- \mu)^2 + \phi^2}$, for $e'=+$
the integral over $q$ gives rise to a term
\begin{equation} \label{BCSlog}
\int_{\mu - \delta}^{\mu + \delta} \frac{{\rm d}q}{\epsilon_{\bf q}^+} 
\equiv 2 \int_0^\delta \frac{{\rm d} \xi}{\sqrt{\xi^2 + \phi^2}}
= 2 \, \ln\left( \frac{ \delta + \sqrt{\delta^2 + \phi^2}}{\phi} \right)
\simeq 2 \, \ln \left( \frac{2 \, \delta}{\phi} \right)
\sim \ln \left[ \frac{ g}{\exp(-1/g)} \right] \sim \frac{1}{g}\,\, ,
\end{equation}
where I have
substituted the variable $\xi \equiv q - \mu$
and used the fact that $\delta \sim M \sim g \mu \gg \phi \sim \mu \exp(-1/g)$.
The logarithm appearing in the estimate (\ref{BCSlog})
is called the ``BCS-logarithm'', because it also appears in standard
BCS theory \cite{fetwal}.
For $e'= -$, the BCS-logarithm does not occur,
as $\epsilon_{\bf q}^- \simeq q + \mu$, such that the integral
is parametrically only of order $\delta/\mu \sim M/\mu \sim g$
(provided that the gap function $\phi^-$ falls off sufficiently
fast that one may restrict the integral to a narrow range around
the Fermi surface).

For the leading-order contribution to the gap equation,
we need another term which is also $\sim 1/g$, such that this term
and the BCS-logarithm combine to give a contribution $\sim 1/g^2$ which
cancels the prefactor $g^2$ in front of the integral in Eq.~(\ref{a1}).
To estimate the order of magnitude of the
remaining terms, one notes that the coefficients
$\eta_{2m}^{\ell, t}$ are parametrically at most of order $O(1)$ 
\cite{schmitt},
such that they can be neglected for the purpose of power counting.
The term $\sim 1/g$ which we are looking for
arises from the term $m=0$ in the sum over $m$
in conjunction with the Landau-damped magnetic gluon propagator.
Abbreviating $\omega_\pm \equiv \epsilon_{{\bf q},s}^{e'} \pm
\epsilon_{{\bf k},r}^e$, one estimates
\begin{equation} \label{ll}
\int_{k-q}^{k+q} {\rm d}p \, p \, \Delta_t^{\rm LDM} (\omega_\pm,p)
= \int_{k-q}^{M} {\rm d}p \, \frac{p^5}{p^6 + M^4 \omega_\pm^2}
= \frac{1}{6} \, \ln \left[ \frac{M^6 + M^4 \omega_\pm^2}{
(k-q)^6 + M^4 \omega_\pm^2} \right]
\sim \ln \left( \frac{M^2}{\omega_\pm^2} \right)\,\, ,
\end{equation}
where the approximation $k \simeq q$ was used. 
(Only when $k - q \simeq 0$, the logarithm may become large,
see argument below. Otherwise, if $k - q \sim M$, the
logarithm is only of order $O(1)$, not $O(1/g)$.)
If either $e= -$ or
$e'=-$, or both $e = e'=-$, the logarithm is parametrically of order $O(1)$,
and not $\sim 1/g$. Consequently, the only case of interest is if
both $e = e' =+$. In this case, $\omega_\pm \sim \phi$, and
the logarithm is large, $\ln (\mu/\phi) \sim 1/g$.

One readily convinces oneself that
the $p$ integral over the other terms in Eq.~(\ref{a1}) gives at most
a contribution of order $O(1)$. In combination with the BCS-logarithm, this
leads to a subleading contribution in the gap equation.
From the quasi-antiparticle poles, $e'=-$, one does not obtain a large
BCS-logarithm, but a term $\sim g$. With the prefactor
and a factor of $\phi$ from the gap function under the integral,
their contribution to the gap equation
is of order $g^3 \phi$, i.e., even beyond sub-subleading order.
In the following, one may therefore safely neglect the contribution 
from quasi-antiparticles
when computing the gap for the quasiparticle excitations, $e=+$.

\begin{table}
\begin{center}
\begin{minipage}[t]{16.5 cm}
\caption{The normalization factors $a_s$, the
coefficients $\eta_{2m}^{\ell,t}$, and the constant $d$
from Eq.~(\ref{b}) in various color-superconducting phases.
In the polar phase, $\vartheta$ is the angle
between the direction of the color-superconducting order parameter
and the momentum of the quarks in the Cooper pair.
\vspace*{5mm}}
\label{tableeta}
\end{minipage}
\begin{tabular}{|l|c|c|c|c|c|c|c|c|c|}
\hline
      &      &      &     &      &      &        &       &    &\\[-2mm]
phase &  $a_1$ & $a_2$  & $\eta_0^\ell$ &  $\eta_2^\ell $ & $\eta_4^\ell$ 
      &  $\eta_0^t$ &  $\eta_2^t $ & $\eta_4^t$ & $ d$    \\
      &      &      &     &      &      &        &       &    & \\[-2mm]
\hline
      &      &      &     &      &      &        &       &    & \\[-2mm]
2SC   &  1   &  0   & $\frac{2}{3}$ &  $-\frac{1}{6}$ &  $0$           
      &  $\frac{2}{3}$ &  $\frac{1}{6}$  &  $0$          & $0$       \\
      &      &      &     &      &      &        &       &    & \\[-2mm]\hline
      &      &      &     &      &      &        &       &    & \\[-2mm]
CFL   &  $\frac{1}{3}$ &  $\frac{2}{3} $ & $\frac{2}{3}$ & $-\frac{1}{6}$ 
      &  $0$  &  $\frac{2}{3}$ &  $\frac{1}{6}$  &  $0$  & $0$   \\
      &      &      &     &      &      &        &       &    & \\[-2mm]\hline
      &      &      &     &      &      &        &       &    & \\[-2mm]
CSL   &  $\frac{2}{3}$ &  $\frac{1}{3} $ & $\frac{2}{3}$ &  $-\frac{7}{18}$ 
      &  $\frac{1}{18}$ & $\frac{2}{3}$ & $-\frac{5}{18}$ & $0$ & $5$ \\
      &      &      &     &      &      &        &       &    &  \\[-2mm]\hline
      &      &      &     &      &      &        &       &    &  \\[-2mm]
polar & 1   &  0   & $\frac{2}{3}$ &  $-\frac{2+\cos^2 \vartheta}{6}$ 
              &  $\frac{1+ \cos^2 \vartheta}{24}$           
      &  $\frac{2}{3}$ &  $-\frac{2-\cos^2 \vartheta}{6}$  
              &  $\frac{1- 3 \cos^2 \vartheta}{24}$        & 
                 $ \frac{3}{2}\, (3+ \cos^2 \vartheta)$       \\
      &      &      &     &      &      &        &       &    &  \\[-2mm]
\hline   
\end{tabular}
\end{center}
\end{table}

In order to proceed, one performs the $p$ integrals,
which can be done exactly \cite{schmitt}, but only need to
be known to order $O(1)$.
One furthermore analyzes the coefficients $\eta_{2m}^{\ell,t}$
and realizes \cite{schmitt} that,
to subleading order, one may approximate $k \simeq q \simeq \mu$ in 
the expressions for these coefficients. 
Thus, they become pure
numbers of order $O(1)$. Moreover, to subleading order the coefficients
$\eta_{-2}^{\ell , t} \equiv 0$ and need not be considered further.
I list the coefficients for $m=0,1,2$ in Table~\ref{tableeta} 
together with the normalization factors $a_s$ for the
four color-superconducting phases considered here.
The final result for the gap equation can be written in the concise
form \cite{schmitt} (let us omit the superscript ``+'' for the sake
of simplicity)
\begin{equation}
\phi(\epsilon_{{\bf k},r},k)=\bar{g}^2
\int_0^\delta {\rm d}(q-\mu)\sum_s a_s \; Z(\epsilon_{{\bf q},s})\,
\frac{\phi(\epsilon_{{\bf q},s},q)}{\epsilon_{{\bf q},s}}\;
\tanh\left(\frac{\epsilon_{{\bf q},s}}{2T}\right)\;\frac{1}{2}
\ln\left(\frac{b^2\mu^2}{|\epsilon_{{\bf q},s}^2-\epsilon_{{\bf k},r}^2|}
\right) \,\, ,
\label{gap1}
\end{equation}
which is exact to subleading order. In Eq.~(\ref{gap1}) we
have introduced 
\begin{equation} \label{b}
b\equiv \tilde{b}\,\exp(-d) \,\,, \;\;\;\;
\tilde{b} \equiv 256 \pi^4 \left(\frac{2}{N_f g^2} \right)^{5/2}\,\, , \;\;\;\;
d = -\frac{6}{\eta_0^t}\, \left[\eta_2^\ell+\eta_2^t+2(\eta_4^\ell+\eta_4^t)
\right] \,\, .
\end{equation}
The $N_f$-dependence of $b$ arises from the corresponding dependence
of the gluon mass parameter $m_g$, cf.~Eq.~(\ref{m_g}).
The values for the constant $d$ are also listed in Table~\ref{tableeta}.
For the spin-zero color-superconducting phases, $d = 0$, due to an
accidental cancellation of the coefficients $\eta_2^\ell$
and $\eta_2^t$. This does not happen in the spin-one phases and,
consequently, $d \neq 0$.

In order to solve Eq.~(\ref{gap1}), one makes the following approximation 
which was first proposed by Son \cite{sonetal} and is
valid to subleading order,
\begin{equation}
\frac{1}{2}\,\ln\left(\frac{b^2\mu^2}{|\epsilon_{{\bf q},s}^2-
\epsilon_{{\bf k},r}^2|}\right)
\simeq\Theta(\epsilon_{{\bf q},s}-\epsilon_{{\bf k},r})\, 
\ln \left(\frac{b\mu}{\epsilon_{{\bf q},s}}\right)
+\Theta(\epsilon_{{\bf k},r}-\epsilon_{{\bf q},s})\, 
\ln \left(\frac{b\mu}{\epsilon_{{\bf k},r}}\right) \,\, .
\end{equation}
The remainder of the calculation is technical, but straightforward
and given in detail in Ref.~\cite{schmitt}. To summarize the
steps, a suitable substitution
of variables allows to rewrite the gap equation (\ref{gap1}), which
is an integral equation, in terms of Airy's differential equation
\cite{schmitt,qwangdhr}.
The result for the gap function has the form
\begin{equation} \label{solution}
\phi(x_r) \equiv \phi_0 \, F(x_r)\,\, ,
\end{equation}
where $\phi_0$ is the value of the gap function at the Fermi surface,
i.e., the color-superconducting gap parameter or ``gap'', and
$F(x_r)$ parametrizes
the momentum dependence of the gap function. The variable $x_r$ is
defined as 
\begin{equation}
x_r \equiv \bar{g}\, \ln \left( \frac{2 b \mu}{k-\mu + \epsilon_{{\bf k},r}}
\right)\,\, .
\end{equation}
At the Fermi surface, $k= \mu$, one has $x_r \equiv x_r^*
= \bar{g} \ln [2 b \mu / (\sqrt{\lambda_r} \phi_0)]= \pi/2 + O(\bar{g})
\sim O(1)$. 
If one moves away from the Fermi surface, $x_r$ stays of order $O(1)$,
as long as the momentum difference from the Fermi surface is
$|k - \mu| \sim O(\phi)$. When $|k - \mu| \sim M$ or larger, 
$x_r \sim O(\bar{g})$.
The precise form of the function $F(x_r)$ is not very illuminating
(it consists of a combination of Airy functions \cite{schmitt,qwangdhr}),
and thus will not be discussed here.
All one needs to know is that it has a narrow
peak in an interval $|k - \mu| \sim O(\phi)$ around the Fermi surface.
At the Fermi surface, $x_r \equiv x_r^*$,
the function $F(x_r)$ assumes the value $F(x_r^*) \equiv 1 + O(\bar{g}^2)$.
At a distance $|k- \mu| \sim M$ from the Fermi surface, 
$F(x_r) \sim O(\bar{g})$.
If one neglects the factor $Z(\epsilon_{{\bf q},s})$ in
Eq.~(\ref{gap1}),
the differential equation satisfied by the gap function
is that of the harmonic oscillator and, consequently, the
solution of the gap equation becomes simpler and more amenable 
to interpretation: $F(x_r) \equiv \sin x_r$ \cite{pr2,sonetal}.

The value of the gap function at the Fermi surface is
\begin{equation}
\label{phi0}
\phi_0 = 2\,\, b \, b_0' \, \mu \, \exp\left(- \frac{\pi}{2 \, \bar{g}}
\right)\, \left( \lambda_1^{a_1} \, \lambda_2^{a_2} \right)^{-1/2}\,\, .
\end{equation}
The constant $b_0' \equiv \exp[-(\pi^2 + 4)/8]$ arises from the
wave function renormalization factor $Z(\epsilon_{{\bf q},s})$ in
Eq.~(\ref{gap1}) \cite{rockefeller,qwangdhr}.
The result (\ref{phi0}) differs from the standard
BCS result in the power of the coupling constant $g$
in the exponent. In weak-coupling BCS theory, $\phi_0 \sim \exp(-1/g^2)$,
while here $\phi_0 \sim \exp(-1/g)$. The difference in the
parametric dependence on $g$ arises from the long-range nature
of magnetic gluon exchange. 
In BCS theory, the attractive interaction is assumed to
be short-range (point-like or at least exponentially screened).
On the other hand, in QCD static magnetic gluon exchange is 
not screened \cite{lebellac}. Almost static
magnetic gluons are dynamically screened, but the screening is rather
weak. It gives rise to the large logarithm (\ref{ll})
in addition to the BCS logarithm (\ref{BCSlog}). This reduces
the power of $g$ in the exponent. 

In Table~\ref{tablephi} I list the value of the gap $\phi_0$ in units
of its value in the 2SC phase. For the spin-one gaps, the nonzero
value of the constant $d$ leads to a strong suppression $\sim e^{-d}
\sim 10^{-2} - 10^{-3}$ as compared to the spin-zero gaps. In the
CFL and CSL phases, the second gapped excitation leads to a nontrivial
factor $\left( \lambda_1^{a_1} \, \lambda_2^{a_2} \right)^{-1/2} < 1$,
which reduces the gap as compared to the 2SC and polar phases where
there is only a single gapped excitation.

\begin{table}
\begin{center}
\begin{minipage}[t]{16.5 cm}
\caption{The value of the gap function at the Fermi surface,
$\phi_0$ in units of its value in the 2SC phase,
and the critical temperature, in units of its value
expected from BCS theory, Eq.~(\ref{BCS}),
and in units of the critical temperature in the 2SC phase,
$T_c^{\rm 2SC}$.
\vspace*{5mm}}
\label{tablephi}
\end{minipage}
\begin{tabular}{|l|c|c|c|}
\hline
      &                           &                      &           \\[-2mm]
phase &  $\phi_0/\phi_0^{\rm 2SC}$ & $T_c/T_c^{\rm BCS}$  & 
                                                $T_c/T_c^{\rm 2SC}$    \\
      &                           &                      &           \\[-2mm]
\hline
      &                           &                      &           \\[-2mm]
2SC   &                1          &       1              &   1       \\
      &                           &                      &    \\[-2mm]\hline
      &                           &                      &           \\[-2mm]
CFL   &  $2^{-1/3}$               &  $2^{1/3}$           &   1       \\
      &                           &                      &   \\[-2mm]\hline
      &                           &                      &           \\[-2mm]
CSL   &  $2^{-2/3}\, e^{-d}$      &  $2^{2/3}$           & $e^{-d}$  \\
      &                           &                      &   \\[-2mm]\hline
      &                           &                      &           \\[-2mm]
polar & $e^{-d}$           &   1                  & $e^{-d}$  \\
      &                           &                      &           \\[-2mm]
\hline   
\end{tabular}
\end{center}
\end{table}

\begin{figure}[tb]
\begin{center}
\begin{minipage}[t]{14cm}
\vspace*{1cm}
\epsfig{file=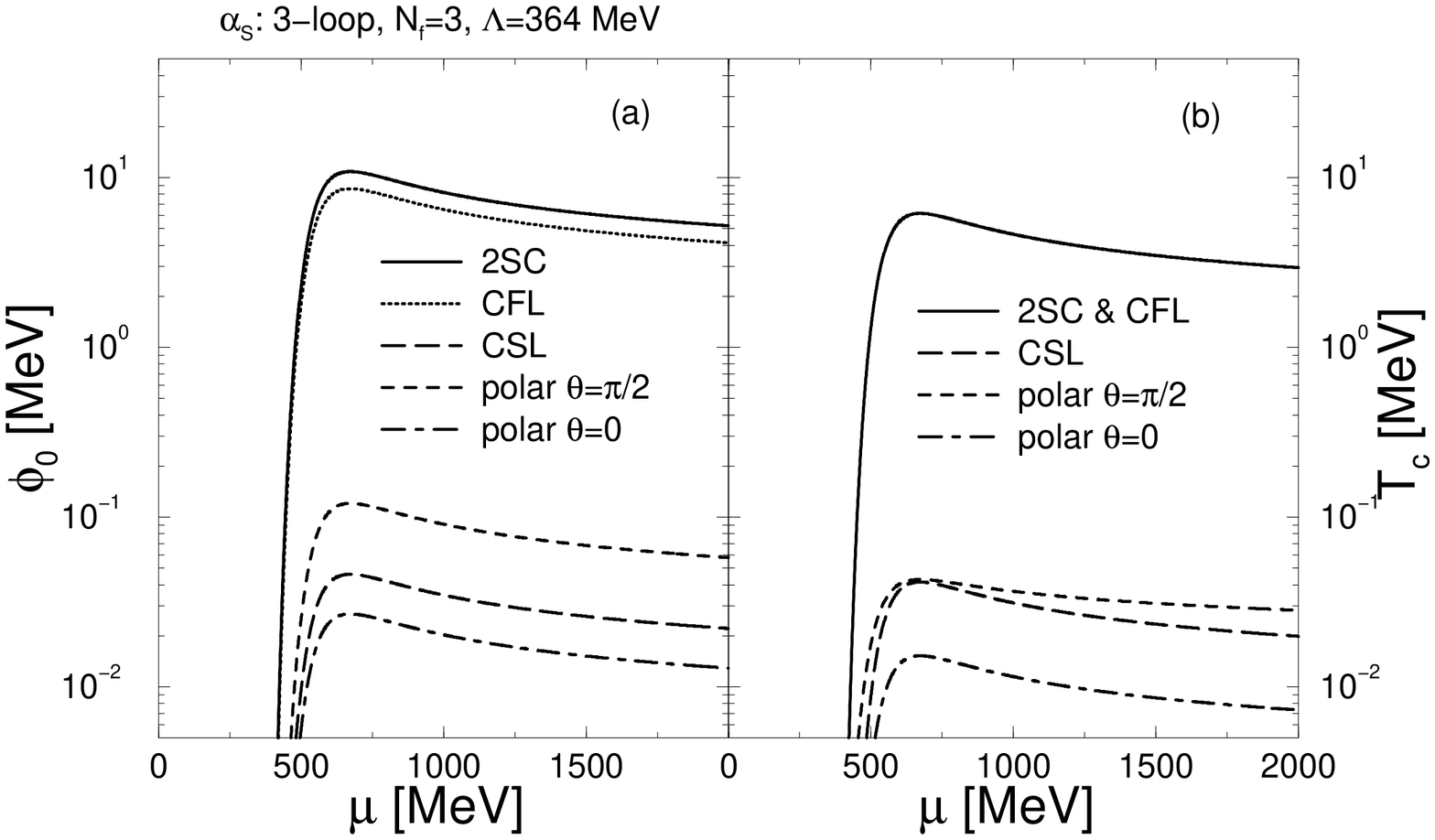,height=140mm,angle=-90}
\end{minipage}
\begin{minipage}[t]{16.5 cm}
\caption{(a) The gap and (b) the critical temperature 
as a function of the quark chemical potential. Solid curves
are for the 2SC phase, and dotted curves for the CFL phase.
(In the case of the critical temperature, both curves
coincide). The long-dashed curve is for the CSL phase, and the dashed
and dash-dotted curves are for the polar phase with $\vartheta = \pi/2$
and $\vartheta = 0$, respectively.
\label{phimu}}
\end{minipage}
\end{center}
\end{figure}

The result (\ref{phi0}) is rigorously valid in weak coupling, i.e.,
for asymptotically large quark chemical potentials, where
the value of the strong coupling constant evaluated at the
scale $\mu$ is small, $g(\mu) \ll 1$. However, for
phenomenology it is of considerable interest to determine
the gap also at values of $\mu$ which might occur in nature,
for instance in the core of compact stellar objects.
To this end, one extrapolates the weak-coupling result (\ref{phi0})
to large values of $g(\mu) \sim O(1)$. Such an extrapolation
has to be considered under the caveat that the sub-subleading terms 
are not really small for $g(\mu) \sim 1$ and could lead to large
deviations of the actual value of the gap from the subleading
result (\ref{phi0}). Nevertheless, the computation
of $\phi_0$ at $g(\mu) \ll 1$ is a well-posed problem with a
definite result, and so is its extrapolation to large values of $g(\mu)$.
In this sense, this approach should be considered to
be more reliable than {\em ad hoc}
calculations within NJL-type models which are very popular in 
the description of color-superconducting quark matter \cite{rajwil}.

For the running of the strong coupling constant $g$ with
$\mu/\Lambda$, where $\Lambda$ is the QCD scale parameter,
I take the standard 3-loop formula \cite{PDB}.
I assume that there are only $N_f=3$ active quark flavors involved
in the running of the coupling constant, so that
in order to obtain the correct value of $g(\mu)$ at the
mass of the $Z$ boson, one has to adjust the QCD scale parameter,
$\Lambda = 364$ MeV. I also take $N_f =3$ in the factor $b$
in Eq.~(\ref{b}). Physically, this means that, independent of
the number of quark flavors which form Cooper pairs, there are
always three (massless) quark flavors which screen color charges. 
The result of extrapolating Eq.~(\ref{phi0}) to
realistic values of $\mu$ is shown in Fig.~\ref{phimu} (a)
for the various color-superconducting phases considered here.

One observes that the 2SC phase has the largest
gap, $\phi_0^{\rm 2SC} \simeq 10$ MeV followed by the CFL phase. 
The spin-one phases have gaps which
are about $2 -3$ orders of magnitude smaller,
$\phi_0^{J=1} \sim 10^{-2} - 10^{-1}$ MeV. The gap 
is approximately zero for chemical potentials below 500 MeV, 
rapidly rises to assume a maximum around $\mu \simeq 600$ MeV and
then decreases. (For larger values of $\mu$ it will eventually
increase again.) This behavior is due to the dependence of $\phi_0$
on $g$. For large values of $g$ (small values of $\mu$), the 
power-law behavior $g^{-5}$ from the prefactor $b$ leads
to a suppression, while at small values of $g$ (large values
of $\mu$), the exponential suppression $\sim \exp(-1/g)$ dominates.
This leads to a maximum for intermediate values of $\mu$. 
(For asymptotically large values of $\mu$, the gap increases again,
because the prefactor $\mu$ dominates the $\mu$ dependence of
the remaining factors.)

One can also solve the gap equation at nonzero temperature. One
finds that the shape $F(x_r)$ of the gap function hardly
changes with $T$, but that the value of the gap decreases \cite{pr2}.
The gap equation (\ref{Phi+}) is equivalent to the
one obtained in the mean-field approximation \cite{prscalar}, and
therefore the temperature dependence of the gap follows the
predictions from mean-field theory. In particular, the transition
to the normal-conducting phase is of second order, irrespective
of the symmetries of the order parameter. The critical temperature
$T_c$ for this transition can be computed analytically, for details
see Refs.~\cite{pr2,schmitt,qwangdhr}. The result is \cite{schmitt}
\begin{equation} \label{Tc}
T_c = \frac{e^\gamma}{\pi}\, \phi_0 \, 
\left( \lambda_1^{a_1} \, \lambda_2^{a_2} \right)^{1/2}
\left[ 1 + O(g) \right]\,\, .
\end{equation}
This result is surprising for two reasons. First,
in a phase with a single gapped excitation, like the 2SC phase or the polar
phase, where 
$\left( \lambda_1^{a_1} \, \lambda_2^{a_2} \right)^{1/2} \equiv 1$,
the critical temperature in QCD, measured in units of the gap,
is the {\em same\/} as in BCS theory \cite{pr2}, at least to leading order
in weak coupling. This is unexpected, since we have seen
that the dependence of the gap itself on $g$ is parametrically very different
than in BCS theory. Second, in a phase with two different
nontrivial excitation branches, like the CFL and CSL phase, the
factor $\left( \lambda_1^{a_1} \, \lambda_2^{a_2} \right)^{1/2} \neq 1$
{\em violates\/} the expectation from BCS theory \cite{schafer,schmitt}.
In Table~\ref{tablephi} I show $T_c$ in units
of the critical temperature expected from BCS theory, 
$T_c^{\rm BCS} \equiv (e^\gamma/\pi) \phi_0$,
to demonstrate this violation. In physical units, say the value
of the critical temperature in the 2SC phase, 
$T_c^{\rm 2SC} = (e^\gamma/\pi) \phi_0^{\rm 2SC}$, 
the factor $\left( \lambda_1^{a_1} \, \lambda_2^{a_2} \right)^{1/2}$
cancels against its inverse in Eq.~(\ref{phi0}). This leads
to the conclusion that, in the mean-field type approach
pursued here, the critical temperatures in the 2SC and
CFL phases are actually identical.
The critical temperatures in the spin-one phases are just a factor
$e^{-d}$ smaller than in the 2SC and CFL phases \cite{rockefeller}.

The critical temperature (\ref{Tc}) is shown as a function
of $\mu$ in Fig.~\ref{phimu} (b). These curves also define the
boundaries of the color-superconducting phases in the phase
diagram of nuclear matter, cf.~Fig.~\ref{phasediagramNM}. The subleading
result (\ref{Tc}) for $T_c$ implies that one would have to cool quark
matter below temperatures of order 5 MeV, before one enters
a color-superconducting quark matter phase (the 2SC or CFL phase). This means
that, unless sub-subleading corrections to the gap (\ref{phi0})
(and thus to $T_c$) are large, color-superconductivity is 
irrelevant in the context of heavy-ion physics, but that it may
play a large role for compact stellar objects which have a sufficiently
dense core. While spin-zero color-superconducting matter may occur already
quite early in the evolution of such a compact stellar object, i.e.,
while it is still comparatively hot, matter in
a spin-one color-superconducting state only occurs after the core of
the stellar object has cooled below a temperature of order 10 keV,
i.e., in the later stage of its evolution \cite{prakash}.

\subsection{\it Gluon and photon properties} \label{gluephoton}

In this section, I take a first step towards a self-consistent
solution of the Dyson-Schwinger equations (\ref{DSNG}) and
compute gluon properties in a color superconductor.
Within the two-loop approximation to $\Gamma_2$, the gluon
self-energy consists of the diagrams shown in Fig.~\ref{gluonselfenergy}.
At temperatures of relevance for color superconductivity,
$T \leq T_c \sim \phi_0 \sim \mu\, \exp(-1/g) \ll \mu$, we
may neglect the contributions from the gluon (and ghost) loops to
the gluon self-energy: they are $\sim g^2 T^2$, while
the quark loop is $\sim g^2 \mu^2 \gg g^2 T^2$.
Thus, the gluon self-energy in momentum space is
\begin{equation} \label{glueNG}
\Pi^{\mu \nu}_{ab} (P) = \frac{g^2}{2} \, \int_K 
{\rm Tr} \left[ \Gamma^\mu_a
\, {\cal S} (K) \, \Gamma^\nu_b \, {\cal S}(K-P) \right]
\,\,,
\end{equation}
where the trace runs over color, flavor, Dirac, and Nambu-Gor'kov space.
By introducing the Nambu-Gor'kov basis one has effectively
doubled the degrees of freedom by introducing charge-conjugate
quarks in addition to quarks \cite{rischke2SC}. The factor
$1/2$ in Eq.~(\ref{glueNG}) prevents overcounting these
degrees of freedom.

Similarly to the gluon self-energy one can compute
the photon self-energy $\Pi^{\mu \nu}_{\gamma \gamma}$ replacing
the quark-gluon vertices $\Gamma^\mu_a,\, \Gamma^\nu_b$ in
Eq.~(\ref{glueNG}) by the corresponding ones for
the coupling between quarks and photons,
\begin{equation} \label{quarkphotonNG}
\Gamma^\mu_\gamma \equiv \frac{e}{g} \left( \begin{array}{cc}
                           \gamma^\mu\, Q & 0 \\
                            0 & - \gamma^\mu\, Q  \end{array} \right)\,\, ,
\end{equation}
where $Q \equiv {\rm diag} (2/3, - 1/3, - 1/3)$ is the quark electric
charge matrix.
Furthermore, as discussed in Sec.~\ref{classcsc}, 
in a color superconductor gluons can mix with the photon,
leading to a ``rotated'' electromagnetic
$[\tilde{U}(1)]$ symmetry in the 2SC and CFL phases.
This fact manifests itself in a nonvanishing ``mixed''
gluon-photon self-energy $\Pi^{\mu \nu}_{a \gamma}$, which follows
from Eq.~(\ref{glueNG}) by replacing just one of the quark-gluon
vertices with the quark-photon vertex (\ref{quarkphotonNG}).
In order to determine the gluon and photon properties in a color
superconductor, one has to compute all these different
self-energies. For the sake of convenience, in the following let us
set the index $\gamma \equiv 9$ and consider 
Eq.~(\ref{glueNG}) for $a, b = 1, 2, \ldots, 9$.
I also introduce $T_9 \equiv (e/g) Q$ as the appropriate
generator for $[U(1)_{\rm em}]$.

\begin{figure}[tb]
\begin{center}
\begin{minipage}[t]{13cm}
\epsfig{file=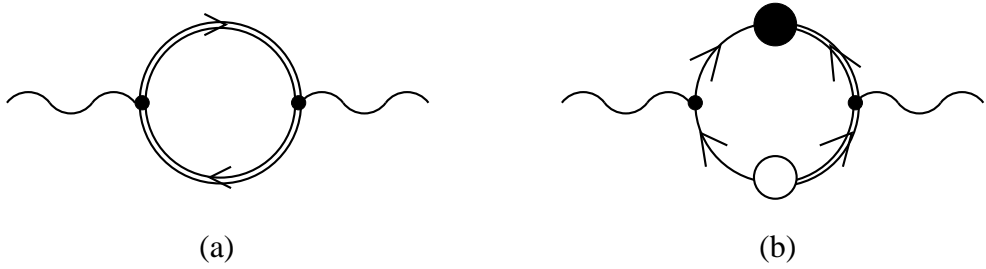,height=35mm}
\end{minipage}
\begin{minipage}[t]{16.5 cm}
\caption{Contributions from (a) normal and (b) anomalous quark propagation
to the self-energies of gluons and photons, and the mixed
gluon-photon self-energy. The notation follows that of 
Fig.~\ref{quarkselfenergy}.
\label{gluonPiNG}}
\end{minipage}
\end{center}
\end{figure}

Taking the trace over Nambu-Gor'kov space in Eq.~(\ref{glueNG}), 
one realizes that 
the quark loop consists of four contributions, two ``regular'' ones
with normal propagators $G^{\pm}$ for quarks and charge-conjugate quarks
and two with anomalous propagators $\Xi^\pm$, 
\begin{eqnarray}
\Pi^{\mu \nu}_{ab} (P) \!\! & = & \!\! \frac{g^2}{2} \, 
\int_K
{\rm Tr} \left[ \gamma^\mu T_a \, G^+ (K) \, \gamma^\nu T_b 
\, G^+(K-P) + \gamma^\mu T_a^T \, G^- (K) \, \gamma^\nu T_b^T 
\, G^-(K-P) \right. \nonumber \\
&   & \left. \hspace*{1.3cm}
- \gamma^\mu T_a \, \Xi^- (K) \, \gamma^\nu T_b^T 
\, \Xi^+(K-P) - \gamma^\mu T_a^T \, \Xi^+ (K) \, \gamma^\nu T_b 
\, \Xi^-(K-P)  \right] . 
\label{PiPscf}
\end{eqnarray}
The two different topologies corresponding to these
contributions are shown in Fig.~\ref{gluonPiNG}.

To further evaluate the trace one has to specify which
color-superconducting phase one would like to consider.
The form of the propagators $G^\pm, \Xi^\pm$ can then be
determined following the method outlined in Sec.~\ref{secexspec},
see Eqs.~(\ref{G+}), (\ref{Xi+}). After inserting these propagators
into Eq.~(\ref{PiPscf}), one performs the Matsubara
sum. The resulting expressions for the self-energies
are rather unwieldy and will not be shown here. For the
2SC phase they were first derived in Ref.~\cite{rischke2SC} and for
the CFL phase in Ref.~\cite{rischkeCFL}. For 
the spin-one color-superconducting phases, this was done in
Ref.~\cite{schmitt3}. To further evaluate these expressions,
one has to compute the integral over ${\rm d}^3 {\bf k}$.
For an arbitrary gluon 4-momentum $P^\mu$, this has not yet been done.
However, in the static, homogeneous limit, the self-energy
$\Pi^{\mu \nu}_{ab}(0)$ was computed in the aforementioned
references in order to derive the Debye and Meissner masses
in the respective color-superconducting phases.
The results will be discussed in more detail in the following.
The gluon self-energy in the 2SC phase was also evaluated for
nonzero energies and momenta $p_0, p$, which are small
compared to the quark chemical potential.
This calculation is rather technical, and I simply refer
to Ref.~\cite{rischkeshovkovy,rischke2SC2} for the details. The main result
was that the modification of the gluon self-energy in a
color superconductor does not influence the value for
the gap parameter at leading or subleading order in weak coupling.
For the other color-superconducting phases, a similar calculation
has yet to be done.

In general, in a medium at nonzero temperature and/or density
static, long-wavelength (color-) electric fields
are screened. The screening length is determined by the
(inverse) Debye mass. If the medium is normal-conducting, 
static, long-wavelength (color-) magnetic fields are not screened.
This changes in a superconductor, where the Meissner effect
expels (color-) magnetic fields. They can only penetrate a certain
distance into the superconducting medium. For static, long-wavelength
(color-) magnetic fields, the (inverse) penetration length
is determined by the so-called Meissner mass.
The acquisition of a Meissner mass by a gauge boson
indicates that the corresponding
gauge symmetry is broken via the Anderson-Higgs mechanism.
The Debye and Meissner masses are defined as
\begin{equation}
m_{D, ab}^2 \equiv - \lim_{p \rightarrow 0} \Pi^{00}_{ab}(0,{\bf p})\,\, ,
\;\;\;\;
m_{M, ab}^2 \equiv \lim_{p \rightarrow 0} \Pi^{ii}_{ab}(0,{\bf p})\,\, .
\end{equation}
I present the values for the Debye masses in Table~\ref{debyemass}
and for the Meissner masses in Table~\ref{meissnermass} for 
various color-superconducting phases.

\begin{table}
\begin{center}
\begin{minipage}[t]{16.5 cm}
\caption{The Debye masses for gluons, photons and from
the mixed gluon-photon polarization tensor for
various color-superconducting phases. The results are given
in units of $N_f \mu^2/(6 \pi^2)$, where $N_f=2$ in the 2SC phase,
$N_f=3$ in the CFL phase, and $N_f = 1$ in the spin-one color-superconducting
phases. The constants are $\zeta \equiv (21 - 8 \ln 2)/54$,
$\alpha \equiv ( 3 + 4 \ln 2)/27$, and $\beta \equiv ( 6 - 4 \ln 2)/9$.
\vspace*{5mm}}
\label{debyemass}
\end{minipage}
\begin{tabular}{|l|cccccccc|cc|c|}
\hline
      &     &     &     &     &     &     &     &     &      &     &   \\[-2mm]
      &  gluons  &    &    &    &    &    &    &    &  mixed  &    & photon \\
      &     &     &     &     &     &     &     &     &      &     &   \\[-2mm]
\hline 
      &     &     &     &     &     &     &     &     &      &     &   \\[-2mm]
 $a$  &  1  &  2  &  3  &  4  &  5  &  6  &  7  &  8  &  1-7 &  8  & 9 \\
      &     &     &     &     &     &     &     &     &      &     &   \\[-2mm]
\hline \hline
      &     &     &     &     &     &     &     &     &      &     &   \\[-2mm]
2SC   &  0  &  0  &  0  &  $\frac{3}{2}\, g^2$ &  $\frac{3}{2}\, g^2$ & 
                        $\frac{3}{2} \, g^2$ & $\frac{3}{2}\, g^2$ & 
                                       $3\, g^2$   & 0   & 0   & $2\, e^2$ \\
      &     &     &     &     &     &     &     &     &      &     &   \\[-2mm]
\hline
      &     &     &     &     &     &     &     &     &      &     &   \\[-2mm]
CFL   &  $3\, \zeta g^2$ &  $3\, \zeta g^2$ &  $3\, \zeta g^2$ &  
         $3\, \zeta g^2$ &  $3\, \zeta g^2$ &  $3\, \zeta g^2$ &  
         $3\, \zeta g^2$ &  $3\, \zeta g^2$ &
                   0 & $-\sqrt{12}\, \zeta e g$ &  $ 4 \,\zeta e^2$ \\
      &     &     &     &     &     &     &     &     &      &     &   \\[-2mm]
\hline
      &     &     &     &     &     &     &     &     &      &     &   \\[-2mm]
CSL   &  $3\, \beta g^2$ &  $3\, \alpha g^2$ &  $3\, \beta g^2$ &  
         $3\, \beta g^2$ &  $3\, \alpha g^2$ &  $3\, \beta g^2$ &  
         $3\, \alpha g^2$ &  $3\, \beta g^2$ &
                   0 &  0 &  $ 18 \, q^2 e^2$ \\
      &     &     &     &     &     &     &     &     &      &     &   \\[-2mm]
\hline
      &     &     &     &     &     &     &     &     &      &     &   \\[-2mm]
polar & 0  &  0  &  0  &  $\frac{3}{2}\, g^2$ &  $\frac{3}{2} \, g^2$ & 
                        $\frac{3}{2}\, g^2$ & $\frac{3}{2}\, g^2$ & 
                                    $3\, g^2$   & 0   & 0   & $18\, q^2 e^2$ \\
      &     &     &     &     &     &     &     &     &      &     &   \\[-2mm]
\hline
\end{tabular}
\end{center}
\end{table}

\begin{table}
\begin{center}
\begin{minipage}[t]{16.5 cm}
\caption{The Meissner mass
for gluons, photons and from
the mixed gluon-photon polarization tensor for
various color-superconducting phases. The results are given 
in units of $N_f \mu^2/(6 \pi^2)$, where $N_f=2$ in the 2SC phase,
$N_f=3$ in the CFL phase, and $N_f = 1$ in the spin-one color-superconducting
phases. The constants are $\zeta \equiv (21 - 8 \ln 2)/54$,
$\alpha \equiv ( 3 + 4 \ln 2)/27$, and $\beta \equiv ( 6 - 4 \ln 2)/9$.
\vspace*{5mm}}
\label{meissnermass}
\end{minipage}
\begin{tabular}{|l|cccccccc|cc|c|}
\hline
      &     &     &     &     &     &     &     &     &      &     &   \\[-2mm]
      &  gluons  &    &    &    &    &    &    &    &  mixed  &    & photon \\
      &     &     &     &     &     &     &     &     &      &     &   \\[-2mm]
\hline 
      &     &     &     &     &     &     &     &     &      &     &   \\[-2mm]
 $a$  &  1  &  2  &  3  &  4  &  5  &  6  &  7  &  8  &  1-7 &  8  & 9 \\
      &     &     &     &     &     &     &     &     &      &     &   \\[-2mm]
\hline \hline
      &     &     &     &     &     &     &     &     &      &     &   \\[-2mm]
2SC   &  0  &  0  &  0  &  $\frac{1}{2} \, g^2$ &  $\frac{1}{2}\, g^2$ & 
                        $\frac{1}{2}\, g^2$ & $\frac{1}{2}\, g^2$ & 
               $\frac{1}{3}\, g^2$   & 0   & $\frac{1}{\sqrt{27}}\,eg$  &
                        $\frac{1}{9} \, e^2$ \\
      &     &     &     &     &     &     &     &     &      &     &   \\[-2mm]
\hline
      &     &     &     &     &     &     &     &     &      &     &   \\[-2mm]
CFL   &  $ \zeta g^2$ &  $ \zeta g^2$ &  $ \zeta g^2$ &  $ \zeta g^2$ &
          $ \zeta g^2$ &  $ \zeta g^2$ &  $ \zeta g^2$ &  $ \zeta g^2$ &
                   0 & $-\frac{2}{\sqrt{3}} \, \zeta e g$ &  
                       $\frac{4}{3}\, \zeta e^2$ \\
      &     &     &     &     &     &     &     &     &      &     &   \\[-2mm]
\hline
      &     &     &     &     &     &     &     &     &      &     &   \\[-2mm]
CSL   &  $ \beta g^2$ &  $ \alpha g^2$ &  $ \beta g^2$ &  $ \beta g^2$ &
          $ \alpha g^2$ &  $ \beta g^2$ &  $ \alpha g^2$ &  $ \beta g^2$ &
                   0 &  0 &  $ 6\, q^2 e^2$ \\
      &     &     &     &     &     &     &     &     &      &     &   \\[-2mm]
\hline
      &     &     &     &     &     &     &     &     &      &     &   \\[-2mm]
polar & 0  &  0  &  0  &  $\frac{1}{2}\, g^2$ &  $\frac{1}{2}\, g^2$ & 
                        $\frac{1}{2}\, g^2$ & $\frac{1}{2}\, g^2$ & 
                   $\frac{1}{3}\, g^2$   & 0   & $\frac{2}{\sqrt{3}}\, q e g$ &
                    $4\, q^2 e^2$ \\
      &     &     &     &     &     &     &     &     &      &     &   \\[-2mm]
\hline
\end{tabular}
\end{center}
\end{table}     

In the 2SC and polar phases, the Meissner mass of the first three gluons 
vanishes. These gluons correspond to the unbroken $[SU(2)]_c$ subgroup,
cf.~Table~\ref{table1}. What is interesting is that they
also have a vanishing Debye mass, indicating that the corresponding
color-electric fields are unscreened. Implications of this result 
were discussed in Ref.~\cite{ssr}. The other five gluons acquire both
a Debye as well as a Meissner mass. Electric and magnetic fields
are always screened in these color-superconducting phases.
Only the eighth gluon mixes with the photon.
Another interesting aspect is, however, that this mixing only occurs
in the magnetic sector, electric and color-electric fields remain
unmixed. In order to obtain the eigenmodes of the gauge bosons,
one has to diagonalize the mass matrices for electric and magnetic
gluons. A zero eigenvalue in this mass matrix indicates the
presence of an unbroken ``rotated'' $[\tilde{U}(1)]$ gauge symmetry.

In the CFL phase, all gluons acquire a Debye as well as a Meissner
mass, indicating that the $[SU(3)]_c$ color symmetry is completely
broken. Photons are Debye- as well as Meissner-screened, and there
is again mixing between the eighth gluon and the photon. In contrast
to the 2SC and polar phases,
however, this mixing extends also to the electric sector.
In the CSL phase, all gluons and the photon obtain Debye and
Meissner masses. There is no mixing between the gluons and the photon.
This means that the mass matrix of the (former gauge) bosons
is already diagonal, and it has no zero eigenvalue. 
Consequently, there is no unbroken residual symmetry,
and no room for a rotation that could generate one. This is in
agreement with the general arguments presented in Sec.~\ref{classcsc}
and summarized in Table~\ref{table1}. The particular pattern
of gluon masses reflects the residual $SO(3)_{c+J}$ symmetry 
in the CSL phase, cf.~Table~\ref{table1}:
the gluons corresponding to the three antisymmetric generators
of $[SU(3)_c]$ (which are simultaneously generators of $SO(3)$)
assume a different mass than the ones corresponding
to the symmetric generators. 

\begin{table}
\begin{center}
\begin{minipage}[t]{16.5 cm}
\caption{The diagonal elements of the
electric and magnetic gluon-photon mass matrix and
the (square of the) cosine of the rotation angles for
Debye and Meissner masses.
The results are given 
in units of $N_f \mu^2/(6 \pi^2)$, where $N_f=2$ in the 2SC phase,
$N_f=3$ in the CFL phase, and $N_f = 1$ in the spin-one color-superconducting
phases. The constants are $\zeta \equiv (21 - 8 \ln 2)/54$
and $\beta \equiv ( 6 - 4 \ln 2)/9$.
\vspace*{5mm}}
\label{tableunmix}
\end{minipage}
\begin{tabular}{|l|c|c|c|c|c|c|}
\hline
      &     &     &     &     &     &        \\[-2mm]
phase &  $\tilde{m}_{D,8}^2 $  &  $\tilde{m}_{D,\gamma}^2 $  &  
        $\cos^2 \theta_D $  &   $\tilde{m}_{M,8}^2 $  &  
        $\tilde{m}_{M,\gamma}^2 $  &   $\cos^2 \theta_M $  \\
      &     &     &     &     &     &        \\[-2mm]
\hline \hline
      &     &     &     &     &     &        \\[-2mm]
2SC   &  $3\, g^2 $ &  $2\, e^2 $  &  1  &  
      $\frac{1}{3}\, g^2 + \frac{1}{9}\, e^2$ & 0 & 
      $3\, g^2/(3\, g^2 + e^2)$  \\
      &     &     &     &     &     &        \\[-2mm]
\hline
      &     &     &     &     &     &        \\[-2mm]
CFL   &  $ (3\, g^2 + 4\, e^2)\, \zeta $ &  0 & 
         $3\,g^2/(3\,g^2 + 4\, e^2)$  
      &  $(g^2 + 4\, e^2 /3) \, \zeta $  &  0  &
      $3\,g^2/(3\,g^2 + 4\, e^2)$  \\
      &     &     &     &     &     &        \\[-2mm]
\hline
      &     &     &     &     &     &        \\[-2mm]
CSL   &  $ 3\, \beta g^2$ &  $ 18\, q^2 \, e^2$ &  1  &
         $ \beta g^2$ &  $ 6\, q^2 \, e^2$ &  1 \\
      &     &     &     &     &     &        \\[-2mm]
\hline
      &     &     &     &     &     &        \\[-2mm]
polar &  $3\, g^2$ &  $18\, q^2 \, e^2$ &  1 &
     $\frac{1}{3}\, g^2 + 4\, q^2 \, e^2$ & 0 & 
     $ g^2/(g^2 + 12\, q^2 \, e^2)$ \\
      &     &     &     &     &     &        \\[-2mm]
\hline
\end{tabular}
\end{center}
\end{table}

The final step is to diagonalize the mass matrices $m^2_{D,ab}$,
$m^2_{M,ab}$ for electric and magnetic gluons. Since only the 
eighth gluon mixes with the photon,
this diagonalization is achieved by a simple 
orthogonal rotation in the $2 \times 2$ block corresponding
to the indices $a = 8,9$.
The resulting diagonal (squared) Debye masses $\tilde{m}^2_{D,a}$ and 
(squared) Meissner masses $\tilde{m}_{M,a}^2$, as well as the
(square of the) cosine of the rotation angles $\theta_D$, $\theta_M$
are shown in Table~\ref{tableunmix}.
In the case of an unbroken $[\tilde{U}(1)]$ symmetry, cf.~Table~\ref{table1},
the ``rotated'' (magnetic) photon is massless. The ``rotated'' gluon
remains massive, but its degeneracy with the other massive
gluons is lifted.

The case of the polar phase is special. If there is only
one quark flavor, or all quark flavors have the same
electric charge, the results shown in 
Tables~\ref{debyemass} -- \ref{tableunmix}
hold. In this case, the rotated photon is massless, and
there is indeed an unbroken $[\tilde{U}(1)]$ symmetry.
However, in the case of two or more quark flavors with
different electric charges, the results change \cite{schmitt2}.
Let us assume that the chemical potentials of all quark flavors
are identical. Then the (squared) gluon masses are the same,
but in the mixed masses, the factor $q$ has to be replaced
by $\sum_f q_f$, while in the photon masses, the factor
$q^2$ is replaced by $\sum_f q_f^2$. In this case, it is not
hard to realize that a diagonalization
of the Meissner mass matrix does not lead to a massless 
rotated photon. There is therefore no $[\tilde{U}(1)]$ symmetry.
Consequently, spin-one color-superconducting quark matter exhibits
a Meissner effect, while color superconductors with spin-zero
Cooper pairs do not. Due to the smallness of the ratio $\phi_0/m_g$
for spin-one color superconductors, these are very likely
of type I, i.e., the magnetic field is completely expelled.
This is in contrast to the standard model of a neutron star,
where the core is assumed to be a type-II superconductor and
thus threaded by magnetic flux tubes.
It was recently argued in Ref.~\cite{link} that the short
precession period of some pulsars contradicts this assumption
and requires the core of the pulsar to be a type-I superconductor.
The question then is whether the core could be made of
spin-one color-superconducting quark matter \cite{schmitt2}.


\section{Conclusions and Outlook} \label{concloutl}

In this review, I have presented the current
knowledge of the equilibrium properties of strongly interacting matter
at large temperatures and/or densities.
In particular, I have qualitatively discussed
the phase diagram. I have presented calculations of
thermodynamic properties of strongly interacting matter,
both via lattice QCD, as well as within analytic approaches.
Finally, I have given an overview of color superconductivity
in weak coupling.

Our knowledge of the QCD phase transition and the QGP at
zero quark chemical potential has tremendously increased over the
last few years. Lattice QCD
calculations are well under control for the pure
$[SU(3)_c]$ gauge theory, and the quality of the 
data is such that an extrapolation to the
continuum limit as well as to the thermodynamic limit
has become possible. Lattice calculations with dynamical fermions 
are more challenging and, consequently, the data are
not of the same quality as for the pure gauge theory.
The main problem is that, with present methods of putting
fermions on the lattice, the pion comes out too heavy. Since pions
dominate the equation of state in the hadronic
phase, calculations of the pressure below the chiral restoration temperature
do not yet reflect the correct physics.
The challenge for the future is to improve the methods such
that the pion mass on the lattice is close to the value in nature.
Besides a reliable computation of the equation of state, 
this will also allow to decide
the question about the order of the QCD phase transition in nature.

Another important development in lattice QCD is to extend
the investigation of thermodynamic properties to 
nonzero quark chemical potentials. For many years, 
the fermion sign problem has impeded progress in this direction.
Recent attempts, like multiparameter
reweighting, Taylor expansion around $\mu = 0$, or analytic
continuation from imaginary values of $\mu$, have made an attempt
to work around this problem. Much work remains to be done
to improve these methods in order to correctly determine
the location of the critical point in the $(T,\mu)$ plane.
This is of great phenomenological importance: in order to
find a signal for the first-order phase transition to the QGP
in nuclear collisions, one has to tune the bombarding energy
such that one probes the region of the phase diagram,
which is to the right of the critical point.
Nuclear collisions at very high energies most likely probe
the region to the left, i.e., the crossover region of the 
quark-hadron transition. By definition,
there is no qualitative
difference between hadronic and QGP phase in this region, and
a clear signal for the QGP will be hard to identify.

Although the question about the location of the critical point
is important and can be investigated with
the above mentioned lattice methods, 
these methods circumvent the fermion sign problem
rather than solving it. Moreover, they are only applicable for
quark chemical potentials $\mu$ from zero up to 
values of order $T$. Therefore, one will ultimately have to find a true
solution which also works at
small temperatures and large chemical potential, so that
the color-superconducting quark matter phase can be explored.

Analytic approaches to compute the equation of state of
strongly interacting matter at high temperature have advanced
rapidly in recent years. The equation of state is now known to all 
orders which are perturbatively computable. Work is in progress to determine
the remaining nonperturbative contribution of order $O(g^6)$.
Resummation techniques have been applied to compute the 
thermodynamic properties of strongly interacting matter. 
At large temperatures $T \gg T_c$, they 
suggest that the QGP is a gas of weakly interacting quasiparticles.
However, when approaching the critical temperature from above,
the approaches based on resummation techniques fail to describe 
lattice QCD data. At the moment, one seems to be forced to either abandon
field-theoretical rigor in favor of simple quasiparticle models with
sufficiently many fit parameters to reproduce the data, or
turn to an alternative description, such
as the Polyakov loop model, which is physically less intuitive. 
It remains to be shown how this model is related to
the quasiparticle picture at large temperatures.

Color superconductivity is a rapidly evolving field. It is fairly
likely that color-superconducting quark matter can be found
in the core of compact stellar objects. It remains to explore
how this phase influences the properties of the star. Much work has
still to be done, for instance to compute the transport properties
of color-superconducting matter and the phase diagram
under the conditions of electric
and color neutrality. Although NJL-type models 
may give a qualitative picture of possible scenarios, they are 
unreliable when one wants to draw quantitative conclusions. 
The task is to improve existing
weak-coupling calculations or apply nonperturbative techniques
to obtain further knowledge about this interesting, exotic phase
of strongly interacting matter.

\section*{Acknowlegdment}

I would like to thank
A. Dumitru, Z. Fodor, S. Hands,
O. Kaczmarek, F. Karsch, E. Laermann, P. Levai, R. Pisarski,
A. Rebhan, T. Sch\"afer, C. Schmidt, A. Schmitt, Y. Schr\"oder, and
A. Steidl for valuable discussions and for help in preparing the
figures.
I greatly appreciate the hospitality of the INT Seattle, where part
of this work was done.


\end{document}